\documentstyle[11pt,epsfig]{report}
\newcommand{\beq}{\begin{equation}}
\newcommand{\eeq}{\end{equation}}
\newcommand{\beqa}{\begin{eqnarray}}
\newcommand{\eeqa}{\end{eqnarray}}

\newtheorem{theorem}{Theorem}[chapter]
\newtheorem{lemma}{Lemma}[chapter]
\newtheorem{corollary}{Corollary}[chapter]

\title{\bf
Distinguishability and Accessible Information in Quantum Theory\\ \bigskip
\bigskip
\large{Dissertation \\ submitted in partial fulfillment of the
requirements for the degree of \\ Doctor of Philosophy in Physics}}
\author{\Large\bf Christopher A.~Fuchs\\
\footnotesize\it Center for Advanced Studies, Department of Physics and 
Astronomy,\\ 
\footnotesize\it University of New Mexico, Albuquerque, NM 87131--1156}
\date{22 December 1995}

\begin{document}

\maketitle

\def\openone{\leavevmode\hbox{\small1\kern-3.8pt\normalsize1}}
\def\RR{{\rm I\kern-.2emR}}
\def\tr{{\rm tr}}


\pagenumbering{roman}






\vspace*{3.7in}

\begin{center}
\copyright~1995, Christopher A. Fuchs
\end{center}
\pagebreak

\vspace*{3in}

\begin{center}
To\medskip\\
my mother Geraldine, who has been with me every day of my life\medskip\\
and to\medskip\\
my brother Mike, who gave me a frontier to look toward
\end{center}

\pagebreak


\begin{center}
{\large ACKNOWLEDGEMENTS}
\end{center}
\bigskip

No one deserves more thanks for the success of this work than my advisor and
friend Carlton Caves.  Carl is the model of the American work ethic applied
to physical thought.  The opportunity to watch him in action has fashioned my 
way of thought, both in the scientific and the secular.
He has been a valued teacher, and I hope my three years in Albuquerque have 
left me with even a few of his qualities.

Special thanks go to Greg Comer, my philosophical companion.  Greg's
influence on this dissertation was from a distance, but no less great because
of that.  Much of the viewpoint espoused here was worked out in conversation
with him. 

I thank the home team, Howard Barnum, Sam Braunstein, Gary Herling, Richard
Jozsa, R\"udiger Schack, and Ben Schumacher for patiently and 
critically listening to so many of my ideas---the few successful ones and
the many failed ones.

Thanks also go to my stepfather Travis Spears for telling me long ago that
travel is one of the best forms of education.  If it had not been for my
travels the last two years, I would have never come into contact with so many helpful colleagues.  Each of the following have in one way or other shaped
this dissertation, even if only by prompting an explanation or turn of phrase.
I thank:
Adriano Barenco,
Charlie Bennett,
Herb Bernstein,
Gilles Brassard,
Colston Chandler,
Claude Cr\'epeau,
Ivan Deutsch,
Matthew Donald,
Artur Ekert,
Salman Habib,
Michael Hall,
Richard Hughes,
Bruno Huttner,
Kingsley Jones,
Arkady Kheyfets,
Raymond LaFlamme,
Norbert Luetkenhaus,
Hideo Mabuchi,
Chiara Macchiavello,
G\"unter Mah\-ler,
Dominic Mayers,
Gerard Milburn,
Warner Miller,
Tal Mor,
Michael Nielsen,
Asher Peres,
David Rosen,
Barry Sanders,
John Smolin,
Kalle-Antti Suominen,
Bill Unruh,
Jeroen van de Graaf,
Harald Weinfurter,
David Wolpert,
Bill Wootters, and
Wojciech Zurek.
In the same respect, John Archibald Wheeler holds a singular position; his
vision of ``it from bit'' has been my inspiration for many years.

I thank the following fellas for a lot of beer:
Mark Endicott,
Jack Glassman,
Chris Hains,
Jeff Nicholson,
Nora Pencola,
John Racette,
Charlie Rasco, and
Amy Sponzo.  If it weren't for these guys, this dissertation
most certainly would have never been written.

Finally, I thank my wife Kiki for giving me purpose and strength and the will
to see this dissertation through.

\pagebreak


%


\begin{center}
{\Large\bf Distinguishability and Accessible Information in Quantum Theory}
\\ \medskip
{\large by \\ \medskip
Christopher Alan Fuchs}\\ \medskip
\baselineskip=14pt
B.~S., Physics, The University of Texas at Austin, 1987\\
B.~S., Mathematics, The University of Texas at Austin, 1987\\
Doctor of Philosophy in Physics, The University of New Mexico, 1996
\end{center}
\bigskip
\bigskip
\bigskip

\begin{center}
{\large ABSTRACT}
\end{center}
\bigskip

Quantum theory forbids physical measurements from giving observers enough
evidence to distinguish nonorthogonal quantum states---this is an expression of
the indeterminism inherent in all quantum phenomena.  As a consequence, 
operational measures of how distinct one quantum state is from another must
depend exclusively on the statistics of measurements and nothing else.  This
compels the use of various information-theoretic measures of distinguishability
for probability distributions as the starting point for quantifying the
distinguishability of quantum states.  The process of translating these
classical distinguishability measures into quantum mechanical ones is the focus
of this dissertation.  The measures so derived have important applications in 
the young fields of quantum cryptography and quantum computation and, also,
in the more established field of quantum communication theory.

Three measures of distinguishability are studied in detail.  The first---{\it
the statistical overlap\/} or {\it fidelity\/}---upper bounds the decrease
(with the number of measurements on identical copies) of the probability of 
error in guessing a quantum state's identity.  The second---{\it the 
Kullback-Leibler relative information}---quantifies the distinction between the 
frequencies of measurement outcomes when the true quantum state is one or the 
other of two fixed possibilities.  The third---{\it the mutual information}---is
the amount of information that can be recovered about a state's identity from a 
measurement; this quantity dictates the amount of redundancy required to
reconstruct reliably a message whose bits are encoded by quantum systems
prepared in the specified states.  For each of these measures, an optimal
quantum measurement is one for which the measure is as large or as small
(whichever is appropriate) as it can possibly be.  The ``quantum 
distinguishability'' for each of the three measures is its value when an
optimal measurement is used for defining it.  Generally all these
quantum distinguishability measures correspond to different optimal
measurements.

The results reported in this dissertation include the following.  An exact
expression for the quantum fidelity is derived, and the optimal measurement
that gives rise to it is studied in detail.  The techniques required for
proving this result are very useful and may be applied to other, quite
different problems.  Several upper and lower bounds on the quantum mutual
information are derived via similar techniques and compared to each other.  Of 
particular note is a simplified derivation for the important upper bound first 
proved by Holevo in 1973.  An explicit expression is given for another (tighter) upper bound that appears implicitly in the same derivation.  Several upper and
lower bounds to the quantum Kullback information are also derived.  Particular 
attention is paid to a technique for generating successively tighter lower 
bounds, contingent only upon one's ability to solve successively higher order 
nonlinear matrix equations.

In the final Chapter, the distinguishability measures developed here are
applied to a question at the foundation of quantum theory: to what extent must 
quantum systems be disturbed by information gathering measurements?  This is 
tackled in two ways.  The first is in setting up a general formalism for 
ferreting out the tradeoff between inference and disturbance.  The main
breakthrough in this is that it gives a way of expressing the problem so that
it appears as algebraic as that of the problem of finding quantum
distinguishability measures.  The second result on this theme is the
proof of a theorem that prohibits ``broadcasting'' an unknown quantum 
state.  That is to say, it is proved that there is no way to replicate an 
unknown quantum state onto two separate quantum systems {\it when\/} each
system is considered without regard to the other (though there may well be 
correlation or quantum entanglement between the systems).  This result is a 
significant extension and generalization of the standard ``no-cloning'' theorem 
for pure states.


\tableofcontents
\listoffigures

\chapter{Prolegomenon}

\begin{flushright}
\baselineskip=13pt
\parbox{2.3in}{\baselineskip=13pt
``The information's unavailable to the mortal man.''}\medskip\\
---{\it Paul Simon}\\
Slip Slidin' Away
\end{flushright}

\pagenumbering{arabic}

\section{Introduction}

Suppose a single photon is secretly prepared in one of two known but 
{\it nonorthogonal\/} linear polarization states $\vec{e}_0$ and $\vec{e}_1$.
A fundamental consequence of the indeterminism inherent in all quantum phenomena
is that there is no measurement whatsoever that can discriminate
which of the two states was actually prepared.  For instance, imagine
an attempt to ascertain the polarization value by sending the photon through a
beam splitter; the photon will either pass straight through
or be deflected in a direction dependent upon the crystal's orientation
$\vec{n}$.  In this example, the only means available for the photon to
``express'' the value of its polarization is through the quantum mechanical
{\it probability\/} law for it to go straight
\beq
p_i=\left|\vec{e}_i\cdot\vec{n}\right|^2,\;\;\;\;
i=0,1\;.
\eeq
Since the polarization vectors $\vec{e}_i$ are nonorthogonal, there
is no orientation $\vec{n}$ that can assure that only one preparation will pass 
straight through while the other is deflected.  To the extent that one can gain 
information by sampling a probability distribution, one can also gain 
information about the preparation, but indeed only to that extent. If the
photon goes straight through, one might conjecture that
$p_i$ is closer to 1 than not, and thus that
the actual polarization is the particular $\vec{e}_i$ most closely aligned with
the crystal orientation $\vec{n}$, but there is no clean-cut certainty here.  
Ultimately one must make do with a guess.  The necessity of this guess is the 
unimpeachable signature of quantum indeterminism.

Fortunately, quantum phenomena are manageable enough that we are
allowed at least the handle of a probability assignment in predicting their
behavior.  The world is certainly not the higgledy-piggledy place it would be
if we were given absolutely no predictive power over the phenomena we
encounter.  This fact is the foundation for this dissertation.  It
provides a starting point for building several notions of how distinct one
quantum state is from another.

Since there is no completely reliable way to
identify a quantum state by measurement, one cannot simply reach into the space
of quantum states, place a yardstick next to the line connecting two of them, 
and read off a distance.  Similarly one cannot reach into the
ethereal space of probability assignments.  Nevertheless classical information
theory does give several ways to gauge the distinction between two probability 
distributions.  The idea of the game of determining a {\it quantum\/}
distinguishability measure is to start with one of the ones specified
by classical information theory.  The probabilities appearing in it are
assumed to be generated by a measurement on a system described by the quantum 
states that one wants to distinguish.  The quantum distinguishability is
defined by varying over all possible measurements to find the one that
makes the classical distinguishability the best it can possibly be.  The
best classical distinguishability found in this search is dubbed the quantum 
measure of distinguishability.

Once reasonable measures of quantum distinguishability are in hand, there
are a great number of applications in which they can be used. To give an
example---one to which we will return in the final chapter---step back to
1927, the year quantum mechanics became a relatively
stable theory.  The question that was the rage was why an electron 
could not be ascribed a classical state of motion.  The answer---so the standard
story of that year and thereafter goes---is that whenever one tries to discern
the position of an electron, one necessarily disturbs its momentum in an 
uncontrollable way.  Similarly, whenever one tries to discern the momentum,
the position is necessarily disturbed.  One can never get at both quantities
simultaneously.  Thus there are no means to specify operationally a phase space
trajectory for the electron.

Yet if one looks carefully at the standard textbook Heisenberg relation
of Robertson \cite{Robertson29}, the first one derived without recourse
to semiclassical thought experiments, one finds nothing even remotely 
resembling this picture.
What is found instead is that when many copies of a system are all prepared in
the same quantum state $\psi(x)$, if one makes enough position
measurements on the copies to get an estimate of 
$\Delta x$, and similarly makes enough momentum measurements on
(different!)\ copies to get $\Delta p$, then one can be assured
that the product of the numbers so found will be no smaller than $\hbar/2$.
Any attempt to extend the meaning of this relation beyond this is 
dangerous and unwarranted.

Nevertheless there is most certainly truth to the idea that information
gathering measurements in quantum theory necessarily cause disturbances.
This is one of the greatest distinctions between classical physics and
quantum physics, and is---after all---the ultimate reason that quantum systems
cannot be ascribed classical states of motion.  It is just that this is not
captured properly by the Heisenberg relation.  
What is needed, {\it among other things}, are two ingredients considered in 
detail in this dissertation.  The first is a notion of the information that can
be gained about the identity of quantum state from a given measurement model.
The second is a way of comparing the quantum state that describes a system before measurement to the quantum state that describes it after
measurement---in short, one of the quantum distinguishability measures
described above.

This gives some hint of the wealth that may lie at the end of the rainbow
of quantum distinguishability and accessible information.  First, however,
there are many rainbows to be made [See Fig.~\ref{snouter}], and this is the
focus of the work reported here.  In the remainder of this Chapter, we
summarize the main results of our research and describe the salient features of 
probabilities, quantum states, and quantum measurements that will be the 
starting point for the later chapters.  In Chapter 2, we describe in great
detail the motivations for and derivations of several classical measures of
distinguishability for probability distributions.  Many of the things presented
there are not well known to the average physicist, but little of the work
represents original research. In Chapter 3---the main chapter of new
results---we report everything we know about the quantum mechanical versions
of the classical distinguishability measures introduced in Chapter 2.  Chapter
4 applies the methods and measures developed in Chapter 3 to the deeper
question of the tradeoff between information gathering and disturbance in
quantum theory.  The first section of Chapter 4 is devoted to developing a 
general formalism for tackling questions of this ilk.  The second section
(which represents a collaboration with H.~Barnum, C.~M. Caves,
R.~Jozsa, and B.~Schumacher) proves that there is a very general sense in which
it is impossible to make a copy of an unknown quantum state.  This is a result
that extends the now-standard ``no-cloning'' theorem for pure quantum states \cite{Wootters82,Dieks82}.  Chapter 5 caps off the dissertation with a 
comprehensive bibliography of 527 books and articles relevant to quantum
distinguishability and quantum state disturbance.

\section{Summary of Results}

\subsubsection{Mutual Information}

As stated already, the main focus of this work is in deriving distinguishability
measures for quantum mechanical states.  First and foremost in importance among 
these is the one defined by the question of {\it accessible information\/}.  In
this problem's simplest form, one considers a binary communication
channel---i.e., a channel in which the alternative transmissions are 0 and 1.
The twist in the quantum mechanical case is that the 0 and 1 are encoded
physically as distinct states $\hat\rho_0$ and $\hat\rho_1$ of some quantum
system (described on a $D$-dimensional Hilbert space, $D$ being finite); the
states $\hat\rho_0$ and $\hat\rho_1$ need not be orthogonal nor pure states for
that matter.  The idea here is to think of the transmissions more literally as 
``letters'' that are mere components in much longer words and sentences, i.e., 
the meaningful messages to be sent down the channel.  When the quantum states 
occur with frequencies $\pi_0$ and $\pi_1$, the channel encodes, according to 
standard information theory,
\beq
H(\pi)=-(\pi_0\log_2\pi_0+\pi_1\log_2\pi_1)
\eeq
bits of information per transmission.  This 
scenario is of interest precisely because of the way in which quantum 
indeterminism bars any receiver from retrieving the full information encoded in 
the individual transmissions: there is generally no measurement with outcomes 
uniquely corresponding to whether 0 or 1 was transmitted.

The amount of information that is recoverable in this scheme is quantified by
an expression called the mutual information.  This quantity can be understood
as follows.  When the receiver performs a measurement to recover the message,
he initially sees the quantum system neither in state $\hat\rho_0$ nor in state 
$\hat\rho_1$ but rather in the mean quantum state
\beq
\hat\rho=\pi_0\hat\rho_0+\pi_1\hat\rho_1\;;
\eeq
this is an expression of the fact that he does not know which message was sent.
Therefore the raw information he gains 
upon measuring some nondegenerate Hermitian operator $\hat M$, say, 
\beq
\hat M=\sum_b m_b |b\rangle\langle b|\;, 
\eeq
is the Shannon information
\beq
H(p)=-\sum_b\langle b|\hat\rho|b\rangle\log_2\langle b|\hat\rho|b\rangle
\eeq
of the distribution $p(b)=\langle b|\hat\rho|b\rangle$ for the outcomes $b$.
This raw information, however, is not solely about the transmission; for, even 
if the receiver knew the system to be in one state over the other, measuring 
$\hat M$ would still give him a residual information gain quantified by the 
Shannon formula for the appropriate distribution,
\beq
p_0(b)=\langle b|\hat\rho_0|b\rangle
\;\;\;\;\;\;\mbox{or}\;\;\;\;\;\;
p_1(b)=\langle b|\hat\rho_1|b\rangle\;.
\eeq
The residual information gain is due to the 
fact that the measurement outcome $b$ is not deterministically predictable even
with the transmission, 0 or 1, known. 
Given this, the mutual information $J$, or information concerning the 
transmission, is just what one would expect intuitively: the raw information 
gain minus the average residual gain.  That is to say,
\beq
J=H(p)-\pi_0H(p_0)-\pi_1H(p_1)\;.
\eeq
The uniquely quantum mechanical question that arises
in this context is which measurement maximizes the mutual information for this 
channel and what is that maximal amount; this is the question of accessible 
information in quantum theory.  The value of the maximal information, $I$, is
called the {\it accessible information}.

Until recently, little has been known about the accessible information of a 
general quantum communication channel.  The trouble is in not having 
sufficiently powerful techniques for searching over all possible quantum 
measurements, i.e., not only those measurements corresponding to Hermitian 
operators, but also the more general positive-operator-valued measures (POVMs)
with any number of outcomes.  Aside from a few contrived examples in which
the accessible information can be calculated explicitly, the most notable 
results have been in the form of bounds---bounds on the number of outcomes in an optimal measurement \cite{Davies78} and bounds on the accessible information 
itself \cite{Holevo73d,Jozsa94b}.

Much of the goal of my own research on this
question, reported in Section~\ref{SnaffleDoo}, has been to improve upon these 
bounds and 
ultimately to find a procedure for successively approximating, to any desired degree of accuracy, the question's solution.  In this respect, some progress
has been made.  The Holevo upper bound \cite{Holevo73d} to the accessible 
information
\beq
I\le S(\hat\rho)-\pi_0S(\hat\rho_0)-\pi_1S(\hat\rho_1)\;,
\eeq
where $S(\hat\rho)=-\tr(\hat\rho\log_2\hat\rho)$ is the von Neumann entropy
of the density operator $\hat\rho$, and the Jozsa-Robb-Wootters lower bound \cite{Jozsa94b}
\beq
I\ge Q(\hat\rho)-\pi_0Q(\hat\rho_0)-\pi_1Q(\hat\rho_1)\;,
\eeq
where $Q(\hat\rho)$ is a complicated quantity called the sub-entropy of
$\hat\rho$, are both the best bounds expressible solely in terms of the
channel's mean density operator $\hat\rho$ {\it when\/} $\hat\rho_0$ and 
$\hat\rho_1$ are pure states. To go beyond them, one must include more details 
about the actual ensemble from which the signals are drawn, i.e., details about 
the prior probabilities and the density operators $\hat\rho_0$ and $\hat\rho_1$ 
themselves.

Along these lines, several new tighter ensemble-dependent bounds are reported.
Two of these---an upper and a lower bound---are found in the
process of simplifying the derivation of the Holevo upper bound (an exercise
useful in and of itself).  The lower bound of this pair takes on a
particularly pleasing form when $\hat\rho_0$ and $\hat\rho_1$ have a common
null subspace:
\beq
\pi_0\,\tr\Big(\hat\rho_0\log_2{\cal L}_{\hat\rho}(\hat\rho_0)\Big)\,+\,
\pi_1\,\tr\Big(\hat\rho_1\log_2{\cal L}_{\hat\rho}(\hat\rho_1)\Big)\,\le\, I\;,
\label{mica}
\eeq
where the ${\cal L}_{\hat\rho}(\hat\rho_i)$, $i=0,1$, are operators that 
satisfy the anti-commutator equation 
\beq
\hat\rho\,{\cal L}_{\hat\rho}(\hat\rho_i)\,+\,{\cal L}_{\hat\rho}
(\hat\rho_i)\,\hat\rho\,=\,2\,\hat\rho_i\;,
\eeq
which has a solution in this case.  It turns out that there is actually a 
measurement that generates this bound and---at least for two-dimensional
Hilbert spaces---it is often very close to being optimal.  When $\hat\rho_0$
and $\hat\rho_1$ are pure states this lower bound actually equals the accessible information.

Another significant bound on the accessible information comes about by thinking
of the states $\hat\rho_0$ and $\hat\rho_1$ as arising from a
partial trace over some larger Hilbert space with two possible pure states
$|\psi_0\rangle$ and $|\psi_1\rangle$ on it.  For any two such
{\it purifications\/} the corresponding accessible information will be larger
than that for the original mixed states; this is because by expanding a
Hilbert space one can only add distinguishability.  However, the accessible 
information for two pure states can be calculated exactly; it is given by
the appropriate analog to Eq.~(\ref{mica}).  Using a result for
the maximal possible overlap between two purifications due to Uhlmann
\cite{Uhlmann76}, one arrives at the tightest possible upper bound of this
form.

\subsubsection{Statistical Overlap}

The second most important way of defining a notion of statistical 
distinguishability (see Section~\ref{MachoMan}) concerns the following scenario.  One imagines a finite number of copies, $N$, of a quantum system secretly
prepared in the same quantum state---either the state $\hat\rho_0$ or the state 
$\hat\rho_1$.  It is the task of an observer to perform the same measurement
on each of these copies and then, based on the collected data, to make a guess
as to the identity of the quantum state.  Intuitively, the more
``distinguishable'' the quantum states, the easier the observer's task, but how 
does one go about quantifying such a notion?  The idea is to rely on the 
probability of error $P_e(N)$ in this inference problem to point the way. For 
instance, one might take the probability of error itself as a measure of statistical distinguishability and define the quantum distinguishability to be 
that quantity minimized over all possible quantum measurements.  Appealing as 
this may be, a difficulty crops up in that the optimal quantum measurement in 
this case explicitly depends on the number of measurement repetitions, $N$.

To get past this blemish, one can ask instead which measurement will cause the 
error probability to decrease exponentially as fast as possible in $N$; that is
to say, what is the smallest $\lambda$ for which
\beq
P_e(N)\le\lambda^N\;.
\eeq
Classically, the best bound of this form is called the Chernoff bound, and is
given by
\beq
\lambda=\min_{0\le\alpha\le1}\sum_b p_0(b)^\alpha p_1(b)^{1-\alpha}\;,
\eeq
where $p_0(b)$ and $p_1(b)$ are the probability distributions for the
measurement outcomes.  The exponential rate of decrease in error probability,
$\lambda$, optimized over all measurements must be, by definition, independent
of the number of measurement repetitions, and thus makes a natural
(operationally defined) candidate for a measure of quantum distinguishability.  
Unfortunately the search for an explicit expression for this quantity remains a 
subject for future research.  On the brighter side, however, there is a closely
related upper bound on the Chernoff bound that is of interest in its own 
right---the {\it statistical overlap} \cite{Wootters81}.

If a measurement generates
probabilities $p_0(b)$ and $p_1(b)$ for its outcomes, then the statistical
overlap between these distributions is defined to be
\beq
F(p_0,p_1)=\sum_b\sqrt{p_0(b)}\sqrt{p_1(b)}\;.
\label{ShuffleBlood}
\eeq
This quantity, as stated, gives a more simply expressed upper bound on 
$\lambda$.
This is quite important because this expression is manageable enough that it
actually can be optimized to give a useful quantum distinguishability measure
(even though it may not be as well-founded conceptually as $\lambda$ itself).
The optimal statistical overlap is given by 
\begin{equation}
F(\hat\rho_0,\hat\rho_1)={\rm 
tr}\,\sqrt{\hat\rho_1^{1/2}\hat\rho_0\hat\rho_1^{1/2}\,}
={\rm tr}\,\sqrt{\hat\rho_0^{1/2}\hat\rho_1\hat\rho_0^{1/2}\,}\;,
\end{equation}
a quantity known as the {\it fidelity for quantum states\/}, which has 
appeared in other mathematical physics contexts
\cite{Uhlmann76,Jozsa94a}.  (Here we start the trend in notation that the same
function is used to denote both the classical distinguishability and its
quantum version; notice that the former has the probability distributions as
its argument and the latter has the density operators themselves.)  This 
measure of distinguishability has many useful properties and is crucial to 
the proof of ``no-broadcasting'' in Chapter 4. 

Of particular note is the way the technique used in finding the quantum
fidelity reveals the actual measurement by which Eq.~(\ref{ShuffleBlood}) is 
minimized.  This measurement (i.e., orthonormal basis) is the one specified by 
the Hermitian operator
\beq
\hat M=\hat\rho_1^{-1/2}\sqrt{\hat\rho_1^{1/2}\hat\rho_0\hat\rho_1^{1/2}}
\hat\rho_1^{-1/2}\;,
\eeq
when $\hat\rho_1$ is invertible.  This technique comes in very handy
for the problems considered in Chapter 4.

\subsubsection{Kullback-Leibler Relative Information}

The final information theoretic measure of distinguishability we consider can
be specified by the following problem (see Section~\ref{MadameClue}).  Suppose
$N\!\gg\!1$ copies of a quantum system are all prepared in the same state 
$\hat\rho_1$.  If some observable is measured on each of these, the most likely 
{\it frequencies\/} for its various outcomes $b$ will be those given by the probabilities $p_1(b)$ assigned by quantum theory.   All other frequencies
beside this ``natural'' set will become less and less likely for large $N$ as
statistical fluctuations in the frequencies eventually damp away.  In fact, any
set of outcome frequencies $\{f(b)\}$---distinct from the ``natural'' ones
$\{p_1(b)\}$---will become exponentially less likely with the number of
measurements according to \cite{Cover91}
\beq
\mbox{PROB}\Bigl(\Bigl.\mbox{freq}=\{f(b)\}\,\Bigr|\,\mbox{prob}=\{p_1(b)\}
\Bigr)\,\approx\,e^{-NK(f/p_1)}\;,
\eeq
where
\beq
K(f/p_1)=\sum_b f(b)\ln\!\left({f(b)\over p_1(b)}\right)
\eeq
is the Kullback-Leibler relative information \cite{Kullback51} between the
frequency distribution $f(b)$ and the probability distribution $p_1(b)$.  
Therefore the quantity $K(f/p_1)$, which controls the behavior of this 
exponential decline, says something about how dissimilar the frequencies 
$\{f(b)\}$ are from the ``natural'' ones $\{p_1(b)\}$.  This gives an easy 
way to gauge the distinguishability of two probability distributions, as will
be seen presently.

Suppose now that the same measurements as above are performed on quantum systems
all prepared in the state $\hat\rho_0$.  The outcome frequencies most
likely to appear in this scenario are again those specified by the probability
distribution given by quantum theory---in this case $p_0(b)$.  This simple fact
points to a natural way to define an optimal measurement for this 
problem.  An optimal measurement is one for which the natural frequencies of 
outcomes for state $\hat\rho_0$ are maximally {\it improbable\/}, given that 
$\hat\rho_1$ is actually controlling the statistics.  That is to say, an
optimal measurement is one for which
\beq
K(p_0/p_1)=\sum_b p_0(b)\ln\!\left({p_0(b)\over p_1(b)}\right)
\eeq
is as large as it can be.  The associated quantum measure of
distinguishability, called the {\it quantum Kullback information}, is just that 
maximal value \cite{Donald86}.

The quantum Kullback information is much like the accessible information in that
it contains a nasty logarithm in its definition.  As such we have only been
able to find bounds on this quantity.  Two of the lower bounds take on
particularly pretty forms:
\beq
K_{\rm F}(\hat\rho_0/\hat\rho_1)\equiv{\rm tr}\biggl(\hat\rho_0\ln
\Bigl({\cal L}_{\hat\rho_1}(\hat\rho_0)\Bigr)\biggr)\;,
\eeq
and
\beq
K_{\rm B}(\hat\rho_0/\hat\rho_1)\equiv2\:{\rm tr}\Biggl(\hat\rho_0\ln\!
\left(\hat\rho_1^{-1/2}\sqrt{\hat\rho_1^{1/2}\hat\rho_0\hat\rho_1^{1/2}}
\hat\rho_1^{-1/2}\right)\Biggr)\;.
\eeq
A very general procedure has also been found for generating successively tighter
lower bounds than these.  This procedure is, however, contingent upon one's
ability to solve higher and higher order nonlinear matrix equations.  Several 
upper bounds to the quantum Kullback information are also reported.

\subsubsection{Inference--Disturbance Tradeoff for Quantum States}

With some results on the quantum distinguishability measures in hand, we turn
our attention to the sorts of problems in which they may be used.  A typical
one is that already described in the Introduction:  how might one gauge the
necessary tradeoff between information gain and disturbance in quantum 
measurement?  Not many results have yet poured from this direction,
but we are able to go a long way toward defining the problem in its most
general setting.  The main breakthrough is in realizing a way to express the
problem so that it becomes as algebraic as that of finding explicit formulae
for the quantum distinguishability measures.  

\subsubsection{The No-Broadcasting Theorem}

Suppose a quantum system, secretly prepared in one state from the set
${\cal A}\!=\!\{\hat\rho_0,\hat\rho_1\!\}$, is dropped into a ``black box''
whose purpose is to {\it broadcast\/} or replicate that quantum state 
onto two separate quantum systems.  That is to say, a state identical
to the original should appear in each system {\it when\/} it is considered
without regard to the other (though there may be correlation or quantum
entanglement between the systems).  Can such a black box be
built?

The ``no-cloning theorem'' \cite{Wootters82,Dieks82} insures that the answer
to this question is {\it no\/} when the states in ${\cal A}$ are pure and 
nonorthogonal; for the only way to have each of the broadcast systems described 
separately by a pure state $|\psi\rangle$ is for their joint state to be
$|\psi\rangle\otimes|\psi\rangle$.  When the states are mixed, however, things
are not so clear.  There are many ways each broadcast system can be described
by $\hat\rho$ without the joint state being $\hat\rho\otimes\hat\rho$, the
mixed state analog of cloning. The systems may also be entangled or correlated
in such a way as to give the correct marginal density operators.  

For instance, consider the limiting case in which $\hat\rho_0$ and $\hat\rho_1$
commute and so may be thought of as probability distributions 
$p_0(b)$ and $p_1(b)$ for their eigenvectors.  In this case, one easily sees 
that the states can be broadcast; the broadcasting device need merely perform a
measurement of the eigenbasis and prepare two systems, each in the state
corresponding to the outcome it finds.  The resulting joint state is not of the
form $\hat\rho\otimes\hat\rho$ but still reproduces the correct marginal 
probability distributions and thus, in this case, the correct marginal density
operators.

It turns out that two states $\hat\rho_0$ and $\hat\rho_1$ can be
broadcast if and only if they commute.\footnote{This finding represents a
collaboration with H.~Barnum, C.~M. Caves, R.~Jozsa, and B.~Schumacher.}
The way this is demonstrated is via a use of the quantum fidelity derived in 
Chapter 3.  One can show that the most general process or ``black box'' allowed
by quantum mechanics can never increase the distinguishability of quantum
states (as measured by fidelity); yet broadcasting requires that 
distinguishability actually increase unless the quantum states commute.

This theorem is important because it draws a communication theoretic distinction
between commuting and noncommuting density operators.  This is a distinction
that has only shown up before in the Holevo bound to accessible information:
the bound is achieved by the accessible information if and only if the
signal states commute.  The no-broadcasting result also has implications for 
quantum cryptography.

\section{Essentials}

What are probability distributions?  What are quantum systems?  What are
quantum states?  What are quantum measurements?  We certainly cannot answer
these questions in full in this short treatise.  However we need at least
working definitions for these concepts to make any progress at all in our
endeavors.  This Section details some basic ideas and formalism used
throughout the remainder of the dissertation.  Also it lays the groundwork
for some of the ideas presented in the Postscript.

\subsection{What Is a Probability?}

In this document, we hold fast to the Bayesian view of probability
\cite{Earman92,Bernardo94}.  This is that a probability assignment summarizes 
what one does and does not know about a particular situation.  A probability 
represents a state of knowledge; its numerical value quantifies the plausibility one is willing to give a hypothesis given some background information.

This point of view should be contrasted with the idea that probability must
be {\it identified\/} with the relative frequency of the various outcomes in
an infinite repetition of a given experiment.  The difficulties with the
{\it frequency theory\/} of probabilities are numerous \cite{Howson89} and
need not be repeated here.  Suffice it to point out that if one takes the
frequency idea seriously then one may never apply the probability calculus
to situations where an experiment cannot---by definition---be repeated more
than once.  For instance, I say that the probability that my heart will stop
beating at 10:29 this evening is about one in a million.  There can be no
real infinite ensemble of repetitions for this experiment.  Alternatively, if
I must construct an ``imaginary'' conceptual ensemble to understand the 
probability statement, then why bother?  Why not call it a degree of belief to 
begin with?

The Bayesian point of view should also be contrasted with the {\it propensity
theory\/} of probability \cite{Popper82}.  This is the idea that a probability
expresses an objective tendency on the part of the experimental situation to 
produce one outcome over the other.  If this were the case, then one could
hardly apply the probability calculus to situations where the experimental
outcome already exists at the time the experiment is performed.  For instance,
I give it an 98\% chance there will be a typographical error in this manuscript
the night I defend it.  But surely there either will or will not be such an
error on the appropriate night, independently of the probability I have
assigned.

One might argue that probabilities in quantum mechanics are something different
from the Bayesian sort \cite{Benioff73,Benioff74,Benioff77a}, perhaps more aligned with the propensity idea \cite{Giere73,Strauss72}.  This idea is taken 
to task in Ref.~\cite{Caves96}.  The argument in a nutshell is that if it
looks like a Bayesian probability and smells like a Bayesian probability, then 
why draw a distinction where none is to be found.

With all that said and done, what is a probability?  Formally, it is the
{\it plausibility\/} $P(H|S)$ of a hypothesis $H$ given some background 
information $S$ and satisfies the probability calculus:
\begin{itemize}
\item
$P(H|S)\ge0$,
\item
$P(H|S)+P(\neg H|S)=1$, where $\neg H$ signifies the negation of $H$, and
\item
plausibilities are updated according to Bayes' rule upon the acquisition of new 
evidence $E$, i.e.
\beq
P(H|E,S)=\frac{P(H|S)\times P(E|H,S)}{P(E|S)}\;,
\eeq
{\it when it is clear\/} that the background information is not changed by the
process of acquiring the evidence.
\end{itemize}
These properties are essential for the work presented here and will be used
over and over again.

\subsection{What Is a Quantum State?}

We take a very pragmatic approach to the meaning of a quantum state in this
dissertation.  Quantum states are the formal devices we use to describe what
we do and do not know about a given quantum system---nothing more, nothing
less \cite{Hartle68,Peres84,Peres93b,Caves96}.  In this sense, quantum states
are quite analogous to the Bayesian concept of probability outlined above.
(Though more strictly speaking, they form the background information $S$ that
may be used in a Bayesian probability assignment.)

What is a quantum system?  To say it once evocatively, a line drawn in the
sand.  A quantum system is any part of the physical world that we may wish
to conceptually set aside for consideration.  It need not be microscopic and
it certainly need not be considered in complete isolation from the remainder
of our conceptual world---it may interact with an {\it environment}.

Mathematically speaking, in this dissertation what we call a quantum
state is any {\it density operator\/} $\hat\rho$ over a $D$-dimensional
Hilbert space, where $D$ is finite.  The properties of a density operator are 
that it be Hermitian with nonnegative eigenvalues and that its trace be equal
to unity.  A very special class of density operators are the one-dimensional
projectors $\hat\rho=|\psi\rangle\langle\psi|$.  These are called {\it pure
states\/} as opposed to density operators of higher rank, which are often
called {\it mixed states}.

Pure states correspond to states of maximal
knowledge in quantum mechanics \cite{Caves96}.  Mixed states, on the other hand,
correspond to less than maximal knowledge.  This can come about in at least
two ways.  The first is simply by not knowing---to the extent that one could
in principle---the precise preparation of a quantum system.  The second is in
having maximal knowledge about a composite quantum system, i.e., describing it
via a pure state $|\psi\rangle\langle\psi|$ on some tensor-product Hilbert space
${\cal H}_1\otimes{\cal H}_2$, but restricting one's attention completely to a
subsystem of the larger system: quantum theory generally requires that one's
knowledge of a subsystem be less than maximal even though maximal knowledge
has been attained concerning the composite system.  Formally, the states
corresponding to the subsystems are given by tracing out the irrelevant
Hilbert space.  That is to, if $|1_i\rangle|2_\alpha\rangle$ is a basis for
${\cal H}_1\otimes{\cal H}_2$, and
\beq
|\psi\rangle=\sum_{i\alpha} c_{i\alpha}|1_i\rangle|2_\alpha\rangle
\eeq
then the state on subsystem 1 is
\beqa
\hat\rho_1
&=&
\tr_2(|\psi\rangle\langle\psi|)
\nonumber\\
&=&
\sum_\beta\langle2_\beta|\psi\rangle\langle\psi|2_\beta\rangle
\rule{0mm}{8mm}
\nonumber\\
&=&
\sum_{i\alpha}c_{i\alpha}c_{i\alpha}^*|1_i\rangle\langle 1_i|\;.
\rule{0mm}{6mm}
\eeqa
A similar mixed state comes about when attention is restricted to subsystem 2.
Calling the reduced density operator the quantum state of the subsystem
ensures that the form of the probability formula for measurement outcomes
will remain the same in going from composite system to subsystem.

\subsection{What Is a Quantum Measurement?}

We shall be concerned with a very general notion of quantum measurement
throughout this dissertation, and it will serve us well to make the ideas
plain at the outset.  We take as a {\it quantum measurement\/} any physical
process that can be used on a quantum system to generate a probability
distribution for some ``outcomes.''

To make this idea rigorous, we recall two standard axioms for quantum theory.
The first is that when the conditions and environment of a quantum system are
completely specified and no measurements are being made, then the system's state
evolves according to the action of a unitary operator $\hat U$,
\beq
\hat\rho\;\longrightarrow\;\hat U\hat\rho\hat U^\dagger\;.
\eeq
The second is that a {\it repeatable\/} measurement corresponds to some
complete set of projectors $\hat\Pi_b$ (not necessarily one-dimensional) onto
orthogonal subspaces of the quantum system's Hilbert space.  The probabilities
of the outcomes for such a measurement are given according to the von Neumann
formula,
\beq
p(b)=\tr(\hat\rho\hat\Pi_b)\;.
\eeq
With these two basic facts, we can lay out the structure of a {\it general\/}
quantum measurement.

The most general action that can be performed on a quantum system 
to generate a set of ``outcomes'' is \cite{Kraus83}
\begin{enumerate}
\item
to allow the system to be placed in contact with an auxiliary
system or {\it ancilla\/} \cite{Helstrom76} prepared in a standard state,
\item
to have the two evolve unitarily so as to become correlated or entangled, and 
then
\item
to perform a repeatable measurement on the ancilla.
\end{enumerate}
One might have thought that a measurement on the composite system as a
whole---in the last stage of this---could lead to a more  general set of 
measurements, but, in fact, it cannot.  For this can always be accomplished
in this scheme by a unitary operation that first swaps the system's state
into a subspace of the ancilla's Hilbert space and then proceeds as
above.

More formally, these steps give rise to a probability distribution in the
following way.  Suppose the system and ancilla are initially described by 
the density operators $\hat\rho_{\rm s}$ and $\hat\rho_{\rm a}$ respectively.
The conjunction of the two systems is then described by the initial quantum 
state
\begin{equation}
\hat\rho_{\rm sa}=\hat\rho_{\rm s}\otimes\hat\rho_{\rm a}\;.
\label{MushMush}
\end{equation}
Then the unitary time evolution leads to a new state,
\begin{equation}
\hat\rho_{\rm sa}\;\longrightarrow\;\hat U\hat\rho_{\rm sa}\hat U^\dagger\;.
\end{equation}
Finally, a reproducible measurement on the ancilla is described via a set of
orthogonal projection operators $\{\hat{\openone}\otimes\hat\Pi_b\}$ acting on 
the ancilla's Hilbert space, where $\hat{\openone}$ is the identity operator.
Any particular outcome $b$ is found with probability
\begin{equation}
p(b)={\rm tr}\!\left(\hat U(\hat\rho_{\rm s}\otimes\hat\rho_{\rm a})\hat U^\dagger(\hat{\openone}\otimes\hat\Pi_b)\right)\;.
\label{RibTie}
\end{equation}
The number of outcomes in this generalized notion of measurement is limited
only by the dimensionality of the ancilla's Hilbert space---in principle,
there can be arbitrarily many.  There are no limitations on the number of 
outcomes due to the dimensionality of the system's Hilbert space.

It turns out that this formula for the probabilities can be re\"expressed in
terms of operators on the system's Hilbert space alone.  This is easy to see.
If we let $|s_\alpha\rangle$ and $|a_c\rangle$ be an orthonormal basis for
the system and ancilla respectively, $|s_\alpha\rangle|a_c\rangle$ will be a
basis for the composite system.  Then using the cyclic property of the trace
in Eq.~(\ref{RibTie}), we get
\beqa
p(b)
&=&
\sum_{\alpha c}\langle s_\alpha|\langle a_c|\Big(
(\hat\rho_{\rm s}\otimes\hat\rho_{\rm a})
\hat U^\dagger(\hat{\openone}\otimes\hat\Pi_b)\hat U\Big)|s_\alpha
\rangle|a_c\rangle
\nonumber\\
&=&
\sum_\alpha\langle s_\alpha|\,\hat\rho_{\rm s}\!\left(\sum_c\langle a_c|\Big(
(\hat{\openone}\otimes\hat\rho_{\rm a})\hat U^\dagger(\hat{\openone}
\otimes\hat\Pi_b)\hat U\Big)|
a_c\rangle\!\right)\!|s_\alpha\rangle\;.
\rule{0mm}{8mm}
\eeqa
It follows that we may write
\beq
p(b)={\rm tr}_{\rm s}(\hat\rho_{\rm s}\hat E_b)\;,
\eeq
where
\beq
\hat E_b={\rm tr}_{\rm a}\!\left(
(\hat{\openone}\otimes\hat\rho_{\rm a})
\hat U(\hat{\openone}\otimes\hat\Pi_b)\hat U^\dagger\right)
\label{oncogene}
\eeq
is an operator that acts on the Hilbert space of the original system only.
Here ${\rm tr}_{\rm a}$ and ${\rm tr}_{\rm s}$ denote partial traces over the
system and ancilla Hilbert spaces, respectively.

Note that the $\hat E_b$ are positive operators, i.e., Hermitian operators
with {\it nonnegative\/} eigenvalues, usually denoted
\beq
\hat E_b\ge0\;,
\label{HeartBurn}
\eeq
because they are formed from the partial trace of the product of two positive
operators.  Moreover, these automatically satisfy a completeness relation of a
sort,
\beq
\sum_b\hat E_b=\hat{\openone}\;.
\label{AppleSauce}
\eeq
These two properties taken together are the defining properties of something
called a {\bf P}ositive {\bf O}perator-{\bf V}alued {\bf M}easure or POVM.  Sets
of operators $\{\hat E_b\}$ satisfying this are so called because they give an 
obvious (mathematical) generalization of the probability concept.  As opposed
to a complete set of orthogonal projectors, the POVM elements $\hat E_b$ need
not commute with each other.

A theorem, originally due to Neumark \cite{Peres90}, that is very useful for
our purposes is that {\it any\/} POVM can be realized in the fashion of
Eq.~(\ref{oncogene}).  This allows us to make full use of the defining
properties of POVMs in setting up the optimization problems considered here.
Namely, whenever we wish to optimize something like the mutual information, say,
over all possible quantum measurements, we just need write the expression in 
terms of a POVM $\{\hat E_b\}$ and optimize over all operator sets satisfying
Eqs.~(\ref{HeartBurn}) and (\ref{AppleSauce}).

Why need we go to such lengths to describe a more general notion of measurement
than the standard textbook one?  The answer is simple: because there are
situations in which the repeatable measurements, i.e., the orthogonal
projection-valued measurements $\{\hat\Pi_b\}$, are just not general enough to 
give the optimal measurement.  The paradigmatic example of such a thing
\cite{Holevo73c} is that of a quantum system with a 2-dimensional Hilbert
space---a spin-$\frac{1}{2}$ particle say---that can be prepared in one of three pure states, all with equal prior probabilities.  If the possible states are 
each $120^\circ$ apart on the Bloch sphere, then the measurement that 
optimizes the information recoverable about the quantum state's 
identity is one with three outcomes.  Each outcome excludes one of the 
possibilities, narrowing down the state's identity to one of the other two.
Therefore this measurement cannot be described by a standard two-outcome
orthogonal projection-valued measurement.


\chapter{The Distinguishability of Probability Distributions}

\begin{flushright}
\baselineskip=13pt
\parbox{2.8in}{\baselineskip=13pt
``\ldots~and as these are the only ties of our thoughts, they are really
{\it to us\/} the cement of the universe, and all the operations of the mind
must, in a great measure, depend on them.''}\medskip\\
---{\it David Hume}\\
An Abstract of a \\ Treatise of Human Nature
\end{flushright}

\section{Introduction}

The idea of distinguishing probability distributions is slippery business.
What one means in saying ``two probability distributions are
distinguishable'' depends crucially upon one's prior state of knowledge and the
context in which the probabilities are applied.  This chapter is devoted to
developing three quantitative measures of distinguishability, each
tailor-made to a distinct problem.

To make firm what we strive for in developing these measures, one should keep
at least one or the other of two models in mind.  The first, very concrete,
one is the model of a noisy communication channel.  In this model, things
are very simple.  A sender prepares a simple message, 0 or 1, encoded in
distinct preparations of a physical system, with probability $\pi_0$ and 
$\pi_1$.  A receiver performs some measurement on the system he receives for 
which the measurement outcomes $b$ have conditional probability $p_0(b)$ or 
$p_1(b)$, depending upon the preparation of the system.  The idea is that a 
notion of distinguishability of probability distributions should tell us 
something about what options are available to the receiver once he collects his
data---what sort of inferences or estimates the receiver may make about the 
sender's actual preparation.  The more distinguishable the probability
distributions, the more the receiver's data should give him some insight (in
senses to be defined) about the actual preparation.

The second, more abstract, model is that of the ``quantum-information
channel'' \cite{Schumacher95,Jozsa94c}---though at this level applied within a
purely classical context. Here we start with a physical system that we describe 
via some probability distribution $p_0(b)$, and we wish to transpose that state 
of knowledge onto another physical system in the possession of someone with 
which we may later communicate.  This transposition process must by necessity be carried out by some physical means, even if only by actually transporting our 
system to the other person.  The only difference between this and a standard 
communication transaction is that no mention will be made as to what the 
receiver of the ``signal'' will actually find if he makes an observation on it.  Rather we shall be concerned with what we, the senders, can {\it predict\/} 
about what he {\it would\/} find in an observational situation.  This 
distinction is crucial, for it leads to the following point.

If the state of knowledge is left truly intact during the transposition process,
then the system possessed by the receiver will be described by the sender
via the initial probability distribution $p_0(b)$.
If on the other hand---through lack of absolute control
of the physical systems concerned or through the intervention of a third
party---the state of knowledge changes during the transposition attempt, the
receiver's system will have to be described by some different distribution 
$p_1(b)$.  We might even imagine a certain probability $\pi_0$
that the transposition will be faithful and a probability $\pi_1$ that it will
be imperfect.  The question we should like to address is to what extent 
the states of knowledge $p_0(b)$ and $p_1(b)$ can be distinguished one from the 
other given these circumstances.  In what quantifiable and operational sense
can one state be said to be distinct from the other?

This language, it should be noted, makes no mention of probability
distributions being either true or false.  Neither $p_0(b)$ nor $p_1(b)$ need
have anything to do with the receiver's subjective expectation for his
observational outcomes; the focus here is completely upon the distinction in
predictions the sender may make under the opposing circumstances and how well
he himself might be able to check that something did or did not go amiss during
the transposition process.

In the following chapters these measures will be applied to the quantum
mechanical context, where states of knowledge are more completely represented
by density operators $\hat\rho_0$ and $\hat\rho_1$.  For the time being, 
however, it may be useful---though not necessary---to think of the distributions
as arising from some {\it fixed\/} measurement POVM $\hat E_b$ via
\beq
p_0(b)={\rm tr}(\hat\rho_0\hat E_b)
\;\;\;\;\;\;\;\;\mbox{and}\;\;\;\;\;\;\;\;
p_1(b)={\rm tr}(\hat\rho_1\hat E_b)\;.
\eeq
This representation, of course, will be the starting point for the quantum
mechanical considerations of the next chapter.

\section{Distinguishability via Error Probability and the Chernoff Bound}
\label{PECBsec}

Perhaps the simplest way to build a notion of distinguishability for probability
distributions is through a simple decision problem.  In this problem, an 
observer blindly samples {\it once\/} from either the 
distribution $p_0(b)$ or the distribution $p_1(b)$, $b=1,\ldots,n$; at most he
might know prior probabilities $\pi_0$ and  $\pi_1$ for which distribution 
he samples.  If the distributions are distinct, the outcome of the sampling will reveal something about the identity of the distribution from which it was drawn.  An easy way to quantify how distinct the distributions are comes from imagining 
that the observer must make a guess or inference about the identity after 
drawing his sample.  The observer's best-possible probability of error in this
game says something about the distinguishability of the distributions $p_0(b)$
and $p_1(b)$ with respect to his prior state of knowledge (as encapsulated in 
the distribution $\{\pi_0,\pi_1\}$).  This idea we develop as our first 
quantitative measure of distinguishability.  Afterward we generalize
it to the possibility of many samplings from the same distribution.  This
gives rise to a measure of distinguishability associated with the exponential
decrease of error probability with the number of samplings, the Chernoff
bound.

\subsection{Probability of Error}

What is the best possible probability of error for the decision problem?  This
can be answered easily enough by manipulating the formal definition of error
probability.  Let us work at this immediately.

A {\it decision function\/} is any function
\beq
\delta:\{1,\ldots,n\}\rightarrow\{0,1\}
\eeq
representing the method of guess an observer might use in this problem.  The
probability that such a guess will be in error is
\beq
P_e(\delta)\,=\,\pi_0P(\delta=1\,|\,0)\,+\,\pi_1P(\delta=0\,|\,1)\;,
\label{maltese}
\eeq
where $P(\delta=1\,|\,0)$ denotes the probability that the guess is 
$p_1(b)$ when, in fact, the distribution drawn from is really $p_0(b)$.
Similarly $P(\delta=0\,|\,1)$ denotes the probability that the guess is
$p_0(b)$ when, in fact, the distribution drawn from is really $p_1(b)$.

A natural decision function is the one such that 0 or 1 is
chosen according to which has the highest posterior probability, given the
sampling's outcome $b$.  Since the posterior probabilities are given by Bayes'
Rule,
\beq
p(i|b)=\frac{\pi_i p_i(b)}{p(b)}=
\frac{\pi_i p_i(b)}{\pi_0p_0(b)+\pi_1p_1(b)}\;,
\eeq
where $i=0,1$ and
\beq
p(b)=\pi_0p_0(b)+\pi_1p_1(b)
\eeq
is the total probability for outcome $b$, this decision function is called
Bayes' decision function.  Symbolically, Bayes' decision
function translates into
\beq
\delta_{\rm B}(b)=\left\{\begin{array}{ll}
0 & \;\;\mbox{if}\:\;\;\;\; \pi_0p_0(b)>\pi_1p_1(b)\\
1 & \;\;\mbox{if}\:\;\;\;\; \pi_1p_1(b)>\pi_0p_0(b)
\rule{0mm}{5mm}\\
\mbox{anything} & \;\;\mbox{if}\:\;\;\;\; \pi_0p_0(b)=\pi_1p_1(b)
\rule{0mm}{5mm}
\end{array}\right]\;.
\eeq
(When the posterior probabilities are equal, it makes no difference which guessing method is used.)  It turns out, not unexpectedly, that this decision 
method is optimal as far as error probability is concerned \cite{Renyi66}.  This is seen easily.  (In this Chapter, we denote the beginning and ending of proofs
by $\bigtriangleup$ and $\Box$, respectively.)

$\bigtriangleup$ Note that for any decision procedure $\delta$, 
Eq.~(\ref{maltese}) can be rewritten as
\beq
P_e(\delta)\,=\,\pi_0\sum_{b=1}^n\delta(b)p_0(b)\;+\;
\pi_1\sum_{b=1}^n\,[1-\delta(b)]p_1(b)\;,
\eeq
because $\sum_b\delta(b)p_0(b)$ is the total probability of guessing 1 when
the answer is 0 and $\sum_b[1-\delta(b)]p_1(b)$ is the total probability of 
guessing 0 when the answer is 1.  Then it follows that
\beq
P_e(\delta)-P_e(\delta_{\rm B})\,=\,\sum_{b=1}^n\Bigl(\delta(b)-
\delta_{\rm B}(b)\Bigr)\!\Bigl(\pi_0p_0(b)-\pi_1p_1(b)\Bigr)\;.
\eeq
Suppose $\delta\ne\delta_{\rm B}$. Then the only nonzero terms in this sum
occur when $\delta(b)\ne\delta_{\rm B}(b)$
and $\pi_0p_0(b)\ne\pi_1p_1(b)$.  Let us consider these terms.  When
$\delta(b)=0$ and $\delta_{\rm B}(b)=1$, $\pi_0p_0(b)-\pi_1p_1(b)<0$; thus
the term in the sum is positive.  When $\delta(b)=1$ and $\delta_{\rm B}(b)=0$,
we have $\pi_0p_0(b)-\pi_1p_1(b)>0$, and again the term in the sum is positive.
Therefore it follows that
\beq
P_e(\delta)>P_e(\delta_{\rm B})\;,
\eeq
for any decision function $\delta$ other than Bayes'.  This proves Bayes'
decision function to be optimal. $\Box$

We shall hereafter denote the probability of error with respect to
Bayes' decision method simply by $P_e$.  This quantity is expressed more
directly by noticing that, when the outcome $b$ is found, the probability of
a correct decision is just $\max\{p(0|b),p(1|b)\}$.  Therefore
\beqa
P_e
&=&
\sum_{b=1}^np(b)\Bigl(1-\max\{p(0|b),p(1|b)\}\Bigr)
\nonumber\\
&=&
\sum_{b=1}^np(b)\min\{p(0|b),p(1|b)\}
\label{svetlich}
\\
&=&
\sum_{b=1}^n\min\{\pi_0p_0(b),\pi_1p_1(b)\}\;.
\label{blatherboy}
\eeqa
Eq.~(\ref{svetlich}) follows because, for any $b$, $p(0|b)+p(1|b)=1$.
Equation~(\ref{blatherboy}) gives a concrete expression for the sought
after measure of distinguishability: the smaller this expression is numerically,
the more distinguishable the two distributions are.

Notice that Eq.~(\ref{blatherboy}) depends explicitly on the observer's
subjective prior state of knowledge through $\pi_0$ and $\pi_1$ and is
{\it not\/} solely a function of the probability distributions to be 
distinguished.  This dependence is neither a good thing nor a bad thing, for,
after all, Bayesian probabilities are subjective notions to begin with: they
are always defined with respect to someone's state of knowledge.  One need
only be aware of this extra dependence.

The main point of interest for this measure of distinguishability is that
it is operationally defined and can be written in terms of a fairly 
simple expression---one expressed in terms of the first power of the 
probabilities.  However, one should ask a few simple immediate questions to
test the robustness of this concept.  For instance, why did we not consider
two samplings before a decision was made?  Indeed, why not three or four or 
more?  If the error probabilities in these scenarios lead to nothing new and
interesting, then one's work is done.  On the other hand, if such cases lead
to seemingly different measures of distinguishability, then the foundation of
this approach might require examination.

These questions are settled by an example due to Cover \cite{Cover74}. Consider
the following four different probability distributions over two outcomes:
$p_0=\{.96,\,.04\}$, $p_1=\{.04,\,.96\}$, $q_0=\{.90,\,.10\}$, and
$q_1=\{0,\,1\}$.  Let us compare the distinguishability of $p_0$ and $p_1$
via Eq.~(\ref{blatherboy}) to that of $q_0$ and $q_1$, both under the assumption
of equal prior probabilities:
\beq
P_e(p_0,p_1)=\frac{1}{2}\min\{.96,\,.04\}+\frac{1}{2}\min\{.04,\,.96\}=.04\;,
\eeq
and
\beq
P_e(q_0,q_1)=\frac{1}{2}\min\{.90,\,0\}+\frac{1}{2}\min\{.10,\,1\}=.05\;.
\eeq
Therefore
\beq
P_e(p_0,p_1)<P_e(q_0,q_1)\;,
\eeq
and so, by this measure, the distributions $p_0$ and $p_1$ are more
distinguishable from each other than the distributions $q_0$ and $q_1$.

On the other hand, consider modifying the scenario so that two samples are
taken before a guess is made about the identity of the distribution.  This
scenario falls into the same framework as before, only now there are four
possible outcomes to the experiment.  These must be taken into account in
calculating the Bayes' decision rule probability of error.  Namely, in
obvious notation,
\beqa
P_e(p_0^2,p_1^2)
&=&
\frac{1}{2}\min\{.96\times.96,\,.04\times.04\}+
\frac{1}{2}\min\{.96\times.04,\,.04\times.96\}
\nonumber\\
& &\;\;\;\;+\frac{1}{2}\min\{.04\times.96,\,.96\times.04\}+
\frac{1}{2}\min\{.04\times.04,\,.96\times.96\}
\nonumber\\
&=&
\rule{0mm}{5mm}
.04\;,
\eeqa
and
\beqa
P_e(q_0^2,q_1^2)
&=&
\frac{1}{2}\min\{.90\times.90,\,0\times0\}+
\frac{1}{2}\min\{.90\times.10,\,0\times1\}
\nonumber\\
& &\;\;\;\;+\frac{1}{2}\min\{.10\times.90,\,1\times0\}+
\frac{1}{2}\min\{.10\times.10,\,1\times1\}
\nonumber\\
&=&
\rule{0mm}{5mm}
.005\;.
\eeqa
Therefore
\beq
P_e(q_0^2,q_1^2)<P_e(p_0^2,p_1^2)\;.
\eeq
The distributions $q_0$ and $q_1$ are actually more distinguishable from each 
other than the distributions $p_0$ and $p_1$, {\it when\/} one allows two 
samplings into the decision problem.

This example suggests that the probability of error, though a perfectly
fine measure of distinguishability for the particular problem of
decision-making after one sampling, still leaves something to be desired.
Even though it is operationally defined, it is not a measure that adapts
easily to further data acquisition.  For this one needs a measure that is not
explicitly tied to the exact number of samplings in the decision problem.

\subsection{The Chernoff Bound}
\label{YidDish}

The optimal probability of error in the decision problem---the one given by
using Bayes' decision rule---must decrease toward zero as the the number of
samplings increases.  This is intuitively clear.  The exact form of that
decrease, however, may not be so obvious.  It turns out that the decrease
asymptotically approaches an exponential in the number of samples
drawn before the decision.  The particular value of this exponential is called
the Chernoff {\it bound\/} \cite{Chernoff52,Hellman70,Cover91} because it is
not only achieved asymptotically, but also envelopes the true decrease from
above.

The Chernoff bound thus forms an attractive notion of distinguishability for
probability distributions:  the faster two distributions allow the probability
of error to decrease to zero in the number of samples, the more distinguishable 
the distributions are.  It is operationally defined by being intimately tied to
the decision problem.  Yet it neither depends on the prior probabilities
$\pi_0$ and $\pi_1$ nor on the number of samples drawn before the decision.
The formal statement of the Chernoff bound is given in the following theorem.

\begin{theorem}[Chernoff]
Let $P_e(N)$ be the probability of error for Bayes' decision rule after
sampling $N$ times one of the two distributions $p_0(b)$ or $p_1(b)$.  Then
\beq
P_e(N)\le\lambda^N
\eeq
where
\beq
\lambda=\min_\alpha F_\alpha(p_0/p_1)\;,
\label{vodka}
\eeq
and
\beq
F_\alpha(p_0/p_1)=\sum_{b=1}^n p_0(b)^\alpha p_1(b)^{1-\alpha}\;,
\eeq
for $\alpha$ restricted to be between 0 and 1.
Moreover this bound is approached asymptotically in the limit of large $N$.
\label{ChernoffThm}
\end{theorem}

$\bigtriangleup$
Let us demonstrate the first part of this theorem.  Denote the outcome of
the $k$'th trial by $b_k$.  Then the two probability distributions for the
outcomes of a string of $N$ trials can be written
\beq
p_0(b_1b_2\ldots b_N)=p_0(b_1)p_0(b_2)\cdots p_1(b_N)\;,
\eeq
and
\beq
p_1(b_1b_2\ldots b_N)=p_1(b_1)p_1(b_2)\cdots p_1(b_N)\;.
\eeq
Now note that, for any two positive numbers $a$ and $b$ and any 
$0\le\alpha\le1$,
\beq
\min\{a,b\}\le a^\alpha b^{1-\alpha}\;.
\eeq
This follows easily. First suppose $a\le b$.  Then, because $1-\alpha\ge0$, we
know
\beq
\left(\frac{b}{a}\right)^{1-\alpha}\ge1\;.
\eeq
So
\beq
\min\{a,b\}=a\le a\!\left(\frac{b}{a}\right)^{1-\alpha}=a^\alpha b^{1-\alpha}\;
\eeq
Alternatively, suppose $b\le a$; then
\beq
\left(\frac{a}{b}\right)^\alpha\ge1\;,
\eeq
and
\beq
\min\{a,b\}=b\le b\!\left(\frac{a}{b}\right)^\alpha=a^\alpha b^{1-\alpha}\;.
\eeq

Putting the notation and the small mathematical fact from the last paragraph
together, we obtain that for any $\alpha\in[0,1]$,
\beqa
P_e(N)
&=&
\sum_{b_1b_2\ldots b_N}\min\left\{\pi_0p_0(b_1b_2\ldots b_N),\,
\pi_0p_0(b_1b_2\ldots b_N)\right\}
\nonumber\\
&\le&
\pi_0^\alpha\pi_1^{1-\alpha}\!\sum_{b_1b_2\ldots b_N}
p_0(b_1b_2\ldots b_N)^\alpha\,p_0(b_1b_2\ldots b_N)^{1-\alpha}
\rule{0mm}{6mm}
\nonumber\\
&=&
\pi_0^\alpha\pi_1^{1-\alpha}\!\sum_{b_1b_2\ldots b_N}\!\left(\,\prod_{k=1}^N
p_0(b_k)^\alpha p_1(b_k)^{1-\alpha}\right)
\rule{0mm}{7mm}
\nonumber\\
&=&
\pi_0^\alpha\pi_1^{1-\alpha}\prod_{k=1}^N\left(\,\sum_{b_k=1}^n p_0(b_k)^\alpha 
p_1(b_k)^{1-\alpha}\right)
\rule{0mm}{9mm}
\nonumber\\
&=&
\pi_0^\alpha\pi_1^{1-\alpha}\left(\,\sum_{b=1}^n p_0(b)^\alpha 
p_1(b)^{1-\alpha}\right)^{\! N}
\rule{0mm}{8mm}
\nonumber\\
&\le&
\left(\,\sum_{b=1}^n p_0(b)^\alpha p_1(b)^{1-\alpha}\right)^{\! N}\;.
\rule{0mm}{7mm}
\label{Cobain}
\eeqa
The tightest bound of this form is found by further minimizing the right
hand side of Eq.~(\ref{Cobain}) over $\alpha$.  This completes the proof
that the Chernoff bound is indeed a bound on $P_e(N)$.  The remaining part
of the theorem, that the bound is asymptotically achieved, is more difficult
to demonstrate and we shall not consider it here; see instead
Ref.~\cite{Cover91}. $\Box$

The value $\alpha^*$ of $\alpha$ that achieves the minimum in Eq.~(\ref{vodka}) 
generally cannot be expressed any more explicitly than there.  This is because 
to find it in general one must solve a transcendental equation.  A notable 
exception is when the probability distributions $p_0\equiv\{q,\,1-q\}$ and 
$p_1\equiv\{r,\,1-r\}$ are distributions over two outcomes.  Let us
write out this special case.  Here
\beq
F_\alpha(p_0,p_1)=q^\alpha r^{1-\alpha}+(1-q)^\alpha (1-r)^{1-\alpha}\;.
\eeq
Setting the derivative of this (with respect to $\alpha$) equal to zero, we
find that the optimal $\alpha$ must satisfy
\beq
q^\alpha r^{1-\alpha}\ln\!\left(\frac{q}{r}\right)=-
(1-q)^\alpha (1-r)^{1-\alpha}\ln\!\left(\frac{1-q}{1-r}\right)\;;
\eeq
hence
\beq
\alpha^*=\left[\,\ln\!\left(\frac{q(1-r)}{r(1-q)}\right)\right]^{-1}\!
\ln\!\left(-\frac{1-r}{r}\,\frac{\ln(1-q)-\ln(1-r)}{\ln q-\ln r}\right)
\eeq
With this and a lot of algebra, one can show
\beqa
\ln F_{\alpha^*}(p_0,p_1)
&=&
-x\ln\!\left(\frac{x}{q}\right)-(1-x)\ln\!\left(\frac{1-x}{1-q}\right)
\nonumber\\
&=&
-x\ln\!\left(\frac{x}{r}\right)-(1-x)\ln\!\left(\frac{1-x}{1-r}\right)
\rule{0mm}{7mm}
\eeqa
where
\beq
x=\left[\,\ln\!\left(\frac{q(1-r)}{r(1-q)}\right)\right]^{-1}\!
\ln\!\left(\frac{1-r}{1-q}\right)\;.
\eeq
That is to say, using an expression to be introduced in Section~\ref{KEHsec},
the optimal $\ln F_\alpha(p_0,p_1)$ is given by the Kullback-Leibler relative
information \cite{Kullback51} between the distribution $p_0$ (or $p_1$) and the
``distribution'' $\{x,\,1-x\}$.

This property, relating the Chernoff bound to a Kullback-Leibler relative
information is more generally true and worth mentioning.  The precise statement
is the following theorem; details of its proof may be found in 
Ref.~\cite{Cover91}.
\begin{theorem}[Chernoff]
The constant $\lambda$ in the Chernoff bound can also be expressed as
\beq
\lambda=K(p_{\alpha^*}/p_0)=K(p_{\alpha^*}/p_1)\;,
\eeq
where $K(p_\alpha/p_0)$ is the Kullback-Leibler relative information between
the distributions $p_\alpha(b)$ and $p_0(b)$,
\beq
K(p_\alpha/p_0)=\sum_{b=1}^n p_\alpha(b)\ln\!\left(\frac{p_\alpha(b)}{p_0(b)}
\right)\;,
\eeq
and the distribution $p_\alpha(b)$---depending upon the parameter $\alpha$---is
defined by
\beq
p_\alpha(b)=\frac{p_0(b)^\alpha p_1(b)^{1-\alpha}}
{\sum_b p_0(b)^\alpha p_1(b)^{1-\alpha}}\;.
\eeq
The particular value of $\alpha$ used in this, i.e., $\alpha^*$, is the
one for which
\beq
K(p_{\alpha^*}/p_0)=K(p_{\alpha^*}/p_1)\;.
\eeq
\end{theorem}

\subsubsection{Statistical Overlap}

Because the Chernoff bound is generally hard to get at analytically, it is
worthwhile to reconsider the other bounds arising in the derivation of
Theorem~\ref{ChernoffThm}.  These are all quantities of the form
\beq
F_\alpha(p_0/p_1)=\sum_{b=1}^n p_0(b)^\alpha p_1(b)^{1-\alpha}\;,
\label{SteakGuru}
\eeq
for $0<\alpha<1$.  We shall call the functions appearing in 
Eq.~(\ref{SteakGuru}) the {\it R\'enyi overlaps\/} of order $\alpha$ because
of their close connection to the relative information of order $\alpha$
introduced by R\'enyi \cite{Renyi61,Renyi70},
\beq
K_\alpha(p_0/p_1)\,=\,\frac{1}{\,\alpha-1\,}\,\ln\!\left(\,\sum_{b=1}^n\,
p_0(b)^\alpha p_1(b)^{1-\alpha}\right)\;.
\eeq

Each $F_\alpha(p_0/p_1)$ forms a notion of distinguishability in its own
right, albeit not as operationally defined as the Chernoff bound.  All these
quantities are bounded between 0 and 1---reaching the minimum of 0 if and only
if the distributions do not overlap at all, and reaching 1 if and only if the
the distributions are identical.  For a fixed $\alpha$, the smaller
$F_\alpha(p_0/p_1)$ is, the more distinguishable the distributions are.
Moreover each of these can be used for generating a valid notion of 
nearness---more technically, a topology \cite{Vaidyanathaswamy60}---on the set 
of probability distributions \cite{Csiszar62,Csiszar64}.

A particular R\'enyi overlap that is quite useful to study because of its
many pleasant properties is the one of order $\alpha=\frac{1}{2}$,
\beq
F(p_0,p_1)\equiv\sum_{b=1}^n\sqrt{p_0(b)}\sqrt{p_1(b)}\;.
\eeq
We shall dub this measure of distinguishability the {\it statistical overlap\/}
or {\it fidelity}.  It has had a long and varied history, being rediscovered
in different contexts at different times by Bhattacharyya 
\cite{Bhattacharyya43,Bhattacharyya46}, Jeffreys \cite{Jeffreys46}, 
Rao \cite{Rao47}, R\'enyi \cite{Renyi61}, Csisz\'ar \cite{Csiszar62}, and
Wootters \cite{Wootters80b,Wootters81,Wootters83}.  Perhaps its most compelling
foundational significance is that the quantity
\beq
D(p_0/p_1)=\cos^{-1}\!\left(\,\sum_{b=1}^n\sqrt{p_0(b)}\sqrt{p_1(b)}\right)
\eeq
corresponds to the geodesic distance between $p_0(b)$ and $p_1(b)$ on the
probability simplex when its geometry is specified by the Riemannian metric
\beq
ds^2\,=\,\sum_{b=1}^n{\left(\delta p(b)\right)^{2}\over p(b)}\;.
\label{LineEl}
\eeq
This metric is known as the Fisher information metric
\cite{Fisher22,Cramer46,Cover91} and is useful because it appears in
expressions for the decrease (with the number of samplings) of an estimator's 
variance in maximum likelihood parameter estimation.

Unfortunately, $F(p_0/p_1)$ does not appear to be strictly tied to any
statistical inference problem or achievable resource bound as are most other
measures of distinguishability studied in this Chapter (in particular
the probability of error, the Chernoff bound, the Kullback-Leibler relative 
information, the mutual information, and the Fisher information).\footnote{Imre
Csisz\'ar, private communication.}  However, it nevertheless remains an
extremely useful quantity mathematically as will be seen in the following
chapters.  Moreover it remains of interest because of its intriguing resemblance
to a quantum mechanical inner product \cite{Wootters80b,Wootters81,Wootters83}:
it is equal to the sum of a product of ``amplitudes'' (square roots of
probabilities) just as the quantum mechanical inner product is.

\section{Distinguishability via ``Keeping the Expert Honest''}
\label{KEHsec}

We have already agreed that probabilities must be interpreted as
subjective states of knowledge.  How then can we ever verify whether a
probability assignment is ``true'' or not?  Simply put, we cannot.
We can never know whether a source of probability assignments, such as 
a weatherman or economic forecaster, is telling the truth or not.
The best we can hope for in a situation where we must rely on someone else's
probability assignment is that there is an effective strategy for inducing
him to tell the truth, an effective strategy to steer him to be as
true to his state of knowledge as possible when disseminating what he knows.  
This is the problem of ``keeping the expert honest'' 
\cite{Good52,Aczel66b,Buehler71,Aczel73,Aczel84}. The 
resolution of this problem gives rise to another measure of 
distinguishability for probability distributions, the Kullback-Leibler relative
information \cite{Kullback51}.

Let us start this section by giving a precise statement of the honest-expert 
problem.  Suppose an expert's knowledge of some state of affairs is quantified 
by a probability distribution $p_0(b)$, $b=1,\ldots,n$, and he is willing to
communicate that distribution for a price.  If we agree to pay for his services, then, barring the use of lie detector tests and truth serums, we can never know
whether we got our money's worth in the deal.  There is no way to tell just
by looking at the outcome of an experiment whether the distribution $p_0(b)$
represents his true state of knowledge or whether some other distribution
$p_1(b)$ does.  The only thing we can do to safeguard against dishonesty is to
agree exclusively to payment schemes that somehow build in an incentive for the
expert to be honest.

Imagine the expert agrees to the following payment scheme.  If the expert gives 
a probability distribution $p_1(b)$, then after we perform an experiment to 
elicit the actual state of affairs, he will be paid an amount that depends upon 
which outcome occurs {\it and\/} the distribution $p_1(b)$.  This particular
type of payment is proposed because, though probabilities do not dictate the
outcomes of an experiment, the events themselves nevertheless do give us an
objective handle on the problem.

Say, for instance, we pay an amount
$F_b\bigl(p_1(b)\bigr)$ if outcome $b$ actually occurs, where 
\beq
F_b\!:\![0,1]\!\rightarrow\!\RR\;\,,\;\;b=1,\ldots,n
\eeq
is some 
fixed set of functions independent of the probabilities under consideration.
Depending upon the form of the functions $F_b$,
it may well be in the expert's best interest to lie in reporting his
probabilities.  That is to say, if the expert's true state of knowledge is
captured by the distribution $p_0(b)$, his expected earnings for reporting
the distribution $p_1(b)$ will be
\beq
\overline{F}=\sum_{b=1}^n p_0(b)F_b\bigl(p_1(b)\bigr)\;.
\eeq
Unless his expected earnings turn out to be less upon lying than in telling
the truth, i.e.,
\beq
\sum_{b=1}^n p_0(b)F_b\bigl(p_1(b)\bigr)\,\le\,\,\sum_{b=1}^n
p_0(b)F_b\bigl(p_0(b)\bigr)\;,
\label{HonestExpert}
\eeq
there is no incentive for him to be honest (that is, if the expert acts
rationally!).  In this context, the problem of ``keeping the expert honest'' is 
that of trying to find a set of functions $F_b$ for which
Eq.~(\ref{HonestExpert}) remains true for all distributions $p_0(b)$ and 
$p_1(b)$, $b=1,\ldots,n$.

If such a program can be carried out, then it will automatically give a
measure of distinguishability for probability distributions.  Namely, the
difference between the maximal expected payoff and the expected payoff for a
dishonesty,
\beq
K(p_0/p_1)\equiv\sum_{b=1}^n p_0(b)\!\left[\tilde F_b\bigl(p_0(b)\bigr)-
\tilde F_b\bigl(p_1(b)\bigr)\right]\;,
\label{RoughKullback}
\eeq
becomes an attractive notion of distinguishability.  (The tildes over the
$F_b$ in this formula signify that they are functions optimal for
keeping the expert honest.)  This quantity captures the idea that the more
distinguishable two probability distributions are, the harder it should be
for an expert to pass one off for the other.  In this case, the bigger the 
expert's lie, the greater the expected loss he will have to take in giving it.

Of course, one could consider using any payoff function whatsoever in
Eq.~(\ref{RoughKullback}) and {\it calling\/} the result a measure of 
distinguishability, but that would be rather useless.  Only a
payment scheme optimal for this problem is relevant for distilling
a probability distribution's identity.  Only this sort of payment scheme
has a useful interpretation.

Nevertheless, one would be justified in expecting even more from a measure of distinguishability.  For instance, the honest-expert problem does not, at first
sight, appear to restrict the class of optimal functions very tightly at
all.  For instance, one might further want a measure of distinguishability
that attains its minimal value of zero {\it if and only if\/} $p_1(b)=p_0(b)$.
Or one might want it to have certain concavity properties.  Interestingly
enough, these sorts of things are already assured---though not explicit---in
the posing of the honest-expert problem.

\subsection{Derivation of the Kullback-Leibler Information}

Let us use the work of Acz\'{e}l \cite{Aczel80} to demonstrate
an exact expression for Eq.~(\ref{RoughKullback}).  When $n\ge3$ it turns
out that, up to various constants, there is a unique function satisfying 
Eq.~(\ref{HonestExpert}) for all distributions $p_0(b)$ and $p_1(b)$.  We shall
consider this case first by proving the following theorem.
\begin{theorem}[Acz\'{e}l]
Let $n\ge3$.  Then the inequality
\beq
\sum_{k=1}^n p_k F_k(q_k)\le\sum_{k=1}^n p_k F_k(p_k)
\eeq
is satisfied for all
$n$-point probability distributions $(p_1,\,\ldots\,,p_n)$
and $(q_1,\,\ldots\,,q_n)$ if and only if there exist constants $\alpha$ and
$\gamma_1,\ldots,\gamma_n$ such that
\beq
F_k(p)=\alpha\ln p+\gamma_k\;,
\eeq
for all $k=1,2,\ldots,n$.
\label{AczelTheorem}
\end{theorem}

$\bigtriangleup$
The main point of this theorem is that Eq.~(\ref{HonestExpert}), though
appearing quite imprecise, is tremendously restrictive.  We shall presently
establish the ``only if'' part of the theorem.  To do this we assume the
inequality to hold and focus on the functions $F_1$ and $F_2$.  This can be
carried out by restricting the distributions $(p_1,\,\ldots\,,p_n)$ and 
$(q_1,\,\ldots\,,q_n)$ to be such that $q_i=p_i$ for all $i\ge3$ while
$p_1\equiv p$, $q_1\equiv q$, $p_2$ and $q_2$ otherwise remain free.  Then we
can define $r\equiv p+p_2=q+q_2$ where also
\beq
r=1-\sum_{i=3}^n p_k=1-\sum_{i=3}^n q_k\;.
\eeq
Note that because $n\ge3$, $r$ is a
number strictly less than unity.  With these definitions, 
Eq.~(\ref{HonestExpert}) reduces to
\beq
pF_1(q)+(r-p)F_2(r-q)\le pF_1(p)+(r-p)F_2(r-p)\;,
\label{TwoPointIneq}
\eeq
since all the other terms for $k\ge3$ cancel.  This new inequality already
contains within it enough to show that $F_1$ and $F_2$ are monotonically
nondecreasing and differentiable at every point in their domain.  Let us work
on showing these properties straight away.

Eq.~(\ref{TwoPointIneq}) can be rearranged to become
\beq
p\left[F_1(p)-F_1(q)\right]\ge(r-p)\left[F_2(r-q)-F_2(r-p)\right]\;.
\label{IneqOne}
\eeq
Upon interchanging the symbols $p$ and $q$, the same reasoning also gives
rise to the inequality,
\beq
q\left[F_1(q)-F_1(p)\right]\ge(r-q)\left[F_2(r-p)-F_2(r-q)\right]\;.
\label{IneqTwo}
\eeq
Multiplying Eq.~(\ref{IneqOne}) by $(r-q)$, Eq.~(\ref{IneqTwo}) by $(r-p)$,
and adding the resultants, we get
\beq
(r-q)\,p\left[F_1(p)-F_1(q)\right]\,+\,(r-p)\,q\left[F_1(q)-F_1(p)\right]
\,\ge\,0\;,
\eeq
which implies
\beq
r\,(p-q)\left[F_1(p)-F_1(q)\right]\ge\,0\;.
\eeq
Then if $p\ge q$, it must be the case that $F_1(p)\ge F_1(q)$ so that this 
inequality is maintained. It follows
that $F_1$ is a monotonically nondecreasing function.

Now we must show the same property for $F_2$.  To do this, we instead multiply Eq.~(\ref{IneqOne}) by $q$, Eq.~(\ref{IneqTwo}) by $p$, and add the
results of these operations.  This gives
\beq
0\,\ge\,q\,(r-p)\left[F_2(r-q)-F_2(r-p)\right]\,+\,
p\,(r-q)\left[F_2(r-p)-F_2(r-q)\right]\;,
\eeq
or, after rearranging,
\beq
0\le r\,[(r-p)-(r-q)]\,[F_2(r-p)-F_2(r-q)]\;.
\eeq
So that, if $(r-p)\ge(r-q)$, we must have in like manner $F_2(r-p)\ge F_2(r-q)$
to maintain the inequality.  Therefore, $F_2$ must also be a monotonically
nondecreasing function.

Putting these two facts to the side for the moment, we presently seek a tighter
relation between the functions $F_1$ and $F_2$.  To this end, we multiply
Eq.~(\ref{IneqOne}) by $q$, Eq.~(\ref{IneqTwo}) by $-p$.  This gives
\beq
p\,q\left[F_1(p)-F_1(q)\right]\ge q\,(r-p)\left[F_2(r-q)-F_2(r-p)\right]\;,
\eeq
and
\beq
p\,q\left[F_1(p)-F_1(q)\right]\le p\,(r-q)\left[F_2(r-q)-F_2(r-p)\right]\;.
\eeq
These two inequalities together imply
\beq
{r-p\over p}\,[F_2(r-q)-F_2(r-p)]\,\le\, F_1(p)-F_1(q)\;,
\eeq
and
\beq
F_1(p)-F_1(q)\,\le\,{r-q\over q}\,[F_2(r-q)-F_2(r-p)]\;.
\eeq
Dividing this through by $(p-q)$, we get finally
\beq
{r-p\over p}\left({F_2(r-q)-F_2(r-p)\over (r-q)-(r-p)}\right)\le\,
{F_1(p)-F_1(q)\over p-q}\;,
\eeq
and
\beq
{F_1(p)-F_1(q)\over p-q}\,\le\,
{r-q\over q}\left({F_2(r-q)-F_2(r-p)\over (r-q)-(r-p)}\right)\;.
\eeq
From this expression we know, by the Pinching Theorem of elementary calculus,
that if the limits
\beq
\lim_{q\rightarrow p}
{r-p\over p}\!\left({F_2(r-q)-F_2(r-p)\over (r-q)-(r-p)}\right)\!\equiv
{r-p\over p}F^\prime_2(r-p)
\eeq
and
\beq
\lim_{q\rightarrow p}
{r-q\over q}\!\left({F_2(r-q)-F_2(r-p)\over (r-q)-(r-p)}\right)\!\equiv
{r-p\over p}F^\prime_2(r-p)
\eeq
exist, then so does
\beq
\lim_{q\rightarrow p}{F_1(p)-F_1(q)\over p-q}\equiv F^\prime_1(p)\;,
\eeq
and the limits must be identical.  In other words, if $F_2$ is
differentiable at $(r-p)$, then $F_1$ is differentiable at $p$ and
\beq
p F^\prime_1(p)=(r-p) F^\prime_2(r-p)\;.
\label{DerivRel}
\eeq
Recall, however, that $p_2$ is not uniquely fixed by $p$ since $n\ge3$.  Thus
neither is $r$; it can range anywhere between $p$ and $1$.  This allows us to
write the statement preceding Eq.~(\ref{DerivRel}) in the converse form: if 
$F_1$ is {\it not\/} differentiable at the point $p$, then $F_2$ is {\it not\/}
differentiable at any point $(r-p)$ in the open set $(0,1-p)$.  This is the
sought after tight relation between $F_1$ and $F_2$.

This statement can be combined with the fact that $F_1$ and $F_2$ are monotonic
for the final thrust of the proof.  For this we will rely on a theorem from 
elementary real analysis sometimes called Lebesgue's theorem
\cite[page 96]{Royden63}:
\begin{lemma}
Let $f$ be an increasing real-valued function on the interval $[a,b]$.  Then
$f$ is differentiable almost everywhere.  The derivative $f^\prime$ is
measurable, and $\int_a^b f^\prime(x)dx\le f(b)-f(a)$.
\end{lemma}
Actually we are only concerned with the first conclusion of this.  It can
be seen in a qualitative manner as follows.  The points where $f$ is not
differentiable can only correspond to kinks or jumps in its graph that are at
best never decreasing in height.  Thus one can easily imagine that, in
travelling from points $a$ to $b$, a continuous infinity of such kinks and
jumps (as would be required for a measurable set) would cause the graph to
blow up to infinity before the end of the interval were ever reached.  A more
rigorous demonstration of this theorem will not be given here.

From this theorem we immediately have that $F_1$ must be differentiable 
everywhere on the closed interval $[0,1]$.  For suppose there were a point $p$
at which it were {\it not\/} differentiable.  Then $F_2$ would not be
differentiable anywhere within the measurable set $(0,1-p)$, contradicting the
fact that it is a monotonically nondecreasing 
function.  Now, using the Pinching Theorem again, we have that $F_2$ is
differentiable everywhere.

Therefore, let $s=r-p$.  Since $p$ and $s$ are independent and 
Eq.~(\ref{DerivRel}) must always hold, it follows that there must be a constant
$\alpha$ such that
\beq
p F^\prime_1(p)=s F^\prime_2(s)=\alpha\;.
\eeq
Because $F_1$ and $F_2$ are monotonically nondecreasing, $\alpha\ge 0$.  So
\beq
F^\prime_1(p)={\alpha\over p}\;\;\;\;\;\;\;\;\mbox{and}\;\;\;\;\;\;\;\;
F^\prime_2(p)={\alpha\over p}\; ;
\eeq
integrating this we get
\beq
F_1(p)=\alpha\ln p+\gamma_1\;\;\;\;\;\;\;\;\mbox{and}\;\;\;\;\;\;\;\;
F_2(p)=\alpha\ln p+\gamma_2
\eeq
where $\gamma_1$ and $\gamma_2$ are integration constants.

Running through the same argument but assuming $q_i=p_i$ for all $i$ except
$i=1$ and $i=k$, it follows in like manner that
$F_k(p)=\alpha\ln p+\gamma_k$ for all $k=1,\ldots,n$.  This concludes the proof
of the ``only if'' side of the theorem.

To establish the ``if'' part of the theorem, we need only demonstrate the
verity of the Shannon inequality,
\beq
\sum_{k=1}^n p_k\ln q_k\,\le\,\sum_{k=1}^n p_k\ln p_k\;,
\label{ShaIneq}
\eeq
with equality if and only if $p_k=q_k$ for all $k$.  To see this note that
the function $f(x)=\ln x$ is convex (since $f^{\prime\prime}=-x^{-2}\le0$) and
consequently always lies below its tangent line at $x=1$.  In other words,
$f(x)$ always lies below the line
\beqa
y
&=&
f^\prime(1)x+[f(1)-f^\prime(1)]
\nonumber\\
&=&
x-1\;.
\eeqa
Hence, $\ln x\le x-1$ with equality holding if and only if $x=1$.  Therefore
it immediately follows that
\beq
\ln\!\left({q_k\over p_k}\right)\le\,{q_k\over p_k}-1\;,
\eeq
with equality if and only if $p_k=q_k$.  This, in turn, implies
\beq
p_k(\ln q_k-\ln p_k)\,=\,p_k\ln\!\left({q_k\over p_k}\right)\le\, q_k-p_k\;,
\eeq
so that
\beq
\sum_{k=1}^n p_k(\ln q_k-\ln p_k)\,\le\,0\;,
\label{LastExpr}
\eeq
with equality holding in the last expression if and only if $p_k=q_k$ for all
$k$.  Rearranging Eq.~(\ref{LastExpr}) concludes the proof. $\Box$

When $n=2$, the above method of proof for the ``only if'' side of
Theorem~(\ref{AczelTheorem}) fails.  This follows technically because, in that 
case, the quantity $r$ must be fixed to the value $1$.  Moreover,
it turns out that Eq.~(\ref{HonestExpert}), independent of proof method,
no longer specifies a relatively unique solution.  For instance, even
if one were to make the restriction $F_1(p)=F_2(p)\equiv f(p)$, a theorem due
to Musz\'{e}ly \cite{Fischer72,Muszely73} states that any function of the 
following form will satisfy the honest-expert inequality for all distributions:
\beq
f(p)\,=\,(1-p)\,U\!\!\left(p-\frac{1}{2}\right)\,+\,
\int_0^{p-\frac{1}{2}}U(t)\,dt\,+\,C\;,
\eeq
where $C$ is constant and $U(t)$ is any continuous, increasing, odd function
defined on the open interval $(-\frac{1}{2},\frac{1}{2})$.  Thus there are
many, many ways of keeping the expert honest when talking about probability
distributions over two outcomes.

We, however, shall not let a glitch in the two-outcome case deter us in 
defining a new measure of distinguishability.  Namely, using the robust payoff 
function defined by Theorem~(\ref{AczelTheorem}) in conjunction with 
Eq.~(\ref{RoughKullback}), we obtain
\beqa
K(p_0/p_1)
&=&
\sum_{b=1}^n p_0(b)\left[\,\ln p_0(b)-\ln p_1(b)\,\right]
\nonumber\\
&=&
\sum_{b=1}^n p_0(b)\,\ln\!\left({p_0(b)\over p_1(b)}\right)\;.
\eeqa
This measure of distinguishability has appeared in other contexts and is known
by various names: the Kullback-Leibler relative information, cross entropy, 
directed divergence, update information, and information gain.

\subsection{Properties of the Relative Information}

Though a probability assignment can never be confirmed as true or false,
one's trust in a state of knowledge or probability assignment for some
phenomenon can be either strengthened or weakened through observation.  The
simplest context where this can be seen is when the phenomenon in question
is repeatable.  Consider, as an example, a coin believed to be unusually
weighted so that the probability for heads in a toss is 25\%.  If, upon
tossing the coin 10,000 times, one were to find roughly 50\% of the actual 
tosses giving rise to heads, one might indeed be compelled to re\"evaluate his 
assessment.  The reason for this is that probability assignments can be used to 
make relatively sharp predictions about the frequencies of outcomes in 
repeatable experiments; the standard Law of Large Numbers \cite{Feller68}
specifies that relative frequencies of outcomes in a set of repeated trials
approach the pre-assigned probabilities with probability 1.

Not surprisingly, the probability for an ``incorrect'' frequency in a set of
repeated trials has something to do with our distinguishability measure based
on keeping the expert honest.  This is because all the payment schemes
considered in its definition were explicitly tied to the observational context.
It turns out that the Kullback-Leibler relative information
controls the exponential rate to zero forced upon the probability of an 
``incorrect'' frequency by the Law of Large Numbers \cite{Csiszar81,Cover91}. 
This gives us another useful operational meaning for the Kullback-Leibler information and gives more trust that it is a quantity worthy of study.
This subsection is devoted to fleshing out this fact in detail.

Let us start by demonstrating that the most probable frequency 
distribution in many trials is indeed essentially the probability distribution 
for the outcomes of a single trial.  For this we suppose that an experiment of
$B$ outcomes, $b\in{\cal B}=\{1,\ldots,B\}$, will be repeated $n$ times.  The 
probability distribution for the outcomes of a single trial will be denoted 
$p_0(b)$.  The 
$n$ outcomes of the $n$ experiments can be denoted by a vector
$\vec{b}=(b_1,b_2,\ldots,b_n)\in{\cal B}^n$ where $b_i\in{\cal B}$ for each
$i$.  The probability of any event $E$ on the space ${\cal B}^n$---that is to
say, any set $E$ of outcome strings---will be denoted by $P(E)$; the probability
of the special case in which $E$ is a single string $\vec{b}$ will be denoted
$P(\vec{b})$.

The empirical frequency distribution of outcomes in $\vec{b}$
will be written as $F_{\vec{b}}(b)$,
\beq
F_{\vec{b}}(b)={1\over n}
\left(\mbox{\# of occurrences of $b$ in $\vec{b}$}\,\right)\;,
\eeq
and the set of all possible frequencies will be denoted by ${\cal F}$,
\beq
{\cal F}=\left\{\left(F_{\vec{b}}(1),F_{\vec{b}}(2),\ldots,F_{\vec{b}}(B)\right)
:\vec{b}\in{\cal B}^n\right\}\;.
\eeq
Note that the cardinality of the set ${\cal F}$, which we shall write as
$|{\cal F}|$, is bounded above by $(n+1)^B$.  This follows because there are
$B$ components in the vector specifying any particular frequency $F_{\vec{b}}$,
i.e.,
\beq
F_{\vec{b}}=\left(\frac{n_1}{n},\frac{n_2}{n},\ldots,\frac{n_B}{n}\right)\;,
\eeq
and each of the numerators of these components can take on any value between
$0$ and $n$ (subject only to the constraint that $\sum_i n_i=n$).  Thus there
are less than $(n+1)^B$ choices for the vectors $F_{\vec{b}}$.  The fact that
$|{\cal F}|$ grows only polynomially in $n$ turns out to be quite important
for these considerations.

Now to get started, we shall also need a notation for the equivalence class of 
outcome strings with the same empirical frequency distribution $p_1(b)$; for
this we adopt
\beq
{\cal T}(p_1)=\left\{\vec{b}\in{\cal B}^n:F_{\vec{b}}(b)=p_1(b)\;\,\forall\;\,
b\in{\cal B}\right\}\;.
\eeq
In this notation, we then have the following theorem.

\begin{theorem}
Suppose the experimental outcomes are described by a probability distribution 
$p_0(b)$ that is in fact an element of ${\cal F}$.  Let $P({\cal T}(p_1))$
denote the probability that an outcome sequence will be in ${\cal T}(p_1)$.  
Then
\beq
P({\cal T}(p_0))\ge P({\cal T}(p_1))\;.
\eeq
That is to say, the most likely frequency distribution in $n$ trials is actually
the pre-assigned probability distribution.
\label{MaxProbTheo}
\end{theorem}

Note that this theorem is restricted to probability assignments that are
numerically equal to frequencies in $n$ trials.  We make this restriction
to simplify the techniques involved in proving it and because it is all we
will really need for the later considerations.

$\bigtriangleup$ To see how this theorem comes about, note that
\beqa
P({\cal T}(p_0))
&=&
\sum_{\vec{b}\in{\cal T}(p_0)}P(\vec{b})
\nonumber\\
&=&
\sum_{\vec{b}\in{\cal T}(p_0)}\,\prod_{i=1}^n p_0(b_i)
\rule{0mm}{8mm}
\nonumber\\
&=&
\sum_{\vec{b}\in{\cal T}(p_0)}\,\prod_{b=1}^B p_0(b)^{np_0(b)}
\rule{0mm}{8mm}
\nonumber\\
&=&
|{\cal T}(p_0)|\,\prod_{b=1}^B p_0(b)^{np_0(b)}
\rule{0mm}{8mm}
\label{p0Prob}
\eeqa
and similarly
\beq
P({\cal T}(p_1))\,=\,|{\cal T}(p_1)|\,\prod_{b=1}^B p_0(b)^{np_1(b)}\;.
\label{p1Prob}
\eeq
Expressions~(\ref{p0Prob}) and (\ref{p1Prob}) can be compared if one realizes
that $|{\cal T}(p_0)|$ is---essentially by definition---identically equal to the
number of ways of inserting $np_0(1)$ objects of type 1, $np_0(2)$ objects of 
type 2, etc., into a total $n$ slots.  That is to say, $|{\cal T}(p_0)|$ is a 
multinomial coefficient \cite{Feller68},
\beq
|{\cal T}(p_0)|={n!\over(np_0(1))!\;(np_0(2))!\;\cdots\;(np_0(B))!}\;.
\eeq
With this, we have
\beqa
{P({\cal T}(p_0))\over P({\cal T}(p_1))}
&=&
{\left(\prod_{b=1}^B (np_1(b))!\right)\!\left(\prod_{b=1}^B 
p_0(b)^{np_0(b)}\right)\over
\left(\prod_{b=1}^B (np_0(b))!\right)\!\left(\prod_{b=1}^B
p_0(b)^{np_1(b)}\right)}
\nonumber\\
&=&
\prod_{b=1}^B{(np_1(b))!\over(np_0(b))!}\,p_0(b)^{n[p_0(b)-p_1(b)]}\;.
\rule{0mm}{9mm}
\label{ProbRatio}
\eeqa

The desired result comes about through the inequality
\beq
{m!\over n!}\ge n^{m-n}\;.
\label{FactorialIneq}
\eeq
Let us quickly demonstrate this before moving on.  First, suppose $m\ge n$.
Then,
\beq
{m!\over n!}=m(m-1)\cdots(n+1)\ge \underbrace{n\cdot n\cdots n}_{(m-n)\;
\mbox{\scriptsize times}}=n^{m-n}\;.
\eeq
Now suppose $m<n$.  Then,
\beq
{m!\over n!}=[n(n-1)\cdots(m+1)]^{-1}\ge[\underbrace{n\cdot
n\cdots n}_{(n-m)\;\mbox{\scriptsize times}}]^{-1}=(n^{n-m})^{-1}=n^{m-n}\;.
\eeq

With Eq.~(\ref{FactorialIneq}) in hand, Eq.~(\ref{ProbRatio}) gives
\beqa
{P({\cal T}(p_0))\over P({\cal T}(p_1))}
&\ge&
\prod_{b=1}^B(np_0(b))^{[np_1(b)-np_0(b)]}p_0(b)^{n[p_0(b)-p_1(b)]}
\nonumber\\
&=&
\prod_{b=1}^B n^{n[p_1(b)-p_0(b)]}
\rule{0mm}{9mm}
\nonumber\\
&=&
n^{\left(n\!\sum_b[p_1(b)-p_0(b)]\right)}\;=\;n^0\;=\;1\;.
\rule{0mm}{9mm}
\eeqa
Therefore,
\beq
P({\cal T}(p_0))\ge P({\cal T}(p_1))\;,
\eeq
and this completes the proof that the most likely frequency distribution in
$n$ trials is just the pre-assigned probability distribution. $\Box$

The next step toward our goal of demonstrating the Kullback-Leibler information
in this new context is to work out an expression for the probability of an
outcome string $\vec{b}$ with an ``incorrect'' frequency distribution.
\begin{theorem}
Suppose the experimental outcomes are described by a probability distribution 
$p_0(b)$.  The probability for a particular string of outcomes
$\vec{b}\in{\cal T}(p_1)$ with the ``wrong'' frequencies is
\beq
P(\vec{b})=e^{-n[H(p_1)+K(p_1/p_0)]}\;,
\eeq
where
\beq
H(p_1)=-\sum_{b=1}^B p_1(b)\ln p_1(b)
\eeq
is the Shannon entropy of the
distribution $p_1(b)$.
\end{theorem}

$\bigtriangleup$ This can be seen with a little algebra:
\beqa
P(\vec{b})
&=&
\prod_{i=1}^n p_0(b_i)\;=\;\prod_{b=1}^B p_0(b)^{np_1(b)}\;=\;
\prod_{b=1}^B e^{np_1(b)\ln p_0(b)}
\nonumber\\
&=&
\prod_{b=1}^B\exp\left\{n[p_1(b)\ln p_0(b)+p_1(b)\ln p_1(b)-p_1(b)\ln p_1(b)]
\right\}
\rule{0mm}{9mm}
\nonumber\\
&=&
\prod_{b=1}^B\exp\left\{n\left[p_1(b)\ln p_1(b)+p_1(b)\ln
\left({p_0(b)\over p_1(b)}\right)\right]\right\}
\rule{0mm}{9mm}
\nonumber\\
&=&
\exp\left\{-n\sum_{b=1}^B\left[-p_1(b)\ln p_1(b)+p_1(b)\ln
\left({p_1(b)\over p_0(b)}\right)\right]\right\}
\rule{0mm}{9mm}
\nonumber\\
&=&
\exp\{-n[H(p_1)+K(p_1/p_0)]\}\;.\;\;\Box
\rule{0mm}{7mm}
\eeqa
A simple corollary to this is,
\begin{corollary}
If the experimental outcomes are described by $p_0(b)$, the probability for a particular string $\vec{b}\in{\cal T}(p_0)$ with the ``correct'' frequencies is
\beq
P(\vec{b})=e^{-nH(p_0)}\;.
\eeq
\label{FreqCor}
\end{corollary}
This follows because $K(p_0/p_0)=0$.

Using this corollary, in conjunction with Theorem~\ref{MaxProbTheo}, we can
derive a relatively good estimate of the number of elements in ${\cal T}(p_1)$.
This estimate is necessary for connecting the relative information to the
probability of an ``incorrect'' frequency.
\begin{lemma}
For any $p_1(b)\in{\cal F}$,
\beq
(n+1)^{-B}e^{nH(p_1)}\le|{\cal T}(p_1)|\le e^{nH(p_1)}\;.
\eeq
\label{UMPH}
\end{lemma}

$\bigtriangleup$
This lemma can be seen in the following way.  Suppose the probability is
actually $p_1(b)$.  Then, using Corollary~\ref{FreqCor}, we have
\beq
P({\cal T}(p_1))=|{\cal T}(p_1)|\,e^{-nH(p_1)}\;.
\eeq
This probability, however, must be less than 1.  Hence,
$|{\cal T}(p_1)|\le e^{nH(p_1)}$---proving the right-hand side of the lemma.
The left-hand side follows similarly by considering the probability of all
possible frequencies and using Theorem~\ref{MaxProbTheo},
\beqa
1\;=\;\sum_{F_{\vec{b}}\in{\cal F}}P\!\left({\cal T}(F_{\vec{b}})\right)
&\le&
\sum_{F_{\vec{b}}\in{\cal F}}\max_{F_{\vec{b}}\in{\cal F}}
P\!\left({\cal T}(F_{\vec{b}})\right)
\nonumber\\
&=&
P({\cal T}(p_1))\left(\sum_{F_{\vec{b}}\in{\cal F}}1\right)\;=\;
|{\cal F}|\,P({\cal T}(p_1))
\nonumber\\
&\le&
\rule{0mm}{6mm}
(n+1)^B P({\cal T}(p_1))
\nonumber\\
&=&
\rule{0mm}{6mm}
(n+1)^B\,|{\cal T}(p_1)|\,e^{-nH(p_1)}\;.
\eeqa
Thus $(n+1)^{-B}e^{nH(p_1)}\le|{\cal T}(p_1)|$ and this completes the proof
of the lemma. $\Box$

Let us now move quickly to our sought after theorem and its proof.
\begin{theorem}
Suppose the experimental outcomes are described by the probability distribution
$p_0(b)$.  The probability that the outcome string will have some ``incorrect''
frequency distribution $p_1(b)$ is $P({\cal T}(p_1))$.  This number is bounded
in the following way,
\beq
(n+1)^{-B}e^{-nK(p_1/p_0)}\le P({\cal T}(p_1))\le e^{-nK(p_1/p_0)}\;.
\eeq
\end{theorem}

$\bigtriangleup$
This is now easy to prove.  We just need write
\beqa
P({\cal T}(p_1))
&=&
\sum_{\vec{b}\in{\cal T}(p_1)}P(\vec{b})
\nonumber\\
&=&
\sum_{\vec{b}\in{\cal T}(p_1)}P(\vec{b})e^{-n[H(p_1)+K(p_1/p_0)]}
\nonumber\\
&=&
|{\cal T}(p_1)|e^{-n[H(p_1)+K(p_1/p_0)]}\;,
\rule{0mm}{6mm}
\eeqa
to see that we can use Lemma~\ref{UMPH} to give the desired result. $\Box$

\section{Distinguishability via Mutual Information}

Consider what happens when one samples a known probability distribution $p(b)$, 
$b=1,\ldots,n$.  The probability quantifies the extent to which the outcome 
can be predicted, but it generally {\it does not\/} pin down the precise
outcome itself.  Upon learning the outcome of a sampling, one, in a very
intuitive sense, ``gains information'' that he does not possess beforehand.
For instance, if all the outcomes of the sampling are equally probable, then
one will generally gain a lot of information from the sampling; there is
essentially nothing that can be predicted about the outcome beforehand.  On the
other hand, if the probability distribution is highly peaked about one 
particular outcome, then one will generally gain very little information from 
the sampling; there will be almost no point in carrying out the sampling---its
outcome can be predicted at the outset.  This simple idea provides the 
starting point for building our last notion of distinguishability.

Let us sketch the idea briefly before attempting to make it precise.  Suppose
there is a reasonable way of quantifying the average information gained when
one samples a distribution $q(b)$; denote that quantity, whatever it may be, by
$H(q)$.  Then, if two probability distributions $p_0(b)$ and
$p_1(b)$ are distinct in the sense that one has more unpredictable outcomes
than the other, the average information gained upon sampling them
will also be distinct, i.e., either $H(p_0)\ge H(p_1)$ or vice versa.  For, in 
sampling the distribution with the more unpredictable outcomes, one can expect 
to gain a larger amount of information.  Thus, immediately, the notion of
information gain in a sampling provides a means for distinguishing probability
distributions.  At this level, however, one is no better off than in simply
comparing the probabilities themselves.  To get somewhere with this idea, a
more interesting scenario must be developed.

The problem in comparing distributions through the information one gains
upon sampling them is that the information gain has nothing to say about the
distribution itself---that quantity is assumed already known.  What would
happen, however, if one were to randomly choose between sampling the two
different distributions, say with probabilities $\pi_0$ and $\pi_1$?  We can 
make a case for two distinct possibilities.  First, perhaps trivially, suppose
that in spite of choosing the sampling distribution randomly, one still knows 
which distribution is in front of him at any given moment.  Then an average
information gain $H(p_0)$ will ensue when the actual distribution is $p_0(b)$ and $H(p_1)$ will ensue when the actual distribution is $p_1(b)$; that is to
say, the expected information gain in a sampling a {\it known\/}
distribution is just $\pi_0 H(p_0)+\pi_1 H(p_1)$.  Notice that this quantity,
being an average, is greater than the lesser of the two information gains and 
less than the greater of the two gains, i.e.,
\beq
\min\{H(p_0),H(p_1)\}\le\pi_0 H(p_0)+\pi_1 H(p_1)\le\max\{H(p_0),H(p_1)\}\;.
\eeq

Now consider the opposing case where the 
identity of the distribution to be sampled remains unknown.  In this case, the
most one can say about which outcome will occur in the sampling is that it
is controlled by the probability distribution $p(b)=\pi_0 p_0(b)+\pi_1 p_1(b)$.
In other words, the sampling outcome will be even more unpredictable than it
was in either of the two individual cases; some of the unpredictability will be
due to the indeterminism $p_0(b)$ and $p_1(b)$ describe and some of the
unpredictability will be due to the fact that the individual distribution from 
which the sample is drawn remains unknown.  Hence it must be the case that
$H(p)\ge H(p_0)$ and $H(p)\ge H(p_1)$.

The excess of $H(p)$ over $\pi_0 H(p_0)+\pi_1 H(p_1)$ is the average gain of 
information one can expect about the distribution itself.
This quantity, called the {\it mutual information\/} \cite{Shannon48,Lindley56},
\beq
J(p_0,p_1;\pi_0,\pi_1)=H(\pi_0 p_0+\pi_1 p_1)-\Bigl(
\pi_0 H(p_0)+\pi_1 H(p_1)\Bigr)\;,
\label{AbsMutInf}
\eeq
is the natural candidate for distinguishability that we seek in this section.
If the two distributions $p_0(b)$ and $p_1(b)$ are completely distinguishable,
then all the information gained in a sampling should be solely about the 
identity of the distribution; the quantity $J(p_0,p_1;\pi_0,\pi_1)$ should 
reduce to $H(\pi)$, the information that can be gained by sampling the prior 
distribution $\pi=\{\pi_0,\pi_1\}$.  If the distributions $p_0(b)$ and $p_1(b)$
are completely indistinguishable, then $J(p_0,p_1;\pi_0,\pi_1)$ should reduce 
to zero; this signifies that in sampling one learns nothing whatsoever about
the distribution from which the sample is drawn.

Notice that this distinguishability measure depends crucially on the observer's
prior state of knowledge, quantified by $\pi=\{\pi_0,\pi_1\}$, about whether
$p_0(b)$ or $p_1(b)$ is actually the case.  Thus it is a measure of
distinguishability relative to a given state of knowledge.  There is, of course,
nothing wrong with this, just as there was nothing wrong with the
error-probability distinguishability measure; one just needs to recognize it as
such.

These are the ideas behind taking mutual information as a measure of 
distinguishability.  In the remainder of this section, we work toward
justifying a precise expression for Eq.~(\ref{AbsMutInf}) and showing in a 
detailed way how it can be interpreted in an operational context.

\subsection{Derivation of Shannon's Information Function}
\label{ShannonDer}

The function $H(p)$ that quantifies the average information gained upon
sampling a distribution $p(b)$ will ultimately
turn out to be the famous Shannon information function 
\cite{Shannon48,Slepian74}
\beq
H(p)=-\sum_b p(b)\ln p(b)\;.
\label{ShannonInf}
\eeq
What we should like to do here is justify this expression from first
principles.  That is to say, we shall build up a theory of ``information gain''
based solely on the probabilities in an experiment and find that that theory
gives rise to the expression (\ref{ShannonInf}).

To start with our most basic assumption, we 
reiterate the idea that the information gained in performing an experiment or 
observation is a function of how well the outcomes to that experiment or 
observation can be predicted in the first place.  Other characteristics of an
outcome that might convey ``information'' in the common sense of the
word, such as shape, color, smell, feel, etc., will be considered
irrelevant; indeed, we shall assume any such properties
already part of the very definition of the outcome events.
Formally this means that if a set of events $\{x_1,x_2,\,\ldots\,,x_n\}$ has
a probability distribution $p(x)$, not only is the expected information gain
in a sampling, $H(p)$, exclusively a function of the numbers
$p(x_1)$, $p(x_2)$, \ldots, $p(x_n)$, but also it must be independent of the
labelling of that set.  In other words,
$H(p)\equiv H(p(x_1),p(x_2),\,\ldots\,,p(x_n))$ is required to be invariant 
under permutations of its arguments.  This is called the requirement of
``symmetry.''

The most important technical property of $H(p)$ is that, even though information
gain is a subjective concept depending on the observer's prior state of
knowledge, it should at least be objective enough that it not depend on the
method by which knowledge of the experimental outcomes is acquired.  We can
make this idea firm with a simple example.  Consider an experiment with three
mutually exclusive outcomes $x$, $y$, and $z$.  Note that the probability that
$z$ does not occur is 
\beq
p(\neg z)=1-p(z)=p(x)+p(y)\;.
\eeq
The probabilities for $x$ and $y$ given that $z$ does not occur are
\beq
p(x|\neg z) = {p(x)\over p(x) + p(y)}
\;\;\;\;\;\;\;\;\;\;\mbox{and}\;\;\;\;\;\;\;\;\;\;
p(y|\neg z) = {p(y)\over p(x) + p(y)}\;.
\eeq

There are at least two methods by which an observer can gather the result of 
this experiment.  The first method is by the obvious tack of simply finding 
which outcome of the three possible ones actually occurred.  In this case, the
expected information gain is, by our convention, 
\beq
H\Bigl(p(x),\,p(y),\,p(z)\Bigr)\;.
\label{RegEnt}
\eeq
The 
second method is more roundabout.  One could, for instance, first check whether
$z$ did or did not occur, and {\it then\/} in the event that it did {\it not\/}
occur, further check which of $x$ and $y$ {\it did\/} occur.  In the first
phase of this method, the expected information gain is
\beq
H\Bigl(p(\neg z),\,p(z)\Bigr)\;.
\eeq
For those cases in which the second phase of the method must be carried out, a
further gain of information can be expected.  Namely,
\beq
H\Bigl(p(x|\neg z),\,p(y|\neg z)\Bigr)\;.
\eeq
Note, though, that this last case is only expected to occur a fraction
$p(\neg z)$ of the time.  Thus, in total, the expected information gain by this
more roundabout method is
\beq
H\Bigl(p(\neg z),\,p(z)\Bigr)\,+\,
p(\neg z)\,H\Bigl(p(x|\neg z),\,p(y|\neg z)\Bigr)\;.
\label{SecEnt}
\eeq
The assumption of ``objectivity'' is that the quantities in Eqs.~(\ref{RegEnt})
and (\ref{SecEnt}) are identical.  That is to say, upon changing the notation
slightly to $p_x=p(x)$, $p_y=p(y)$, $p_z=p(z)$
\beq
H(p_x,\,p_y,\,p_z)\,=\,H(p_x+p_y\,,\,p_z)\,+\,(p_x+p_y)\,
H\!\left({p_x\over p_x + p_y}\,,\,{p_y\over p_x+p_y}\right)\;.
\eeq
In the event that we are instead concerned with $n$ mutually exclusive events,
the same assumption of ``objectivity'' leads to the identification,
\beq
H(p_1,\,\ldots\,,p_n)\,=\,H(p_1+p_2\,,p_3\,,\,\ldots\,,p_n)\,+\,(p_1+p_2)\,
H\!\left({p_1\over p_1+p_2}\,,\,{p_2\over p_1+p_2}\right)\;.
\label{Objectivity}
\eeq

It turns out that the requirements of symmetry and objectivity (as embodied in
Eq.~(\ref{Objectivity})\,) are enough to uniquely determine the form of $H(p)$
(up to a choice of units) provided we allow ourselves one extra convenience
\cite{Aczel81}, namely, that we allow the introduction of an arbitrary positive
parameter $\alpha\ne1$ into Eq.~(\ref{Objectivity}) in the following way,
\beq
H_\alpha(p_1,\,\ldots\,,p_n)\,=\,H_\alpha(p_1+p_2\,,p_3\,,\,\ldots\,,p_n)\,+
\,(p_1+p_2)^\alpha\, H_\alpha\!\left({p_1\over p_1+p_2}\,,\,
{p_2\over p_1+p_2}\right)\;,
\label{ObjectivityAlpha}
\eeq
and define $H(p)$ to be the limiting value of $H_\alpha(p)$ as
$\alpha\!\rightarrow\!1$.  (The introduction of the subscript on $H_\alpha(p)$ 
is made simply to remind us that the solutions to Eq.~(\ref{ObjectivityAlpha}) 
depend upon the parameter $\alpha$.)  This idea is encapsulated in the following
theorem.
\begin{theorem}[Dar\'{o}czy]
Let
\beq
\Gamma_n=\left\{(p_1,\,\ldots\,,p_n)\;|\;p_k\ge0,\;k=1,\ldots,n,\;\mbox{and}
\;\sum_{i=1}^n p_i=1\right\}
\eeq
be the set of all $n$-point probability 
distributions and let $\Gamma=\bigcup_n\Gamma_n$ be the set of all discrete
probability distributions.  Suppose the function
$H_\alpha:\Gamma\rightarrow\RR$, $\alpha\ne1$, is symmetric in all its 
arguments and satisfies Eq.~(\ref{ObjectivityAlpha}) for each $n\ge2$.  Then,
under the convention that $0\ln 0=0$, the limiting value of $H_\alpha$ as
$\alpha\rightarrow1$ is uniquely specified up to a constant $C$ by
\beq
H(p_1,\,\ldots\,,p_n)\,=\,-\,{C\over\ln 2}\,\sum_{i=1}^n p_i\ln p_i\;.
\eeq
\label{InfoTheorem}
\end{theorem}
The constant $C$ in this expression fixes the ``units'' of information.  If
$C=1$, information is said to be measured in {\it bits\/}; if $C=\ln 2$,
information is said to be measured in {\it nats\/}.  (A relatively obscure
measure of information is the case $C=\log_{10} 2$, where the
units are called {\it Hartleys\/} \cite{Hamming80}.)
In this document, we will generally take $C=\ln 2$.  On the occasion, however, 
that we do consider information in units of bits we shall write $\log()$ for
the base-2 logarithm, rather than the more common $\log_2()$.

$\bigtriangleup$
The proof of Theorem~\ref{InfoTheorem}, deserving wider recognition, is due
to Dar\'{o}czy 
\cite{Daroczy70} and proceeds as follows.  Define $s_i=p_1+\cdots+p_i$
and let
\beq
f(x)=H_\alpha(x,1-x)\;\;\;\;\;\;\;\;\mbox{for}\;\;\;\;\;\;\;\;0\le x\le1\;.
\eeq
Then, by repeated application 
of condition~(\ref{ObjectivityAlpha}), it follows immediately that
\beq
H_\alpha(p_1,\,\ldots\,,p_n)\,=\,\sum_{i=2}^n s_i^\alpha\,f\!\left(
{p_i\over s_i}\right)\;.
\label{EntropyAlpha}
\eeq
Thus all we need do is focus on finding an explicit expression for the function
$f$.

We have from the symmetry requirement that $H_\alpha(x,1-x)=H_\alpha(1-x,x)$
and hence,
\beq
f(x)=f(1-x)\;.
\eeq
In particular, $f(0)=f(1)$.  Furthermore, if $x$ and
$y$ are two nonnegative numbers such that $x+y\le1$, we must also have
\beq
H_\alpha(x,y,1-x-y)=H_\alpha(y,x,1-x-y)\;.
\eeq
However, by Eq.~(\ref{ObjectivityAlpha})
\beqa
H_\alpha(x,y,1-x-y)
&=&
H_\alpha(x,1-x)\,+\,(1-x)^\alpha
H_\alpha\!\left({y\over1-x},{1-x-y\over1-x}\right)
\nonumber\\
&=&
H_\alpha(x,1-x)\,+\,(1-x)^\alpha
H_\alpha\!\left({y\over1-x},1-{y\over1-x}\right)
\rule{0mm}{8mm}
\nonumber\\
&=&
f(x)\,+\,(1-x)^\alpha f\!\left({y\over1-x}\right)\;.
\rule{0mm}{8mm}
\eeqa
Thus it follows that $f$ must satisfy the functional equation
\beq
f(x)\,+\,(1-x)^\alpha f\!\left({y\over1-x}\right)\,=\,
f(y)\,+\,(1-y)^\alpha f\!\left({x\over1-y}\right)\;,
\label{FundInfEq}
\eeq
for $x,y\in[0,1)$ with $x+y\le1$.  (In the case $\alpha=1$, 
Eq.~(\ref{FundInfEq}) is known commonly as {\it the fundamental equation of
information\/} \cite{Aczel75}.)

We base the remainder of our conclusions on the study of Eq.~(\ref{FundInfEq}).
Note first that if $x=0$, it reduces to,
\beq
f(0)+f(y)=f(y)+(1-y)^\alpha f(0)\;.
\eeq
Since $y$ is still arbitrary, it follows from this that $f(0)=0$; thus
$f(1)=0$, too.  Now let $p=1-x$ for $x\ne1$ and let $q=y/(1-x)=y/p$.  With
this, the information equation~(\ref{FundInfEq}) becomes
\beq
f(p)+p^\alpha f(q)=f(pq)+(1-pq)^\alpha f\!\left({1-p\over1-pq}\right)\;.
\label{NewInfEq}
\eeq
We can use this equation to show that
\beq
F(p,q)\equiv f(p)+\left[p^\alpha+(1-p)^\alpha\right]f(q)
\eeq
is symmetric in $q$ and $p$, i.e., $F(p,q)=F(q,p)$.
From that fact, a unique expression for $f(p)$ follows trivially.  Let us just
show this before going further:
\beqa
0
&=&
F\!\left(p\,,\,\frac{1}{2}\right)-F\!\left(\frac{1}{2}\,,\,p\right)
\nonumber\\
&=&
f(p)+\left[p^\alpha+(1-p)^\alpha\right]f\!\left(\frac{1}{2}\right)-
f\!\left(\frac{1}{2}\right)-\left[\left(\frac{1}{2}\right)^\alpha
+\left(\frac{1}{2}\right)^\alpha\right]f(p)
\rule{0mm}{8mm}
\nonumber\\
&=&
\left(1-2^{1-\alpha}\right)f(p)+f\!\left(\frac{1}{2}\right)
\left[\,p^\alpha+(1-p)^\alpha-1\,\right]\;,
\rule{0mm}{8mm}
\eeqa
which implies that
\beq
f(p)=C\left(2^{1-\alpha}-1\right)^{-1}\left[\,p^\alpha+
(1-p)^\alpha-1\,\right]\;,
\label{InfoTypeAlpha}
\eeq
where the constant $C=f(\frac{1}{2})$.
Because $f(0)=f(1)=0$, Eq.~(\ref{InfoTypeAlpha}) also holds for $p=0$ and $p=1$.

To cap off the derivation of Eq.~(\ref{InfoTypeAlpha}), let us demonstrate
that $F(p,q)$ is symmetric.  Just expanding and regrouping, we have, by
Eq.~(\ref{NewInfEq}), that
\beqa
F(p,q)
&=&
\left[f(p)+p^\alpha f(q)\right]+(1-p)^\alpha f(q)
\nonumber\\
&=&
f(pq)+(1-pq)^\alpha f\!\left({1-p\over1-pq}\right)+(1-p)^\alpha f(q)
\rule{0mm}{9mm}
\nonumber\\
&=&
f(pq)+(1-pq)^\alpha\left[f\!\left({1-p\over1-pq}\right)+
\left({1-p\over1-pq}\right)^{\!\alpha} f(q)\right]\;.
\rule{0mm}{9mm}
\eeqa
If we can show that the last term in this expression is symmetric in $q$ and
$p$, then we will have shown that $F(p,q)$ is symmetric.  To this end, let us
define
\beq
A(p,q)=f\!\left({1-p\over1-pq}\right)+
\left({1-p\over1-pq}\right)^{\!\alpha} f(q)\;.
\eeq
Also, to save a little room, let
\beq
z=\frac{1-p}{1-pq}\;.
\eeq
Then,
\beq
1-zq={1-q\over1-pq}
\;\;\;\;\;\;\;\;\;\;\;\;\;\mbox{and}\;\;\;\;\;\;\;\;\;\;\;\;\;
1-z=p\left({1-q\over1-pq}\right)\;.
\eeq
So that, upon using Eq.~(\ref{NewInfEq}) again, we get
\beqa
A(p,q)
&=&
f(z)+z^\alpha f(q)
\nonumber\\
&=&
f(zq)+(1-zq)^\alpha f\!\left({1-z\over1-zq}\right)
\rule{0mm}{9mm}
\nonumber\\
&=&
f(1-zq)+(1-zq)^\alpha f\!\left({1-z\over1-zq}\right)
\rule{0mm}{9mm}
\nonumber\\
&=&
f\!\left({1-q\over1-pq}\right)+
\left({1-q\over1-pq}\right)^{\!\alpha} f(p)
\rule{0mm}{9mm}
\nonumber\\
&=&
A(q,p)\;.
\rule{0mm}{6mm}
\eeqa
Thus $F(p,q)$ is symmetric.  This completes the demonstration of
Eq.~(\ref{InfoTypeAlpha}).

We just need plug the expression for $f(p)$ into Eq.~(\ref{EntropyAlpha})
to get a nearly final result,
\beqa
H_\alpha(p_1,\,\ldots\,,p_n)
&=&
\sum_{i=2}^n s_i^\alpha
C\left(2^{1-\alpha}-1\right)^{-1}\left[
\left({p_i\over s_i}\right)^\alpha + \left(1-{p_i\over s_i}\right)^\alpha-
1\right]
\nonumber\\
&=&
C\left(2^{1-\alpha}-1\right)^{-1}\sum_{i=2}^n\left[\,
p_i^\alpha+(s_i-p_i)^\alpha-s_i^\alpha\,\right]
\rule{0mm}{8mm}
\nonumber\\
&=&
C\left(2^{1-\alpha}-1\right)^{-1}\sum_{i=2}^n\left(
p_i^\alpha+s_{i-1}^\alpha-s_i^\alpha\right)
\rule{0mm}{8mm}
\nonumber\\
&=&
C\left(2^{1-\alpha}-1\right)^{-1}\left(\sum_{i=2}^n
p_i^\alpha+s_1^\alpha-s_n^\alpha\right)
\rule{0mm}{8mm}
\nonumber\\
&=&
C\left(2^{1-\alpha}-1\right)^{-1}\left(\sum_{i=1}^n
p_i^\alpha-1\right)\;.
\rule{0mm}{8mm}
\label{EntropyTypeAlpha}
\eeqa
Now in taking the limit $\alpha\!\rightarrow\!1$, note that both the numerator
and denominator of this expression vanishes.  Thus we must use l'Hospital's
rule in the calculating limit, i.e., first take the derivative with respect to
$\alpha$ of the numerator and denominator separately and then take the limit:
\beqa
\lim_{\alpha\rightarrow0}H_\alpha(p_1,\,\ldots\,,p_n)
&=&
\lim_{\alpha\rightarrow0}C\!\left(-2^{1-\alpha}\ln 2\right)^{-1}
\left(\sum_{i=1}^n p_i^\alpha\ln p_i\right)
\nonumber\\
&=&
-\,{C\over\ln 2}\,\sum_{i=1}^n p_i\ln p_i\;.
\rule{0mm}{8mm}
\eeqa
This completes our derivation of the Shannon information 
formula~(\ref{ShannonInf}). It is to be hoped that this has conveyed something 
of the austere origin of the information-gain concept. $\Box$
  
We finally mention that the Dar\'{o}czy informations of type-$\alpha$, i.e., 
Eq.~(\ref{EntropyTypeAlpha}), appearing in this derivation are of interest in 
their own right.  First of all, there is a simple relation between these and
the Renyi informations of degree-$\alpha$ introduced in Section~\ref{PECBsec};
namely,
\beqa
H^\alpha(p)
&\equiv&
{1\over\alpha-1}\,\ln\!\left(\,\sum_{i=1}^n p_i^\alpha\,\right)
\nonumber\\
&=&
{1\over\alpha-1}\,\ln\!\left(\frac{1}{C}\!
\left(2^{1-\alpha}-1\right)\!H_\alpha(p)+1\right)\;.
\rule{0mm}{8mm}
\eeqa
Secondly, they share many properties with the Shannon information \cite{Aczel75}
while being slightly more tractable for some applications, there being no
logarithm in their expression.

\subsection{An Interpretation of the Shannon Information}

\begin{figure}
\begin{center}
\leavevmode
\epsfig{figure=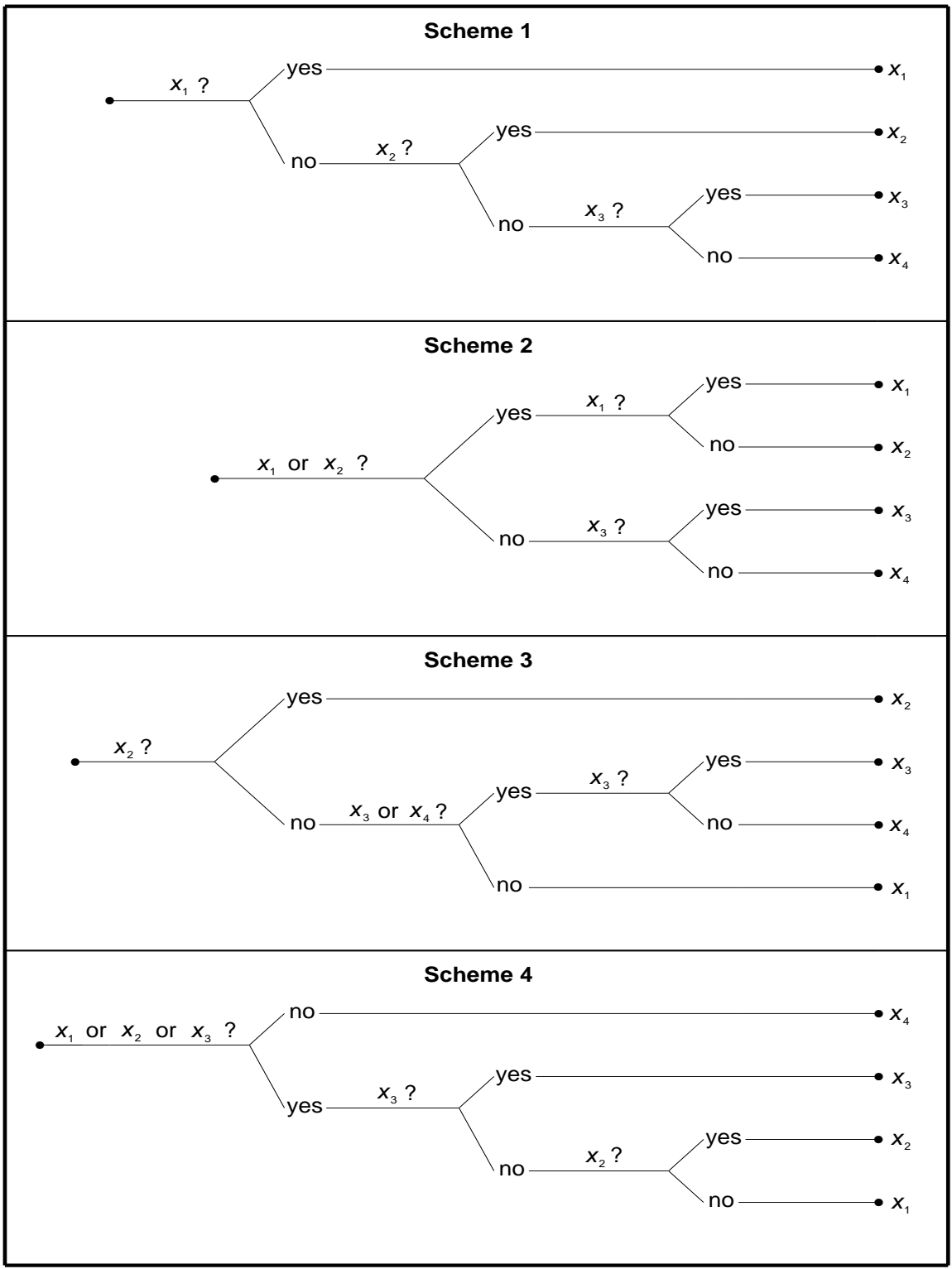,width=6in,height=8in}
\caption{Binary Question Schemes}
\label{BQS}
\end{center}
\end{figure}

The justification of the information-gain concept can be strengthened through
an operational approach to the question.  To carry this out, let us develop the 
following example.  Suppose we were to perform an experiment with four 
possible outcomes $x_1,\,x_2,\,x_3,\,x_4$, the respective probabilities being
$p(x_1)=\frac{1}{20}$, $p(x_2)=\frac{1}{5}$, $p(x_3)=\frac{1}{4}$, and
$p(x_4)=\frac{1}{2}$.  The expected gain of information in this experiment is 
given by Eq.~(\ref{ShannonInf}) and is numerically approximately 1.68 bits.
By the fundamental postulate of Section~\ref{ShannonDer}, we know that this
information gain will be independent of the method of questioning used in 
discerning the outcome.  In particular, we could consider all possible ways of
determining the outcome by way of binary yes/no questions.  For instance, we 
could start by asking, ``Is the outcome $x_1$?''  If the answer is yes, then
we are
done.  If the answer is no, then we could further ask, ``Is the outcome 
$x_2$?,'' and proceed in similar fashion until the identity of the outcome is
at hand.  This and three other such binary-question methodologies are depicted
schematically in Figure~\ref{BQS}.

The point of interest to us here is that each such questioning scheme
generates, by its very nature, a {\it code\/} for the possible
outcomes to the experiment.  That code can be generated by writing down,
in order, the yes's and no's encountered in traveling from the {\it root\/} to
each {\it leaf\/} of these schematic {\it trees\/}.  For instance, by 
substituting 0 and 1 for yes and no, respectively, the four trees depicted in
Figure~\ref{BQS} give rise to the codings:
\begin{center}
\vspace{.2in}
\begin{tabular}{||l|l|l|l||} \hline
Scheme 1 & Scheme 2 & Scheme 3 & Scheme 4 \\ \hline
$x_1\leftrightarrow0$ & $x_1\leftrightarrow00$ & $x_1\leftrightarrow11$ &
$x_1\leftrightarrow011$ \\
$x_2\leftrightarrow10$ & $x_2\leftrightarrow01$ & $x_2\leftrightarrow0$ &
$x_2\leftrightarrow010$ \\
$x_3\leftrightarrow110$ & $x_1\leftrightarrow10$ & $x_1\leftrightarrow100$ &
$x_3\leftrightarrow00$ \\
$x_4\leftrightarrow111$ & $x_1\leftrightarrow11$ & $x_1\leftrightarrow101$ &
$x_4\leftrightarrow1$ \\
\hline
\end{tabular}
\vspace{.2in}
\end{center}
Codes that can be generated from trees in this way are called
{\it instantaneous\/} or {\it prefix-free\/} and are noteworthy for the property
that concatenated strings of their codewords can be uniquely deciphered just
by reading from left to right.  This follows because no codeword in such a
coding can be the ``prefix'' of any other codeword. As a case in point, using
the coding generated by Scheme 1, the ``message'' $1111100010100110$ uniquely 
corresponds to the concatenation $x_4x_3x_1x_1x_2x_2x_1x_3$; there are no
other possibilities.

From these four examples, one can see that not all such questioning schemes are 
equally efficient.  For Scheme 1 the expected codeword length, i.e.,
\beq
\overline{l}=\sum p(x_i)l(x_i)\;,
\eeq
where $l(x_i)$ is the number of digits in the
code word for $x_i$, is 2.70 binary digits.  Those for Schemes 2--4 are
2.00, 2.55, and 1.75 binary digits, respectively.  To see where this example is
going, note that each expectation value is greater than $H(p)$, the average
information gained in sampling the distribution $p(x_i)$.  This inequality is no
accident.  As we shall see, the Shannon noiseless coding theorem 
\cite{Shannon48} specifies that the expected codeword length of any 
instantaneous code must be greater than $H(p)$.  Moreover, the minimal 
average codeword length is bounded above by $H(p)+1$.
Reverting back to the language of questioning schemes, we have that the
minimum average number of binary questions required to discern the outcome of 
sampling $p(x_i)$ is approximately equal to $H(p)$.

This we take as a new starting point for interpreting the information
function: it is approximately the minimal effort (quantified in terms of
expected number of binary questions) required to discern the outcome of an 
experiment.  To make this precise, we presently set out to demonstrate how the 
noiseless coding theorem comes about within this context.

Our first step in doing this is to demonstrate an elementary lemma of 
information theory, known as the Kraft inequality \cite{Kraft49}, giving a 
useful analytic characterization of all possible instantaneous codes.
\begin{lemma}[Kraft]
The codeword lengths $l_1\le l_2\le\cdots\le l_n$ of any binary instantaneous
code for a set of messages $\{x_1,\,x_2,\,\ldots\,,\,x_n\}$ must satisfy the 
inequality
\beq
\sum_{k=1}^n 2^{-l_i}\le1\;.
\eeq
Moreover, for any set of integers $k_1\le k_2\le\cdots\le k_n$ satisfying
this inequality, there exists a instantaneous code with these as codeword lengths.
\end{lemma}

\begin{figure}
\begin{center}
\leavevmode
\epsfig{figure=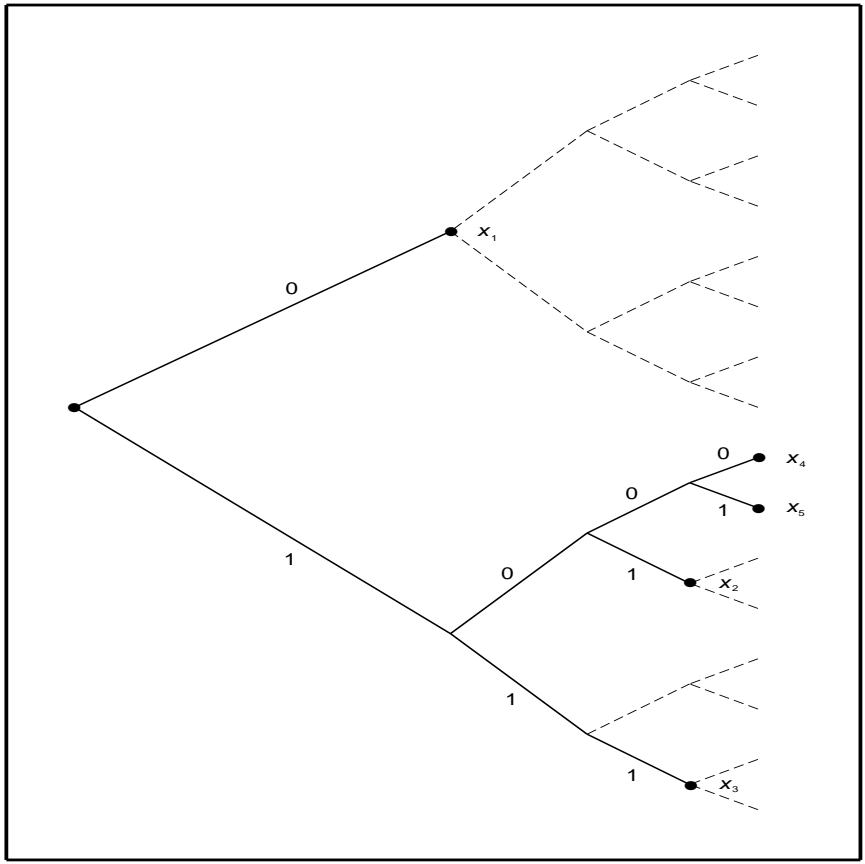,width=4.5in,height=4.5in}
\caption{Coding Tree Imbedded in a Full Tree}
\end{center}
\label{FullTree}
\end{figure}

$\bigtriangleup$
The derivation of this lemma is really quite simple.  Start with the coding
tree generating the instantaneous code and imbed it in a {\it full\/} tree of
$2^{l_n}$ leaves.  A full tree is a tree for which each direct path leading
from the root to a terminal leaf is of the same length; see 
Figure~\ref{FullTree}.
With this, one sees easily that the number of terminal leaves in the full
tree stemming from the node associated with the codeword of length $l_i$ is
just $2^{l_n-l_i}$, $i=1,\ldots,n$.  Therefore it follows that the total number
of terminal leaves in this tree not associated with codewords must be
\beq
\sum_{i=1}^n 2^{l_n-l_i}\;.
\eeq
This number, however, can be no larger than the number of leaves in the full
tree.  Thus
\beq
\sum_{i=1}^n 2^{l_n-l_i}\le2^{l_n}\;.
\eeq
This proves the first statement of the lemma.

For the second statement of the lemma, start with a full tree of $2^{k_n}$
leaves.  First place the symbol $x_1$ at some node of depth $k_1$ and delete
all further descendents of that node.  Then place the symbol $x_2$ at any
remaining node of depth $k_2$ and remove its descendents.  Iterating this
procedure will produce the appropriate coding tree and thus a instantaneous
code with the specified codeword lengths. $\Box$

With the Kraft inequality in hand, it is a simple matter to derive the
Shannon noiseless coding theorem.
\begin{theorem}[Shannon]
Suppose messages $x_1$, \ldots, $x_n$ occur with 
probabilities $p(x_1)$, \ldots, $p(x_n)$.  Then the minimal
average codeword length 
$\overline{l}_{\min}$ for a binary instantaneous coding of these messages
satisfies
\beq
H(p)\le\overline{l}_{\min}\le H(p)+1\;,
\eeq
where $H(p)$ is the Shannon information of the distribution $p(x_i)$ measured
in bits.
\label{NoiselessCode}
\end{theorem}

$\bigtriangleup$
To show the left-hand inequality, we just need note that for any instantaneous 
code the Kraft inequality specifies
\beqa
\overline{l}-H(p)
&=&
\sum_i p(x_i)l(x_i)+\sum_i p(x_i)\log p(x_i)
\nonumber\\
&=&
\rule{0mm}{6mm}
-\sum_i p(x_i)\log 2^{-l(x_i)}+\sum_i p(x_i)\log p(x_i)
\nonumber\\
&=&
\sum_i p(x_i)\log\!\left({p(x_i)\over q(x_i)}\right)-
\log\!\left(\sum_i2^{-l(x_i)}\right)
\rule{0mm}{8mm}
\nonumber\\
&\ge&
\rule{0mm}{8mm}
\sum_i p(x_i)\log\!\left({p(x_i)\over q(x_i)}\right)\;,
\label{NoisCodDer}
\eeqa
where
\beq
q(x_i)={2^{-l(x_i)}\over\sum_i2^{-l(x_i)}}
\eeq
is a probability distribution constructed for the purpose at hand.  Namely,
the final quantity on the right-hand side of Eq.~(\ref{NoisCodDer}) is then
positive by the Shannon inequality, Eq.~(\ref{ShaIneq}), already demonstrated.  
Thus it follows that
\beq
\overline{l}-H(p)\ge0\;,
\eeq
and so the minimal average instantaneous codeword length must be at least as
large as $H(p)$.

Now to show the right hand side of Theorem~\ref{NoiselessCode}, we need only
note that there is an instantaneous code with codeword lengths given by
\beq
l(x_i)=\lceil-\log p(x_i)\rceil\;,
\eeq
where $\lceil x\rceil$ denotes the smallest integer greater than or equal
to $x$, because these integers satisfy the Kraft inequality.  Therefore
it must hold that
\beqa
\overline{l}_{\min}
&\le&
\sum_i p(x_i)\lceil-\log p(x_i)\rceil
\nonumber\\
&\le&
\sum_i p(x_i)\Bigl(-\log p(x_i)+1\Bigr)
\nonumber\\
&=&
H(p)+1\;.
\eeqa
This concludes our proof of the Shannon noiseless coding theorem. $\Box$

This is one precise sense in which $\overline{l}_{\min}\approx H(p)$.  Actually,
the exact value of $\overline{l}_{\min}$ can be calculated given
the message probabilities $p(x_1)$, \ldots, $p(x_n)$.  This comes about by an
optimal coding procedure known as Huffman coding 
\cite{Huffman52,Ash65,Gallager68,Cover91}.
Therefore, one might have wished to choose $\overline{l}_{\min}$ as the
appropriate measure of information under the present interpretation.  This,
however, encounters an immediate objection: the Huffman coding algorithm gives
no explicit {\it analytical\/} expression for $\overline{l}_{\min}$.  Thus
using $\overline{l}_{\min}$ as a measure of information would be operationally
difficult at best.  Also, though, there are even tighter upper bounds on
$\overline{l}_{\min}$ than given by the noiseless coding theorem.  For instance
if $p(x_1)\ge p(x_2)\ge\cdots\ge p(x_n)$, then \cite{Gallager78,Capocelli91}
\beq
\overline{l}_{\min}-H(p)\le\left\{
\begin{array}{ll}
p(x_1)+\sigma & \mbox{if $p(x_1)<\frac{1}{2}$} \\
2-h(p(x_1))-p(x_1)\le p(x_1) & \mbox{if $p(x_1)\ge\frac{1}{2}$}
\rule{0mm}{6mm}
\end{array}
\right]
\eeq
where $\sigma=1-\log e+\log\log e\approx .086$ and
\beq
h(x)=-x\log x-(1-x)\log(1-x)\;.
\eeq
Another bound is
\beq
\overline{l}_{\min}-H(p)\,\le\,1-h(p(x_n))\,\le\,1-2p(x_n)\;.
\eeq
Tighter bounds than this, in terms of $p(x_1)$ {\it and\/} $p(x_n)$, are
known \cite{Capocelli91}, but are not so easily expressible.  The upshot is 
that these generally force $\overline{l}_{\min}$ closer to $H(p)$ than the
noiseless coding theorem and thus strengthen the notion of ``approximate''
here.

Finally, in this context, we note that a direct consequence of
Theorem~\ref{NoiselessCode} is the following.  If we were to repeat the 
experiment described by $p(x_k)$, say, $N$ times before asking a set of yes--no
questions to discern all $N$ outcomes, the minimum expected number of questions
will be some $\overline{L}_{\min}$ that satisfies
\beq
H(P)\le\overline{L}_{\min}\le H(P)+1\;,
\label{BigNCT}
\eeq
where $P(x_{i_1},x_{i_2},\ldots,x_{i_N})=p(x_{i_1})p(x_{i_2})\cdots p(x_{i_N})$
is the (product) probability distribution describing all $N$ outcomes.  Using
the fact that
\beqa
H(P)
&=&
-\sum_{x_{i_1},\ldots,x_{i_N}}P(x_{i_1},\ldots,x_{i_N})\log
P(x_{i_1},\ldots,x_{i_N})
\nonumber\\
&=&
-\sum_{x_{i_1},\ldots,x_{i_N}}P(x_{i_1},\ldots,x_{i_N})\left(
\sum_{k=1}^n\log p(x_{i_k})\right)
\rule{0mm}{7mm}
\nonumber\\
&=&
-N\sum_{k=1}^n p(x_k)\log p(x_k)
\nonumber\\
&=&
N H(p)\;,
\rule{0mm}{6mm}
\eeqa
Eq.~(\ref{BigNCT}) reduces to
\beq
H(p)\,\le\,\frac{1}{N}\overline{L}_{\min}\,\le\, H(p)+\frac{1}{N}\;.
\eeq
Therefore, if one is willing to collect data on multiple experiments before
asking the yes--no questions required to discern the outcomes, then one can
make the expected number of questions {\it per\/} experiment as close to
$H(p)$ as one wishes---just by choosing $N$ sufficiently large.  This is
another, strong sense in which $H(p)$ quantifies the minimal effort required
to discern the outcome of an experiment.

\subsection{The Mutual Information}

The notion of information gain has now been analyzed from two very different
perspectives, one axiomatic and one operational.  With a firm expression for
this notion finally at hand, we may return to the real object of this
section, the measure of distinguishability known as {\it mutual information\/}.
Using Eq.~(\ref{ShannonInf}) in conjunction with Eq.~(\ref{AbsMutInf}), we
obtain various formulations of this quantity
\beqa
J(p_0,p_1;\pi_0,\pi_1)
&=&
-\sum_b p(b)\ln p(b)+\pi_0\sum_b p_0(b)\ln p_0(b)+\pi_1\sum_b p_1(b)\ln p_1(b)
\nonumber\\
&=&
\pi_0\sum_b p_0(b)\ln\!\left({p_0(b)\over p(b)}\right)+
\pi_1\sum_b p_1(b)\ln\!\left({p_1(b)\over p(b)}\right)
\nonumber\\
&=&
\pi_0K(p_0/p)+\pi_1K(p_1/p)\;,
\rule{0mm}{6mm}
\eeqa
where $p(b)=\pi_0 p_0(b)+\pi_1 p_1(b)$ and $K(/)$ denotes the Kullback-Leibler
relative information of Section~\ref{KEHsec}.  The last form gives a secondary
interpretation to the mutual information: in an honest expert problem, it is
the expert's expected loss for trying to pass off the mean distribution $p(b)$
in place of either actual distribution $p_0(b)$ or $p_1(b)$.


\chapter{The Distinguishability of Quantum States}

\begin{flushright}
\baselineskip=13pt
\parbox{2.8in}{\baselineskip=13pt
``\ldots~a priori one should expect a chaotic world which cannot
be grasped by the mind in any way.  One could (yes one should) expect the world
to be subjected to law only to the extent that we order it through our 
intelligence.''}\medskip\\
---{\it Albert Einstein}\\
Letter to Maurice Solovine\\30 March 1952
\end{flushright}

\section{Introduction}

For any given quantum state, the outcomes of most possible measurements are
completely lawless in their determination.  Quantum theory, however, provides
the means for calculating probabilities for the outcomes.  It is by this
handle that the measures of distinguishability for probability distributions
can be used to say something about quantum mechanical states.
In this Chapter, we work toward ``quantizing'' the classical distinguishability
measures introduced in Chapter 2.  The problem of statistically 
distinguishing quantum states $\hat\rho_0$ and $\hat\rho_1$ on a
$D$-dimensional Hilbert space via quantum measurement is that of using some 
measurement with $n$ outcomes to generate the probability distributions
$p_0(b)$ and $p_1(b)$ used in the classical measures.  The number $n$ here
should be thought of as a free variable that remains to be fixed; certainly
it need not equal $D$. An {\it optimal\/} quantum measurement with respect to
any of these measures is just a measurement that makes these quantities as
large or as small as they can possibly be.  The ``quantized'' measures
are simply the numerical values of the classical measures when an optimal
measurement is used.

The strategy for making progress toward precise expressions of these
quantities is to use the formalism of positive-operator-valued measures or
POVMs \cite{Jauch67,Davies70,Kraus83} introduced in Chapter~1.
As a quick reminder, a POVM is a set of positive operators $\hat E_b$ which
is complete, i.e.,
\beq
\langle\psi|\hat E_b|\psi\rangle\ge 0\mbox{\ \ \ for all }b\mbox{ and all
vectors\ }|\psi\rangle\;,
\eeq
and
\beq
\sum_b\hat 
E_b=\hat{\openone}=\left({\mbox{identity}\atop\mbox{operator}}\right)\;.
\eeq
A consequence of this definition is that the $\hat E_b$ are Hermitian operators
with nonnegative eigenvalues \cite{Schatten60}.
The subscript $b$ here, as before, indexes the possible outcomes of the
measurement.  Again, the conditions on the $\hat E_b$ are those 
necessary and sufficient for the expression
$p(b)={\rm tr}\bigl(\hat\rho\hat E_b\bigr)$
to be a valid probability distribution for the $b$.
As described in Chapter 1, it turns out that a measurement corresponding to a 
POVM can always be interpreted as an ``ordinary'' orthogonal projection-valued 
measurement (i.e., one for which the outcomes correspond to eigenvalues of a 
Hermitian operator) on an extended system consisting of the given one along
with an independently prepared auxiliary system; the labels $b$ in that 
interpretation stand for the various outcomes to the ordinary measurement on
the composite system \cite{Peres90}.

The quantum measures of distinguishability we shall focus upon in this chapter
are, listed in order of increasing unwieldiness:
\begin{itemize}
\item
the Quantum Error Probability
\beq
P_e(\hat\rho_0|\hat\rho_1)\equiv\min_{\{\hat E_b\}}\sum_b\,\min\!\left\{
\pi_0{\rm tr}(\hat\rho_0\hat E_b),\,\pi_1{\rm tr}(\hat\rho_1\hat E_b)\right\}
\label{QError}
\eeq
\item
the Quantum Fidelity
\beq
F(\hat\rho_0,\hat\rho_1)\equiv\min_{\{\hat E_b\}}\sum_b
\sqrt{{\rm tr}(\hat\rho_0\hat E_b)}
\sqrt{{\rm tr}(\hat\rho_1\hat E_b)}
\label{QOwwtters}
\eeq
\item
the Quantum R\'enyi Overlaps
\beq
F_\alpha(\hat\rho_0/\hat\rho_1)\equiv\min_{\{\hat E_b\}}\sum_b
\left({\rm tr}(\hat\rho_0\hat E_b)\right)^{\!\alpha}\!
\left({\rm tr}(\hat\rho_1\hat E_b)\right)^{\!{1-\alpha}},
\;\;\;\;\;\;\;0<\alpha<1
\label{QRenyi}
\eeq
\item
the Quantum Kullback Information
\beq
K(\hat\rho_0/\hat\rho_1)\equiv\max_{\{\hat E_b\}}\,
\sum_b{\rm tr}(\hat\rho_0\hat E_b)\,
\ln\!\left({{\rm tr}(\hat\rho_0\hat E_b)\over
{\rm tr}(\hat\rho_1\hat E_b)}\right)
\label{QKullback}
\eeq
\item
the Accessible Information
\end{itemize}
\beq
I(\hat\rho_0|\hat\rho_1)\equiv\max_{\{\hat E_b\}}\,
\sum_b\!\left(\pi_0\,{\rm tr}(\hat\rho_0\hat E_b)\,
\ln\!\left({{\rm tr}(\hat\rho_0\hat E_b)\over
{\rm tr}(\hat\rho\hat E_b)}\right)\,+\,
\pi_1\,{\rm tr}(\hat\rho_1\hat E_b)\,
\ln\!\left({{\rm tr}(\hat\rho_1\hat E_b)\over
{\rm tr}(\hat\rho\hat E_b)}\right)\rule{0mm}{7mm}\!\right)
\label{AInformation}
\eeq
where
\beq
\hat\rho=\pi_0\hat\rho_0+\pi_1\hat\rho_1.
\eeq
Notice again that the number of measurement outcomes in these definitions
has not been fixed at the outset as must be the case in the classical 
expressions.  The notation used here is meant to convey the following.  The
comma separating $\hat\rho_0$ and $\hat\rho_1$ in $F(\hat\rho_0,\hat\rho_1)$ is
meant to convey that this function is symmetric upon their interchange.  The
slash in $F_\alpha(\hat\rho_0/\hat\rho_1)$ and $K(\hat\rho_0/\hat\rho_1)$ is
signifies that these functions are explicitly asymmetric in the two
density operators.  The bar in $P_e(\hat\rho_0|\hat\rho_1)$ and
$I(\hat\rho_0|\hat\rho_1)$ is used to emphasize that these may or may not be
symmetric, depending upon the value of the prior probabilities $\pi_0$ and
$\pi_1$.

The difficulty that crops up in extremizing quantities like these is that, so
far at least, there seems to be no way to make the problem amenable to a 
variational approach: the problems associated with allowing $n$ to be arbitrary 
while enforcing the constraints on positivity and completeness for the
$\hat E_b$ appear to be intractable.  Moreover variational techniques generally
only lead to the assurance of local extrema, perhaps never revealing the one
that is globally best.  New methods are required.

Fortunately, the error-probability and statistical overlap distinguishability 
measures appear to be ``algebraic'' enough that one could well imagine using 
standard operator inequalities, such as the Schwarz Inequality for operator 
inner products, to aid in finding explicit expressions for 
$P_e(\hat\rho_0|\hat\rho_1)$ and $F(\hat\rho_0,\hat\rho_1)$.  That, in fact,
is the case.  The expression for $F_\alpha(\hat\rho_0/\hat\rho_1)$ appears
less tractable, but one might still hope that something like a H\"older 
Inequality can be of use in this context.  This remains an open question.
On the other hand, when it comes to finding useful expressions for 
$K(\hat\rho_0/\hat\rho_1)$ and $I(\hat\rho_0|\hat\rho_1)$, for the same reason,
one should be less optimistic.  Progress toward explicit expressions for 
Eqs.~(\ref{QKullback}) and (\ref{AInformation}) are necessarily impeded by 
the ``transcendental'' character of the logarithm appearing in their 
definitions.  Generally only bounds for these quantities may be found.

This Chapter is devoted to fleshing out what is known about the quantum
measures of distinguishability.

\section{The Quantum Error Probability}
\label{GiveEmHell}

\subsection{Single Sample Case}

An interesting particular case of the general quantum decision 
problem \cite{Helstrom76} is connected to the one introduced in
Section~\ref{PECBsec}.  A given quantum mechanical system is secretly
prepared either in the (pure or mixed) state $\hat\rho_0$ or in the state
$\hat\rho_1$.  These two possibilities are described by the prior probabilities $\pi_0$ and $\pi_1$ respectively.  It is an observer's task to perform any
quantum measurement he pleases on this system and then to make the ``best'' 
possible guess as to the state's true identity.  The word ``best'' is in
quotes because there are many things it can mean---for instance, it may depend
upon the observer's various personal costs for being right or wrong.  Here we 
shall specialize the notion of ``best'' measurement and ``best'' guess to be
those which, when combined, minimize the expected error probability of the
decision.  The question is this: what quantum measurement should the observer
use so that his expected probability of error is indeed as small as it can 
possibly be?  The answer to this gives rise to an explicit expression for the 
measure of distinguishability called the Quantum Error Probability:
\beq
P_e(\hat\rho_0|\hat\rho_1)\equiv\min_{\{\hat E_b\}}\sum_b\,\min\!\left\{
\pi_0{\rm tr}(\hat\rho_0\hat E_b),\,\pi_1{\rm tr}(\hat\rho_1\hat E_b)\right\}\;.
\eeq

This problem can be much simplified by noticing the following.  
Any ``measurement $\,+\,$ guess'' the observer can make can be summed up neatly
as the measurement of a binary-valued POVM $\{\hat E_0,\hat E_1\}$, 
i.e., two nonnegative definite operators $\hat E_0$ and $\hat E_1$ such that
$\hat E_0+\hat E_1=\hat{\openone}$.  When outcome
0 occurs, the observer chooses the state $\hat\rho_0$; when
outcome 1 occurs, he chooses state $\hat\rho_1$.  Therefore the expected
probability of error for a decision based on {\it this\/} measurement can
be written as
\beq
P_e=\pi_0\,{\rm tr}(\hat\rho_0\hat E_1)\,+\,\pi_1\,{\rm tr}
(\hat\rho_1\hat E_0)\;.
\label{ErrorProb}
\eeq
That is to say, the expected probability of error is just the probability that
$\hat\rho_0$ is the true state times the conditional probability that the 
decision will be wrong when this is the case {\it plus\/} a similar term for
$\hat\rho_1$.  So, it must be the case that
\beq
\min_{\{\hat E_b\}}\sum_b\,\min\!\left\{
\pi_0{\rm tr}(\hat\rho_0\hat E_b),\,\pi_1{\rm tr}(\hat\rho_1\hat E_b)\right\}
=\min_{\{\hat E_0,\hat E_1\}}\!\left(\pi_0\,{\rm tr}(\hat\rho_0
\hat E_1)\,+\,\pi_1\,{\rm tr}(\hat\rho_1\hat E_0)\right)\;.
\eeq

Helstrom's minimal error-probability measurement is just the POVM 
$\{\hat E^{\rm o}_0,\hat E^{\rm o}_1\}$ that minimizes Eq.~(\ref{ErrorProb}).
In Ref.~\cite{Helstrom76}, he showed that this POVM possesses the
following explicit form.  Both $\hat E^{\rm o}_0$ and $\hat E^{\rm o}_1$
are diagonal in a basis diagonalizing the Hermitian operator
\beq
\hat\Gamma\equiv \pi_1\hat\rho_1-\pi_0\hat\rho_0\;.
\label{Schmuck}
\eeq
With respect to this basis, the diagonal elements $\lambda^0_j$ of
$\hat E^{\rm o}_0$ are
assigned values according to the diagonal elements $\gamma_j$ of $\hat\Gamma$ 
via the rule:
\beqa
\lambda^0_j = 1
&\;\;\;\;\;\;\;\;\;\mbox{when}\;\;\;\;\;\;\;\;\;&
\gamma_j<0\;,
\nonumber\\
\lambda^0_j = 0
&\;\;\;\;\;\;\;\;\;\mbox{when}\;\;\;\;\;\;\;\;\;&
\gamma_j>0\;.
\eeqa
For $j$ such that $\gamma_j=0$, $\lambda^0_j$ may be assigned any value between
0 and 1; we take it to be 0 for definiteness.  The operator $\hat E^{\rm o}_1$ 
is formed simply by working out
$\hat E^{\rm o}_1=\hat{\openone}-\hat E^{\rm o}_0$.

One way the observer can implement this POVM is just to perform a standard von
Neumann measurement of the Hermitian operator $\hat\Gamma$ and bin the outcomes 
according to whether they correspond to positive, negative, or zero eigenvalues.
If a positive eigenvalue results, outcome 1 of the POVM is said to be found and
the observer makes a decision appropriately.  If a negative eigenvalue results, 
outcome 0 of the POVM is said to be found.  If a zero eigenvalue results, the
posterior information is that either of the density operators is just as likely 
as the other.  So in that case, any strategy for a decision will do.

In the remainder of this Section, we shall rederive the explicit form of
Helstrom's measurement in an elementary way that does not depend upon the 
variational techniques of Ref.~\cite{Helstrom76}.  This derivation is
closely connected to the one appearing in Ref.~\cite{Helstrom67}.

Let $\{\hat E_0,\hat E_1\}$ be an arbitrary binary-valued POVM.
Using the fact that
\beq
\hat E_0+\hat E_1=\hat{\openone}\;,
\eeq
Eq.~(\ref{ErrorProb}) becomes
\beqa
P_e
&=&
\pi_0{\rm tr}\Bigl(\hat\rho_0(\hat{\openone}-\hat E_0)\Bigr)+
\pi_1{\rm tr}(\hat\rho_1\hat E_0)
\nonumber\\
&=&
\pi_0{\rm tr}\hat\rho_0-\pi_0{\rm tr}(\hat\rho_0\hat E_0)+
\pi_1{\rm tr}(\hat\rho_1\hat E_0)
\rule{0mm}{4.5mm}
\nonumber\\
&=&
\pi_0 + {\rm tr}\Bigl((\pi_1\hat\rho_1-\pi_0\hat\rho_0)\hat E_0\Bigr)\;.
\rule{0mm}{6mm}
\label{ErrorArray}
\eeqa
Therefore finding the minimum of $P_e$ reduces to finding the minimum of
${\rm tr}(\hat\Gamma\hat E_0)$ over all operators
$\hat E_0$ such that $0\le\hat E_0\le\hat{\openone}$.

To do this, suppose the operator $\hat\Gamma$ has a spectral decomposition
given by
\beq
\hat\Gamma=\sum_j \gamma_j\,|j\rangle\langle j|\;.
\eeq
Then
\beq
{\rm tr}(\hat\Gamma\hat E_0)=\sum_j\gamma_j\langle j|\hat E_0|j\rangle\;.
\eeq
Because $\hat\Gamma$ is neither positive- nor negative-definite,
this quantity is bounded below by the sum of its negative terms and, moreover,
\beq
{\rm tr}(\hat\Gamma\hat E_0)\ge{\sum_j}^{\,\prime}\gamma_j\;,
\label{Bound}
\eeq
where the 
prime on the summation sign signifies that the sum is {\it restricted\/} to 
those $j$ for which $\gamma_j\le 0$.  This follows because
$0\le\langle j|\hat E_0|j\rangle\le1$ for all $j$.  Note that the right hand
side of Eq.~(\ref{Bound}) is independent of $\hat E_0$.  Hence, if we can find
any $\hat E_0$ that satisfies this inequality via a strict equality, that POVM
element must be optimal.

To construct such an optimal $\hat E^{\rm o}_0$, i.e., one that achieves
this lower bound $\sum_j^\prime\gamma_j$, we may start by  
specifying its diagonal elements in this basis: 
\beqa
\langle j|\hat E^{\rm o}_0|j\rangle = 1
&\;\;\;\;\;\;\;\;\;\mbox{when}\;\;\;\;\;\;\;\;\;&
\gamma_j<0
\nonumber\\
\langle j|\hat E^{\rm o}_0|j\rangle = 0
&\;\;\;\;\;\;\;\;\;\mbox{when}\;\;\;\;\;\;\;\;\;&
\gamma_j>0\;.
\label{Conditions}
\eeqa
For $j$ such that $\gamma_j=0$, we make take
$\langle j|\hat E^{\rm o}_0|j\rangle$ to be any value between 0 and 1; again
we take it to be 0 for definiteness.

Now, since no mention has yet been made of the off-diagonal elements,
it might at first appear that there are {\it many} optimal measurements
for this problem.  That, however, is incorrect; for it turns out
that any measurement operator $\hat E^{\rm o}_0$ satisfying 
Eq.~(\ref{Conditions}) must also be diagonal in this basis.
To see this, suppose the operator $\hat E^{\rm o}_0$ has the 
spectral decomposition
\beq
\hat E^{\rm o}_0=\sum_k e_k\,|e_k\rangle\langle e_k|\;.
\eeq
Then, first consider a $j$ such that $\gamma_j\ge0$.  For that,
\beq
0=\langle j|\hat E^{\rm o}_0|j\rangle=\sum_k e_k\,|\langle e_k|j\rangle|^2\;.
\eeq
Hence, because $|\langle e_k|j\rangle|^2\ge0$ in general, it must be the case
that $\langle e_k|j\rangle=0$ whenever $e_k\ne0$. So
\beq
\langle k|\hat E^{\rm o}_0|j\rangle=\sum_l e_l
\langle k|e_l\rangle\langle e_l|j\rangle=0\;,
\eeq
for {\it any\/} $k\ne j$ such that $\gamma_j\ge0$ or $\gamma_k\ge0$.  To see
that all other off-diagonal terms must vanish, one just need return to
Eq.~(\ref{ErrorArray}) and run through exactly the same argument as above to
find that $\langle k|\hat E^{\rm o}_1|j\rangle=0$ for any $k\ne j$ such that
$\gamma_j\le0$ or $\gamma_k\le0$.  Then because
$\hat E^{\rm o}_0+\hat E^{\rm o}_1=\hat{\openone}$, it follows that
$\langle k|\hat E^{\rm o}_1|j\rangle=0$ {\it for all} $k\ne j$.

This completes the proof.  The measurement operator $\hat E^{\rm o}_0$ we have 
specified is unique up to the arbitrary choice of the diagonal elements for which $\gamma_j=0$, though here we have chosen them to vanish for definiteness.
It follows that $\hat E^{\rm o}_1$ is also unique to the 
same extent, i.e., in the basis diagonalizing $\hat\Gamma$, its $(j,j)$ matrix 
element is 1 whenever $\gamma_j>0$---all other matrix elements either vanish or 
are set by the condition $\hat E^{\rm o}_0+\hat E^{\rm o}_1=\hat{\openone}$.

Thus we have an explicit form for the quantum error probability:
\beq
P_e(\hat\rho_0|\hat\rho_1)\,=\,\pi_0\,+\sum_{\gamma_j\le 0}\gamma_j\;,
\eeq
where $\gamma_j$ are eigenvalues of the operator $\hat\Gamma$ defined in
Eq.~(\ref{Schmuck}).

\subsection{Many Sample Case}

What happens to this criterion of distinguishability when there are $M\!>\!1$
copies of the quantum state upon which measurements can be performed?  There
are at least two ways it loses its unique standing in this situation.  The
first is that one could imagine making a sophisticated measurement on all $M$ 
quantum systems at once, i.e.\ on the $N^M$-dimensional Hilbert space
describing the complete ensemble of quantum states.  This measurement is very 
likely to be more useful than any set of measurements on the systems separately
\cite{Peres91}.  In particular, the optimal error-probability measurement on the big Hilbert space is a Helstrom measurement operator (\ref{Schmuck}),
except that the density operators of concern now are
$\hat\Upsilon_0=\hat\rho_0\otimes\hat\rho_0\otimes\cdots\otimes\hat\rho_0$ and
$\hat\Upsilon_1=\hat\rho_1\otimes\hat\rho_1\otimes\cdots\otimes\hat\rho_1$, 
where each expression contains $M$ terms and $\otimes$ is the direct or 
Kronecker product for matrices \cite{Bellman70,Marcus92}.  Thus the optimal 
measurement on the big Hilbert space can be written explicitly as the Hermitian 
operator,
\beq
\hat\Gamma_M=\pi_0\!\left(\,\bigotimes_{k=1}^M\hat\rho_0\right)-
\pi_1\!\left(\,\bigotimes_{k=1}^M\hat\rho_1\right)\;.
\eeq 
Clearly the distinguishability $P_e(\hat\Upsilon_0|\hat\Upsilon_1)$ will be no
simple function of $P_e(\hat\rho_0|\hat\rho_1)$.

\begin{figure}
\begin{center}
\leavevmode
\epsfig{figure=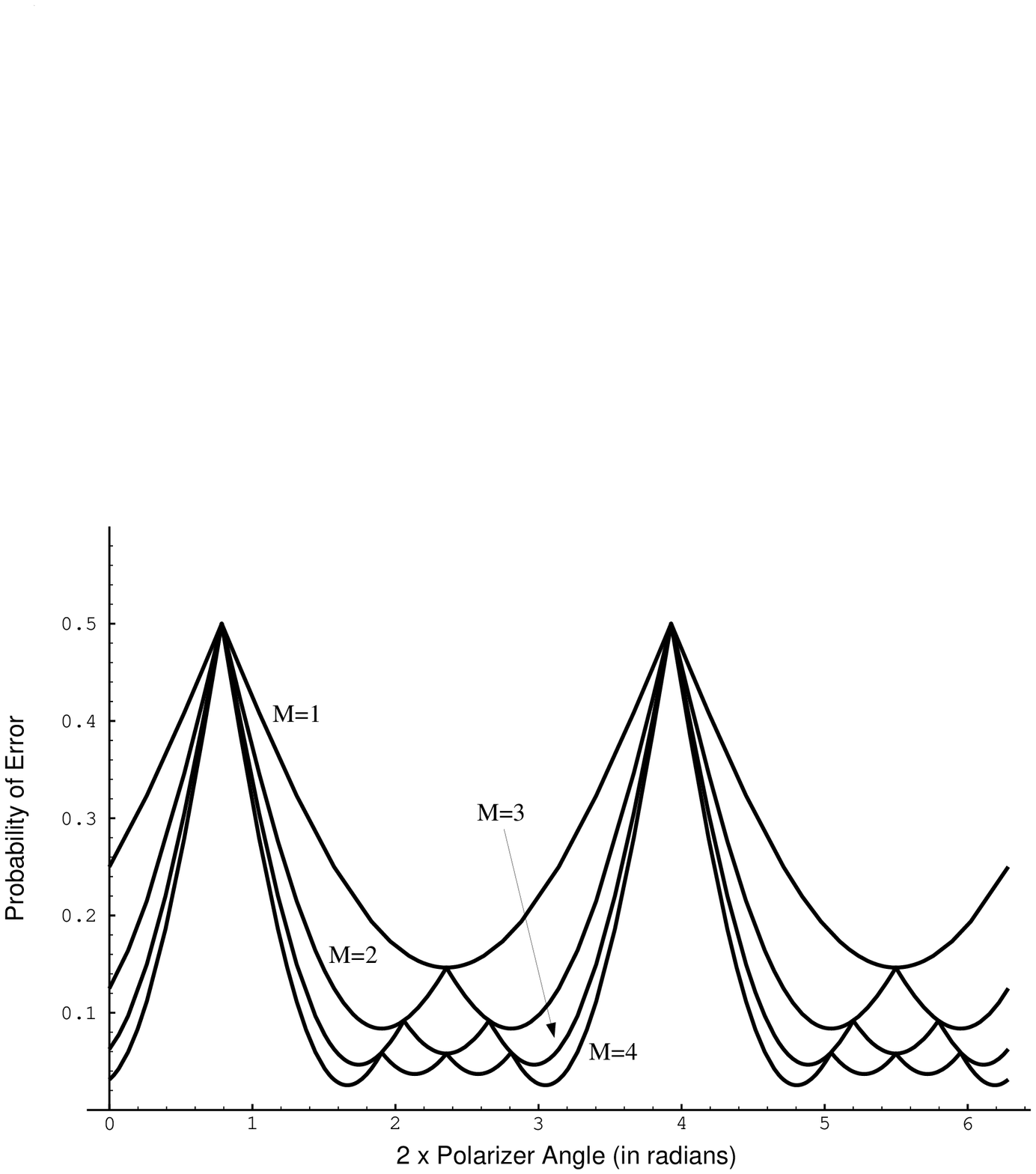,width=6in,height=7in}
\caption{Probability of error in guessing a photon's polarization that is
either horizontal or $45^\circ$ from the horizontal.  Error probability is 
plotted here as a function of measurement (i.e., radians from the horizontal)
and number of measurement repetitions $M$ before the guess is made.}
\label{ManualLabor}
\end{center}
\end{figure}

A second way for $P_e(\hat\rho_0|\hat\rho_1)$ to lose its unique standing in 
the decision problem comes about even when all the measurements are restricted
to the individual quantum systems.  For instance, suppose that $\hat\rho_0$ 
and $\hat\rho_1$ are equally probable pure linear polarization states of a 
photon, one along the horizontal and the other $45^\circ$ from the horizontal.  
The optimal error-probability measurement for the case $M\!=\!1$ is given by
the Helstrom operator $\hat\Gamma$ just derived, i.e., the measurement of the
yes/no question of 
whether the photon is polarized $67.50^\circ$ from the horizontal.  On the other hand, if $M\!=\!2$, the expression that must be optimized over all (Hermitian 
operator) measurements is no longer Eq.~(\ref{ErrorProb}), but rather
\beqa
P_e
&=&
\frac{1}{2}\min\{p_0(\uparrow)p_0(\uparrow),\,p_1(\uparrow)
p_1(\uparrow)\}\,+\,
\min\{p_0(\uparrow)p_0(\downarrow),\,p_1(\uparrow)p_1(\downarrow)\}\,
\nonumber\\
& &
\;\;\;\;\;\;+\frac{1}{2}\min\{p_0(\downarrow)p_0(\downarrow),\,p_1(\downarrow)
p_1(\downarrow)\}\;.
\label{BinErProb2}
\eeqa
This reflects the fact that this experiment has four possible outcomes:
$\uparrow\uparrow$, $\uparrow\downarrow$, $\downarrow\uparrow$, and
$\downarrow\downarrow$, with $\uparrow$ and $\downarrow$ denoting yes and
no outcomes, respectively.  The measurement that minimizes 
Eq.~(\ref{BinErProb2}) can be found easily by numerical means; it turns out to
be a polarization measurement $54.54^\circ$ from the horizontal.  In similar 
fashion, if $M\!=\!3$, the optimal measurement is along the axis $49.94^\circ$ 
from the horizontal.  See Fig.~\ref{ManualLabor}.
The lesson to be learned from this is that the optimal error-probability
measurement is expressly dependent upon the number of repetitions $M$
expected.  This phenomenon has already been encountered in the classical
example of Chapter~2.  If $M$ is to be left undetermined, then something 
beside the minimal error probability itself is required for an adequate measure 
of statistical distinguishability within the context of the decision problem.

\section{The Quantum Fidelity}
\label{MachoMan}

The solution of Section~\ref{YidDish} to remedy this predicament was to shift
focus to the optimal exponential decrease in error probability in the number
of samples $M$.  Translated into the quantum context, that would mean we should
use the Quantum Chernoff Bound
\beq
C(\hat\rho_0,\hat\rho_1)\,\equiv\,\min_{0\le\alpha\le1}\,
\min_{\{\hat E_b\}}\,\sum_b
\left({\rm tr}(\hat\rho_0\hat E_b)\right)^{\!\alpha}\!
\left({\rm tr}(\hat\rho_1\hat E_b)\right)^{\!{1-\alpha}}
\eeq
as the appropriate measure of distinguishability.  Instead, in this Section
we shall focus on optimizing a particular upper bound to this measure of
distinguishability, the statistical overlap introduced in Section~\ref{YidDish}.

The ``quantized'' version of the statistical overlap is called the {\it quantum
fidelity\/} and is defined by
\beq
F(\hat\rho_0,\hat\rho_1)=\min_{\{\hat E_b\}}\sum_b
\sqrt{{\rm tr}\hat\rho_0\hat E_b}\sqrt{{\rm tr}\hat\rho_1\hat E_b}
\label{WannaProve}
\eeq
We shall show in this Section that
\beq
F(\hat\rho_0,\hat\rho_1)=
{\rm tr}\,\sqrt{\hat\rho_1^{1/2}\hat\rho_0\hat\rho_1^{1/2}}\;,
\label{QWootters2}
\eeq
where for any nonnegative operator $\hat A$ we mean by $\hat A^{1/2}$
$\Bigl(\mbox{or }\sqrt{\hat A}\,\Bigr)$ the unique nonnegative operator such 
that $\hat A^{1/2}\hat A^{1/2}=\hat A$.

The quantity on the right hand side of Eq.~(\ref{QWootters2})
has appeared before as the distance function
\beq
d_{\rm B}^2(\hat\rho_0,\hat\rho_1)=
2-2F(\hat\rho_0,\hat\rho_1)\;
\label{butter}
\eeq
of Bures \cite{Bures69,Hubner92},
the generalized transition probability for mixed states
\beq
{\rm prob}(\hat\rho_0\!\rightarrow\!\hat\rho_1)=
\Big(F(\hat\rho_0,\hat\rho_1)\Big)^{\!2}
\label{BurpStool}
\eeq
of Uhlmann \cite{Uhlmann76}, and---in the same form as Uhlmann's---Jozsa's
criterion \cite{Jozsa94a} for ``fidelity'' of signals in a quantum
communication channel.  Note that Jozsa's ``fidelity'' \cite{Jozsa94a} is
actually the square of the quantity called fidelity here.  The fidelity was
found to be of use within that context because it is symmetric in 0 and 1,
because it is invariant under unitary operations, i.e.,
\beq
F(\hat U\hat\rho_0\hat U^\dagger,\hat U\hat\rho_1\hat U^\dagger)=
F(\hat\rho_0,\hat\rho_1)\;,
\eeq
for any unitary operator $\hat U$, because
\beq
0\le F(\hat\rho_0,\hat\rho_1)\le 1
\eeq
reaching 1 if and only if $\hat\rho_0=\hat\rho_1$, and because
\beq
\Big(F(\hat\rho_0,\hat\rho_1)\Big)^{\!2}=\langle\psi_1|\hat\rho_0|\psi_1\rangle
\eeq
when $\hat\rho_1=|\psi_1\rangle\langle\psi_1|$ is a pure state.

The notion defined by Eq.~(\ref{BurpStool}) is particularly significant because
of the auxiliary interpretation it gives the quantity in Eq.~(\ref{QWootters2}).
Imagine another system B, described by an $D$-dimensional Hilbert space, 
attached to our given system.  There are many pure states $|\psi_0\rangle$ and 
$|\psi_1\rangle$ on the composite system such that
\beq
{\rm tr}_{\rm B}(|\psi_0\rangle\langle\psi_0|)=\hat\rho_0
\;\;\;\;\;\;\mbox{and}\;\;\;\;\;\;
{\rm tr}_{\rm B}(|\psi_1\rangle\langle\psi_1|)=\hat\rho_1\;,
\eeq
where ${\rm tr}_{\rm B}$ denotes a partial trace over System B's Hilbert
space.  Such pure states are called ``purifications'' of the density operators
$\hat\rho_0$ and $\hat\rho_1$.  For these, the following theorem can be
shown \cite{Uhlmann76,Jozsa94a}
\begin{theorem}[Uhlmann]
For all purifications $|\psi_0\rangle$ and $|\psi_1\rangle$ of $\hat\rho_0$ and
$\hat\rho_1$, respectively,
\beq
|\langle\psi_0|\psi_1\rangle|\le{\rm tr}\,
\sqrt{\hat\rho_1^{1/2}\hat\rho_0\hat\rho_1^{1/2}}\;.
\eeq
Moreover, equality is achievable in this expression by an appropriate choice
of purifications.
\end{theorem}
That is to say, of all purifications of $\hat\rho_0$ and $\hat\rho_1$, the ones
with the maximal modulus for their inner product have it actually equal to
the quantum fidelity as defined here.

We should note that, in a roundabout way through the mathematical-physics 
literature (cf., for instance, in logical order \cite{Wootters81}, 
\cite{Hadjisavvas82}, \cite{Araki82}, \cite{Hadjisavvas86}, and
\cite{Uhlmann76}), one can put together a result quite similar in spirit to
Eq.~(\ref{QWootters2})---that is, a maximization like (\ref{QOwwtters}) but,
instead of over all POVMs, restricted to orthogonal projection-valued measures.  What is novel here is the explicit statistical interpretation, the simplicity 
and generality of the derivation, and the fact that it pinpoints the
measurement by which Eq.~(\ref{QWootters2}) is attained.

\subsection{The General Derivation}

Before getting started, let us note that if
$\hat\rho_0=|\psi_0\rangle\langle\psi_0|$ and
$\hat\rho_1=|\psi_1\rangle\langle\psi_1|$ are pure states, the expression in
Eq.~(\ref{QWootters2}) reduces to
\beqa
F(\hat\rho_0,\hat\rho_1)
&=&
{\rm tr}\,\sqrt{
|\psi_1\rangle\langle\psi_1|\psi_0\rangle\langle\psi_0|\psi_1\rangle\langle
\psi_1|}
\nonumber\\
&=&
|\langle\psi_0|\psi_1\rangle|\;{\rm tr}\sqrt{
|\psi_1\rangle\langle\psi_1|}
\nonumber\\
&=&
|\langle\psi_0|\psi_1\rangle|\;.
\eeqa
This already agrees with the expression derived by Wootters
\cite{Wootters80a,Wootters80b,Wootters81} for the optimal statistical overlap
between pure states. Moreover, it indicates that Eq.~(\ref{QWootters2}) has a
chance of being a general solution to Eq.~(\ref{WannaProve}).
  
The method we use for deriving Eq.~(\ref{QWootters2}) is to apply the Schwarz
inequality to the statistical overlap in such a way that its specific
conditions for equality can be met by a suitable measurement.
First, however, it is instructive to consider a quick and dirty---and for this
problem inappropriate---application of the Schwarz inequality;
the difficulties encountered therein point naturally toward the correct proof.
The Schwarz inequality for the operator inner product
${\rm tr}(\hat A^\dagger\hat B)$ is given by
\beq
|{\rm tr}(\hat A^\dagger\hat B)|^2\le{\rm tr}(\hat A^\dagger\hat A)\,
{\rm tr}(\hat B^\dagger\hat B)\;,
\eeq
where equality is achieved if and only if $\hat B = \mu\hat A$ for some constant
$\mu$.

Let $\{\hat E_b\}$ be an arbitrary POVM and
\beq 
p_0(b)={\rm tr}(\hat\rho_0\hat E_b)
\;\;\;\;\;\;\;\;\mbox{and}\;\;\;\;\;\;\;\;
p_1(b)={\rm tr}(\hat\rho_1\hat E_b)\;.
\eeq
By the cyclic property of the trace and this inequality, we must have for any 
$b$,
\beqa
\sqrt{p_0(b)}\sqrt{p_1(b)}
&=&
\sqrt{{\rm tr}\!\left(\hat\rho_0^{1/2}\hat E_b\hat\rho_0^{1/2}\right)}
\sqrt{{\rm tr}\!\left(\hat\rho_1^{1/2}\hat E_b\hat\rho_1^{1/2}\right)}
\nonumber\\
&=&
\sqrt{{\rm tr}\!\left(\Bigl(\hat E_b^{1/2}\hat\rho_0^{1/2}\Bigr)^\dagger
\Bigl(\hat E_b^{1/2}\hat\rho_0^{1/2}\Bigr)\right)}\:
\sqrt{{\rm tr}\!\left(\Bigl(\hat E_b^{1/2}\hat\rho_1^{1/2}\Bigr)^\dagger
\Bigl(\hat E_b^{1/2}\hat\rho_1^{1/2}\Bigr)\right)}
\rule{0mm}{9mm}
\nonumber\\
&\ge&
\left|
{\rm tr}\!\left(\Bigl(\hat E_b^{1/2}\hat\rho_0^{1/2}\Bigr)^\dagger
\Bigl(\hat E_b^{1/2}\hat\rho_1^{1/2}\Bigr)\right)
\right|
\rule{0mm}{8mm}
\nonumber\\
&=&
\left|{\rm tr}\!\left(\hat\rho_0^{1/2}\hat E_b\hat\rho_1^{1/2}\right)\right|\;.
\rule{0mm}{8mm}
\label{TermIneq}
\eeqa
The condition for attaining equality here is that
\beq
\hat E_b^{1/2}\hat\rho_1^{1/2}=\mu_b\hat E_b^{1/2}\hat\rho_0^{1/2}\;.
\label{EqualCond}
\eeq
A subscript $b$ has been placed on the constant $\mu$ as a reminder of its
dependence on the particular $\hat E_b$ in this equation.  From inequality
(\ref{TermIneq}), it follows by the linearity of the trace and the completeness
property of POVMs that
\beqa
\sum_b\sqrt{p_0(b)}\sqrt{p_1(b)}
&\ge&
\sum_b\left|{\rm tr}\!\left(\hat\rho_0^{1/2}\hat E_b\hat\rho_1^{1/2}\right)
\right|
\nonumber\\
&\ge&
\left|\sum_b{\rm tr}\!\left(\hat\rho_0^{1/2}\hat E_b\hat\rho_1^{1/2}\right)
\right|
\rule{0mm}{8mm}
\label{Herbert}
\\
&=&
\left|{\rm tr}\!\left(\hat\rho_1^{1/2}\hat\rho_0^{1/2}\right)\right|
\rule{0mm}{8mm}
\label{SumIneq}
\eeqa

The quantity \cite{Albrecht94}
\beq
F_{\rm A}(\hat\rho_0,\hat\rho_1)=
{\rm tr}\Bigl(\hat\rho_0^{1/2}\hat\rho_1^{1/2}\Bigr)
\eeq
is thus a lower bound to $F(p_0,p_1)$.  For it actually to be the minimum,
there must be a POVM such that, for all $b$, 
Eq.~(\ref{EqualCond}) is satisfied and
\beq
{\rm tr}\Bigl(\hat\rho_0^{1/2}\hat E_b\hat\rho_1^{1/2}\Bigr)=
\left|{\rm tr}\Bigl(\hat\rho_0^{1/2}\hat E_b\hat\rho_1^{1/2}\Bigr)\right|
e^{i\phi}\;,
\eeq
where $\phi$ is an arbitrary phase independent of $b$,
so that the sum can be taken past the absolute values sign in
Eq.~(\ref{Herbert}) without effect.

These conditions, however, cannot be fulfilled by any POVM $\{\hat E_b\}$
when $\hat\rho_0$ and $\hat\rho_1$ do not commute.  This can be 
seen as follows.  Suppose $[\hat\rho_0,\hat\rho_1]\neq 0$ and, for simplicity,
let us suppose that $\hat\rho_0$ can be inverted. Then
condition~(\ref{EqualCond}) can be written equivalently as
\beq
\hat E_b^{1/2}\!\left(\mu_b\hat{\openone}-\hat\rho_1^{1/2}\hat\rho_0^{-1/2}\right)
=0\;.
\label{WrongEq}
\eeq
The only way this can be satisfied is if we take the $\hat E_b$ to
be proportional to the projectors formed from the {\it left}-eigenvectors
of $\hat\rho_1^{1/2}\hat\rho_0^{-1/2}$ and let the $\mu_b$ be the corresponding 
eigenvalues.  This is seen easily.

The operator $\hat\rho_1^{1/2}\hat\rho_0^{-1/2}$ is non-Hermitian by assumption.
Thus, though it has $D$ linearly independent left- and right-eigenvectors, they
cannot be orthogonal.  Let us denote the left-eigenvectors by
$\langle\psi_r|$ and their corresponding eigenvalues by $\sigma_r$; let us 
denote the right-eigenvectors and eigenvalues by $|\phi_q\rangle$ and 
$\lambda_q$.  Then if Eq.~(\ref{WrongEq}) is to hold, we must have 
\beq
\hat E_b^{1/2}\!\left(\mu_b\hat{\openone}-\hat\rho_1^{1/2}
\hat\rho_0^{-1/2}\right)|\phi_q\rangle=0\;.
\eeq
It follows that
\beq
(\mu_b-\lambda_q)\hat E_b^{1/2}|\phi_q\rangle=0
\eeq
for all $q$ and $b$. 
Now assume---again for simplicity---that all the $\lambda_q$ are distinct.  If 
$\hat E_b$ is not identically zero, then we must have that (modulo
relabeling)
\beq
\hat E_b^{1/2}|\phi_q\rangle=0\;\mbox{ for all $q\neq b$}
\;\;\;\;\;\;\;\;\mbox{\it and}\;\;\;\;\;\;\;\;
\mu_b=\lambda_q\;\mbox{ for $q=b$}\;.
\eeq
This means that $\hat E_b^{1/2}$ is 
proportional to the projector onto the one-dimensional subspace that 
is orthogonal to all the $|\phi_q\rangle$ with $q\ne b$.  But since
\beqa
0
&=&
\langle\psi_r|\hat\rho_1^{1/2}\hat\rho_0^{-1/2}|\phi_q\rangle
-\langle\psi_r|\hat\rho_1^{1/2}\hat\rho_0^{-1/2}|\phi_q\rangle
\nonumber\\
&=&
(\sigma_r-\lambda_q)\langle\psi_r|\phi_q\rangle\;,
\eeqa
we have that 
(again modulo relabeling) $|\psi_r\rangle$ is orthogonal to 
$|\phi_q\rangle$ for $q\ne r$ and $\sigma_r=\lambda_q$ for $q=r$. 
Therefore
\beq
\hat E_b^{1/2}\propto|\psi_b\rangle\langle\psi_b|\;.
\eeq
The reason Eq.~(\ref{WrongEq}) cannot be satisfied by {\it any\/} POVM is
now apparent; it is just that the $|\psi_b\rangle$ are nonorthogonal.  When the 
$|\psi_b\rangle$ are nonorthogonal, there are {\it no\/} positive constants 
$\alpha_b$ $(b=1,\ldots,n)$ such that
\beq
\sum_{b=1}^n\alpha_b|\psi_b\rangle\langle\psi_b|=\hat{\openone}\;.
\eeq
For if there were, then the completeness relation would give rise to
the equation
\beq
\sum_{b=1}^n\alpha_b\langle\psi_b|\psi_c\rangle|\psi_b\rangle=|\psi_c\rangle
\eeq
so that
\beq
\sum_{b\ne c}\alpha_b\langle\psi_b|\psi_c\rangle|\psi_b\rangle=(1-\alpha_c)
|\psi_c\rangle
\eeq
contradicting the fact that the $|\psi_b\rangle$ are linearly independent but
nonorthogonal.  If $\hat\rho_0$ and $\hat\rho_1$ were commuting operators so 
that $\hat\rho_1^{1/2}\hat\rho_0^{-1/2}$ were Hermitian, there would be no 
problem; for then all the eigenvectors would be mutually orthogonal.  A
complete set of orthonormal projectors necessarily sum to the identity operator.

The lesson from this example is that the na\"{\i}ve Schwarz inequality is not
enough to prove Eq.~(\ref{QWootters2}); one must be careful to ``build in''
a way to attain equality by at least one POVM.  
Plainly the way to do this is to take advantage of the invariances of
the trace operation.  In particular, in the set of
inequalities~(\ref{TermIneq}), we could just as well have written
\beq
p_0(b)={\rm tr}(\hat\rho_0\hat E_b)=
{\rm tr}\!\left(\hat U\hat\rho_0^{1/2}
\hat E_b\hat\rho_0^{1/2}\hat U^\dagger\right)
\eeq
for any unitary operator $\hat U$ since $\hat U^\dagger\hat U=\hat{\openone}$.
Then, in exact analogy to the previous derivation, it follows that
\beqa
\sqrt{p_0(b)}\sqrt{p_1(b)}
&=&
\sqrt{{\rm tr}\!\left(\Bigl(\hat E_b^{1/2}\hat\rho_0^{1/2}\hat U^\dagger
\Bigr)^\dagger
\Bigl(\hat E_b^{1/2}\hat\rho_0^{1/2}\hat U^\dagger\Bigr)\right)}\:
\sqrt{{\rm tr}\!\left(\Bigl(\hat E_b^{1/2}\hat\rho_1^{1/2}\Bigr)^\dagger
\Bigl(\hat E_b^{1/2}\hat\rho_1^{1/2}\Bigr)\right)}
\rule{0mm}{8mm}
\nonumber\\
&\ge&
\left|{\rm tr}\!\left(\Bigl(\hat E_b^{1/2}\hat\rho_0^{1/2}\hat U^\dagger
\Bigr)^\dagger\Bigl(\hat E_b^{1/2}\hat\rho_1^{1/2}\Bigr)\right)\right|
\rule{0mm}{8mm}
\nonumber\\
&=&
\left|{\rm tr}\!\left(\hat U\hat\rho_0^{1/2}\hat E_b
\hat\rho_1^{1/2}\right)\right|\;,
\rule{0mm}{8mm}
\label{trailor}
\eeqa
where the condition for equality is now
\beq
\hat E_b^{1/2}\hat\rho_1^{1/2}=\mu_b\hat E_b^{1/2}\hat\rho_0^{1/2}
\hat U^\dagger\;.
\label{stogie}
\eeq
This equation, it turns out, can be satisfied by an appropriate choice for
the unitary operator $\hat U$.

To see this, let us first suppose that $\hat\rho_0$ and $\hat\rho_1$ are 
invertible.  Then Eq.~(\ref{stogie}) is equivalent to
\beq
\hat E_b^{1/2}\!\left(\hat{\openone}-\mu_b\hat\rho_0^{1/2}
\hat U^\dagger\hat\rho_1^{-1/2}
\right)=0\;.
\label{CorrEq}
\eeq
Summing Eq.~(\ref{trailor}) on $b$, we get
\beq
\sum_b\sqrt{p_0(b)}\sqrt{p_1(b)}\ge
\left|{\rm tr}\!\left(\hat U\hat\rho_0^{1/2}\hat\rho_1^{1/2}\right)\right|\;.
\label{FinalIneq}
\eeq
The final conditions for equality in this is that the $\hat E_b$ satisfy both
Eq.~(\ref{CorrEq}) and the requirement
\beq
{\rm tr}\Bigl(\hat U\hat\rho_0^{1/2}\hat E_b\hat\rho_1^{1/2}\Bigr)=
\left|{\rm tr}\Bigl(\hat U\hat\rho_0^{1/2}\hat E_b\hat\rho_1^{1/2}\Bigr)\right|
e^{i\phi}
\label{TiredBoy}
\eeq
for all $b$, where again $\phi$ is an arbitrary phase.

As in the last example, there can be no POVM $\{\hat E_b\}$ that satisfies 
condition~(\ref{CorrEq}) {\it unless\/} the operator
$\hat\rho_0^{1/2}\hat U^\dagger\hat\rho_1^{-1/2}$ is Hermitian.  An easy way
to find a unitary 
$\hat U$ that makes a solution to Eq.~(\ref{CorrEq}) possible is to note a 
completely different point about inequality~(\ref{FinalIneq}).  The unitary 
operator $\hat U$ there is arbitrary; if there is to be a chance of attaining 
equality in (\ref{FinalIneq}), $\hat U$ had better be chosen so as to maximize
$\left|{\rm tr}\Bigl(\hat U\hat\rho_0^{1/2}\hat\rho_1^{1/2}\Bigr)\right|$.
It turns out that that particular $\hat U$ forces 
$\hat\rho_0^{1/2}\hat U^\dagger\hat\rho_1^{-1/2}$ to be Hermitian.

To demonstrate the last point, we need a result from the mathematical
literature \cite{Jozsa94a,Fan51,Schatten60}:  for any operator $\hat A$,
\beqa
\max_{\hat U}\left|{\rm tr}\Bigl(\hat U\hat A\Bigr)\right|
&=&
\max_{\hat U}\,{\rm Re}\!\left[{\rm tr}\Bigl(\hat U\hat A\Bigr)\right]
\nonumber\\
&=&
{\rm tr}\sqrt{\hat A^\dagger\hat A}\;,
\rule{0mm}{7mm}
\label{Gorba}
\eeqa
where the maximum is taken over all unitary operators $\hat U$. The set
of operators $\hat U$ that gives rise to the maximum must satisfy
\beq
\hat U\hat A=\sqrt{\hat A^\dagger\hat A}\;.
\eeq
At least one such unitary operator is assured to exist by the so called
{\it polar decomposition theorem}.  When $\hat A$ is invertible, it is easy
to see that
\beq
\hat U=\sqrt{\hat A^\dagger\hat A}\,\hat A^{-1}
\eeq
has the desired properties and is unique.  When $\hat A$ is not invertible,
$\hat U$ is no longer unique, but can still be shown to exist
\cite[pp.\ 74--75]{Marcus92}.

So a unitary operator $\hat U_{\rm c}$ that gives rise to the tightest 
inequality in Eq.~(\ref{FinalIneq}) can be taken to satisfy
\beqa
\hat U_{\rm c}\hat\rho_0^{1/2}\hat\rho_1^{1/2}
&=&
\sqrt{\left(\hat\rho_0^{1/2}\hat\rho_1^{1/2}\right)^\dagger\!
\left(\hat\rho_0^{1/2}\hat\rho_1^{1/2}\right)}
\nonumber\\
&=&
\sqrt{\hat\rho_1^{1/2}\hat\rho_0\hat\rho_1^{1/2}}
\rule{0mm}{8mm}
\label{Bert}
\eeqa
When $\hat\rho_0$ and $\hat\rho_1$ are both invertible, this equation is
uniquely satisfied by
\beq
\hat U_{\rm c}\equiv\sqrt{\hat\rho_1^{1/2}\hat\rho_0\hat\rho_1^{1/2}}
\hat\rho_1^{-1/2}\hat\rho_0^{-1/2}\;.
\eeq
With this, Eq.~(\ref{FinalIneq}) clearly takes the form needed to prove 
Eq.~(\ref{WannaProve}):
\beq
\sum_b\sqrt{p_0(b)}\sqrt{p_1(b)}
\ge\left|{\rm tr}\!\left(\hat U_{\rm c}
\,\hat\rho_0^{1/2}\hat\rho_1^{1/2}\right)\right|
={\rm tr}\,\sqrt{\hat\rho_1^{1/2}\hat\rho_0\hat\rho_1^{1/2}}\;.
\eeq
Inserting this choice for $\hat U$ into Eq.~(\ref{CorrEq}) gives the condition
\beq
\hat E_b^{1/2}\!\left(\hat{\openone}-\mu_b\hat M\right)=0\;.
\label{MeasCond}
\eeq
where the operator
\beq
\hat M\equiv\hat\rho_0^{1/2}\hat U_{\rm c}^\dagger\hat\rho_1^{-1/2}=
\hat\rho_1^{-1/2}\sqrt{\hat\rho_1^{1/2}\hat\rho_0\hat\rho_1^{1/2}}
\hat\rho_1^{-1/2}
\eeq
is indeed Hermitian and also nonnegative (as can be seen
immediately from its symmetry).  Thus there is a POVM 
$\{\hat E_b^{\rm B}\}$ that satisfies Eq.~(\ref{CorrEq}) for each $b$: the 
$\hat E_b^{\rm B}$ can be taken be projectors onto a basis $|b\rangle$
that diagonalizes $\hat M$.  Here the $\mu_b$ must be taken to be reciprocals
of $\hat M$'s eigenvalues.

With the POVM $\{\hat E_b^{\rm B}\}$, Eq.~(\ref{TiredBoy}) is automatically
satisfied.  Since the eigenvalues $1/\mu_b$ of $\hat M$ are all nonnegative, one finds that
\beq
{\rm tr}\Bigl(\hat U_{\rm c}\,\hat\rho_0^{1/2}
\hat E_b^{\rm B}\hat\rho_1^{1/2}\Bigr)
={\rm tr}\Bigl(\hat\rho_1\hat M\hat E_b^{\rm B}\Bigr)
={1\over\mu_b}{\rm tr}\Bigl(\hat\rho_1\hat E_b^{\rm B}\Bigr)\ge0\;.
\eeq
This concludes the proof of Eq.~(\ref{QWootters2}) under the restriction that
$\hat\rho_0$ and $\hat\rho_1$ be invertible.

When $\hat\rho_0$ and/or $\hat\rho_1$ are {\it not\/} invertible, things
are only slightly more difficult.  Suppose  $\hat\rho_1$ is not invertible;
then there exists a projector $\hat\Pi_{\rm null}$ onto the null subspace
of $\hat\rho_1$, i.e., $\hat\Pi_{\rm null}$ is a projector of maximal rank
such that
\beq
\hat\Pi_{\rm null}\,\hat\rho_1=0
\;\;\;\;\;\;\;\;\mbox{and}\;\;\;\;\;\;\;\;
\hat\rho_1\,\hat\Pi_{\rm null}=0\;.
\eeq
The projector onto the support of $\hat\rho_1$, i.e., the orthogonal complement
to the null subspace, is defined by
\beq
\hat\Pi_{\rm supp}=\hat{\openone}-\hat\Pi_{\rm null}\;.
\eeq
Clearly $\hat E_{\rm null}=\hat\Pi_{\rm null}$ satisfies
Eq.~(\ref{stogie}) if the associated constant $\mu_{\rm null}$ is chosen to
be zero.  Now let us construct a set of orthogonal projectors
$\hat E_b=|b\rangle\langle b|$ that span the support of $\hat\rho_1$ and
satisfy Eqs.~(\ref{stogie}) and (\ref{TiredBoy}) with (\ref{Bert}).  This is 
done easily enough.  Suppose the support of $\hat\rho_1$ is an $m$-dimensional
subspace; the operator
\beq
\hat R_1=\hat\Pi_{\rm supp}\hat\rho_1\hat\Pi_{\rm supp}
\eeq
is invertible on that subspace.  Now consider any set of $m$ orthogonal
one-dimensional projectors $\hat\Pi_b$ in the support of $\hat\rho_1$.  If they
are to satisfy Eq.~(\ref{stogie}), then they must also satisfy the equations
created by sandwiching Eqs.~(\ref{stogie}) and (\ref{Bert}) by the projector
$\hat\Pi_{\rm supp}$:
\beq
\hat\Pi_b\hat R_1^{1/2}=\mu_b\hat\Pi_b\left(\hat\Pi_{\rm supp}\hat\rho_0^{1/2}
\hat U^\dagger\hat\Pi_{\rm supp}\right),
\eeq
and
\beq
\left(\hat\Pi_{\rm supp}\hat U_{\rm c}\hat\rho_0^{1/2}\hat\Pi_{\rm supp}\right)
\hat R_1^{1/2}=\hat\Pi_{\rm supp}
\sqrt{\hat\rho_1^{1/2}\hat\rho_0\hat\rho_1^{1/2}}\hat\Pi_{\rm supp}\;.
\eeq
Therefore, the $\hat\Pi_b$ must satisfy
\beq
\left[\hat R_1^{-1/2}\!\left(\hat\Pi_{\rm supp}
\sqrt{\hat\rho_1^{1/2}\hat\rho_0\hat\rho_1^{1/2}}\hat\Pi_{\rm supp}\right)\!
\hat R_1^{-1/2}\right]\hat\Pi_b=\frac{1}{\mu_b}\hat\Pi_b\;.
\label{Bilbo}
\eeq
Hereafter we may run through the same steps as in the invertible case.
Because the operator on the left-hand side of Eq.~(\ref{Bilbo}) is a positive
operator, there are indeed $m$ orthogonal projectors on the support of 
$\hat\rho_1$ that satisfy the conditions for optimizing the 
statistical overlap.  Taking the $\hat E_b=\hat\Pi_b$ completes the
proof.

A particular case of noninvertibility is when 
$\hat\rho_0=|\psi_0\rangle\langle\psi_0|$ and
$\hat\rho_1=|\psi_1\rangle\langle\psi_1|$ are nonorthogonal pure states.  Then 
Eq.~(\ref{Bert}) becomes
\beqa
\hat U_c |\psi_0\rangle\langle\psi_0|\psi_1\rangle\langle\psi_1|
&=&
\sqrt{|\psi_1\rangle\langle\psi_1|\psi_0\rangle\langle\psi_0|
\psi_1\rangle\langle\psi_1|}
\nonumber\\
&=&
|\langle\psi_0|\psi_1\rangle|\:|\psi_1\rangle\langle\psi_1|
\eeqa
and Eq.~(\ref{stogie}) becomes
\beq
|b\rangle\langle b|\psi_1\rangle\langle\psi_1|=\mu_b|b\rangle\langle b|
\psi_0\rangle\langle\psi_0|\hat U_c^\dagger\;.
\label{burpie}
\eeq
If we redefine the phases of $|\psi_0\rangle$ and $|\psi_1\rangle$ so that
$\langle\psi_0|\psi_1\rangle$ is positive, we have that
\beq
\hat U_c|\psi_0\rangle=|\psi_1\rangle\;.
\label{AfterBurn}
\eeq
Therefore, Eq.~(\ref{burpie}) implies
\beq
\langle b|\psi_1\rangle=\mu_b\langle b|\psi_0\rangle\;.
\label{briggs}
\eeq
This equation specifies that any orthonormal basis $|b\rangle$ containing 
vectors $|0\rangle$ and $|1\rangle$ lying in the plane spanned by the vectors 
$|\psi_0\rangle$ and $|\psi_1\rangle$ and straddling them will form an optimal 
measurement basis.  This follows because in this case all inner products
$\langle b|\psi_1\rangle$ and $\langle b|\psi_0\rangle$ will be nonnegative;
thus Eq.~(\ref{briggs}) has a solution with nonnegative $\mu_b$.  This supplements the set of optimal measurements found in Ref.~\cite{Wootters81} and
is easily confirmed to be true as follows.  Let $\theta$ be the angle between
$|\psi_0\rangle$ and $|\psi_1\rangle$ and let $\phi$ be the angle between
$|0\rangle$ and $|\psi_0\rangle$.  Then for this measurement,
\beqa
F(p_0,p_1)
&=&
\sqrt{\cos^2\phi\cos^2(\phi+\theta)}\,+\,
\sqrt{\cos^2(\phi-{\textstyle\frac{\pi}{2}})\,\cos^2(\phi+\theta-
{\textstyle\frac{\pi}{2}})}
\nonumber\\
&=&
\cos\phi\,\cos(\phi+\theta)\,+\,\sin\phi\,\sin(\phi+\theta)
\nonumber\\
&=&
\cos\theta\;,
\eeqa
which is completely independent of $\phi$.  This verifies the result.

Equation~(\ref{AfterBurn}) raises the interesting question of, more generally,
what is the action of $\hat U_c$?  Could it be that $\hat U_c$ always takes
a basis diagonalizing $\hat\rho_0$ to a basis diagonalizing $\hat\rho_1$?
This appears not to be the case, unfortunately.  The question of a more
geometric interpretation of $\hat U_c$ is an open one.

\subsection{Properties}
\label{UncleDoug}

In this Subsection, we report a few interesting points about the
measurement specified by $\hat M$ and the quantum distinguishability measure
$F(\hat\rho_0,\hat\rho_1)$.  The equation defining the
statistical overlap is clearly invariant under interchanges of
the labels 0 and 1.  Therefore it must follow that
\beq
F(\hat\rho_0,\hat\rho_1)=F(\hat\rho_1,\hat\rho_0)\;.
\eeq
A neat way to see this
directly is to note that the operators 
$\hat\rho_1^{1/2}\hat\rho_0\hat\rho_1^{1/2}$ and
$\hat\rho_0^{1/2}\hat\rho_1\hat\rho_0^{1/2}$ have the same eigenvalue
spectrum.  For if $|b\rangle$ and $\lambda_b$ are an eigenvector and eigenvalue
of $\hat\rho_1^{1/2}\hat\rho_0\hat\rho_1^{1/2}$, it follows that
\beqa
\lambda_b\Bigl(\hat\rho_0^{1/2}\hat\rho_1^{1/2}|b\rangle\Bigr)
&=&
\hat\rho_0^{1/2}\hat\rho_1^{1/2}\Bigl(\lambda_b|b\rangle\Bigr)
\nonumber\\
&=&
\hat\rho_0^{1/2}\hat\rho_1^{1/2}\Bigl(
\hat\rho_1^{1/2}\hat\rho_0\hat\rho_1^{1/2}|b\rangle\Bigr)
\rule{0mm}{6mm}
\nonumber\\
&=&
\Bigl(\hat\rho_0^{1/2}\hat\rho_1\hat\rho_0^{1/2}\Bigr)
\Bigl(\hat\rho_0^{1/2}\hat\rho_1^{1/2}|b\rangle\Bigr)\:.
\rule{0mm}{6mm}
\eeqa
Hence,
\beq
{\rm tr}\,\sqrt{\hat\rho_1^{1/2}\hat\rho_0\hat\rho_1^{1/2}}=
\sum_b\sqrt{\lambda_b}=
{\rm tr}\,\sqrt{\hat\rho_0^{1/2}\hat\rho_1\hat\rho_0^{1/2}}\;,
\eeq
and so $F(\hat\rho_0,\hat\rho_1)=F(\hat\rho_1,\hat\rho_0)$.

By the same token, the derivation of Eq.~(\ref{QWootters2}) itself must remain
valid if all the 0's and 1's in it are interchanged throughout.  When
$\hat\rho_0$ and $\hat\rho_1$ are invertible, however, this
gives rise to a measurement specified by a basis diagonalizing
\beq
\hat N\equiv\hat\rho_0^{-1/2}
\sqrt{\hat\rho_0^{1/2}\hat\rho_1\hat\rho_0^{1/2}}\hat\rho_0^{-1/2}\;.
\label{wumpus}
\eeq
It turns out that $\hat M$ and $\hat N$ can define the same measurement because
not only do they commute, they are inverses of each other.  This can be seen as
follows.  Let $\hat A$ be any operator and $\hat V$ be a unitary operator such
that
\beq
\sqrt{\hat A^\dagger\hat A}=\hat V^\dagger\hat A\;.
\eeq
Hence
$\hat A=\hat V\sqrt{\hat A^\dagger\hat A}$ and also
$\hat A^\dagger=\sqrt{\hat A^\dagger\hat A}\,\hat V^\dagger$.  So
\beq
\left(\hat V\sqrt{\hat A^\dagger\hat A}\,\hat V^\dagger\right)^{\!2}=
\left(\hat V\sqrt{\hat A^\dagger\hat A}\,\right)
\left(\sqrt{\hat A^\dagger\hat A}\,\hat V^\dagger\right)=
\hat A\hat A^\dagger\;,
\eeq
and therefore, because $\hat V\sqrt{\hat A^\dagger\hat A}\,\hat V^\dagger$ is a
nonnegative operator,
\beq
\sqrt{\hat A\hat A^\dagger}=\hat V\sqrt{\hat A^\dagger\hat A}\,\hat V^\dagger
=\hat V\hat A^\dagger\;.
\eeq
In particular, if 
\beq
\sqrt{\hat\rho_1^{1/2}\hat\rho_0\hat\rho_1^{1/2}}=
\hat U_{\rm c}\,\hat\rho_0^{1/2}\hat\rho_1^{1/2}\;,
\eeq
then
\beq
\sqrt{\hat\rho_0^{1/2}\hat\rho_1\hat\rho_0^{1/2}}=
\hat U_{\rm c}^\dagger\,\hat\rho_1^{1/2}\hat\rho_0^{1/2}\;.
\eeq
Therefore, the desired property follows at once,
\beqa
\hat M\hat N
&=&
\hat\rho_1^{-1/2}\Bigl(\hat U_{\rm c}\,\hat\rho_0^{1/2}
\hat\rho_1^{1/2}\Bigr)\hat\rho_1^{-1/2}\hat\rho_0^{-1/2}
\Bigl(\hat U_{\rm c}^\dagger\,\hat\rho_1^{1/2}\hat\rho_0^{1/2}
\Bigr)\hat\rho_0^{-1/2}
\nonumber\\
&=&
\hat\rho_1^{-1/2}\hat U_{\rm c}\,
\hat U_{\rm c}^\dagger\,\hat\rho_1^{1/2}
\rule{0mm}{6mm}
\nonumber\\
&=&
\hat{\openone}\;.
\rule{0mm}{6mm}
\eeqa

We may also note an interesting expression for $\hat M$'s eigenvalues that
arises from the last result.  Let the eigenvalues and eigenvectors of
$\hat M$ be denoted by $m_b$ and $|b\rangle$; in this notation
$\hat E_b^{\rm B}=|b\rangle\langle b|$.  Then we can write two expressions for
$m_b$:
\beqa
m_b\langle b|\hat\rho_1|b\rangle
&=&
\langle b|\hat\rho_1\hat M|b\rangle
\nonumber\\
&=&
\langle b|\hat\rho_1^{1/2}\hat U_{\rm c}\,\hat\rho_0^{1/2}|b\rangle\;,
\rule{0mm}{7mm}
\label{Last1}
\eeqa
and
\beqa
{1\over m_b}\langle b|\hat\rho_0|b\rangle
&=&
\langle b|\hat\rho_0\hat N|b\rangle
\nonumber\\
&=&
\langle b|\hat\rho_0^{1/2}\hat U_{\rm c}^\dagger\,\hat\rho_1^{1/2}|b\rangle
\rule{0mm}{6mm}
\nonumber\\
&=&
\left(\langle b|\hat\rho_1^{1/2}\hat U_{\rm c}\,\hat\rho_0^{1/2}|b\rangle
\right)^\ast\;.
\rule{0mm}{8mm}
\label{Last2}
\eeqa
Because the left hand sides of these equations are real numbers, so are the
right hand sides; in particular, combining Eqs.~(\ref{Last1}) and (\ref{Last2}),
we get
\beq
m_b=\left({\langle b|\hat\rho_0|b\rangle\over\langle b|\hat\rho_1|b\rangle}
\right)^{\!1/2}
=\left({{\rm tr}\hat\rho_0\hat E_b^{\rm B}\over{\rm tr}\hat\rho_1
\hat E_b^{\rm B}}\right)^{\!1/2}\;.
\eeq

The optimal measurement operator $\hat M$ can be considered a sort of operator
analog to the classical likelihood ratio, for its squared eigenvalues are
the ratio of two probabilities.  This fact gives rise to an interesting 
expression for the Kullback-Leibler relative information between $\hat\rho_0$
and $\hat\rho_1$ with respect to this measurement:
\beqa
K_{\rm B}(\hat\rho_0/\hat\rho_1)
&\equiv&
\sum_b \Bigr({\rm tr}\hat\rho_0\hat E_b^{\rm B}\Bigl)\,
\ln\!\left({
{\rm tr}\hat\rho_0\hat E_b^{\rm B}
\over
{\rm tr}\hat\rho_1\hat E_b^{\rm B}
}\right)
\nonumber\\
&=&
2\,{\rm tr}\!\left(\hat\rho_0\sum_b(\ln\,m_b)\hat E_b^{\rm B}\right)
\rule{0mm}{8mm}
\nonumber\\
&=&
2\:{\rm tr}\Biggl(\hat\rho_0\,\ln\!\left(
\hat\rho_1^{-1/2}
\sqrt{\hat\rho_1^{1/2}\hat\rho_0\hat\rho_1^{1/2}}
\hat\rho_1^{-1/2}\right)\Biggr)\;.
\rule{0mm}{8mm}
\label{KullTrace}
\eeqa
This, of course, will generally {\it not\/} be the maximum of the 
Kullback-Leibler information over all measurements, but it does provide a
lower bound for the maximum value.  Moreover, a quantity quite similar
to this arises naturally in the context of still another measure of quantum
distinguishability studied by Braunstein \cite{Braunstein94c,Braunstein95c}.
In yet another guise, it appears in the work of Nagaoka \cite{Nagaoka94}.

There are two other representations for the quantum fidelity 
$F(\hat\rho_0,\hat\rho_1)$ that can be worked out simply with the techniques
developed here.  The first is \cite{Alberti83a}
\beq
F(\hat\rho_0,\hat\rho_1)=\sqrt{\,\min_{\hat G}\,{\rm tr}(\hat\rho_0\hat G)\,
{\rm tr}(\hat\rho_1\hat G^{-1})\rule{0mm}{4.3mm}\,}\;,
\label{repre}
\eeq
when $\hat\rho_0$ and $\hat\rho_1$ are {\it invertible}, and where
the minimum is taken over all invertible positive operators $\hat G$.
This representation, in analogy to the representation of fidelity as the
optimal statistical overlap, also comes about via the Schwarz inequality.  Let
us show this.

Let $\hat G$ be any invertible positive operator and let $\hat U$ be any
unitary operator.  Using the cyclic property of the trace and the Schwarz 
inequality, we have that
\beqa
{\rm tr}(\hat\rho_0 \hat G)\,{\rm tr}(\hat\rho_1 \hat G^{-1})
&=&
{\rm tr}\left[\left(\hat U\hat\rho_0^{1/2}\hat G^{1/2}\right)^{\!\dagger}\!
\left(\hat U\hat\rho_0^{1/2}\hat G^{1/2}\right)\right]
{\rm tr}\!\left[\left(\hat\rho_1^{1/2}\hat G^{-1/2}\right)^{\!\dagger}\!
\left(\hat\rho_1^{1/2}\hat G^{-1/2}\right)\right]
\nonumber\\
&\ge&
\left|{\rm tr}\!\left[\left(\hat U\hat\rho_0^{1/2}
\hat G^{1/2}\right)^{\!\dagger}\!
\left(\hat\rho_1^{1/2}\hat G^{-1/2}\right)\right]\right|^2
\rule{0mm}{8mm}
\nonumber\\
&=&
\left|{\rm tr}\!\left(
\hat U^\dagger\hat\rho_1^{1/2}\hat\rho_0^{1/2}\right)\right|^2.
\rule{0mm}{8mm}
\eeqa
Since $\hat U$ is arbitrary, we may use this freedom to make the inequality as
tight as possible.  To do this we choose
\beq
\hat U^\dagger\hat\rho_1^{1/2}\hat\rho_0^{1/2}=
\sqrt{\hat\rho_0^{1/2}\hat\rho_1\hat\rho_0^{1/2}\,}\;.
\label{rickets}
\eeq
to get that
\beq
{\rm tr}(\hat\rho_0 \hat G)\,{\rm tr}(\hat\rho_1 \hat G^{-1})\ge 
F(\hat\rho_0,\hat\rho_1)^2\;.
\eeq
To find a particular $\hat G$ that achieves equality in this, we need merely
study the condition for equality in the Schwarz inequality; in this case we
must have
\beq
\hat U\hat\rho_0^{1/2}\hat G^{1/2}=\alpha\hat\rho_1^{1/2}\hat G^{-1/2}\;.
\label{scurvy}
\eeq
The solution $\hat G$ to this equation is unique and easy to find,
\beq
\hat G=\alpha \hat\rho_0^{-1/2}\sqrt{\hat\rho_0^{1/2}\hat\rho_1\hat\rho_0^{1/2}\,}
\hat\rho_0^{-1/2}\;.
\label{gop}
\eeq
The constraint that $\hat G$ be a positive operator further restricts $\alpha$
to be a positive real number.  Thus the optimal operator $\hat G$ is 
proportional to the operator $\hat N$ given by Eq.~(\ref{wumpus}).
The choice for $\hat G$ given by Eq.~(\ref{gop})
demonstrates that Eq.~(\ref{repre}) does in fact hold.  This is easy to verify
by noting the fact that $\hat N^{-1}=\hat M$.

The second representation is more literally concerned with the Bures distance
defined by Eq.~(\ref{butter}).  It is \cite{Uhlmann76}
\beq
d_{\rm B}^2(\hat\rho_0,\hat\rho_1)=\min{\rm tr}\!
\left(\left(\hat W_0-\hat W_1\right)\!\left(\hat W_0-
\hat W_1\right)^{\!\dagger}\right)\;,
\label{Grannie}
\eeq
where the minimum is taken over all operators $\hat W_0$ and $\hat W_1$
such that
\beq
\hat W_0\hat W_0^\dagger=\hat\rho_0
\;\;\;\;\;\;\;\;\mbox{and}\;\;\;\;\;\;\;\;
\hat W_1\hat W_1^\dagger=\hat\rho_1\;.
\label{snooker}
\eeq
This is seen easily by noting that Eq.~(\ref{snooker}) requires, by the
operator polar decomposition theorem, that
\beq
\hat W_0=\hat\rho_0^{1/2}\hat U_0
\;\;\;\;\;\;\;\;\mbox{and}\;\;\;\;\;\;\;\;
\hat W_1=\hat\rho_1^{1/2}\hat U_1
\eeq
for some unitary operators $\hat U_0$ and $\hat U_1$.  Then the right hand
side of Eq.~(\ref{Grannie}) becomes
\beqa
\min_{\hat U_0,\hat U_1}{\rm tr}\!
\left(\left(\hat W_0-\hat W_1\right)\!\left(\hat W_0-
\hat W_1\right)^{\!\dagger}\right)
&=&
\min_{\hat U_0,\hat U_1}
{\rm tr}\!\left(\hat W_0\hat W_0^\dagger-\hat W_0\hat W_1^\dagger-
\hat W_1\hat W_0^\dagger+\hat W_1\hat W_1^\dagger\right)
\rule{0mm}{6mm}
\nonumber\\
&=&
{\rm tr}(\hat\rho_0)+{\rm tr}(\hat\rho_1)-2\max_{\hat U_0,\hat U_1}{\rm Re}\!\left[
{\rm tr}\!\left(\hat W_0\hat W_1^\dagger\right)\right]
\rule{0mm}{6mm}
\nonumber\\
&=&
2-2\max_{\hat V}
{\rm Re}\!\left[{\rm tr}\!\left(\hat V\hat\rho_1^{1/2}\hat\rho_0^{1/2}\right)\right]
\rule{0mm}{6mm}
\nonumber\\
&=&
2-2\,F(\hat\rho_0,\hat\rho_1)\;.
\rule{0mm}{6mm}
\eeqa
The last step in this follows from Eq.~(\ref{Gorba}) and demonstrates the
truth of Eq.~(\ref{Grannie}).

Finally, we should mention something about the measurement specified by
$\hat M$, as it might appear in the larger context.  A very general theory of
{\it operator means\/} has been developed by Kubo and Ando \cite{Kubo80,Ando94b}
in which the notion of a {\it geometric mean\/} \cite{Pusz75} plays a special
role.  The geometric mean between two positive operators $\hat A$ and $\hat B$
is defined by
\beq
\hat A\,\#\,\hat B={\hat A\rule{0mm}{1.5mm}}^{1/2}
\sqrt{{\hat A\rule{0mm}{1.5mm}}^{-1/2}
\hat B{\hat A\rule{0mm}{1.5mm}}^{-1/2}\,}{\hat A\rule{0mm}{1.5mm}}^{1/2}\;.
\eeq
More generally an operator mean is any mapping
$(\hat A, \hat B)\mapsto\hat A\,\sigma\,\hat B$ from operators on a $D$-dimen\-sional
Hilbert space to a $D'$-dimensional space such that
\begin{enumerate}
\item
$\alpha(\hat A\,\sigma\,\hat B)=(\alpha\hat A)\,\sigma\,(\alpha\hat B)$ for any
nonnegative constant $\alpha$,
\item
$\hat A\,\sigma\,\hat A=\hat A$,
\item
$\hat A\,\sigma\,\hat B\ge\hat A'\,\sigma\,\hat B'$ whenever $\hat A\ge\hat A'$ and
$\hat B\ge\hat B'$,
\item
a certain continuity condition is satisfied, and
\item
$(\hat T^\dagger\hat A\hat T)\,\sigma\,(\hat T^\dagger\hat B\hat T)\ge
\hat T^\dagger(\hat A\,\sigma\,\hat B)\hat T$ for every operator $\hat T$.
\end{enumerate}

Another characterization of the geometric mean is that (in any representation)
it is the matrix $\hat X$ such that the matrix (on a $D^2$-dimensional space) 
defined by
\beq
\left(\begin{array}{cc}
\hat A & \hat X\\
\hat X & \hat B\end{array}\right)
\eeq
is maximized in the matrix sense.

In this notation, the optimal measurement operator $\hat M$ for statistical
overlap is
\beq
\hat M=\hat\rho_1^{-1}\,\#\,\hat\rho_0\;.
\eeq
The significance of this correspondence, however, is yet to be determined.

\subsection{The Two-Dimensional Case}

In this Subsection, we derive a useful expression for the measurement
$\{\hat E_b^{\rm B}\}$ in a case of particular interest,
two-dimensional Hilbert spaces.  Here the best strategy for finding the basis 
projectors $\hat E_b^{\rm B}$ is not to directly diagonalize the operator 
$\hat M$, but rather to focus on variational methods.
The great simplification for two-dimensional 
Hilbert spaces is that the signal states $\hat\rho_0$ and $\hat\rho_1$ may be 
represented as vectors within the unit ball of $\RR^3$, the so-called
Bloch sphere:
\beq
\hat\rho_0=\frac{1}{2}\bigl(\hat{\openone}+\vec{a}\cdot\vec{\sigma}\bigr)
\;\;\;\;\;\mbox{and}\;\;\;\;\;
\hat\rho_1=\frac{1}{2}\bigl(\hat{\openone}+\vec{b}\cdot\vec{\sigma}\bigr)\;,
\eeq
where $a\equiv|\vec{a}|\le1$, $b\equiv|\vec{b}|\le1$, and $\vec{\sigma}$ is the
Pauli spin vector.  This follows because the identity operator $\hat{\openone}$
and the (trace-free) Pauli operators
$\vec{\sigma}=(\hat\sigma_x,\hat\sigma_y,\hat\sigma_z)$, i.e.,
\beq
\hat\sigma_x=\left(\begin{array}{cr}
0 & 1\\
1 & 0\end{array}\right),\;\;\;\;
\hat\sigma_1=\left(\begin{array}{cr}
0 & -i\\
i & 0\end{array}\right),\;\;\;\;
\hat\sigma_z=\left(\begin{array}{cr}
1 & 0\\
0 & -1\end{array}\right),
\eeq
form a basis for the vector space of $2\times2$ Hermitian operators. In this 
representation the signal states are pure if
$\vec{a}$ and $\vec{b}$ have unit modulus.  More generally, the eigenvalues of
$\hat\rho_0$ are given by $\frac{1}{2}(1-a)$ and $\frac{1}{2}(1+a)$ and
similarly for $\hat\rho_1$.  Consider an orthogonal projection-valued 
measurement corresponding to the unit Bloch vector
$\vec{n}/n$ and its antipode
$-\vec{n}/n$; for this measurement, the possible outcomes
can be labeled simply $+1$ and $-1$.  It is a trivial matter, using the identity
\beq
(\vec{a}\cdot\vec{\sigma})(\vec{n}\cdot\vec{\sigma})=(\vec{a}\cdot\vec{n})
\hat{\openone}+i\,\vec{\sigma}\cdot(\vec{a}\times\vec{n})\;,
\eeq
to show that
\beq
p_0(b)=\frac{1}{2}\!\left(1\pm\vec{a}\cdot\frac{\vec{n}}{n}\right)
\;\;\;\;\;\mbox{and}\;\;\;\;\;
p_1(b)=\frac{1}{2}\!\left(1\pm\vec{b}\cdot\frac{\vec{n}}{n}\right)\;.
\eeq
The statistical overlap, i.e., $F(p_0,p_1)$, with respect to 
this measurement can thus be written as
\beq
W(\vec{n})\equiv{1\over2}\,\sqrt{\left(1+\vec{a}\cdot\frac{\vec{n}}{n}\right)
\!\!\left(1+\vec{b}\cdot\frac{\vec{n}}{n}\right)}
+{1\over2}\,\sqrt{\left(1-\vec{a}\cdot\frac{\vec{n}}{n}\right)
\!\!\left(1-\vec{b}\cdot\frac{\vec{n}}{n}\right)}\;.
\label{2DWootters}
\eeq
The optimal projector for expression~(\ref{2DWootters}) can be found by varying
it with respect to all vectors $\vec{n}$, i.e., by setting
$\delta W(\vec{n})=0$.  Using
\beq
\delta n=\delta\Bigl(\sqrt{\vec{n}\cdot\vec{n}}\,\Bigr)=
{1\over2n}\delta(\vec{n}\cdot\vec{n})=
{1\over2n}\Bigl( (\delta\vec{n})\cdot\vec{n}+\vec{n}\cdot(\delta\vec{n})\Bigr)=
{\vec{n}\over n}\cdot(\delta\vec{n})\;,
\eeq
one finds after a bit of algebra that the optimal $\vec{n}$ must lie in the
plane spanned by $\vec{a}$ and $\vec{b}$ and satisfy the equation
\beq
(a^2-b^2)\,n^2\,-\,(1-b^2)\Big(\vec{a}\cdot\vec{n}\Big)^2\,+\,(1-a^2)
\Big(\vec{b}\cdot\vec{n}\Big)^2\,=\,0\;.
\eeq
A vector $\vec{n}_{\rm B}$ satisfying these two requirements is
\beq
\vec{n}_{\rm B}={\vec{a}\over\sqrt{1-a^2}}-{\vec{b}\over\sqrt{1-b^2}}\;.
\label{BraunVec}
\eeq
It might be noted that the variational equation $\delta W(\vec{n})=0$ also
generates the measurement with respect to which $\hat\rho_0$ and $\hat\rho_1$
are the least distinguishable; this is the measurement specified by the Bloch
vector $\vec{n}_0$ orthogonal to the vector $\vec{d}=\vec{b}-\vec{a}$.

If $\vec{n}_{\rm B}$ is plugged back into Eq.~(\ref{2DWootters}), one finds
after quite some algebra that
\beq
F(\hat\rho_0,\hat\rho_1)\,=\,\frac{1}{\sqrt{2}}\left(
1\,+\,\vec{a}\cdot\vec{b}\,+\,\sqrt{1-a^2}\sqrt{1-b^2}\,\right)^{\!1/2}\;.
\eeq
(This expression does not match H\"ubner's expression in Ref.~\cite{Hubner92}
because of his convention that the Bloch sphere is of radius $\frac{1}{2}$.)

\section{The Quantum R\'enyi Overlaps}
\label{HopScotch}

Recall that the quantum R\'enyi overlap of order $\alpha$ (for $0<\alpha<1$)
is defined by
\beq
F_\alpha(\hat\rho_0/\hat\rho_1)\equiv\min_{\{\hat E_b\}}\sum_b
\left({\rm tr}(\hat\rho_0\hat E_b)\right)^{\!\alpha}\!
\left({\rm tr}(\hat\rho_1\hat E_b)\right)^{\!{1-\alpha}}\;.
\label{UncleRufus}
\eeq
This measure of quantum distinguishability, though it has no compelling
operational meaning, proves to be a convenient testing ground for optimization
techniques more elaborate than so far encountered.  Equation~(\ref{UncleRufus})
is more unwieldy than the quantum statistical overlap, but not so
transcendental in character as to contain a logarithm like the Kullback-Leibler
and mutual informations.  One can still maintain a serious hope that an
explicit expression for it is possible.

Moreover, if an explicit expression can be found for Eq.~(\ref{UncleRufus}),
then one would have a direct route to the quantum Kullback information through
either the R\'enyi relative information of order
$\alpha$ \cite{Renyi61,Renyi70}
\beq
K_\alpha(p_0/p_1)\,=\,\frac{1}{\,\alpha-1\,}\,\ln\!\left(\,\sum_{b=1}^n\,
p_0(b)^\alpha p_1(b)^{1-\alpha}\right)\;,
\eeq
or the relative information of type $\alpha$ of Rathie and Kannappan
\cite{Rathie72,Aczel75}
\beq
K^\alpha(p_0/p_1)\,=\,\left(e^{\alpha-1}-1\right)^{\!-1}\!\left(\,\sum_{b=1}^n\,
p_0(b)^\alpha p_1(b)^{1-\alpha}-1\right)\;.
\eeq
Both these quantities converge to the Kullback-Leibler relative information
in the limit $\alpha\rightarrow1$, as can be seen easily by using l'Hospital's
rule.  Therefore, one just needs to find $F_\alpha(\hat\rho_0/\hat\rho_1)$
and take the limit $\alpha\rightarrow1$ to get the quantum Kullback information.

The brightest ray of hope for finding an explicit expression for
$F_\alpha(\hat\rho_0/\hat\rho_1)$ is that the quantum fidelity was found via
an application of the Schwarz inequality.  There is a closely related inequality
in the mathematical literature known as the H\"older inequality
\cite{Hardy52,Mitrinovic70} that---at first sight at least---would appear to be 
of use in the same way for this context.  This inequality is that, for any two 
sequences $a_k$ and $b_k$ of complex numbers, $k=1,\ldots,n$,
\beq
\sum_{k=1}^n|a_kb_k|\,\le\,\left(\sum_{k=1}^n|a_k|^p\right)^{1/p}\!
\left(\sum_{k=1}^n|b_k|^q\right)^{1/q}\;,
\label{AuntRuby}
\eeq
when $p,q>1$ with
\beq
\frac{1}{p}+\frac{1}{q}=1\;.
\label{AuntDoris}
\eeq
Equality is achieved in Eq.~(\ref{AuntRuby}) if and only if there exists a 
constant $c$ such that
\beq
|b_k|=c|a_k|^{p-1}\;.
\eeq
The standard Schwarz inequality is the special case of this for $p=2$.

One would like to use {\it some\/} appropriate operator analog to the H\"older 
inequality in much the same way as the operator-Schwarz inequality was used
for optimizing the statistical overlap: use it to bound the quantum R\'enyi
overlap and then search out the conditions for achieving equality---perhaps
again by taking advantage of the invariances of the trace.  In particular, one
would like to find something of the form
\beq
\left({\rm tr}(\hat\rho_0\hat E_b)\right)^{\!\alpha}\!
\left({\rm tr}(\hat\rho_1\hat E_b)\right)^{\!{1-\alpha}}\ge
\tr\Big(f(\hat\rho_0,\hat\rho_1;\alpha)\hat E_b\Big)\;,
\label{UncleDell}
\eeq
with a function $f(\hat\rho_0,\hat\rho_1;\alpha)$ that is independent of the
POVM $\{\hat E_b\}$.  In this way, the linearity of the trace and the 
completeness of the $\hat E_b$ could be used in the same fashion as before.

Unfortunately---even after an {\it exhaustive\/} literature search---an
inequality sufficiently strong to carry through the optimization has yet to be 
found. Nevertheless, for future endeavors, we report the most promising lines
of attack found so far.

For the first demonstration, we need to list the standard
operator-H\"older inequality \cite{Mehta68,Gohberg69,Bhatia95}
\beq
\left|\,\tr\!\left(\hat A\hat B\right)\right|\,\le\,
\left(\tr\!\left(\hat A^\dagger\hat A\right)^{\!p/2}\right)^{\!1/p}\!
\left(\tr\!\left(\hat B^\dagger\hat B\right)^{\!q/2}\right)^{\!1/q}\;,
\eeq
and the Araki inequality \cite{Araki90,Wang93},
\beq
\tr\!\left(\hat C^{1/2}\hat D\hat C^{1/2}\right)^{\!r}\,\le\,
\tr\!\left(\hat C^{r/2}\hat D^r\hat C^{r/2}\right)\;,
\eeq
for $\hat C$ and $\hat D$ positive operators and $r\ge1$.  These two
inequalities can get us to something ``close'' to Eq.~(\ref{UncleDell}), though
not quite linear in the $\hat E_b$.  Let us show how this is done.  Suppose
positive $p$ and $q$ satisfy Eq.~(\ref{AuntDoris}).  Then 
\beqa
\left({\rm tr}(\hat\rho_0\hat E_b)\right)^{\!\frac{1}{p}}\!
\left({\rm tr}(\hat\rho_1\hat E_b)\right)^{\!\frac{1}{q}}\!\!\!
&=&\!\!\!
\left(\!{\rm tr}\!\left(\left(\hat E_b^\frac{1}{p}\right)^{\!\frac{p}{2}}\!\!
\left(\hat\rho_0^\frac{1}{p}\right)^{\!p}\!\!
\left(\hat E_b^\frac{1}{p}\right)^{\!\frac{p}{2}}\right)
\rule{0mm}{8mm}\!\right)^{\!\frac{1}{p}}\!
\left(\!{\rm tr}\!\left(\left(\hat E_b^\frac{1}{q}\right)^{\!\frac{q}{2}}\!\!
\left(\hat\rho_1^\frac{1}{q}\right)^{\!q}\!\!
\left(\hat E_b^\frac{1}{q}\right)^{\!\frac{q}{2}}\right)
\rule{0mm}{8mm}\!\right)^{\!\frac{1}{q}}\!
\rule{0mm}{10mm}
\nonumber\\
&\ge&\!\!\!
\left(\!{\rm tr}\!\left(\hat E_b^\frac{1}{2p}\hat\rho_0^\frac{1}{p}
\hat E_b^\frac{1}{2p}\right)^{\!p}\rule{0mm}{6mm}\!\right)^{\!\frac{1}{p}}\!
\left(\!{\rm tr}\!\left(\hat E_b^\frac{1}{2q}\hat\rho_1^\frac{1}{q}
\hat E_b^\frac{1}{2q}\right)^{\!q}\rule{0mm}{6mm}\!\right)^{\!\frac{1}{q}}
\rule{0mm}{10mm}
\nonumber\\
&\ge&\!\!
\left|\,{\rm tr}\!\left(\hat E_b^\frac{1}{2p}\hat\rho_0^\frac{1}{p}
\hat E_b^\frac{1}{2p}\hat E_b^\frac{1}{2q}\hat\rho_1^\frac{1}{q}
\hat E_b^\frac{1}{2q}\right)\right|
\rule{0mm}{10mm}
\nonumber\\
&\ge&\!\!
\left|\,{\rm tr}\!\left(\hat E_b^\frac{1}{2}\hat\rho_0^\frac{1}{p}
\hat E_b^\frac{1}{2}\hat\rho_1^\frac{1}{q}\right)\right|\;.
\rule{0mm}{10mm}
\eeqa
The upshot is a bound that is ``almost'' linear in the $\hat E_b$.
There are other variations on this theme, but they are lacking in the same
respect.

A second way to tackle this problem via the H\"older inequality is at the
eigenvalue level.  To set this problem up, let us use the notation 
$\lambda_i(\hat A)$ to denote the eigenvalues of any Hermitian operator
$\hat A$, when numbered so that they form a nonincreasing sequence, i.e., 
\beq
\lambda_1(\hat A)\ge\lambda_2(\hat A)\ge\cdots\ge\lambda_D(\hat A)\;.
\eeq
With this, we may write down a theorem, originally due to Richter 
\cite{Mirsky59,Marshall79}, that places a bound on the trace of a product of
two operators
\beq
\sum_{i=1}^D\lambda_i(\hat A)\,\lambda_{D-i+1}(\hat B)\;\le\;
\tr(\hat A\hat B)\;\le\;
\sum_{i=1}^D\lambda_i(\hat A)\,\lambda_i(\hat B)\;.
\eeq
Using this and the H\"older inequality for numbers, one immediately obtains,
\beqa
\left({\rm tr}(\hat\rho_0\hat E_b)\right)^{\!\frac{1}{p}}\!
\left({\rm tr}(\hat\rho_1\hat E_b)\right)^{\!\frac{1}{q}}
&\ge&
\left(\sum_{i=1}^D\lambda_i(\hat\rho_0)\,\lambda_{D-i+1}(\hat E_b)
\right)^{\!\frac{1}{p}}\!
\left(\sum_{i=1}^D\lambda_i(\hat\rho_1)\,\lambda_{D-i+1}(\hat E_b)
\right)^{\!\frac{1}{q}}
\nonumber\\
&\ge&
\sum_{i=1}^D\lambda_i(\hat\rho_0)^\frac{1}{p}\,
\lambda_i(\hat\rho_1)^\frac{1}{q}\,\lambda_{D-i+1}(\hat E_b)
\rule{0mm}{9mm}
\label{UncleClarence}
\eeqa
This is---in a certain sense---linear in the operator
$\hat E_b$.  Regardless of this, however, the bound given by
Eq.~(\ref{UncleClarence}), is far too loose.  For instance, if the $\hat E_b$
are one-dimensional projectors, this bound reduces to the product
$\lambda_D(\hat\rho_0)^\frac{1}{p}\,\lambda_D(\hat\rho_1)^\frac{1}{q}$ and
so gives rise to 
\beq
\sum_{b=1}^D\left({\rm tr}(\hat\rho_0\hat E_b)\right)^{\!\frac{1}{p}}\!
\left({\rm tr}(\hat\rho_1\hat E_b)\right)^{\!\frac{1}{q}}\;\ge\;
D\,\lambda_D(\hat\rho_0)^\frac{1}{p}\,\lambda_D(\hat\rho_1)^\frac{1}{q}\;.
\eeq
This can hardly be a tight bound, disregarding as it does all the other
structure of the operators $\hat\rho_0$ and $\hat\rho_1$ and their relation to
each other.

Could it be that an optimal POVM for the R\'enyi overlap can always be taken
to be an orthogonal projection-valued measurement?  Chances of this are
strong, given that that was the case for the statistical overlap.  In case of
this, let us point out the following.  Suppose $\hat A$ is an invertible 
positive operator with spectral decomposition $\hat A=\sum_b a_b\hat\Pi_b$.
Then applying the H\"older inequality again, we have
\beqa
\left({\rm tr}(\hat\rho_0\hat A^p)\right)^{\!\frac{1}{p}}\!
\left({\rm tr}(\hat\rho_1\hat A^{-q})\right)^{\!\frac{1}{q}}
&=&
\left(\sum_b a_b^p{\rm tr}(\hat\rho_0\hat\Pi_b)\right)^{\!\frac{1}{p}}
\left(\sum_b a_b^{-q}{\rm tr}(\hat\rho_1\hat\Pi_b)\right)^{\!\frac{1}{q}}
\nonumber\\
&\ge&
\sum_b\left({\rm tr}(\hat\rho_0\hat\Pi_b)\right)^{\!\frac{1}{p}}\!
\left({\rm tr}(\hat\rho_1\hat\Pi_b)\right)^{\!\frac{1}{q}}\;.
\eeqa
Therefore, as it was for the case of the quantum statistical overlap, it
may well be that
\beq
\min_{\hat A>0}\left({\rm tr}(\hat\rho_0\hat A^p)\right)^{\!\frac{1}{p}}\!
\left({\rm tr}(\hat\rho_1\hat A^{-q})\right)^{\!\frac{1}{q}}
\eeq
actually equals the quantum R\'enyi overlap of order $\frac{1}{p}$.

Let us point out a bound for this quantity.  Note that, for any positive
invertible operator $\hat A$,
\beq
\lambda_{D-i+1}(\hat A^{-1})\,=\,\Big(\lambda_i(\hat A)\Big)^{\!-1}\;.
\eeq
Using the Richter and H\"older inequalities, we have that
\beqa
\left({\rm tr}(\hat\rho_0\hat A^p)\right)^{\!\frac{1}{p}}\!
\left({\rm tr}(\hat\rho_1\hat A^{-q})\right)^{\!\frac{1}{q}}
&\ge&
\left(\sum_{i=1}^D\lambda_{D-i+1}(\hat\rho_0)\,\lambda_i(\hat A^p)
\right)^{\!\frac{1}{p}}\!
\left(\sum_{i=1}^D\lambda_i(\hat\rho_1)\,\lambda_{D-i+1}(\hat A^{-q})
\right)^{\!\frac{1}{q}}
\rule{0mm}{9mm}
\nonumber\\
&=&
\left(\sum_{i=1}^D\lambda_{D-i+1}(\hat\rho_0)\,\lambda_i(\hat A)^p
\right)^{\!\frac{1}{p}}\!
\left(\sum_{i=1}^D\lambda_i(\hat\rho_1)\,\lambda_i(\hat A)^{-q}
\right)^{\!\frac{1}{q}}
\rule{0mm}{9mm}
\nonumber\\
&\ge&
\sum_{i=1}^D\lambda_{D-i+1}(\hat\rho_0)^\frac{1}{p}
\lambda_i(\hat\rho_1)^\frac{1}{q}\;.
\rule{0mm}{9mm}
\eeqa
This gives a nice lower bound, though it is most certainly not tight---as can
be seen from that fact that it does not reproduce the quantum fidelity when
$p=2$.

Let us now mention another method for lower bounding the quantum R\'enyi
overlap that does not appear to be related to a H\"older inequality at all.
This one relies on an inequality of Ando \cite{Ando94b} concerning the operator
means introduced in Section~(\ref{UncleDoug}).  For any operator mean $\sigma$,
\beq
\Big(\tr(\hat C\hat A)\Big)\,\sigma\,\Big(\tr(\hat C\hat B)\Big)
\,\ge\,\tr\Big(\hat C\,(\hat A\sigma\hat B)\Big)\;,
\eeq
where $\hat A$, $\hat B$, and $\hat C$ are all positive operators.  We can use
this in the following way.  Note that the mapping $\#_\alpha$ defined by
\beq
\hat A\,\#_\alpha\,\hat B\;=\;\hat A^{1/2}\left(\hat A^{-1/2}\hat B\hat A^{-1/2}
\right)^{\!\alpha}\hat A^{1/2}
\eeq
satisfies all the properties of an operator mean \cite{Kubo80}.  When $\hat A$
and $\hat B$ are scalars, and so commute, this operator mean reduces to
\beq
\hat A\,\#_\alpha\,\hat B\;=\;\hat B^\alpha\hat A^{1-\alpha}\;.
\eeq
Therefore, for any POVM $\{\hat E_b\}$, we can write the inequality
\beqa
\sum_b\left({\rm tr}(\hat\rho_0\hat E_b)\right)^{\!\alpha}\!
\left({\rm tr}(\hat\rho_1\hat E_b)\right)^{\!{1-\alpha}}
&\ge&
\sum_b\tr\!\left(\hat E_b\left(\hat\rho_1\,\#_\alpha\,\hat\rho_0\right)
\rule{0mm}{4mm}\right)
\nonumber\\
&=&
\tr\!\left(\hat\rho_1^{1/2}\!\left(\hat\rho_1^{-1/2}\hat\rho_0\hat\rho_1^{-1/2}
\right)^{\!\alpha}\!\hat\rho_1^{1/2}\rule{0mm}{5mm}\right)\;.
\rule{0mm}{7mm}
\label{UncleWalter}
\eeqa
Therefore one obtains a lower bound on the quantum R\'enyi overlap of order
$\alpha$.  Again the bound is not as tight as it might be because it does
not reproduce the result known for $\alpha=\frac{1}{2}$.

Finally, we should point out that Hasegawa \cite{Hasegawa93,Hasegawa95} has
studied the quantity
\beq
H_\alpha(\hat\rho_0/\hat\rho_1)=\frac{4}{1-\alpha^2}\,\tr\!\left(\!
\left(1-\hat\rho_0^{\frac{1+\alpha}{2}}\hat\rho_1^{-\frac{1+\alpha}{2}}\right)
\!\hat\rho_1\rule{0mm}{6mm}\!\right)
\eeq
in the context of distinguishing quantum states.  The connection between this 
and Eq.~(\ref{UncleRufus}) (if there is any) is not known.

\section{The Accessible Information}
\label{SnaffleDoo}

A binary quantum communication channel is defined by its signal states
$\{\hat\rho_0,\hat\rho_1\}$ and their prior probabilities
\beq
\pi_0=1-t
\;\;\;\;\;\;\;\;\mbox{and}\;\;\;\;\;\;\;\;
\pi_1=t\;,
\eeq
for $0\le t\le1$.  Consider a measurement $\{\hat E_b \}$ on the
channel.  The probability of an outcome $b$ when the message state is $k$
($k=0,1$) is
\beq
p_k(b)={\rm tr}(\hat\rho_k\hat E_b)\;.
\eeq
The unconditioned probability distribution for the outcomes is
\beq
p(b)={\rm tr}(\hat\rho\hat E_b)\;, 
\eeq
for
\beqa
\hat\rho
&=&
(1-t)\hat\rho_0+t\hat\rho_1
\nonumber\\
&=&
\hat\rho_0+t\hat\Delta
\nonumber\\
&=&
\hat\rho_1-(1-t)\hat\Delta\;,
\eeqa
where the difference operator $\hat\Delta$ is defined by
\beq
\hat\Delta=\hat\rho_1 - \hat\rho_0\;.
\eeq
The Shannon mutual information \cite{Cover91,Shannon48} for the channel,
with respect to the measurement $\{\hat E_b \}$, is defined to be
\beqa
J(p_0,p_1;t)
&\equiv&
H(p)-(1-t)H(p_0)-tH(p_1)
\nonumber\\
&=&
(1-t)K(p_0/p)+tK(p_1/p)\;,
\label{MutualInf}
\eeqa
where
\beq
H(q)=-\sum_b q(b)\ln q(b)
\label{Shnaaoainf}
\eeq
is the Shannon information 
\cite{Cover91,Aczel75,Shannon48} of the probability
distribution $q(b)$. 
Because a natural logarithm has
been used in this definition, information here is quantified in terms of
``nats'' rather than the more commonly used unit of ``bits.''  The
{\it accessible information\/} \cite{Schumacher90a,Schumacher90b}
$I(\hat\rho_0|\hat\rho_1)$ is the mutual information maximized over all 
measurements $\{\hat E_b \}$:
\beqa
I(\hat\rho_0|\hat\rho_1)
&=&
\max_{\{\hat E_b\}}\,
\sum_b\!\left(\rule{0mm}{5.5mm}-
{\rm tr}(\hat\rho\hat E_b)\ln\!\Big({\rm tr}(\hat\rho\hat E_b)\Big)\right.
\nonumber\\
&&\;\;\;\;\;\;\;
+(1-t)\left.\rule{0mm}{5.5mm}
{\rm tr}(\hat\rho_0\hat E_b)\ln\!\Big({\rm tr}(\hat\rho_0\hat E_b)\Big)+t\,
{\rm tr}(\hat\rho_1\hat E_b)\ln\!\Big({\rm tr}(\hat\rho_1\hat E_b)\Big)\right)
\nonumber\\
&=&
\rule{0mm}{10mm}
\max_{\{\hat E_b\}}\,
\sum_b\!\left(\pi_0\,{\rm tr}(\hat\rho_0\hat E_b)\,
\ln\!\left({{\rm tr}(\hat\rho_0\hat E_b)\over
{\rm tr}(\hat\rho\hat E_b)}\right)\,+\,
\pi_1\,{\rm tr}(\hat\rho_1\hat E_b)\,
\ln\!\left({{\rm tr}(\hat\rho_1\hat E_b)\over
{\rm tr}(\hat\rho\hat E_b)}\right)\right)\;.
\nonumber\\
\eeqa
In this Section we will often use the alternative notations $J(t)$ and $I(t)$
for the mutual and accessible informations; this notation is more compact while
still making explicit the dependence of these quantities on the prior 
probabilities.

The significance of expression~(\ref{MutualInf}), explained in great detail in
Chapter 2, can be summarized in the following way.  Imagine for a moment that 
the receiver actually {\it does\/} know which message $m_k$ was sent, but 
nevertheless performs the measurement $\{\hat E_b \}$ on it.  Regardless of
this knowledge, the receiver will not be able to predict the exact 
outcome of his measurement; this is just because of quantum
indeterminism---the most he can say is that outcome $b$ will occur with
probability $p_k(b)$.  A different way to summarize this is that even with
the exact message known, the receiver will gain information via
the unpredictable outcome $b$.  That information, however, has nothing
to do with the message itself.  The residual information gain is a
signature of quantum indeterminism, now quantified by the Shannon information
$H(p_k)$ of the outcome distribution. Now return to the real scenario, where
the receiver actually does {\it not\/} know which message 
was sent.  The amount of residual information the receiver can {\it expect\/} to gain in this case is $(1-t)H(p_0)+tH(p_1)$.  This quantity, however, is not
identical to the information the receiver can expect to gain {\it in toto\/}.
That is because the receiver must describe the quantum system encoding the 
message by the mean density operator $\hat\rho$; 
this reflects his lack of knowledge about the preparation.  For this state, the 
expected amount of information gain in a measurement of $\{\hat E_b \}$ is 
$H(p)$.  This time some of the information will have to do with the message
itself, rather than being due to quantum indeterminism.  The natural quantity
for describing the information gained exclusively about the message itself 
(i.e., not augmented through quantum indeterminism) is just the mutual 
information Eq.~(\ref{MutualInf}).

\subsection{The Holevo Bound}
\label{ChompaCakes}

The problems associated with actually finding $I(t)$ and the
measurement that gives rise to it are every bit as difficult as those in
maximizing the Kullback-Leibler information, perhaps more so---for
here it is not only the logarithm that confounds things, but also the fact that
$\hat\rho_0$ and $\hat\rho_1$ are ``coupled'' through the mean density operator
$\hat\rho$.  Outside of a very few isolated examples, namely the case where
$\hat\rho_0$ and $\hat\rho_1$ are pure states and the case where they are 
$2\times2$ density operators with equal determinant and equal prior 
probabilities \cite{Levitin93,Levitin95}, explicit expressions for
$I(t)$ have never been calculated.  Moreover no general algorithms
for approximating this quantity appear to exist as yet.  There is a result, due
to Davies \cite{Davies78,Fuchs95d}, stating that there always {\it exists\/}
an optimal measurement $\{\hat E_b^{\rm D}\}$ of the form
\beq
\hat E_b^{\rm D}=\alpha_b|\psi_b\rangle\langle\psi_b|
\;\;\;\;\;\;\;\;\;b=1,\ldots,N\;,
\eeq
where the number of terms in this is bracketed by
\beq 
D\le N\le D^2
\eeq
for signal states on a $D$-dimensional Hilbert space.  This 
theorem, however, does not pin down the measurement any further than that.
The most useful general statements about $I(t)$ have been in the
form of bounds:  an upper bound first conjectured in print by Gordon
\cite{Gordon64} in 1964 (though there may also have been some discussion of it by Forney \cite{Forney62} in 1962) and proven by Holevo \cite{Holevo73d} in 1973 (though Levitin did announce a similar but more restricted result \cite{Levitin69} in
1969), and a lower bound first conjectured by Wootters \cite{Wootters93} and proven by Jozsa, Robb, and Wootters \cite{Jozsa94b}.
These bounds, however, are of little use in
pinpointing the measurement that gives rise to $I(t)$.  In what
follows, we simplify the derivation of the Holevo bound via a variation of the
methods used in the Section~\ref{MachoMan}.  This simplification has the
advantage of specifying a measurement, the use of which immediately gives rise
to a new lower bound to $I(t)$.  We also supply an explicit expression for a
still-tighter upper bound whose existence is required within the original Holevo
derivation.

Since Holevo's original derivation, various improved versions have also appeared \cite{Yuen93,Hall93}.  The improvements until now, however, have been in
proving the upper bound for more general situations: infinite dimensional
Hilbert spaces, infinite numbers of messages states and infinite numbers of
measurement outcomes.  In contrast, in this section and throughout the remainder of the report, we retain the finiteness assumptions made by Holevo; the aim
here is to build a deeper understanding of the issues involved in finding 
approximations to the accessible information.

The Holevo upper bound to $I(t)$ is given by
\beq
I(t)\le S(\hat\rho)-(1-t)S(\hat\rho_0)-t S(\hat\rho_1)\equiv S(t)\;,
\label{HolevoBound}
\eeq
where
\beq
S(\hat\rho)=-{\rm tr}\bigl(\hat\rho\ln\hat\rho\bigr)=-\sum_{j=1}^D\lambda_j
\ln\lambda_j
\eeq
is the von Neumann entropy \cite{vonNeumann32} of the density operator 
$\hat\rho$, whose eigenvalues are $\lambda_j$, $j=1,\ldots,D$.  Equality is
achieved in this bound if and only if $\hat\rho_0$ and $\hat\rho_1$ commute.

The Jozsa-Robb-Wootters lower bound to $I(t)$ is formally quite similar to
the Holevo upper bound.  For later reference, we write it out explicitly:
\beq
I(t)\ge Q(\hat\rho)-(1-t)Q(\hat\rho_0)-t Q(\hat\rho_1)\equiv Q(t)\;,
\label{JozsaBound}
\eeq
where
\beq
Q(\hat\rho)=-\sum_{j=1}^D\left(\prod_{k\ne j}{\lambda_j\over\lambda_j-\lambda_k}
\right)\lambda_j\ln\lambda_j
\eeq
is the ``sub-entropy'' \cite{Wootters90,Jones91d} of the density operator $\hat\rho$.  The formal similarity between
Eqs.~(\ref{HolevoBound}) and (\ref{JozsaBound}) becomes even more apparent if
$S(\hat\rho)$ and $Q(\hat\rho)$ are represented as contour integrals
\cite{Holevo73d,Jozsa94b,Poincare99}:
\beq
S(\hat\rho)=-{1\over 2\pi i}\oint_C
(\ln z)\;{\rm tr}\!\left(\Bigl(\hat{\openone}
-z^{-1}\hat\rho\Bigr)^{-1}\right)dz\;,
\eeq
and
\beq
Q(\hat\rho)=-{1\over 2\pi i}\oint_C
(\ln z)\;{\rm det}\!\left(\Bigl(\hat{\openone}
-z^{-1}\hat\rho\Bigr)^{-1}\right)dz\;,
\eeq
where the contour $C$ encloses all the {\it nonzero\/} eigenvalues of
$\hat\rho$.

The key to deriving the Holevo bound is in realizing the importance of 
properties of $J(t)$ and $S(t)$ as functions of $t$ \cite{Holevo73d}.
Note that
\beq
J(0)=J(1)=S(0)=S(1)=0\;.
\eeq
Moreover, both
$J(t)$ and $S(t)$ are downwardly convex, as can be seen by
working out their second derivatives.  For
$J(t)$ a straightforward calculation gives
\beq
J''(t)=-\sum_{b} {\; \Bigl( {\rm tr} \bigl( \hat\Delta\hat E_b \bigr) 
\Bigr)^2
\over {\rm tr} \bigl( \hat\rho\hat E_b \bigr) }\;.
\label{Fisher}
\eeq

For $S(t)$ it is easiest to proceed by using the contour
integral representation of $S(\hat\rho)$. By
differentiating within the integral and using the operator
identity
\beq
\frac{d\hat A^{-1}}{dt}=-\hat A^{-1}\frac{d\hat A}{dt}\hat A^{-1}
\label{Mizas}
\eeq
(which comes simply from the fact that $\hat A^{-1}\hat A=\hat{\openone}$),
one finds
\beq
\frac{d}{dt}S(\hat\rho)=-\frac{1}{2\pi i}\oint_C
(z\ln z)\;{\rm tr}\!\left(\Bigl(z\hat{\openone}-\hat\rho\Bigr)^{-1}
\hat\Delta\Bigl(z\hat{\openone}-\hat\rho\Bigr)^{-1}\right)dz\;,
\eeq
and
\beq
\frac{d^2}{dt^2}S(\hat\rho)=-\frac{2}{2\pi i}\oint_C
(z\ln z)\;{\rm tr}\!\left(\Bigl(z\hat{\openone}-\hat\rho\Bigr)^{-2}
\hat\Delta\Bigl(z\hat{\openone}-\hat\rho\Bigr)^{-1}
\hat\Delta\right)dz\;.
\eeq
Therefore, if $|j\rangle$ is the eigenvector of $\hat\rho$ with eigenvalue $\lambda_j$ and $\Delta_{jk}=\langle j|\hat\Delta|k\rangle$, we can
write
\beq
S''(t)=-\sum_{\{j,k|\lambda_j +\lambda_k\neq0\}}
\Phi\bigl(\lambda_j,\lambda_k\bigr)
\bigl|\Delta_{jk}\bigr|^2\;,
\label{Balian}
\eeq
where
\beq
\Phi\bigl(\lambda_j,\lambda_k\bigr)=\frac{2}{2\pi i}\oint_C
\frac{z\ln z}{(z-\lambda_k)^2(z-\lambda_j)}\,dz\;.
\eeq
An application of Cauchy's integral theorem gives
\beq
\Phi(x,y)=\frac{\ln x-\ln y}{x-y}\;\;\;\;\;\;\mbox{if}\:\;x\neq y\;,
\eeq
and
\beq
\Phi(x,x)=\frac{1}{x}\;.
\eeq
Expressions~(\ref{Fisher}) and~(\ref{Balian}) are thus clearly nonpositive.

The statement that $S(t)$ is an upper bound to $J(t)$ for {\it any\/} $t$
is equivalent to the property that, when plotted versus $t$, the curve for
$S(t)$ has a more negative curvature than the curve for $J(t)$ regardless of 
which POVM $\{\hat E_b \}$ is used in its definition. This has to be because 
$S(t)$ must climb higher than $J(t)$ to be an upper bound for it.
The meat of the derivation is in showing the inequality
\beq
S''(t) \leq J''(t)\leq 0\hspace{.2in}\mbox{for any POVM $\{\hat E_b\}$}\;.
\eeq
Holevo does this by demonstrating the existence of a function $L''(t)$, 
independent of $\{\hat E_b \}$, such that
\beq
S''(t) \leq L''(t)\;\;\;\;\;\mbox{and}\;\;\;\;\;L''(t)\leq J''(t)\;.
\eeq
From this it follows, upon enforcing the boundary condition
\beq
L(0)=L(1)=0\;,
\eeq
that
\beq
I(t)\leq L(t)\leq S(t)\;.
\eeq
(It should be noted that $L(t)$ is not explicitly computed in
Ref.~\cite{Holevo73d}; it is only shown to exist via the expression for its
second derivative.)

At this point a fairly drastic simplification can be made to the original
proof.  An easy way to get at such a function $L''(t)$ is simply to minimize
$J''(t)$ over all POVMs $\{\hat E_b\}$ and to define the
result to be the function $L''(t)$.  Thereafter one can work to show that
$S''(t) \leq L''(t)$.  This is decidedly more tractable than extremizing
the mutual information $J(t)$ itself because no logarithms appear in 
$J''(t)$; there can be hope for a solution by means of standard algebraic
inequalities such as the Schwarz inequality.  This approach, it turns out, 
generates exactly the same function $L''(t)$ as used by Holevo in the original 
proof, though the two derivations appear to have little to do with each other.
The difference of real importance here is that this approach pinpoints the 
measurement that actually minimizes $I''(t)$. This measurement, though it 
generally does not maximize $J(t)$ itself, necessarily does provide a
{\it lower\/} bound to the accessible information $I(t)$.

The problem of minimizing Eq.~(\ref{Fisher}) is formally identical to the
problem considered by Braunstein and Caves \cite{Braunstein94a}: the
expression for $-I''(t)$ is of the same form as the Fisher information
optimized there.  The steps are as follows.  The idea is to think of the 
numerator within the sum~(\ref{Fisher}) as analogous to the left hand side of
the Schwarz inequality:
\beq
|{\rm tr}(\hat A^\dagger\hat B)|^2\le{\rm tr}(\hat A^\dagger\hat A)\,
{\rm tr}(\hat B^\dagger\hat B)\;.
\eeq
One would like to use this inequality in such a way that the
${\rm tr}\bigl(\hat\rho\hat E_b\bigr)$ term in the denominator is cancelled and 
only an expression linear in $\hat E_b$ is left; for then, upon summing over
the index $b$, the completeness property for POVMs will leave the final 
expression independent of the given measurement.

These ideas are formalized by introducing a ``lowering'' super-operator
${\cal G}_{\hat C}$, (i.e., a mapping from operators to operators that depends 
explicitly on a third operator $\hat C$) with the property that for any 
operators $\hat A$ and $\hat B$ and any positive operator $\hat C$,
\beq
{\rm tr}\bigl(\hat A\hat B\bigr)\le
\left|{\rm tr}\!\left(\hat C\hat B\,{\cal G}_{\hat C}(\hat A)\right)\right|\;.
\label{NewR}
\eeq
There are many examples of such super-operators; perhaps the simplest example
is
\beq
{\cal G}_{\hat C}(\hat A)=\hat A\hat C^{-1}\;,
\eeq
when $\hat C$ is invertible.  In any case, for these super-operators, one can derive---via simple applications of the Schwarz inequality---that
\beqa
\Bigl({\rm tr}\bigl(\hat\Delta\hat E_b\bigr)\Bigr)^2
&\le&
\left|{\rm tr}\!\left(\hat\rho\hat E_b {\cal G}_{\hat\rho}
(\hat\Delta)\right)\right|^2
\nonumber\\
&=&
\left|\,{\rm tr}\!\left(
\left(\hat E_b^{1/2}\hat\rho^{1/2}\right)^{\!\dagger}
\left(\hat E_b^{1/2}{\cal G}_{\hat\rho}(\hat\Delta)\hat\rho^{1/2}\right)
\right)\right|^2
\rule{0mm}{8mm}
\nonumber\\
&\le&
{\rm tr}(\hat\rho\hat E_b)\,{\rm tr}\!\left(
\hat E_b\left({\cal G}_{\hat\rho}(\hat\Delta)\hat\rho\,{\cal G}_{\hat\rho}
(\hat\Delta)^\dagger\right)\rule{0mm}{5mm}\right)
\rule{0mm}{8mm}
\label{TBound1}
\eeqa
and
\beqa
\Bigl({\rm tr}\bigl(\hat\Delta\hat E_b\bigr)\Bigr)^2
&\le&
\left|\,{\rm tr}\!\left(\hat\rho^{1/2}\hat E_b {\cal G}_{\hat\rho^{1/2}}
(\hat\Delta)\right)\right|^2
\nonumber\\
&=&
\left|\,{\rm tr}\!\left(
\left(\hat E_b^{1/2}\hat\rho^{1/2}\right)^{\!\dagger}
\left(\hat E_b^{1/2}{\cal G}_{\hat\rho^{1/2}}(\hat\Delta)\right)
\right)\right|^2
\rule{0mm}{8mm}
\nonumber\\
&\le&
{\rm tr}(\hat\rho\hat E_b)\,{\rm tr}\!\left(
\hat E_b\left({\cal G}_{\hat\rho^{1/2}}(\hat\Delta)\,{\cal G}_{\hat\rho^{1/2}}
(\hat\Delta)^\dagger\right)\rule{0mm}{5mm}\right)\;.
\rule{0mm}{8mm}
\label{TBound2}
\eeqa
By ${\cal G}_{\hat C}(\hat A)^\dagger$, we mean simply the Hermitian conjugate to the operator ${\cal G}_{\hat C}(\hat A)$.
The conditions for equality in these are that the super-operators
${\cal G}_{\hat\rho}$ and ${\cal G}_{\hat\rho^{1/2}}$ give equality in the
first steps and, moreover, saturate the Schwarz inequality via
\beq
\hat E_b^{1/2}{\cal G}_{\hat\rho}(\hat\Delta)\hat\rho^{1/2}=\mu_b
\hat E_b^{1/2}\hat\rho^{1/2}\;,
\label{FEqCond}
\eeq
and
\beq
\hat E_b^{1/2}{\cal G}_{\hat\rho^{1/2}}(\hat\Delta)=
\mu_b\hat E_b^{1/2}\hat\rho^{1/2}\;,
\eeq
respectively.

Using inequalities~(\ref{TBound1}) and (\ref{TBound2}) in Eq.~(\ref{Fisher})
for $J''(t)$ immediately gives the lower bounds
\beq
J''(t)\ge-{\rm tr}\Bigl(
{\cal G}_{\hat\rho}(\hat\Delta)\hat\rho\,{\cal G}_{\hat\rho}
(\hat\Delta)^\dagger\Bigr)\;,
\label{SumBound1}
\eeq
and
\beq
J''(t)\ge-{\rm tr}\!\left(
{\cal G}_{\hat\rho^{1/2}}(\hat\Delta)\,{\cal G}_{\hat\rho^{1/2}}
(\hat\Delta)^\dagger\right)\;.
\label{SumBound2}
\eeq
The problem now, much like in Section~\ref{MachoMan}, is to choose a
super-operator ${\cal G}_{\hat C}$ or in such a way that equality can be 
attained in either Eq.~(\ref{SumBound1}) or Eq.~(\ref{SumBound2}).

The super-operator ${\cal L}_{\hat\rho}$ that does the trick for minimizing
Eq.~(\ref{Fisher}) \cite{Braunstein94a} is defined by its action on an
operator $\hat A$ through
\beq
{1\over2}
\left(\hat\rho{\cal L}_{\hat\rho}(\hat A)+{\cal L}_{\hat\rho}(\hat A)
\hat\rho\right)=\hat A.
\label{Lowering}
\eeq
This equation is a special case of the operator equation known as the
Lyapunov equation \cite{Bellman70}:
\beq
\hat B\hat X + \hat X\hat C=\hat D\;.
\label{GilbertGrape}
\eeq
The Lyapunov equation has a solution {\it for all\/} $\hat D$ if and only if
no eigenvalue of $\hat B$ and no eigenvalue of $\hat C$ sum to zero.  Thus when
$\hat\rho$ has zero eigenvalues, ${\cal L}_{\hat\rho}(\hat A)$ is
not well defined for a general operator $\hat A$.  For the case of interest 
here, however, where $\hat A=\hat\Delta$, ${\cal L}_{\hat\rho}(\hat A)$ does
exist regardless of whether $\hat\rho$ has zero eigenvalues or not.  This can
be seen by constructing a solution.

Let $|j\rangle$ be an orthonormal basis that diagonalizes $\hat\rho$ and let
$\lambda_j$ be the associated eigenvalues.  Note that if $\lambda_j=0$, then
\beq
0=\langle j|\hat\rho|j\rangle=(1-t)\langle j|\hat\rho_0|j\rangle+
t\langle j|\hat\rho_1|j\rangle\;.
\eeq
Therefore, if $0<t<1$, we must have that both $\langle j|\hat\rho_0|j\rangle=0$
and $\langle j|\hat\rho_1|j\rangle=0$.  So if $\lambda_j=0$, then 
$\hat\rho_0^{1/2}|j\rangle=0$ and $\hat\rho_1^{1/2}|j\rangle=0$.  In particular,
sandwiching Eq.~(\ref{Lowering}) between eigenvectors of $\hat\rho$, we find
that
\beq
{1\over2}(\lambda_j+\lambda_k){\cal L}_{\hat\rho}(\hat\Delta)_{jk}=\Delta_{jk}
\eeq
has a solution for
${\cal L}_{\hat\rho}(\hat\Delta)_{jk}=
\langle j|{\cal L}_{\hat\rho}(\hat\Delta)|k\rangle$ because
$\Delta_{jk}=\langle j|\hat\Delta|k\rangle$ vanishes whenever 
$\lambda_j+\lambda_k=0$.  With this ${\cal L}_{\hat\rho}(\hat\Delta)$ becomes
\beq
{\cal L}_{\hat\rho}(\hat\Delta)\equiv
\sum_{\{j,k|\lambda_j+\lambda_k\ne0\}}\,
{2\over \lambda_j+\lambda_k}\Delta_{jk}|j\rangle\langle k|
\;,
\label{Rinverse}
\eeq
where we have conveniently set the terms in the null subspace of $\hat\rho$
to be zero.  (For further discussion of why Eq.~(\ref{Rinverse}) is the 
appropriate extension of ${\cal L}_{\hat\rho}(\hat\Delta)$ to the 
zero-eigenvalue subspaces of $\hat\rho$, see Ref.~\cite{Braunstein94a}; note 
that ${\cal L}_{\hat\rho}$ is denoted there by ${\cal R}^{-1}_{\hat\rho}$.)
Eq.~(\ref{Rinverse}) demonstrates that ${\cal L}_{\hat\rho}(\hat\Delta)$ can be
taken to be a Hermitian operator.

The super-operator ${\cal L}_{\hat\rho}$ is easily seen to satisfy
the defining property of a ``lowering'' super-operator, Eq.~(\ref{Lowering}),  
because, for Hermitian $\hat A$ and $\hat B$,
\beqa
\left|{\rm tr}\!\left(\hat\rho\hat A{\cal L}_{\hat\rho}(\hat B)\right)\right|
&\ge&
{\rm Re}\!\left[{\rm tr}\!\left(\hat\rho\hat A{\cal L}_{\hat\rho}(\hat B)
\right)\right]
\nonumber\\
&=&
\frac{1}{2}\left[\,{\rm tr}\!\left(\hat\rho\hat A{\cal L}_{\hat\rho}
(\hat B)\right)
+{\rm tr}\!\left(\hat\rho\hat A{\cal L}_{\hat\rho}(\hat B)\right)^{\!*}\,\right]
\rule{0mm}{7mm}
\nonumber\\
&=&
\frac{1}{2}\left[{\,\rm tr}\!\left(\hat\rho\hat A{\cal L}_{\hat\rho}
(\hat B)\right)
+{\rm tr}\!\left({\cal L}_{\hat\rho}(\hat B)\hat A\hat\rho\right)\right]
\rule{0mm}{7mm}
\nonumber\\
&=&
{\rm tr}\!\left(\hat A\,\frac{1}{2}\!\left(
{\cal L}_{\hat\rho}(\hat B)\hat\rho+\hat\rho{\cal L}_{\hat\rho}(\hat B)
\right)\right)
\rule{0mm}{7mm}
\nonumber\\
&=&
{\rm tr}(\hat A\hat B)\;.
\rule{0mm}{7mm}
\eeqa
The desired optimization is via Eq.~(\ref{SumBound1}):
\beqa
J''(t)
&\ge&
-{\rm tr}\!\left({\cal L}_{\hat\rho}(\hat\Delta)\hat\rho{\cal L}_{\hat\rho} 
(\hat\Delta)\right)
\nonumber\\
&=&
-{\rm tr}\!\left(\hat\Delta {\cal L}_{\hat\rho}(\hat\Delta)\right)
\rule{0mm}{7mm}
\nonumber\\
&=&
-\sum_{\{j,k|\lambda_j+\lambda_k\ne0\}}\,
{2\over \lambda_j+\lambda_k}\;
\bigl|\Delta_{jk}\bigr|^2\;.
\rule{0mm}{7mm}
\label{FisherBound}
\eeqa
Equality can be satisfied in Eq.~(\ref{FisherBound}) if 
\beq
{\rm Im}\!\left[{\rm tr}\!\left(\hat\rho\hat E_b
{\cal L}_{\hat\rho}(\hat\Delta)
\right)\right]=0\;\;\;
\mbox{for all $b$}\;,
\eeq
and, from Eq.~(\ref{FEqCond}),
\beq
{\cal L}_{\hat\rho}(\hat\Delta)\hat E_b^{1/2}=\mu_b\hat E_b^{1/2}
\;\;\;\mbox{for all $b$}\;.
\label{putate}
\eeq
The second of these can be met easily by choosing the operators
$\hat E_b=\hat E_b^{\rm F}=|b\rangle\langle b|$ to be projectors onto an
eigenbasis for the Hermitian operator ${\cal L}_{\hat\rho}(\hat\Delta)$ 
and choosing the constants $\mu_b$ to be the eigenvalues of
${\cal L}_{\hat\rho}(\hat\Delta)$.  The first condition then follows simply
by Eq.~(\ref{putate}) being satisfied. For
\beqa
{\rm tr}\!\left(\hat\rho\hat E_b{\cal L}_{\hat\rho}(\hat\Delta)\right)
&=&
{\rm tr}\!\left(\hat\rho\hat E_b^{1/2}\hat E_b^{1/2}
{\cal L}_{\hat\rho}(\hat\Delta)\right)
\nonumber\\
&=&
\mu_b\,{\rm tr}\!\left(\hat\rho\hat E_b\right)\;,
\rule{0mm}{6mm}
\eeqa
which is clearly a real number.

The function $L''(t)$ can now be defined as
\beq
L''(t)=-{\rm tr}\!\left(\hat\Delta {\cal L}_{\hat\rho}(\hat\Delta)\right)\;.
\label{GottaGo}
\eeq
This, as stated above, is exactly the function $L''(t)$ used by Holevo, but
obtained there by other means.  Again, among other things, what is new here
is that the derivation above gives a way of associating a measurement with
the function $L''(t)$.  This measurement may be used to define a new lower
bound to the accessible information $I(t)$; this bound we shall call $M(t)$.

The next step in the derivation of 
Eq.~(\ref{HolevoBound}), to show that $S''(t)\le L''(t)$; i.e., that
\beq
\sum_{\{j,k|\lambda_j +\lambda_k\neq0\}}\Phi\bigl(\lambda_j,\lambda_k\bigr)
\bigl|\Delta_{jk}\bigr|^2\;\ge\sum_{\{j,k|\lambda_j+\lambda_k\ne0\}}\,
{2\over \lambda_j+\lambda_k}\;\bigl|\Delta_{jk}\bigr|^2\;.
\label{BungeCord}
\eeq
This can be accomplished by demonstrating the arithmetic inequality 
\cite{Holevo73d,Mitrinovic64,Mitrinovic70} 
\beq
\Phi(x,y)\ge\frac{2}{x+y}\;.
\label{BugleBoy}
\eeq
That is, in words, that the arithmetic mean
of $x$ and $y$ is greater than or equal to their {\it logarithmic mean\/}
\cite{Bullen88}.  We reiterate Holevo's method of proof in particular.  The
case for $\Phi(x,x)$ is easy: equality is satisfied automatically. Suppose 
now that $x\ne y$ for $0<x,y\le1$ and let
\beq
s=\frac{x-y}{x+y}\;.
\eeq
Then
\beq
x=\frac{1}{2}(x+y)(1+s)
\;\;\;\;\;\;\mbox{and}\;\;\;\;\;\;
y=\frac{1}{2}(x+y)(1-s)\;,
\eeq
and $0<|s|<1$.  Hence
\beq
\ln x-\ln y=\ln(1+s)-\ln(1-s)
\eeq
has a convergent Taylor series expansion about $s=0$.  In particular, one has
\beq
\frac{1}{2}(x+y)\,\Phi(x,y)=1+\sum_{n=1}^\infty\frac{1}{(2n+1)}s^{2n}\;.
\eeq
Since this expansion contains only even powers of $s$, it follows immediately
that Eq.~(\ref{BugleBoy}) holds, with equality if and only if $x=y$.  This
completes the demonstration that the Holevo upper bound is indeed a
bound for the accessible information of a binary quantum communication channel.

The only piece remaining to be shown is that the bound is achieved if and
only if $\hat\rho_0$ and $\hat\rho_1$ commute.  The ``if'' side of the
statement is trivial; one need only choose the $\hat E_b$ to be projectors
onto a common basis diagonalizing both $\hat\rho_0$ and $\hat\rho_1$.  Then
one has immediately that
\beq
J(t)=I(t)=S(\hat\rho)-(1-t)S(\hat\rho_0)-t S(\hat\rho_1)\;.
\eeq
For the noncommuting case, if we can show that $L(t)$ is strictly less than $S(t)$, then our work will be done.

We do this by showing that $L(t)=S(t)$ implies that $\hat\rho_0$ and 
$\hat\rho_1$ commute.  Taking two derivatives of the supposition, we must
have that $S''(t)=L''(t)$.  Note, from Eq.~(\ref{BungeCord}) and the properties of $\Phi(x,y)$, that this holds if and only if
\beq
|\Delta_{jk}|^2=0\;\;\;\;\mbox{for all}\;\;k\ne j\;.
\label{Stamford}
\eeq
The latter condition implies that $\hat\rho_0$ and $\hat\rho_1$ commute.  This
is seen easily, for note that Eq.~(\ref{Stamford}) implies
\beqa
0
&=&
\sum_{k,j}(\lambda_j-\lambda_k)^2|\Delta_{jk}|^2
\nonumber\\
&=&
\sum_{k,j}(\lambda_j-\lambda_k)^2{\rm tr}\!
\left(\hat\Pi_k\hat\Delta\hat\Pi_j\hat\Delta\right)
\rule{0mm}{7mm}
\nonumber\\
&=&
\sum_{k,j}{\rm tr}\!\left(\hat\Pi_k\left(
\hat\rho\hat\Delta-\hat\Delta\hat\rho\right)\hat\Pi_j
\left(\hat\Delta\hat\rho-\hat\rho\hat\Delta\right)\rule{0mm}{5mm}\!\right)
\rule{0mm}{7mm}
\nonumber\\
&=&
\rule{0mm}{7mm}
{\rm tr}\!\left(\left(\hat\Delta\hat\rho-\hat\rho\hat\Delta\right)^{\!\dagger}
\!\left(\hat\Delta\hat\rho-\hat\rho\hat\Delta\right)\right)\;.
\eeqa
This implies $\hat\Delta\hat\rho-\hat\rho\hat\Delta=0$ and thus
$[\rho_0,\rho_1]=0$.  This completes the proof.

\subsection{The Lower Bound $M(t)$}

Now we focus on deriving an explicit expression for the lower bound $M(t)$ to
the accessible information.  This bound takes on a surprisingly simple form.
Moreover, as we shall see for the two dimensional case, this lower bound can
be quite close to the accessible information and sometimes actually equal to
it.

For simplicity, let us suppose $\hat\rho_0$ and $\hat\rho_1$ are invertible.
We start by rewriting, in the manner of Eq.~(\ref{KullTrace}), the mutual 
information as
\beq
J(t)=
{\rm tr}\biggl( (1-t)\hat\rho_0\sum_b\bigl(\ln\alpha_b\bigr)\hat E_b
+\;
t\,\hat\rho_1\sum_b\bigl(\ln\beta_b\bigr)\hat E_b\biggr)\,,
\eeq
where
\beq
\alpha_b={{\rm tr}\bigl(\hat\rho_0\hat E_b\bigr)\over
{\rm tr}\bigl(\hat\rho\hat E_b\bigr)}
\;\;\;\;\;\;\;\mbox{and}\;\;\;\;\;\;\;
\beta_b={{\rm tr}\bigl(\hat\rho_1\hat E_b\bigr)\over
{\rm tr}\bigl(\hat\rho\hat E_b\bigr)}\;.
\eeq
The lower bound $M(t)$ is defined by inserting the 
projectors $\hat E_b^{\rm F}$ onto a basis that diagonalizes 
${\cal L}_{\hat\rho}(\hat\Delta)$ into this formula.  This expression simplifies
because of a curious fact:  even though $\hat\rho_0$ and $\hat\rho_1$
need not commute, ${\cal L}_{\hat\rho}(\hat\rho_0)$ and
${\cal L}_{\hat\rho}(\hat\rho_1)$ necessarily {\it do\/} commute.  This follows
from the linearity of the ${\cal L}_{\hat\rho}$ super-operator:
\beqa
{\cal L}_{\hat\rho}(\hat\rho_0)
&=&
{{\cal L}_{\hat\rho}\bigl(\hat\rho-t\hat\Delta\bigr)}
\nonumber\\
&=&
\hat{\openone}-t{\cal L}_{\hat\rho}(\hat\Delta)\;,
\eeqa
and
\beqa
{\cal L}_{\hat\rho}(\hat\rho_1)
&=&
{\cal L}_{\hat\rho}\Big(\hat\rho+(1-t)\hat\Delta\Big)
\nonumber\\
&=&
\hat{\openone}+(1-t){\cal L}_{\hat\rho}(\hat\Delta)\;.
\eeqa
Thus the projectors $\hat E_b^{\rm F}$ that diagonalize
${\cal L}_{\hat\rho}(\hat\Delta)$ also clearly diagonalize both
${\cal L}_{\hat\rho}(\hat\rho_0)$ and ${\cal L}_{\hat\rho}(\hat\rho_1)$.

Therefore, if $|b\rangle$ is such that $\hat E_b^{\rm F}=|b\rangle\langle b|$ 
and $\alpha_{{\rm F}b}$ is the associated eigenvalue
of ${\cal L}_{\hat\rho}(\hat\rho_0)$, then the definition
of ${\cal L}_{\hat\rho}(\hat\rho_0)$, Eq.~(\ref{Lowering}), requires that
\beq
\alpha_{{\rm F}b}={\langle b|\hat\rho_0|b\rangle\over\langle b|\hat\rho
|b\rangle}={{\rm tr}\bigl(\hat\rho_0\hat E^{\rm F}_b\bigr)\over
{\rm tr}\bigl(\hat\rho\hat E^{\rm F}_b\bigr)}\;.
\eeq
Thus the operator ${\cal L}_{\hat\rho}(\hat\rho_0)$, much like the operator
$\hat M=\hat\rho_1^{-1/2}
\sqrt{\hat\rho_1^{1/2}\hat\rho_0\hat\rho_1^{1/2}}\hat\rho_1^{-1/2}$, can be
considered an operator analog to the classical likelihood ratio.  Similarly,
for the eigenvalues $\beta_{{\rm F}b}$ of ${\cal L}_{\hat\rho}(\hat\rho_1)$,
we must have
\beq
\beta_{{\rm F}b}={{\rm tr}\bigl(\hat\rho_1\hat E^{\rm F}_b\bigr)\over
{\rm tr}\bigl(\hat\rho\hat E^{\rm F}_b\bigr)}\;.
\eeq
Hence $M(t)$ takes the simple form
\beq
M(t)={\rm tr}\biggl((1-t)\,\hat\rho_0\ln
\Bigl({\cal L}_{\hat\rho}(\hat\rho_0)\Bigr)+
t\,\hat\rho_1\ln
\Bigl({\cal L}_{\hat\rho}(\hat\rho_1)\Bigr)\biggr)\,.
\label{MForm}
\eeq

As an aside, we note that one can also obtain by this method a lower bound 
to the quantum Kullback information (in the vein of
Eq.~(\ref{KullTrace})).  Using the measurement basis that diagonalizes
${\cal L}_{\hat\rho_1}(\hat\rho_0)$ in the classical Kullback-Leibler relative
information, we have, via the steps above, the expression
\beq
K_{\rm F}(\hat\rho_0/\hat\rho_1)\equiv{\rm tr}\biggl(\hat\rho_0\ln
\Bigl({\cal L}_{\hat\rho_1}(\hat\rho_0)\Bigr)\biggr)\;,
\label{FisherKullback}
\eeq
whenever ${\cal L}_{\hat\rho_1}(\hat\rho_0)$ is well defined.

There is a close relation between the lowering super-operator discussed here
and the optimal measurement operator $\hat M$ for statistical overlap found in
Section~\ref{MachoMan}.  This can be seen by noting that for small $\epsilon$
\beq
\sqrt{\hat A+\epsilon\hat B}\,\approx\,\sqrt{\hat A}\,+\,\frac{1}{2}\,
{\cal L}_{\hat A^{1/2}}(\hat B)\;,
\eeq
when $\hat A$ is invertible (as can be seen by squaring the left and right
hand sides), and for any operator $\hat B$ that commutes with
$\hat\rho$,
\beq
{\cal L}_{\hat\rho}(\hat B\hat A\hat B)=\hat B{\cal L}_{\hat\rho}(\hat A)
\hat B\;.
\eeq
Then, when
\beq
\hat\rho_0=\hat\rho_1+\delta\hat\rho\;,
\eeq
so that the two density operators to be distinguished are close to each other,
\beqa
\hat M
&=&
\hat\rho_1^{-1/2}\sqrt{\hat\rho_1^{1/2}(\hat\rho_1+\delta\hat\rho)
\hat\rho_1^{1/2}\,}\hat\rho_1^{-1/2}
\nonumber\\
&\approx&
\hat{\openone}\,+\,\frac{1}{2}\,{\cal L}_{\hat\rho_0}(\delta\hat\rho)\;.
\rule{0mm}{8mm}
\nonumber\\
&=&
\frac{3}{2}\hat{\openone}\,-\,\frac{1}{2}\,{\cal L}_{\hat\rho_0}(\hat\rho_1)\;.
\rule{0mm}{8mm}
\eeqa
Thus the measurement bases defined by $\hat M$ and
${\cal L}_{\hat\rho_0}(\hat\rho_1)$ in this limit can be taken to be the
same.

Equations~(\ref{FisherKullback}) and (\ref{KullTrace}), it should be noted,
are both distinct from that quantity usually considered the quantum analog to 
the Kullback-Leibler information in the literature \cite{Umegaki62,Ohya93}.  
That quantity, given by
\beq
K_{\rm U}(\hat\rho_0/\hat\rho_1)\equiv{\rm tr}\Bigl(
\hat\rho_0\ln\hat\rho_0-\hat\rho_0\ln\hat\rho_1\Bigr)\;,
\eeq
is not a lower bound to the maximum Kullback-Leibler information.  In fact,
the Holevo upper bound to the mutual information is easily seen to be 
expressible in terms of it:
\beq
I(t)\le{(1-t)}K_{\rm U}(\hat\rho_0/\hat\rho)+
tK_{\rm U}(\hat\rho_1/\hat\rho)\;.
\label{Manotoshi}
\eeq

\subsection{The Upper Bound $L(t)$}

The upper bound $L(t)$ has not yet yielded a form as sleek as the one found
for $M(t)$.  All that need be done in principle, of course, is integrate
Eq.~(\ref{GottaGo}) twice and apply the boundary conditions $L(0)=L(1)=0$. 
The trouble lies in that, whereas general methods exist for differentiating 
operators with respect to parameters
\cite{Feynman51a,Aizu63,Daleckii65a,Daleckii65b,Wilcox67,Horn94},
methods for the inverse problem of integration are nowhere to be found.
The main problem here reduces to finding a tractable representation for
${\cal L}_{\hat\rho}(\hat\Delta)$.  This has turned out to be a difficult
matter in spite of the fact that Eq.~(\ref{Lowering}) is a special case of the 
Lyapunov equation and has been studied widely in the mathematical literature.
Though there are convenient ways for finding numerical expressions for
${\cal L}_{\hat\rho}(\hat\Delta)$ \cite{Smith68} (even when the matrices
involved are 1000-dimensional \cite{Hodel92}), this is of no help to our
integration problem.  On the other hand, there do exist various methods for
obtaining an exact expression for ${\cal L}_{\hat\rho}(\hat\Delta)$ in a basis 
that does not depend on $t$ \cite{Smith66,Ma66,Lancaster70,Muller70,Young80}. 
These expressions can be integrated in principle.  However the
representations so found appear to have no compact form that makes obvious the 
algorithm for their calculation.

Two of the more useful representations for ${\cal L}_{\hat\rho}(\hat A)$
\cite{Heinz51,Smith65} might appear to be, a contour integral representation
({\it when $\hat\rho$ is invertible\/}),
\beq
{\cal L}_{\hat\rho}(\hat A)={2\over 2\pi i}\oint
\bigl(z\hat{\openone}-\hat\rho\bigr)^{-1}\hat A
\bigl(z\hat{\openone}+\hat\rho\bigr)^{-1}dz\;,
\label{ContourRep}
\eeq
where the contour contains the pole at $z=\lambda_j$ for all
eigenvalues $\lambda_j$ of $\hat\rho$, but does not contain the 
pole at $z=-\lambda_j$ for any $j$, and, more generally, a Riemann integral 
representation,
\beq
{\cal L}_{\hat\rho}(\hat A)=\int_{0}^\infty e^{-\hat\rho u/2}\hat A
e^{-\hat\rho u/2}\,du\;.
\label{RiemannRep}
\eeq
Both of these expressions can be checked easily enough by writing them out in
a basis that diagonalizes $\hat\rho$.

However, these two representations really lead nowhere.  Seemingly the best one
can do with them is use them to derive a {\it doubly infinite\/} (to be explained momentarily) Fourier sine series expansion for $L(t)$:
\beq
L(t)=\sum_{m=1}^\infty b_m\sin(m\pi t)\;.
\label{Fourier}
\eeq
This unfortunately does not have the compelling conciseness of 
Eq.~(\ref{MForm}), but at least it does automatically satisfy the boundary
conditions.  Perhaps one can hope that only the first few terms in
Eq.~(\ref{Fourier}) are significant.

Luckily it turns out that there are better, somewhat nonstandard representations
of ${\cal L}_{\hat\rho}$ to work with.  Nevertheless, before working toward a 
better representation, we go through the exercise of of building the Fourier 
expansion to illustrate the difficulties encountered.
The idea is to start off with a Taylor expansion for $L''(t)$ 
about $t=0$ and then use that in finding the coefficients $b_m$---these
being given by the standard Fourier algorithm,
\beq
b_m=-{2\over (m\pi)^2}\int_0^1 L''(t)\sin(m\pi t)\,dt\;.
\label{CoeffDef}
\eeq
The Taylor series expansion for $L''(t)$ is
\beq
L''(t)=-\sum_{n=0}^\infty{t^n\over n!}\left[{\rm tr}\!\left(
\hat\Delta{d^n \over dt^n} {\cal L}_{\hat\rho}(\hat\Delta)\right)
\right]_{t=0}\;.
\eeq
Now using the operator-inverse differentiation formula Eq.~(\ref{Mizas})
within the integral of Eq.~(\ref{ContourRep}), one finds the following.  
Differentiating once with respect to $t$ generates one term of the form
\beq
(z\hat{\openone}-\hat\rho)^{-1}\hat\Delta(z\hat{\openone}-\hat\rho)^{-1}\hat
\Delta(z\hat{\openone}+\hat\rho)^{-1}=\hat\Delta^{-1}\left[\hat\Delta
(z\hat{\openone}-\hat\rho)^{-1}\right]^2\left[\hat\Delta(z\hat{\openone}+
\hat\rho)^{-1}\right]
\eeq
and one term of the form
\beq
-(z\hat{\openone}-\hat\rho)^{-1}\hat\Delta(z\hat{\openone}+\hat\rho)^{-1}\hat
\Delta(z\hat{\openone}+\hat\rho)^{-1}=-\hat\Delta^{-1}\left[\hat\Delta
(z\hat{\openone}-\hat\rho)^{-1}\right]\left[\hat\Delta(z\hat{\openone}+
\hat\rho)^{-1}\right]^2\;.
\eeq
Differentiating again generates two terms of the form
\beq
\hat\Delta^{-1}\!\left[\hat\Delta(z\hat{\openone}-\hat\rho)^{-1}\right]^3\!
\left[\hat\Delta(z\hat{\openone}+\hat\rho)^{-1}\right]\;,
\eeq
two of the form
\beq
-\hat\Delta^{-1}\!\left[\hat\Delta(z\hat{\openone}-\hat\rho)^{-1}\right]^2\!
\left[\hat\Delta(z\hat{\openone}+\hat\rho)^{-1}\right]^2\;,
\eeq
and two of the form
\beq
\hat\Delta^{-1}\!\left[\hat\Delta(z\hat{\openone}-\hat\rho)^{-1}\right]\!
\left[\hat\Delta(z\hat{\openone}+\hat\rho)^{-1}\right]^3\;.
\eeq
The pattern quickly becomes apparent; in general, one has:
\beq
\hat\Delta{d^n \over dt^n} {\cal L}_{\hat\rho}(\hat\Delta)=
2(n!)\sum_{k=0}^{n}(-1)^k D_{\hat\rho}(n;k)\;,
\eeq
where
\beq
D_{\hat\rho}(n;k)=\frac{1}{2\pi i}\oint
\Bigl[\hat\Delta\bigl(z\hat{\openone}-\hat\rho\bigr)^{-1}\Bigr]^{n+1-k}
\Bigl[\hat\Delta\bigl(z\hat{\openone}+\hat\rho\bigr)^{-1}\Bigr]^{k+1}dz\;.
\label{UglyD}
\eeq
Putting Eqs.~(\ref{CoeffDef}) through (\ref{UglyD}) together, one arrives at
an expression for the expansion coefficients,
\beq
b_m={2\over m^3 \pi^3}\sum_{n=0}^\infty (n!)\!
\left[b(n;m) - (-1)^m\sum_{j=0}^n{1\over (n-j)!}b(j;m)\right]\!
\sum_{k=0}^n(-1)^k{\rm tr}\Bigl(D_{\hat\rho_0}(n;k)\Bigr)\;,
\eeq
where
\beq
b(j;m)=(-1)^{j/2}\bigl[1+(-1)^j\bigr](m\pi)^{-j}\;.
\eeq
The calculation of the terms ${\rm tr}\Bigl(D_{\hat\rho_0}(n;k)\Bigr)$ in
this expansion is straightforward but tedious.  Because each $b_m$ itself can
only be written in terms of an infinite series, the Fourier sine series here is
dubbed a doubly infinite series.

\subsubsection{A Better Way}

It turns out that one integration of the operator
${\cal L}_{\hat\rho}(\hat\Delta)$ is enough to give an explicit expression for 
$L(t)$ instead of the two integrations na\"{\i}vely called for in the
definition of $L''(t)$:
\beq
L(t)=-{\rm tr}\!\left(\hat\Delta\int^t\!\!\int^{t'}\!\!
{\cal L}_{\hat\rho(t'')}(\hat\Delta)\,dt''dt'\right)\;+\;c_1t\;+\;c_2\;,
\eeq
where now we are making the integration variable in $\hat\rho(t)$ explicit.
This reduction to one integration greatly simplifies things already, and we
develop it now.  The point can be seen from an integration by parts:
\beqa
\int^t\!\!\hat\rho(t'){\cal L}_{\hat\rho(t')}(\hat\Delta)\,dt'
&=&
\int^t\!\!\hat\rho(t')\left(\frac{d}{dt'}\int^{t'}\!\!
{\cal L}_{\hat\rho(t'')}(\hat\Delta)\,dt''\right)\!dt'
\nonumber\\
&=&
\int^t\!\left\{\,\frac{d}{dt'}\!\left(\hat\rho(t')\int^{t'}\!\!
{\cal L}_{\hat\rho(t'')}(\hat\Delta)\,dt''\right)\,-\,\hat\Delta
\int^{t'}\!\!{\cal L}_{\hat\rho(t'')}(\hat\Delta)\,dt''\,\right\}dt'
\rule{0mm}{9mm}
\nonumber\\
&=&
\hat\rho(t)\int^t\!\!{\cal L}_{\hat\rho(t')}(\hat\Delta)\,dt'\,-\,
\hat\Delta\int^t\!\!\int^{t'}\!\!{\cal L}_{\hat\rho(t'')}(\hat\Delta)\,dt''dt'.
\rule{0mm}{9mm}
\eeqa
Similarly
\beq
\int^t\!\!{\cal L}_{\hat\rho(t')}(\hat\Delta)\hat\rho(t')\,dt'=
\left(\int^t\!\!{\cal L}_{\hat\rho(t')}(\hat\Delta)\,dt'\right)\hat\rho(t)\,-\,
\left(\int^t\!\!\int^{t'}\!\!
{\cal L}_{\hat\rho(t'')}(\hat\Delta)\,dt''dt'\right)\hat\Delta
\;.
\eeq
Adding these expressions together and using the fact that
\beq
\hat\rho{\cal L}_{\hat\rho}(\hat\Delta)+{\cal L}_{\hat\rho}(\hat\Delta)\hat\rho
=2\hat\Delta\;,
\eeq
we get,
\beqa
&\mbox{}&
\int^t\!\!\int^{t'}\!\frac{1}{2}\!
\left(\hat\Delta{\cal L}_{\hat\rho(t'')}(\hat\Delta)
+{\cal L}_{\hat\rho(t'')}(\hat\Delta)\hat\Delta\right)dt''dt'
\nonumber\\
&\mbox{}&\;\;\;\;\;\;\;\;\;\;\;\;\;\;\;\;\;\;\;\;\;\;\;\;=
\frac{1}{2}\!\left(\hat\rho\!\left(\int^t\!\!
{\cal L}_{\hat\rho(t')}(\hat\Delta)\,dt'\right)
\!+\!\left(\int^t\!\!
{\cal L}_{\hat\rho(t')}(\hat\Delta)\,dt'\right)\!\hat\rho\rule{0mm}{7mm}\right)
-t\hat\Delta+\eta\;,
\rule{0mm}{9mm}
\nonumber\\
\eeqa
where $\hat\eta$ is a constant operator.  Therefore, using the linearity and the
cyclic property of the trace, we obtain
\beq
L(t)=-{\rm tr}\!\left(\hat\rho\int^t\!\!
{\cal L}_{\hat\rho(t')}(\hat\Delta)\,dt'\right)\;+\;c_1t\;+\;c_2\;.
\label{BoogerPicker}
\eeq

\subsubsection{Kronecker Product Methods}

There is another way of looking at the equation defining
${\cal L}_{\hat\rho}(\hat\Delta)$ [Eq.~(\ref{Lowering})] which leads to a way
of calculating the operator
\beq
\hat H\equiv\int^t\!\!{\cal L}_{\hat\rho}
(\hat\Delta)\,dt'\;.
\label{SnutBug}
\eeq
The resulting expression is not particularly elegant, but it does give an
operational way of carrying out this integral.

The method is to think of Eq.~(\ref{Lowering}) quite literally as a set of
simultaneous linear equations, the solution of which defines the matrix
elements of ${\cal L}_{\hat\rho}$.  Once one realizes this, then there is
some hope that Eq.~(\ref{Lowering}) may be rearranged to be of a form more
amenable to standard linear algebraic techniques.  This can be accomplished
by introducing the notion of {\it vec}'ing a matrix
\cite{Neudecker69,Graham81,Horn94}.
The operation of vec'ing a matrix is that of forming a vector by
stacking all its columns on top each other.  That is to say, if the elements
of a matrix $A$ are given by $A_{ij}$, $1\le i,j\le D$, i.e.,
\beq
A=\left[
\begin{array}{cccc}
A_{11} & A_{12} & \cdots & A_{1D} \\
A_{21} & A_{22} & \cdots & A_{2D} \\
\vdots & \cdots  & \ddots & \vdots \\
A_{D1} & A_{D2} & \cdots & A_{DD}
\end{array}\right]\;,
\eeq
then the (column) vector associated with the vec operation on $A$ is given by
\beq
{\rm vec}(A)=\left[
\begin{array}{c}
A_{11} \\ A_{21} \\ \vdots \\ A_{D1} \\ A_{12} \\ \vdots \\ A_{D2} \\ \vdots
\\ A_{DD}
\end{array}\right]\;.
\eeq
Note that the vec operation is not a simple mapping from operators to vectors
{\it because it is basis dependent}.

We shall need two simple properties of the vec operation.  The first, that it
is linear, is obvious.  The second is not obvious, but quite simple to derive
\cite[p.\ 255]{Horn94}: for any three matrices $A$, $B$, and $X$,
\beq
{\rm vec}(AXB)=(B^{\rm T}\otimes A){\rm vec}(X)\;,
\label{ToeKnuckle}
\eeq
where $B^{\rm T}$ denotes the transpose of $B$ and $\otimes$ denotes the
Kronecker or direct product defined by
\beq
A\otimes B=\left[
\begin{array}{cccc}
A_{11}B & A_{12}B & \cdots & A_{1D}B \\
A_{21}B & A_{22}B & \cdots & A_{2D}B \\
\vdots & \cdots  & \ddots & \vdots \\
A_{D1}B & A_{D2}B & \cdots & A_{DD}B
\end{array}\right]\;.
\eeq

Choosing a particular representation for the operators in the Lyapunov equation
Eq.~(\ref{GilbertGrape}) and letting $I$ be the identity matrix, we see that
it can be rewritten as
\beq
BXI+IXC=D\;.
\eeq
Using Eq.~(\ref{ToeKnuckle}) on this, we have finally that it is equivalent
to the system of linear equations given by
\beq
\Big(I\otimes B\,+\,C^{\rm T}\otimes I\Big){\rm vec}(X)\,=\,{\rm vec}(D)\;.
\eeq
A combination of matrices like on the left hand side of this is called
the Kronecker sum of $B$ and $C^{\rm T}$ and is denoted by $B\oplus C^{\rm T}$.
Therefore, using this notation, when $B\oplus C^{\rm T}$ is invertible we
have
\beq
{\rm vec}(X)\,=\,\Big(B\oplus C^{\rm T}\Big)^{\!-1}{\rm vec}(D)\;.
\eeq

Let us now apply these facts and notations toward finding an explicit
representation for the operator $\hat H$ in Eq.~(\ref{SnutBug}).  We start
by picking a particular basis $|j\rangle$, independent of $t$, in which to
express the operators $\hat\rho_0$, $\hat\rho_1$, $\hat\rho$, $\hat\Delta$, 
${\cal L}_{\hat\rho}(\hat\Delta)$, and $\hat H$.  Let us denote the matrix 
representations of these by the same symbol but with the hat removed, 
except for ${\cal L}_{\hat\rho}(\hat\Delta)$ which we represent by a
matrix $X$.  Since all these matrices are Hermitian, we have that
$\rho^{\rm T}=\rho^*$, etc., where the $*$ just means complex conjugation.  
(As an aside, note that for any Hermitian matrix $A$, the matrix $A^*$ is also
Hermitian.  Moreover, for any positive semidefinite matrix $B$, the matrix
$B^*$ is positive semidefinite. This follows because any positive semidefinite 
matrix $B$ can be written in the form $B=UDU^\dagger$ for some diagonal matrix 
$D=D^*$ and some unitary matrix $U$.  Therefore $B^*=(U^*)D(U^*)^\dagger$ is 
a positive semidefinite matrix because $U^*$ is a unitary matrix.)
Then, for invertible $\hat\rho_0$ and $\hat\rho_1$, we have immediately that
\beq
{\rm vec}(X)\,=\,2\,\Big(\rho\oplus\rho^*\Big)^{\!-1}{\rm vec}(\Delta)\;,
\eeq
where the matrix $\rho\oplus\rho^*$ is positive definite.
Finding the matrix $H$ simply corresponds to unvec'ing the vector
\beq
2\left(\int^t\!\!\Big(\rho\oplus\rho^*\Big)^{\!-1}dt'\right){\rm vec}(\Delta)\;.
\label{mim}
\eeq

Our problem thus reduces to evaluating the operator integral in Eq.~(\ref{mim}).
For this purpose, we rearrange the integrand as follows:
\beqa
\Big(\rho\oplus\rho^*\Big)^{\!-1}
&=&
\Big((\rho_0\oplus\rho_0^*)\,+\,t\,(\Delta\oplus\Delta^*)\Big)^{\!-1}
\nonumber\\
&=&
\tilde{\rho}_0^{-1/2}\left(\tilde{I}\;+\;t
\left(\tilde{\rho}_0^{-1/2}\tilde{\Delta}\tilde{\rho}_0^{-1/2}\right)
\rule{0mm}{5mm}\right)^{\!-1}\tilde{\rho}_0^{-1/2}\;,
\rule{0mm}{8mm}
\label{samba}
\eeqa
where
\beq
\tilde{\rho}_0=\rho_0\oplus\rho_0^*
\;\;\;\;\;\;\;\;\;\;\mbox{and}\;\;\;\;\;\;\;\;\;\;
\tilde{\Delta}=\Delta\oplus\Delta^*\;,
\eeq
and $\tilde{I}$ is the appropriate sized identity matrix.
If we suppose that $\hat\Delta$ is invertible, Eq.~(\ref{samba}) can be
integrated immediately.  This is because the matrices $\tilde{I}$ and
$\tilde{\rho}_0^{-1/2}\tilde{\Delta}\tilde{\rho}_0^{-1/2}$ commute.  Thus
\beqa
\int^t\!
\left(\tilde{I}\;+\;t'
\left(\tilde{\rho}_0^{-1/2}\tilde{\Delta}\tilde{\rho}_0^{-1/2}\right)
\rule{0mm}{5mm}\right)^{\!-1}dt'
&=&
\left(\tilde{\rho}_0^{-1/2}\tilde{\Delta}\tilde{\rho}_0^{-1/2}\right)^{\!-1}
\ln\!\left(\tilde{I}\;+\;t
\left(\tilde{\rho}_0^{-1/2}\tilde{\Delta}\tilde{\rho}_0^{-1/2}\right)\;,
\rule{0mm}{5mm}\right)
\nonumber\\
\rule{0mm}{6mm}
\eeqa
and so
\beqa
{\rm vec}(H)
&=&
2\,\tilde{\rho}_0^{-1/2}\!
\left(\tilde{\rho}_0^{-1/2}\tilde{\Delta}\tilde{\rho}_0^{-1/2}\right)^{\!-1}
\ln\!\left(\tilde{I}\;+\;t
\left(\tilde{\rho}_0^{-1/2}\tilde{\Delta}\tilde{\rho}_0^{-1/2}\right)
\rule{0mm}{5mm}\right)\tilde{\rho}_0^{-1/2}\,{\rm vec}(\Delta)\;.
\nonumber\\
\rule{0mm}{6mm}
\label{ShoutsAndMurmurs}
\eeqa
The logarithm in this is well defined because the operators in its argument
are invertible \cite[p.\ 474]{Horn94}.

Eq.~(\ref{ShoutsAndMurmurs}) contains the algorithm for calculating $\hat H$
in the given basis.  This then may be used in conjunction with
Eq.~(\ref{BoogerPicker}) to calculate $L(t)$.  An interesting open question
is whether there exists a basis in which to represent the operators
$\hat\rho_0$, $\hat\rho_1$, etc., so that the matrix appearing in the logarithm
above is diagonal.  If so, this would greatly simply the computation of $L(t)$.

\subsection{The Two-Dimensional Case}

In this Subsection we consider the important special case of binary 
communication channels on two-dimensional Hilbert spaces.  Here the new
bounds are readily expressible in terms of elementary functions and, moreover, 
the optimal orthogonal projection-valued measurement can be found via a 
variational calculation (just as was possible with the statistical overlap).  
With this case, one can gain a feel for how tightly the new bounds delimit the 
true accessible information $I(t)$.

Let the signal states $\hat\rho_0$ and $\hat\rho_1$ again be represented
by two vectors within the Bloch sphere, 
i.e.,
\beq
\hat\rho_0=\frac{1}{2}\Bigl(\hat{\openone}+\vec{a}\cdot\vec{\sigma}\Bigr)
\;\;\;\;\;\mbox{and}\;\;\;\;\; 
\hat\rho_1=\frac{1}{2}\Bigl(\hat{\openone}+\vec{b}\cdot\vec{\sigma}\Bigr)\;.
\eeq
The total density matrix for the channel can then be written as
\beq
\hat\rho=\frac{1}{2}\Bigl(\hat{\openone}+\vec{c}\cdot\vec{\sigma}\Bigr)\;,
\label{GrandPoobah}
\eeq
where
\beqa
\vec{c}
&=&
(1-t)\vec{a}+t\vec{b}
\nonumber\\
&=&
\vec{a}+t\vec{d}
\nonumber\\
&=&
\vec{b}-(1-t)\vec{d}
\eeqa
and 
\beq
\vec{d}=\vec{b}-\vec{a}\;.
\eeq
For an orthogonal projection-valued measurement 
specified by the Bloch vector $\vec{n}/n$, the mutual 
information takes
the form
\beq
J(t;\vec{n})=(1-t)\,K\bigl(\hat\rho_0/\hat\rho;\vec{n}\bigr)+
t\,K\bigl(\hat\rho_1/\hat\rho;\vec{n}
\bigr)\;,
\label{binary2D}
\eeq
where
\beq
K\bigl(\hat\rho_0/\hat\rho;\vec{n}\bigr)
={1\over2n}\!\left[
\Big(n+\vec{a}\cdot\vec{n}\Big)\ln\!\left({n+\vec{a}\cdot\vec{n}\over
n+\vec{c}\cdot\vec{n}}\right)+
\Big(n-\vec{a}\cdot\vec{n}\Big)\ln\!\left({n-\vec{a}\cdot\vec{n}\over
n-\vec{c}\cdot\vec{n}}\right)\right]
\label{mutinf2D1}
\eeq
and
\beq
K\bigl(\hat\rho_1/\hat\rho;\vec{n}\bigr)
={1\over2n}\!\left[
\Big(n+\vec{b}\cdot\vec{n}\Big)\ln\!\left({n+\vec{b}\cdot\vec{n}\over
n+\vec{c}\cdot\vec{n}}\right)+
\Big(n-\vec{b}\cdot\vec{n}\Big)\ln\!\left({n-\vec{b}\cdot\vec{n}\over
n-\vec{c}\cdot\vec{n}}\right)\right]\;.
\label{mutinf2D2}
\eeq
The optimal projector is found by varying expression~(\ref{binary2D}) over
all vectors $\vec{n}$. The resulting equation for the optimal $\vec{n}$ is
\beq
0\,=\,(1-t)\ln\!\left({\Big(n+\vec{c}\cdot\vec{n}\Big)\Big(n-\vec{a}\cdot
\vec{n}\Big)\over
\Big(n-\vec{c}\cdot\vec{n}\Big)\Big(n+\vec{a}\cdot\vec{n}\Big)}
\rule{0mm}{8.2mm}\right)\,\vec{a}_\perp\;
+\;t\,\ln\!\left({\Big(n+\vec{c}\cdot\vec{n}\Big)\Big(n-\vec{b}\cdot\vec{n}
\Big)\over
\Big(n-\vec{c}\cdot\vec{n}\Big)\Big(n+\vec{b}\cdot\vec{n}\Big)}
\rule{0mm}{8.2mm}\right)\,\vec{b}_\perp\;,
\label{VaryEq}
\eeq
where
\beq
\vec{a}_\perp=\vec{a}-\!\left(\vec{a}\cdot\frac{\vec{n}}{n}\right)\!
\frac{\vec{n}}{n}
\label{DisBeLloyd}
\eeq
and
\beq
\vec{b}_\perp=\vec{b}-\!\left(\vec{b}\cdot\frac{\vec{n}}{n}\right)\!
\frac{\vec{n}}{n}
\eeq
are vectors perpendicular to $\vec{n}$.
Equation~(\ref{VaryEq}) is unfortunately a transcendental equation and as such 
generally has no explicit solution.  That is really no problem, however, for
given any particular $\vec{a}$, $\vec{b}$, and $t$, a numerical solution for
$\vec{n}$ can easily be computed.
Nevertheless, it is of some interest to classify the cases in which one can 
actually write out an explicit solution to Eq.~(\ref{VaryEq}).  There are four 
nontrivial situations where this is possible:
\begin{enumerate}
\item a classical channel, 
where $\hat\rho_0$ and $\hat\rho_1$ commute (i.e., $\vec{a}$ and 
$\vec{b}$ are parallel),
\item $\hat\rho_0$ and $\hat\rho_1$ are both pure 
states (i.e., $a=b=1$), 
\item $a=b$ and $t=\frac{1}{2}$, and 
\item $t$ is explicitly determined by $\hat\rho_0$ and $\hat\rho_1$ according
to
\beq
t=\left(1+\sqrt{{1-b^2\over 1-a^2}}\,\right)^{\!-1}\;.
\eeq
\end{enumerate}
In case (1), taking $\vec{n}/n$ parallel to $\vec{a}$ and $\vec{b}$
causes $\vec{a}_\perp$ and $\vec{b}_\perp$ to vanish.
When conditions (2)--(4) are fulfilled, Eq.~(\ref{VaryEq}) can be solved by
requiring that the arguments of the logarithms be multiplicative inverses, i.e.,
\beq
{\Big(n+\vec{c}\cdot\vec{n}\Big)\Big(n-\vec{a}\cdot\vec{n}\Big)\over
\Big(n-\vec{c}\cdot\vec{n}\Big)\Big(n+\vec{a}\cdot\vec{n}\Big)}\,=\,
{\Big(n-\vec{c}\cdot\vec{n}\Big)\Big(n+\vec{b}\cdot\vec{n}\Big)\over
\Big(n+\vec{c}\cdot\vec{n}\Big)\Big(n-\vec{b}\cdot\vec{n}\Big)}\;,
\eeq
and choosing $\vec{n}$ such that
\beq
(1-t)\vec{a}_\perp=t\vec{b}_\perp\;.
\label{Rupert}
\eeq
Cases (2) and (3), reported previously by Levitin \cite{Levitin93}, are limits 
of case (4).

The condition of Eq.~(\ref{Rupert}) in the exactly solvable cases 
(2)--(4) is equivalent to the requirement that
\beq
\vec{n}=t\vec{b}-(1-t)\vec{a}\;.
\label{HelMeas}
\eeq
This is of significance because, {\it for arbitrary\/} $t$, the measurement that
minimizes the probability of error measure of distinguishability studied in
Section~\ref{GiveEmHell} is just the measurement specified by
Eq.~(\ref{HelMeas}) \cite{Helstrom67,Helstrom76}.  Thus, in the cases where
the optimal information
gathering measurement can be written explicitly, it coincides with the optimal
error probability measurement.  Now, simply rewriting $t$ and $1-t$ in terms of
the vectors $\vec{a}$, $\vec{b}$, and $\vec{c}$, the vector~(\ref{HelMeas})
becomes,
\beq
\vec{n}\,=\,{\vec{a}\cdot\!\Big(\vec{a}-\vec{c}\,\Big)\over\vec{a}\cdot\!\Big(
\vec{a}-\vec{b}\,\Big)}\,\vec{b}\,-\,
{\vec{b}\cdot\!\Big(\vec{b}-\vec{c}\,\Big)\over\vec{b}\cdot\!\Big(\vec{b}-
\vec{a}\,\Big)}
\,\vec{a}\;
\label{UgMeas}
\eeq
In cases~(1)--(3) this can be seen to reduce to
\beqa
\vec{n}_0
&=&
\Big(1-\vec{a}\cdot\vec{c}\,\Big)\,\vec{b}\,-\,\Big(1-\vec{b}\cdot\vec{c}\,
\Big)\,
\vec{a}
\nonumber\\
&=&
\vec{d}\,+\,\vec{c}\times\!\Big(\vec{c}\times\vec{d}\,\Big)\;.
\rule{0mm}{6mm}
\label{optmeas}
\eeqa
(It should be kept in mind that in going from Eqs.~(\ref{HelMeas}) to 
(\ref{UgMeas}) to (\ref{optmeas}) the length of $\vec{n}$ has been allowed to 
vary.)

Case~(2), and consequently the orthogonal projection-valued measurement 
set by Eq.~(\ref{optmeas}), is of particular interest in communication
theory.  This is because two pure states in {\it any\/} Hilbert space 
span only a two-dimensional subspace of that Hilbert space.  Hence Eq.~(\ref{optmeas}) remains valid as the optimal orthogonal projection-valued 
measurement for a pure-state binary channel in a Hilbert space of any dimension.
Moreover, Levitin~\cite{Levitin95} has shown that in this case the optimal 
orthogonal projection-valued measurement is indeed the actual optimal measurement.

Let us write out this case in more standard notation.  Suppose
$\hat\rho_0=|\psi_0\rangle\langle\psi_0|$ and 
$\hat\rho_1=|\psi_1\rangle\langle\psi_1|$ and let
\beqa
q
&\equiv&
|\langle\psi_0|\psi_1\rangle|^2
\nonumber\\
&=&
{\rm tr}(\hat\rho_0\hat\rho_1)
\nonumber\\
&=&
\frac{1}{2}\left(1+\vec{a}\cdot\vec{b}\,\right)\;.
\eeqa
Then, because $a=b=1$ here, the optimal measurement given by Eq.~(\ref{HelMeas})
has norm
\beq
n=\sqrt{1-4t(1-t)q\,}\;.
\eeq
Using this and some algebra, the expression for the accessible information
$I(t)$ reduces to
\beqa
I(t)\!
&=&\!
\frac{1}{2n}\left\{\rule{0mm}{7mm}(1-t)\!\left[
\Bigl(n+1-2tq\Bigr)\ln\!\left(\frac{1+n}{2(1-t)}\right)\,+\,
\Bigl(n-1+2tq\Bigr)\ln\!\left(\frac{1-n}{2(1-t)}\right)\right]\right.
\nonumber\\
& &\;\;\;+\,
t\!\left.\left[
\Bigl(n+1-2(1-t)q\Bigr)\ln\!\left(\frac{1+n}{2t}\rule{0mm}{6.2mm}\right)\,+\,
\Bigl(n-1+2(1-t)q\Bigr)\ln\!\left(\frac{1-n}{2t}\rule{0mm}{6.2mm}\right)\right]
\rule{0mm}{7mm}\right\}
\rule{0mm}{10mm}
\nonumber\\
\rule{0mm}{4mm}
\label{LessUgly}
\eeqa
This confirms Levitin's expression in Ref.~\cite{Levitin93} though this version
is significantly more compact.

With this much known about the exact solutions, let us return to the question
of how well the bounds $M(t)$, $L(t)$, etc., fare in comparison.  For a
measurement specified by the Bloch vector $\vec{n}/n$, the second derivative of
the mutual information takes the form
\beq
J''(t)=-\,{\Big(\vec{d}\cdot\vec{n}\Big)^{\!2}\over 
n^2-\Big(\vec{c}\cdot\vec{n}\Big)^{\!2}}\;.
\eeq
The vector $\vec{n}$ that minimizes this is again given easily enough by a 
variational calculation; the equation specifying the nontrivial solution is
\beq
\left(n^2-\Big(\vec{c}\cdot\vec{d}\,\Big)^{\!2}\right)\vec{d}\,-\,\Big(\vec{d}
\cdot\vec{n}\Big)\!\left(\rule{0mm}{5.5mm}
\vec{n}-\Big(\vec{d}\cdot\vec{n}\Big)\vec{c}\,\right)\,=\,0\;.
\eeq
After a bit of algebra one finds its solution to be given by none other than 
\beq
\vec{n}_0=\Big(1-\vec{a}\cdot\vec{c}\,\Big)\,\vec{b}-\Big(1-\vec{b}\cdot\vec{c}
\,\Big)\,\vec{a}\;,
\eeq
the vector given by Eq.~(\ref{optmeas}); this time, however, the 
expression is valid for all $\vec{a}$, $\vec{b}$, and $t$.  
Inserting this vector into expressions~(\ref{mutinf2D1}) and (\ref{mutinf2D2}) produces the lower bound $M(t)$, 
\beq
M(t)\,=\,(1-t)\,K\bigl(\hat\rho_0/\hat\rho;\vec{n}_0\bigr)\,+\,
t\,K\bigl(\hat\rho_1/\hat\rho;\vec{n}_0\bigr)\;.
\eeq

\begin{figure}
\begin{center}
\leavevmode
\epsfig{figure=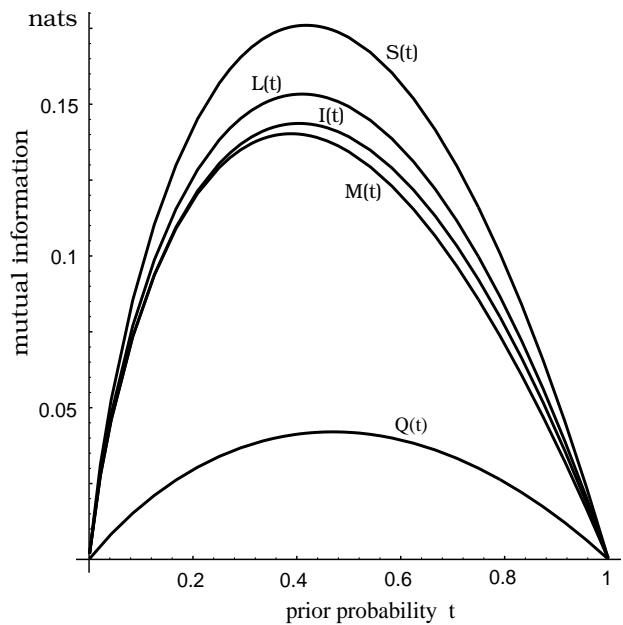,width=5.25in,height=7in}
\caption{The Holevo upper bound $S(t)$, the upper bound 
$L(t)$, the information $I(t)$ extractable by optimal orthogonal 
projection-valued measurement (found numerically), the lower bound 
$M(t)$, and the Jozsa-Robb-Wootters lower bound $Q(t)$, all for the 
case that $\hat\rho_0$ is pure ($a=1$), $\hat\rho_1$ is mixed with 
$b=2/3$, and the angle between the two Bloch vectors is $\pi/3$.}
\label{snouter}
\end{center}
\end{figure}

The upper bound $L(t)$ is found by integrating $J''(t)$ back up but with this 
particular measurement in place; that is to say, by integrating
\beq
L''(t)=-\left(d^2+\frac{1}{1-c^2}\Big(\vec{c}\cdot\vec{d}\,\Big)^{\!2}\right)\;.
\eeq
The result, upon requiring the boundary conditions $L(0)=L(1)=0$, is
\beq
L(t)\,=\,{\delta\over2d^2}\!\left(\rule{0mm}{5mm}
-\Big(\delta-\vec{c}\cdot\vec{d}\,\Big)\ln\!\Big(\delta-\vec{c}\cdot\vec{d}\,
\Big)\,
-\,\Big(\delta+\vec{c}\cdot\vec{d}\,\Big)\ln\!\Big(\delta+\vec{c}\cdot\vec{d}\,
\Big)
\,+\,\beta_1t\,+\,\beta_2\,\right)\;,
\eeq
where
\beq
\beta_2\,=\,\Big(\delta-\vec{a}\cdot\vec{d}\,\Big)\ln\!\Big(\delta-\vec{a}\cdot
\vec{d}\,\Big)\,+\,
\Big(\delta+\vec{a}\cdot\vec{d}\,)\ln\!\Big(\delta+\vec{a}\cdot\vec{d}\,\Big)\;,
\eeq
and
\beq
\beta_1\,=\,\Big(\delta-\vec{b}\cdot\vec{d}\,\Big)\ln\!\Big(\delta-\vec{b}
\cdot\vec{d}\,\Big)+
\Big(\delta+\vec{b}\cdot\vec{d}\,\Big)\ln\!\Big(\delta+\vec{b}\cdot\vec{d}\,
\Big)\,-\,\beta_2\;,
\eeq
and
\beqa
\delta
&=&
\sqrt{\Big(1-\vec{a}\cdot\vec{b}\,\Big)^2-(1-a^2)(1-b^2)}
\nonumber\\
&=&
\sqrt{d^2-\Big|\vec{c}\times\vec{d}\,\Big|^2}
\rule{0mm}{8mm}
\nonumber\\
&=&
\sqrt{d^2-\Big|\vec{a}\times\vec{b}\,\Big|^2}\;.
\rule{0mm}{8mm}
\eeqa
In contrast, the Jozsa-Robb-Wootters lower bound and the Holevo upper bound are 
given by Eqs.~(\ref{JozsaBound}) and (\ref{HolevoBound}), respectively, where in
Bloch vector representation
\beq
Q(\hat\rho)\,=\,\frac{1}{4c}\!\left((1-c)^2\,\ln\!\left(\frac{1-c}{2}\right)
\,-\,(1+c)^2\,\ln\!\left(\frac{1+c}{2}\right)\rule{0mm}{6mm}\right)\;,
\eeq
and
\beq
S(\hat\rho)\,=\,-\frac{1}{2}(1-c)\,\ln\!\left(\frac{1-c}{2}\right)\,-\,
\frac{1}{2}(1+c)\,\ln\!\left(\frac{1+c}{2}\right)\;,
\eeq
and similarly for $Q(\hat\rho_0)$, $S(\hat\rho_0)$, etc.
The extent to which the bounds $M(t)$ and $L(t)$ are tighter than the bounds
$Q(t)$ and $S(t)$ and the degree to which they conform to the exact numerical
answer $I(t)$ is illustrated by a typical example in Fig.~\ref{snouter}.

\subsection{Other Bounds}

\subsubsection{Another Bound from the Schwarz Inequality}

Because of the difficulties in constructing a concise expression for $L(t)$,
it is worthwhile to look at another upper bound to $I(t)$ derived from the
Schwarz inequality.  This is a bound derived from the second usage of
lowering operators ${\cal G}_{\hat\rho}$, i.e., that in Eq.~(\ref{SumBound2}),
but with ${\cal L}_{\hat\rho}$ in particular.  We can immediately write
\beq
J''(t)\ge-{\rm tr}\!\left({\cal L}_{\hat\rho^{1/2}}(\hat\Delta)^2\right)\equiv
N''(t)\;,
\label{SmurfCity}
\eeq
since ${\cal L}_{\hat\rho^{1/2}}(\hat\Delta)$ is Hermitian and guaranteed to
exist for the same reason ${\cal L}_{\hat\rho}(\hat\Delta)$ does.
The right hand side of this, when integrated twice and required to meet the 
boundary conditions
\beq
N(0)=N(1)=1\;,
\eeq
gives a new upper bound $N(t)$.

Since the bound derived from Eq.~(\ref{SmurfCity}) is necessarily worse than
$L(t)$, this would be of little interest if it were not for the following fact.
Consider any positive operator $\hat X$ that depends on a parameter $t$.  Then
\beqa
{d\hat X\over dt}
&=&{d\over dt}\left(\sqrt{\hat X}\sqrt{\hat X}\,\right)
\nonumber\\
&=&
\sqrt{\hat X}\,{d\sqrt{\hat X}\over dt}+
{d\sqrt{\hat X}\over dt}\,\sqrt{\hat X}\;.
\eeqa
Thus,
\beq
{d\sqrt{\hat X}\over dt}=2\,{\cal L}_{\hat X^{1/2}}\!\left(
{d\hat X\over dt}\right)\;,
\eeq
and specifically,
\beq
N''(t)=-\frac{1}{4}\,{\rm tr}\!\left[\left(
{d\displaystyle{\sqrt{\hat\rho\,}}\over dt}\right)^{\!2}\,\,\right]\;.
\label{GooberPea}
\eeq
The simplicity of this expression would apparently give hope that there may be 
some way of integrating it explicitly.

The reason it would be useful to find $N(t)$ is that, though it
cannot be as tight as the $L(t)$ bound, it still beats the Holevo bound $S(t)$.  This can be seen as follows.  In a basis $|j\rangle$ diagonalizing $\hat\rho$,
\beq
N''(t)=-\sum_{\{j,k|\lambda_j+\lambda_k\ne0\}}\,
\left({2\over\sqrt{\lambda_j}+\sqrt{\lambda_k}}\right)^{\!2}\;
\bigl|\Delta_{jk}\bigr|^2\;.
\eeq
Now, as in the proof of the Holevo bound, $S(t)\ge N(t)$ requires that
\beq
\Phi(x,y)\ge\left({2\over\sqrt{x}+\sqrt{y}}\right)^{\!2}\;.
\eeq
This can be shown very simply.  The case for $\Phi(x,x)$ is again automatic.
Suppose $0<x,y\le1$ and $x\ne y$.  We already know that $\Phi(x,y)\ge2/(x+y)$.
Consequently,
\beq
\Phi\!\left(\sqrt{x},\sqrt{y}\right)\ge\frac{2}{\sqrt{x}+\sqrt{y}}\;.
\eeq
But that is to say,
\beq
{\frac{1}{2}(\ln x-\ln y)\over\sqrt{x}-\sqrt{y}}\ge{2\over\sqrt{x}+\sqrt{y}}\;.
\eeq
Multiplying both sides of this by $2/(\sqrt{x}+\sqrt{y})$ gives the desired
result.

Unfortunately, despite the seeming simplicity of Eq.~(\ref{GooberPea}), little progress has been made toward its general integration.  This in itself may be an
indication that the simplicity is only apparent.  This can be seen with the
example of $2\times2$ density operators.  Let us report enough of these results
to make this point convincing.

The operator $\sqrt{\hat\rho\,}$ can represented with the help of the unit 
operator and the Pauli matrices as
\beq
\sqrt{\hat\rho\,}\,=\,r_0\hat{\openone}\,+\,\vec{r}\cdot\vec{\sigma},
\eeq
where the requirement that $\sqrt{\hat\rho\,}$ squares to give the representation of $\hat\rho$ in Eq.~(\ref{GrandPoobah}) forces
\beq
r_0^2\,+\,r^2\,=\,\frac{1}{2}
\eeq
and
\beq
2r_0\,\vec{r}\,=\,\frac{1}{2}\,\vec{c}\;.
\eeq
These two requirements go together to give a quartic equation specifying $r_0$.
Since $\sqrt{\hat\rho\,}$ must be a positive operator, its eigenvalues
$r_0+r$ and $r_0-r$ must both be positive.  Hence, picking the largest value
of $r_0$ consistent with the quartic equation, we find
\beq
r_0\,=\,\frac{1}{2}\left(1+\sqrt{1-c^2}\right)^{\!1/2}
\eeq
and
\beq
\vec{r}\,=\,\frac{1}{4r_0}\,\vec{c}\;.
\eeq
Taking the derivative of these quantities with respect to $t$, we find that
\beq
N''(t)\,=\,-\frac{1}{2}\left( (r'_0)^2\,+\,\vec{r}\,'\cdot\vec{r}\,'\right)\;,
\eeq
which reduces after quite some algebra to
\beq
N''(t)\,=\,-\frac{1}{32r_0^2}\!\left(\,
\frac{(\vec{d}\cdot\vec{c})^{2}}{1-c^2}\!
\left(1-\frac{1}{8r_0^2}\right)\,+\,d^2\right)\;,
\eeq
and really cannot be reduced any further.  Clearly this is a considerable mess
as a function of $t$, and the actual expression for $N(t)$ is far worse.

\subsubsection{A Bound Based on Jensen's Inequality}

Still another upper bound to the accessible information can be built
from the technique developed to find the function $L''(t)$.  This bound makes
crucial use of the concavity of the logarithm function.

For simplicity, let us suppose that both $\hat\rho_0$ and $\hat\rho_1$ are
invertible. Recall the representation, given by Eq.~(\ref{MutualInf}), of the 
mutual information as the average of two Kullback-Leibler relative informations.  For each of these terms, we have two different bounds that come Jensen's 
inequality \cite{Mitrinovic70}, i.e., for any probability distribution $q(b)$,
\beq
\sum_b q(b)\ln x_b\;\le\;\ln\!\left(\sum_b q(b)x_b\right)\;.
\eeq
The first bound is that
\beqa
K(p_i/p)
&=&
\sum_b p_i(b)\ln\!\left(\frac{p_i(b)}{p(b)}\right)
\nonumber\\
&\le&
\ln\!\left(\sum_b\frac{p_i(b)^2}{p(b)}\right)\;.
\label{MooGoo}
\rule{0mm}{8mm}
\eeqa
The second is that
\beqa
K(p_i/p)
&=&
-\sum_b p_i(b)\ln\!\left(\frac{p(b)}{p_i(b)}\right)
\nonumber\\
&=&
-2\sum_b p_i(b)\ln\!\left(\left(\frac{p(b)}{p_i(b)}\right)^{\!1/2}\,\right)
\rule{0mm}{9mm}
\nonumber\\
&\ge&
-2\,\ln\!\left(\sum_b\sqrt{p_i(b)p(b)}\right)
\rule{0mm}{8mm}
\eeqa
Therefore, it follows that the quantum Kullback information
$K(\hat\rho_i/\hat\rho)$ is bounded by
\beq
-2\,\ln\!\left(\min_{\{\hat E_b\}}\sum_b\sqrt{{\rm tr}(\hat\rho_i\hat E_b)}
\sqrt{{\rm tr}(\hat\rho\hat E_b)}\right)\;\le\;K(\hat\rho_i/\hat\rho)\;\le\;
\ln\!\left(\max_{\{\hat E_b\}}\sum_b{\left({\rm tr}(\hat\rho_i\hat E_b)
\right)^{\!2}\over{\rm tr}(\hat\rho\hat E_b)}\right)\;.
\eeq
These bounds can be evaluated explicitly by the techniques of
Sections~\ref{MachoMan} and \ref{ChompaCakes}.  Namely, we have
\beq
-2\,\ln\!\left({\rm tr}\,\sqrt{\hat\rho^{1/2}\hat\rho_i\hat\rho^{1/2}}\,\right)
\;\le\;K(\hat\rho_i/\hat\rho)\;\le\;\ln\!\left({\rm tr}\Big(\hat\rho_i
{\cal L}_{\hat\rho}(\hat\rho_i)\Big)\rule{0mm}{5mm}\right)\;.
\label{FavoriteShadows}
\eeq
(The upper bound is assured to exist in the form stated because
${\cal L}_{\hat\rho}(\hat\rho_i)$ is itself well defined; this follows because
the $\hat\rho_i$ are assumed invertible.)

One of these bounds may be used to upper bound the accessible information.
In particular, since the maximum of a sum is less than or equal to the sum of
the maxima, it follows that
\beq
I(t)\;\le\;(1-t)\,\ln\!\left({\rm tr}\Big(\hat\rho_0{\cal L}_{\hat\rho}
(\hat\rho_0)\Big)\rule{0mm}{5mm}\right)\;+\;t\,\ln\!\left({\rm tr}
\Big(\hat\rho_1{\cal L}_{\hat\rho}(\hat\rho_1)\Big)\rule{0mm}{5mm}\right)
\;\equiv\;R(t)\;.
\label{GuyPan}
\eeq

To see how the bound $R(t)$ looks for 2-dimensional density operators, note
that in this case ${\cal L}_{\hat\rho}(\hat\rho_0)$ can be written in terms
of Bloch vectors as \cite{Braunstein95a}
\beq
{\cal L}_{\hat\rho}(\hat\rho_0)\,=\,S_0\hat{\openone}\,+\,
\vec{R}_0\cdot\vec{\sigma}\;,
\eeq
where
\beq
S_0\,=\,\frac{1}{1-c^2}\Big(1\,-\,\vec{a}\cdot\vec{c}\,\Big)\;,
\eeq
and
\beq
\vec{R}_0\,=\,\vec{a}\,-\,S_0\vec{c}\;.
\eeq
A similar representation holds for ${\cal L}_{\hat\rho}(\hat\rho_1)$; one
need only substitute $\vec{b}$ for $\vec{a}$ above.  Then, we get
\beqa
{\rm tr}\Big(\hat\rho_0{\cal L}_{\hat\rho}(\hat\rho_0)\Big)
&=&
S_0\,+\,\vec{a}\cdot\vec{R}_0
\nonumber\\
&=&
\frac{1}{1-c^2}\Big(1\,-\,\vec{a}\cdot\vec{c}\,\Big)^{\!2}\,+\,a^2\;,
\rule{0mm}{8mm}
\eeqa
and similarly 
\beq
{\rm tr}\Big(\hat\rho_1{\cal L}_{\hat\rho}(\hat\rho_1)\Big)
\,=\,\frac{1}{1-c^2}\Big(1\,-\,\vec{b}\cdot\vec{c}\,\Big)^{\!2}\,+\,b^2\;.
\eeq
Substituting these into Eq.~(\ref{GuyPan}) gives the desired result.

\subsubsection{A Bound Based on Purifications}

Imagine that states $\hat\rho_0$ and $\hat\rho_1$ come from a partial
trace over some subsystem of a larger Hilbert space prepared in either a pure
state $|\tilde{\psi}_0\rangle$ or $|\tilde{\psi}_1\rangle$.  That is to say,
let $|\tilde{\psi}_0\rangle$ and $|\tilde{\psi}_1\rangle$ be
{\it purifications\/} of $\hat\rho_0$
and $\hat\rho_1$, respectively.  Then any measurement POVM $\{\hat E_b\}$ on
the original Hilbert space can be thought of as a measurement POVM
$\{\hat E_b\otimes\hat{\openone}\}$ on the larger Hilbert space that ignores
the extra subsystem.  In particular, we will have that the measurement outcome 
statistics can be rewritten as
\beq
\tr(\rho_s\hat E_b)=
\tr\!\left(|\tilde{\psi}_s\rangle\langle\tilde{\psi}_s|\left(
\hat E_b\otimes\hat{\openone}\right)\rule{0mm}{5mm}\!\right)\;,
\eeq
for $s=0,1$.  (The trace on the left side of this equation is taken over only
the original Hilbert space; the trace on the right side is taken over the
larger Hilbert space in which the purifications live.)  Similarly, we have for
the average density operator $\hat\rho$ that
\beq
\tr(\rho\hat E_b)=\tr\!\left(\hat\rho^{\rm AB}\!\left(
\hat E_b\otimes\hat{\openone}\right)\rule{0mm}{5mm}\!\right)\;,
\eeq
where
\beq
\hat\rho^{\rm AB}\,=\,
(1-t)\,|\tilde{\psi}_0\rangle\langle\tilde{\psi}_0|\,+\,
t\,|\tilde{\psi}_1\rangle\langle\tilde{\psi}_1|
\eeq
is the average density operator of the purifications.

Let us now denote the mutual information for the ensemble consisting of
$\hat\rho_0$ and $\hat\rho_1$ with respect to the measurement $\{\hat E_b\}$
by
\beq
J\!\left(\hat\rho_0,\hat\rho_1;t;\{\hat E_b\}\right)\;.
\eeq
It follows then that
\beqa
I(\hat\rho_0|\hat\rho_1)
&=&
\max_{\{\hat E_b\}}J\!\left(\hat\rho_0,\hat\rho_1;t;\{\hat E_b\}\right)
\nonumber\\
&=&
\max_{\{\hat E_b\}}J\!\left(|\tilde{\psi}_0\rangle,|\tilde{\psi}_1\rangle;t;
\{\hat E_b\otimes\hat{\openone}\}\right)
\rule{0mm}{6mm}
\nonumber\\
&\le&
\max_{\{\hat F_c\}}J\!\left(|\tilde{\psi}_0\rangle,|\tilde{\psi}_1\rangle;t;
\{\hat F_c\}\right)
\rule{0mm}{6mm}
\nonumber\\
&=&
I\!\left(\left.\rule{0mm}{3.5mm}
|\tilde{\psi}_0\rangle\right||\tilde{\psi}_1\rangle\right)\;.
\rule{0mm}{6mm}
\eeqa
That is to say, the accessible information of the original ensemble of states
must be less than or equal to the accessible information of the ensemble of
purifications.

This is of great interest because we already know how to calculate the
accessible information for two pure states. It is given by
Eq.~(\ref{LessUgly}) with
\beq
q=|\langle\tilde{\psi}_0|\tilde{\psi}_1\rangle|^2\;.
\eeq
This observation immediately gives an infinite number of new upper bounds to
the accessible information---one for each possible set of purifications.  The
one of most interest, of course, is the smallest upper bound in this class.

Clearly then, the larger the overlap between the purifications 
$|\tilde{\psi}_0\rangle$ and $|\tilde{\psi}_1\rangle$, the tighter the bound
will be.  For the larger the overlap, the less the distinguishability that
will have been added to the purifications above and beyond that of the original 
states.  We need only recall from Section~\ref{MachoMan} that the largest 
possible overlap between purifications is given by
\beq
q=|\langle\tilde{\psi}_0|\tilde{\psi}_1\rangle|^2=\left(
{\rm tr}\,\sqrt{\hat\rho_1^{1/2}\hat\rho_0\hat\rho_1^{1/2}}\,\right)^{\!2}\;.
\eeq
Use of this $q$ in Eq.~(\ref{LessUgly}) gives the best upper bound based on
purifications.  This bound we shall denote by $P(t)$.  When $\hat\rho_0$ and
$\hat\rho_1$ are both very close to being pure states, this bound can be very
tight.

\subsection{Photo Gallery}

In the following pages, several plots compare the bounds on accessible
information derived in the previous sections.  The plots were generated by
$\mbox{Mathematica}^{\rm TM}$ with the following code.
\bigskip

{\tt
(* Initial Notation *)\\
av = aa \{ 1, 0, 0 \}\\
bv = bb \{ Cos[theta], Sin[theta], 0 \}\\
dv = bv - av\\
cv = av + t*dv\\
a = Sqrt[av.av]\\
b = Sqrt[bv.bv]\\
c = Sqrt[cv.cv]\\
d = Sqrt[dv.dv] \\
(* ********************************************************* *)

(* Holevo Upper Bound *)\\
SC = -( (1-c)*Log[(1-c)/2] + (1+c)*Log[(1+c)/2] )/2\\
SB = -( (1-b)*Log[(1-b)/2] + (1+b)*Log[(1+b)/2] )/2\\
SA = -( (1-a)*Log[(1-a)/2] + (1+a)*Log[(1+a)/2] )/2\\
SS = SC - (1-t)*SA - t*SB\\
(* ********************************************************* *)

(* Jozsa-Robb-Wootters Lower Bound *)\\
QC = ( ((1-c)\^{ }2)*Log[(1-c)/2] - ((1+c)\^{ }2)*Log[(1+c)/2] )/(4*c)\\
QB = ( ((1-b)\^{ }2)*Log[(1-b)/2] - ((1+b)\^{ }2)*Log[(1+b)/2] )/(4*b)\\
QA = ( ((1-a)\^{ }2)*Log[(1-a)/2] - ((1+a)\^{ }2)*Log[(1+a)/2] )/(4*a)\\
QQ = QC - (1-t)*QA - t*QB\\
(* ********************************************************* *)

(* Lower Bound M(t) *)\\
mv = (1 - av.cv)*bv - (1 - bv.cv)*av\\
m = Sqrt[mv.mv] \\
MA = ( (m + av.mv)*Log[(m + av.mv)/(m + cv.mv)] +

       (m - av.mv)*Log[(m - av.mv)/(m - cv.mv)] )/(2*m)\\
MB = ( (m + bv.mv)*Log[(m + bv.mv)/(m + cv.mv)] +

       (m - bv.mv)*Log[(m - bv.mv)/(m - cv.mv)] )/(2*m)\\
MM = (1-t)*MA + t*MB\\
(* ********************************************************* *)

(* Upper Bound L(t) *)\\
Ld = Sqrt[ (1 - av.bv)\^{ }2 - (1 - a\^{ }2)*(1 - b\^{ }2) ]\\
LA = (Ld - av.dv)*Log[Ld - av.dv] + (Ld + av.dv)*Log[Ld + av.dv]\\
LB = (Ld - bv.dv)*Log[Ld - bv.dv] + (Ld + bv.dv)*Log[Ld + bv.dv]

- LA\\
LC = -(Ld - cv.dv)*Log[Ld - cv.dv] - (Ld + cv.dv)*Log[Ld + cv.dv]\\
LL = ( Ld/(2*d\^{ }2) )*( LC + t*LB + LA )\\
(* ********************************************************* *)

(* Upper Bound R(t) Based On Jensen's Inequality *)\\
RA = a\^{ }2 + ((1 - av.cv)\^{ }2)/(1 - c\^{ }2)\\
RB = b\^{ }2 + ((1 - bv.cv)\^{ }2)/(1 - c\^{ }2)\\
RR = (1-t)*Log[RA] + t*Log[RB]\\
(* ********************************************************* *)

(* Upper Bound P(t) Based On Purifications *)\\
qq = (1 + av.bv + Sqrt[1-a\^{ }2]*Sqrt[1-b\^{ }2])/2\\
p = Sqrt[1 - 4*t*(1-t)*qq]\\
PA = (p + 1 - 2*t*qq)*Log[(1 + p)/(2*(1-t))] +

     (p - 1 + 2*t*qq)*Log[(1 - p)/(2*(1-t))]\\
PB = (p + 1 - 2*(1-t)*qq)*Log[(1 + p)/(2*t)] +

     (p - 1 + 2*(1-t)*qq)*Log[(1 - p)/(2*t)]\\
PP = ( (1-t)*PA + t*PB )/(2*p)}

\begin{figure}
\begin{center}
\leavevmode
\epsfig{figure=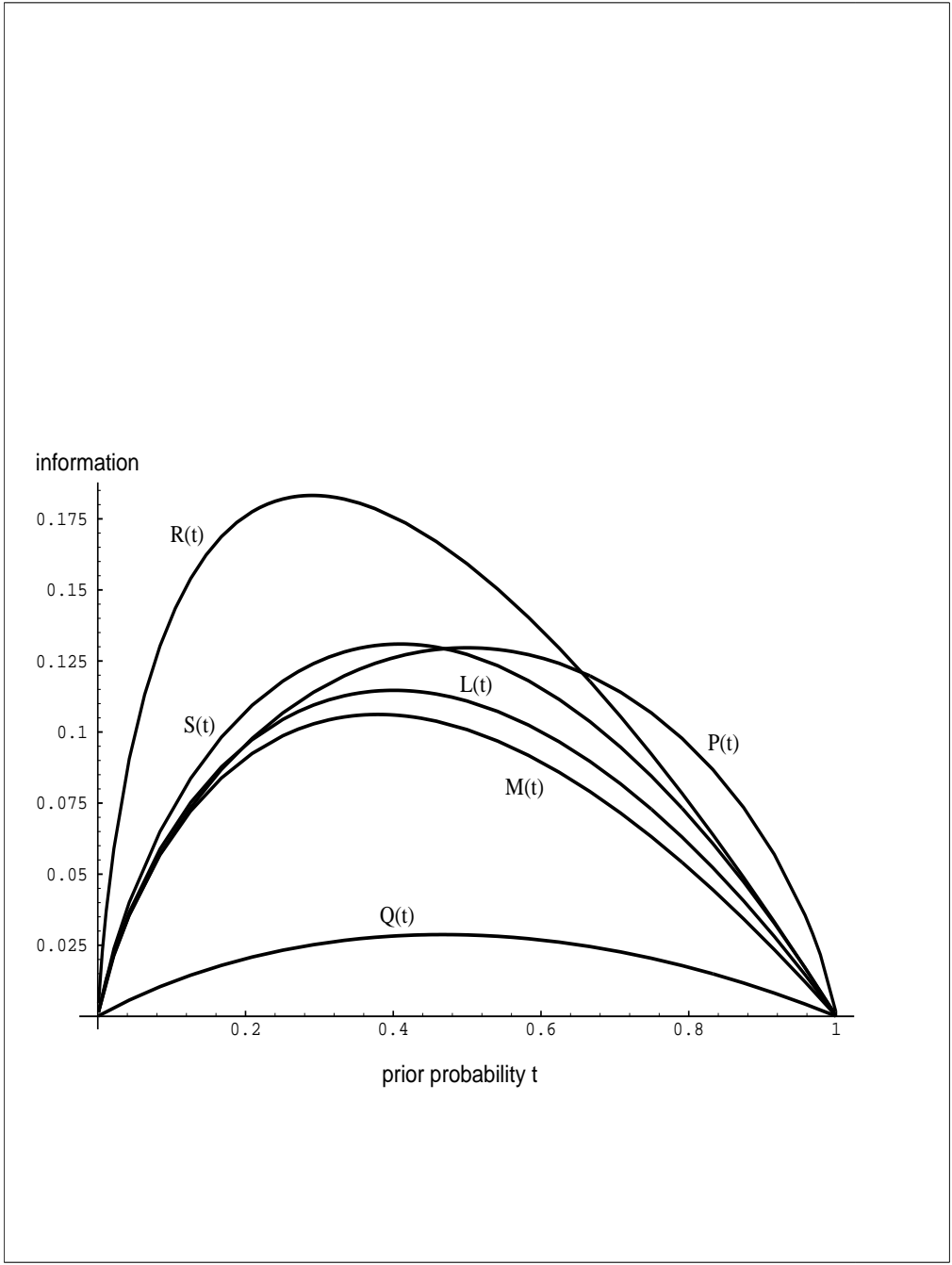,width=5.25in,height=7in}
\caption{All the bounds to accessible information studied here, for the 
case that $\hat\rho_0$ is pure ($a=1$), $\hat\rho_1$ is mixed with 
$b=2/3$, and the angle between the two Bloch vectors is $\pi/4$.}
\label{HoochKooch}
\end{center}
\end{figure}

\begin{figure}
\begin{center}
\leavevmode
\epsfig{figure=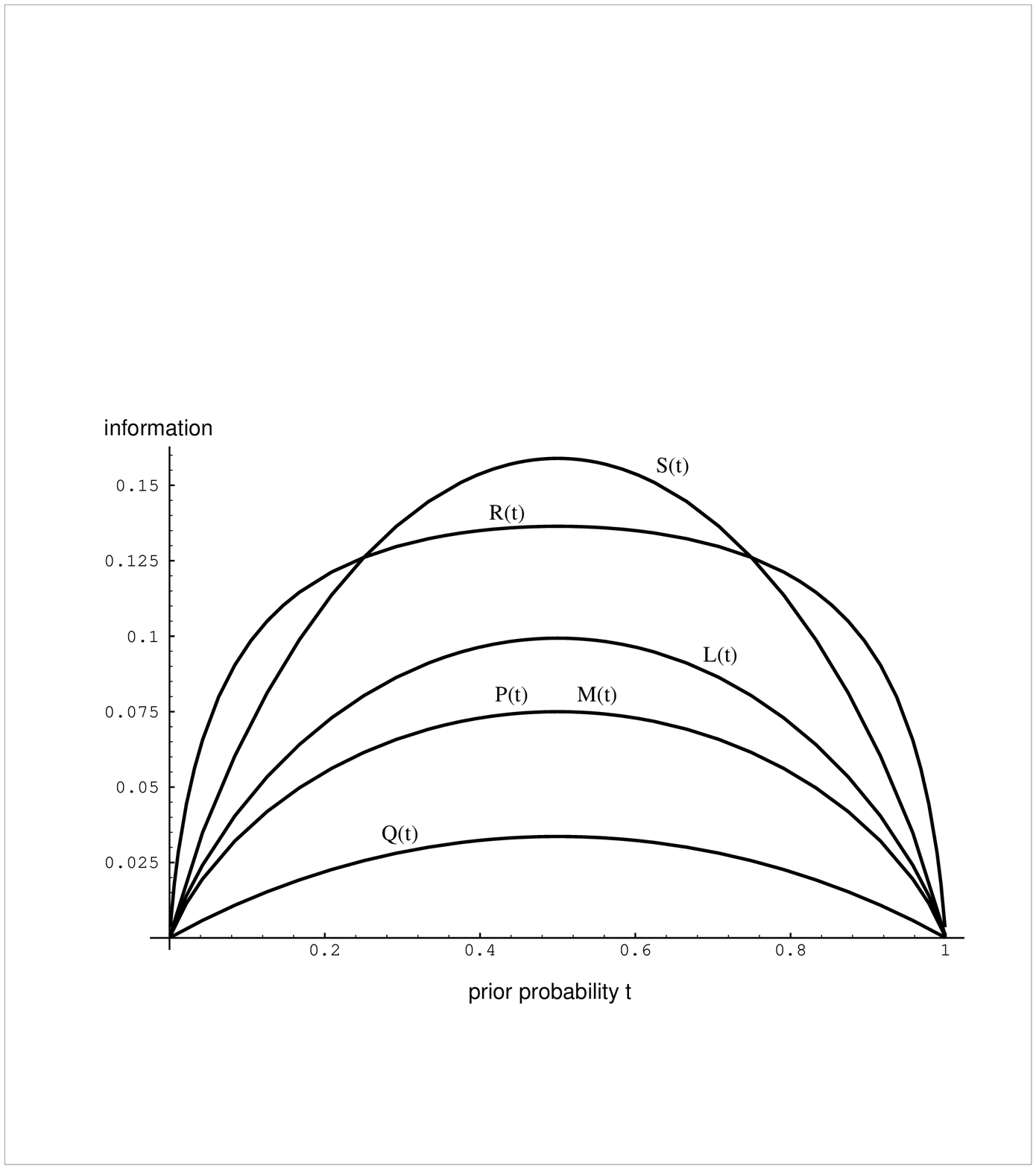,width=5.25in,height=7in}
\caption{All the bounds to accessible information studied here, for the 
case that $\hat\rho_0$ and $\hat\rho_1$ are pure states ($a=b=1$)
and the angle between the two Bloch vectors is $\pi/4$. For this case,
$M(t)=P(t)=I(t)$.}
\label{OshKosh}
\end{center}
\end{figure}

\begin{figure}
\begin{center}
\leavevmode
\epsfig{figure=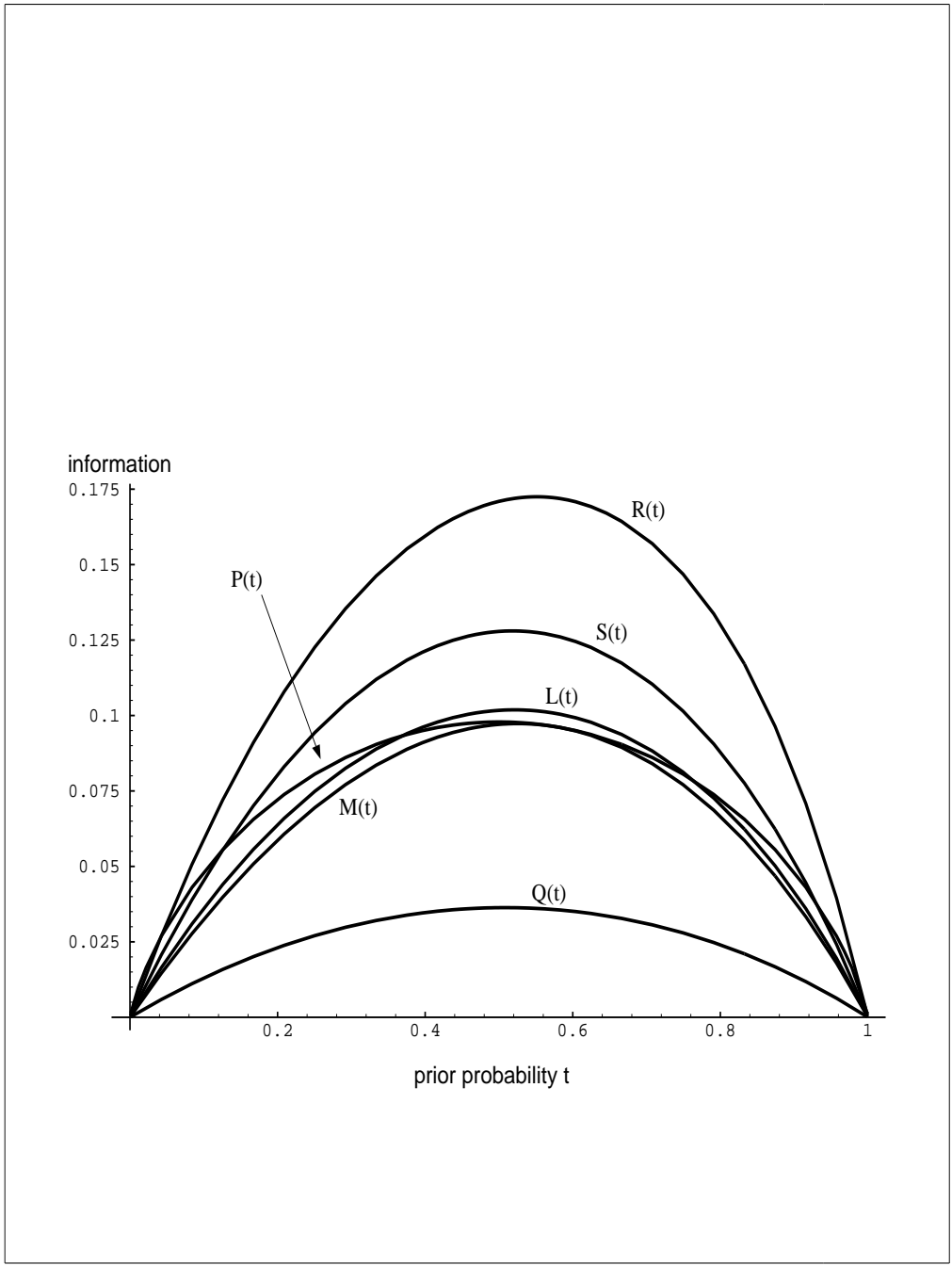,width=5.25in,height=7in}
\caption{All the bounds to accessible information studied here, for the 
case that $\hat\rho_0$ and $\hat\rho_1$ are mixed states with $a=\frac{4}{5}$
and $b=\frac{9}{10}$ and the angle between the two Bloch vectors is $\pi/3$.}
\label{ByGosh}
\end{center}
\end{figure}

\begin{figure}
\begin{center}
\leavevmode
\epsfig{figure=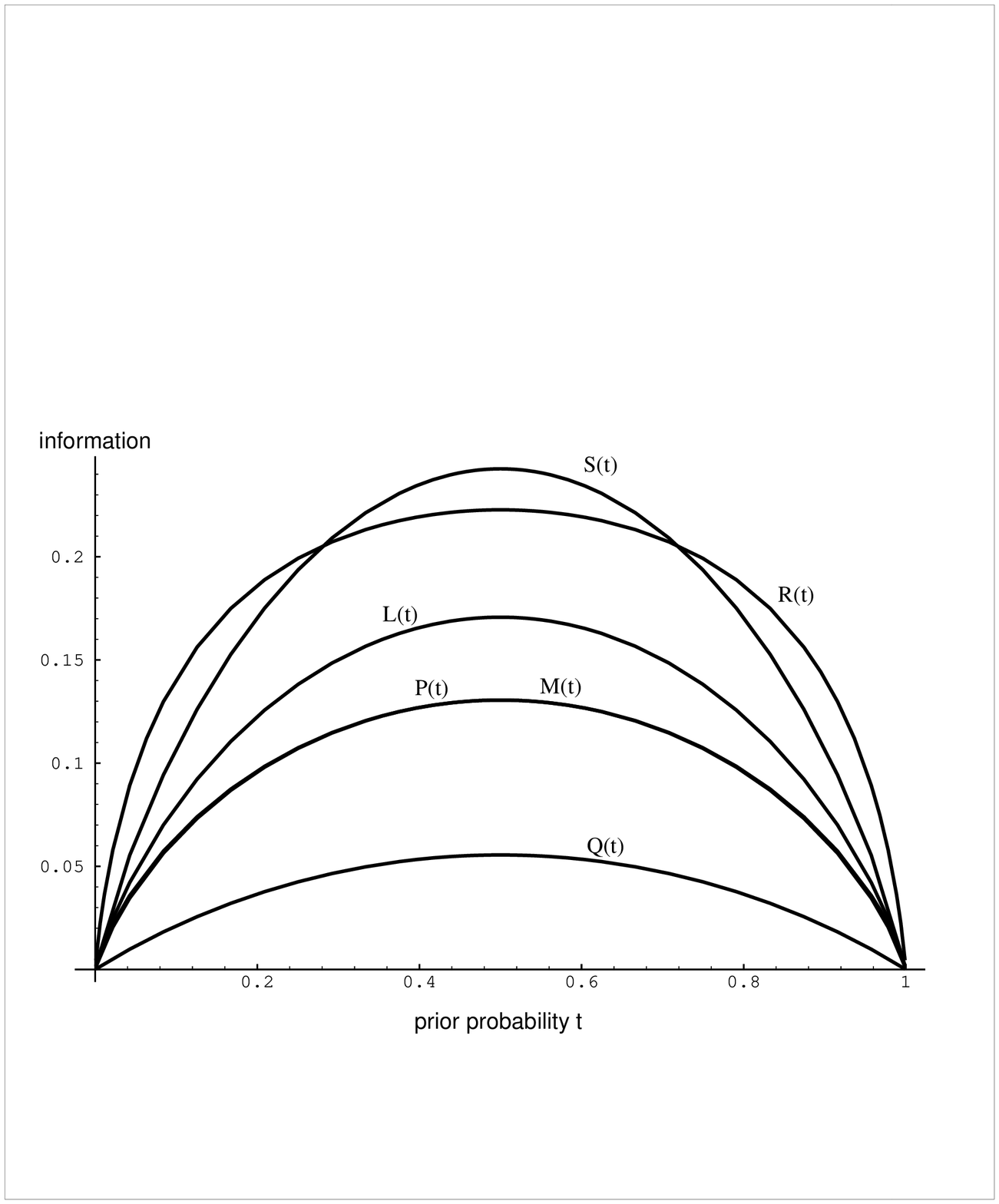,width=5.25in,height=7in}
\caption{All the bounds to accessible information studied here, for the 
case that $\hat\rho_0$ and $\hat\rho_1$ are pure states ($a=b=1$)
and the angle between the two Bloch vectors is $\pi/3$. For this case,
$M(t)=P(t)=I(t)$.}
\label{GrubbingHoe}
\end{center}
\end{figure}

\begin{figure}
\begin{center}
\leavevmode
\epsfig{figure=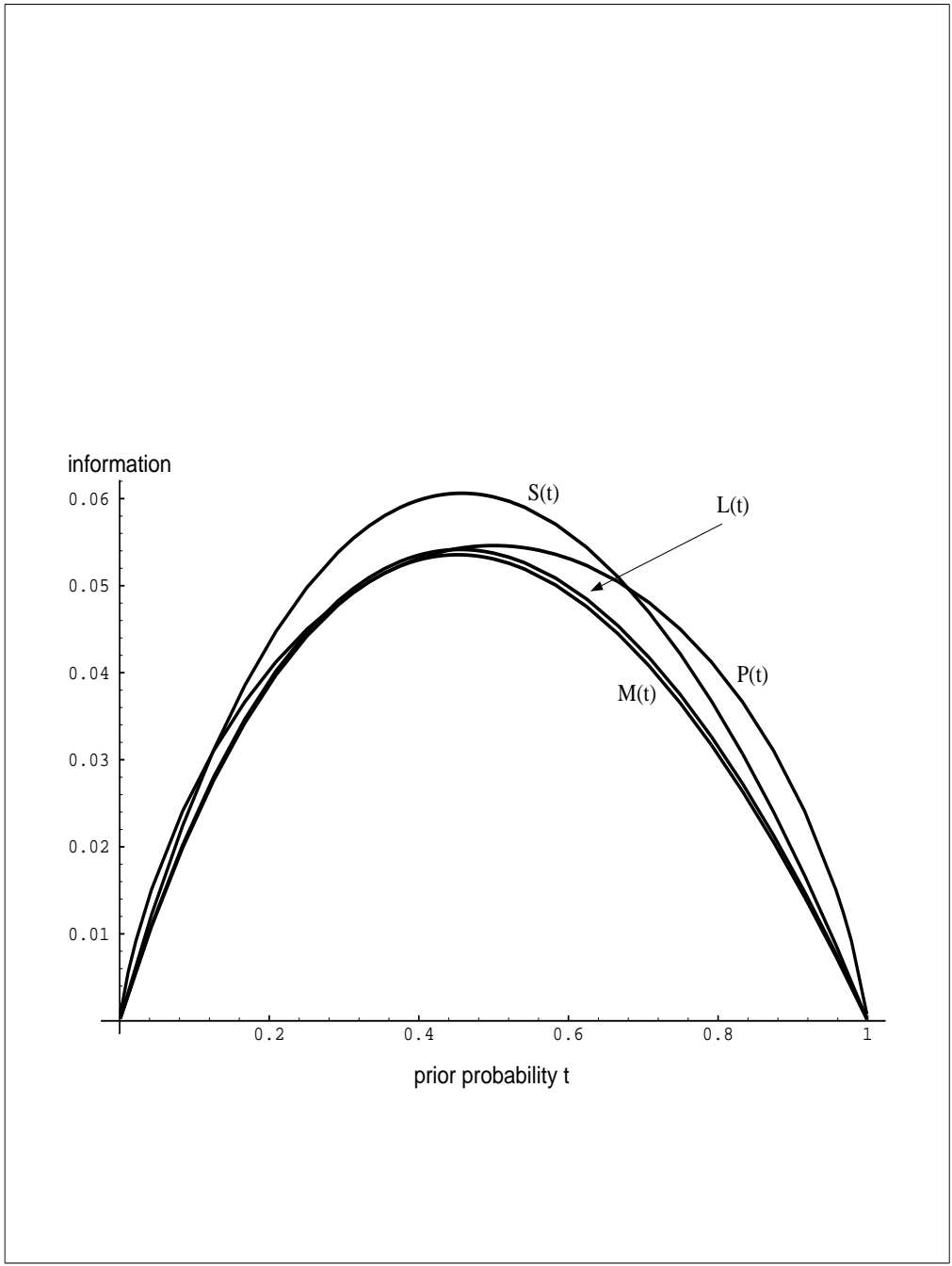,width=5.25in,height=7in}
\caption{The bounds $S(t)$, $L(t)$, $M(t)$, and $P(t)$ for the 
case that $\hat\rho_0$ and $\hat\rho_1$ are mixed states with $a=\frac{9}{10}$
and $b=\frac{3}{5}$ and the angle between the two Bloch vectors is $\pi/5$.}
\label{GardenSpinach}
\end{center}
\end{figure}

\begin{figure}
\begin{center}
\leavevmode
\epsfig{figure=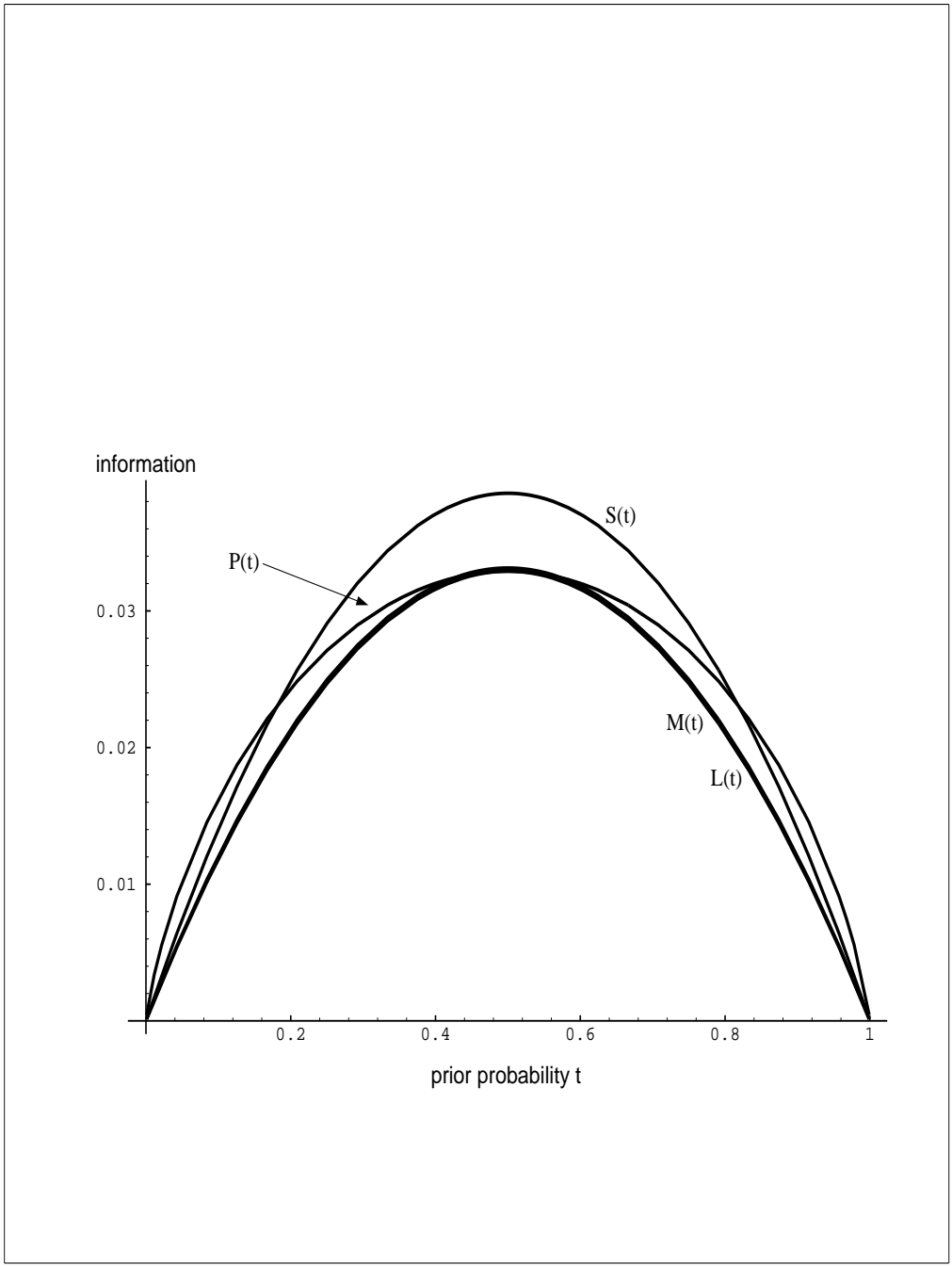,width=5.25in,height=7in}
\caption{The bounds $S(t)$, $L(t)$, $M(t)$, and $P(t)$ for the 
case that $\hat\rho_0$ and $\hat\rho_1$ are mixed states with $a=b=\frac{2}{3}$
and the angle between the two Bloch vectors is $\pi/4$.}
\label{PeteyMoss}
\end{center}
\end{figure}

\section{The Quantum Kullback Information}
\label{MadameClue}

The {\it quantum Kullback information} for a density operator $\hat\rho_0$
relative to density operator $\hat\rho_1$ is defined to be
\beq
K(\hat\rho_0/\hat\rho_1)\equiv\max_{\{\hat E_b\}}\,
\sum_b\,{\rm tr}(\hat\rho_0\hat E_b)\,\ln\!\left({{\rm tr}(\hat\rho_0\hat E_b)
\over{\rm tr}(\hat\rho_1\hat E_b)}\right)\;,
\label{BuggerBoy}
\eeq
where the maximization is taken over all POVMs $\{\hat E_b\}$.  This quantity
is significant because it details a notion of distinguishability for quantum
states most notably in the following way.  (For other interpretations of the
Kullback-Leibler relative information, see Chapter 2.)

Suppose $N\!\gg\!1$ copies of a quantum system
are prepared identically to be in state $\hat\rho_1$.  If a POVM
$\{\hat E_b:b=1,\ldots,n\}$ 
is measured on each of these, the most likely frequencies for the various 
outcomes $b$ will be those given by the probability estimates
$p_1(b)={\rm tr}(\hat\rho_1\hat E_b)$ themselves.   All other frequencies
beside this ``natural'' set will become less and less likely for large $N$ as
statistical fluctuations in the frequencies eventually damp away.  In fact, any
set of outcome frequencies $\{f(b)\}$---distinct from the ``natural'' ones
$\{p_1(b)\}$---will become exponentially less likely with the number of
measurements according to \cite{Cover91}
\beq
e^{-NK(f/p_1)-n\ln(N+1)}\,\le\,
\mbox{PROB}\Bigl(\Bigl.\mbox{freq}=\{f(b)\}\,\Bigr|\,\mbox{prob}=\{p_1(b)\}
\Bigr)\,\le\,e^{-NK(f/p_1)}\;,
\eeq
where
\beq
K(f/p_1)=\sum_{b=1}^n f(b)\,\ln\!\left({f(b)\over p_1(b)}\right)
\eeq
is the Kullback-Leibler relative information \cite{Kullback51} between the
distributions $f(b)$ and $p_1(b)$.  Therefore the quantity $K(f/p_1)$, which
controls the leading behavior of this exponential decline, says something about 
how dissimilar the frequencies $\{f(b)\}$ are from the ``natural'' ones 
$\{p_1(b)\}$.

Now suppose instead that the measurements are performed on quantum systems
identically prepared in the state $\hat\rho_0$.  The outcome frequencies most
likely to appear in this scenario are those specified by the distribution
$p_0(b)={\rm tr}(\hat\rho_0\hat E_b)$.  Therefore the particular POVM
$\hat E^0_b$ satisfying Eq.~(\ref{BuggerBoy}) has the following interpretation.
It is the measurement for which the natural frequencies of outcomes for state
$\hat\rho_0$ are maximally distinct from those for measurements on $\hat\rho_1$,
given that $\hat\rho_1$ is actually controlling the statistics.  In this sense,
Eq.~(\ref{BuggerBoy}) gives an operationally defined (albeit asymmetric) notion
of distinguishability for quantum states.

The main difficulty with Eq.~(\ref{BuggerBoy}) as a notion of distinguishability
is that one would like an explicit, convenient expression for it---not simply
the empty definer given there.  Chances for that,
however, are very slim.  For just by looking at the $2\times2$ density operator
case, one can see that the POVM optimal for this criterion must satisfy a
transcendental equation.  For instance, by setting the variation of
Eq.~(\ref{mutinf2D1}) to zero, we see that an optimal orthogonal
projection-measurement $\vec{n}$ must satisfy,
\beq
\ln\!\left({\Big(n+\vec{a}\cdot\vec{n}\Big)\Big(n-\vec{c}\cdot\vec{n}\Big)\over
\Big(n-\vec{a}\cdot\vec{n}\Big)\Big(n+\vec{c}\cdot\vec{n}\Big)}
\rule{0mm}{8.2mm}\right)\,\vec{a}_\perp\;+\;2\,n\!\left(
\frac{\vec{c}\cdot\vec{n}-\vec{a}\cdot\vec{n}}{n^2-
\Big(\vec{c}\cdot\vec{n}\Big)^2}\right)\,\vec{c}_\perp\;=\;0\;,
\eeq
where the vectors $\vec{a}_\perp$ and $\vec{c}_\perp$ are defined as in
Eq.~(\ref{DisBeLloyd}).
This may well not have an explicit solution in general.  
We are again forced to study bounds rather than exact solutions just as
was the case for accessible information.

So far, several bounds to the quantum Kullback information have come along
incidentally.  There are the lower bounds given by Eqs.~(\ref{KullTrace}) and
(\ref{FisherKullback}), which will soon be revisited in Section~\ref{MTS}.  And
there are the upper and lower bounds due to Jensen's inequality given in
Eq.~(\ref{FavoriteShadows}).  In the remainder of this Section, we shall detail
a few more bounds, both upper and lower.

\subsection{The Umegaki Relative Information}

The Umegaki relative information between $\hat\rho_0$ and $\hat\rho_1$ is
defined by
\beq
K_{\rm U}(\hat\rho_0/\hat\rho_1)\equiv{\rm tr}\Bigl(
\hat\rho_0\ln\hat\rho_0-\hat\rho_0\ln\hat\rho_1\Bigr)\;.
\label{NuffSaid}
\eeq
This concept was introduced by Umegaki in 1962 \cite{Umegaki62}, and a large
literature exists concerning it.  Some authors have gone so far as to label it
the ``proper'' notion of a relative information for two quantum states
\cite{Hiai91,Ohya93}.

As we have already seen, the Holevo upper bound to mutual information is
easily expressible in terms of Eq.~(\ref{NuffSaid}).  This turns out to be no 
surprise because, indeed, this quantity is an upper bound to the quantum 
Kullback information itself \cite{Lindblad74}.  This will be demonstrated 
directly from the Holevo bound in this Section.

For simplicity, let us assume that both $\hat\rho_0$ and $\hat\rho_1$ are
invertible.  Let $\{\hat E_b\}$ be any POVM and $p_0(b)$ and $p_1(b)$ be the 
probability distributions it generates.  Suppose $0<t<1$.  Then the Holevo
bound Eq.~(\ref{Manotoshi}) can be rewritten as
\beq
K(p_0/p)\;+\;\frac{t}{1-t}K(p_1/p)\;\le\;K_{\rm U}(\hat\rho_0/\hat\rho)\;+\;
\frac{t}{1-t}K_{\rm U}(\hat\rho_1/\hat\rho)\;.
\eeq
Taking the limit of this as $t$ approaches 1, we obtain
\beq
K(p_0/p_1)\;+\;\lim_{t\rightarrow1}\frac{t}{1-t}K(p_1/p)\;\le\;
K_{\rm U}(\hat\rho_0/\hat\rho_1)\;+\;\lim_{t\rightarrow1}
\frac{t}{1-t}K_{\rm U}(\hat\rho_1/\hat\rho)\;.
\eeq
It turns out that the limits yet to be evaluated vanish; let us show this.
Note that, trivially,
\beq
\lim_{t\rightarrow1}(1-t)\,=\,\lim_{t\rightarrow1}K(p_1/p)\,=\,
\lim_{t\rightarrow1}K_{\rm U}(\hat\rho_1/\hat\rho)\,=\,0.
\eeq
Therefore we must use l'Hospital's rule to evaluate the desired limits. In the
first case, this readily gives
\beqa
\lim_{t\rightarrow1}\frac{t}{1-t}K(p_1/p)
&=&
-\lim_{t\rightarrow1}\left(K(p_1/p)+t\frac{d}{dt}K(p_1/p)\right)
\nonumber\\
&=&
\sum_b\left(\lim_{t\rightarrow1}\,\frac{\tr(\hat\rho_1\hat E_b)}{\tr(\hat\rho
\hat E_b)}\,\tr(\hat\Delta\hat E_b)\right)
\rule{0mm}{9mm}
\nonumber\\
&=&
\sum_b\tr(\hat\Delta\hat E_b)
\rule{0mm}{8mm}
\nonumber\\
&=&
\tr(\hat\Delta)\;\,=\;\,0\;.
\rule{0mm}{6mm}
\eeqa
The second case is slightly more difficult.  l'Hospital's rule gives
\beqa
\lim_{t\rightarrow1}\frac{t}{1-t}K_{\rm U}(\hat\rho_1/\hat\rho)
&=&
-\lim_{t\rightarrow1}\left(K_{\rm U}(\hat\rho_1/\hat\rho)+t\frac{d}{dt}
K_{\rm U}(\hat\rho_1/\hat\rho)\right)
\nonumber\\
&=&
\lim_{t\rightarrow1}\,\frac{d}{dt}\,\tr(\hat\rho_1\ln\hat\rho)\;,
\rule{0mm}{8mm}
\eeqa
but there is still some work required to evaluate the last expression.

Recall that the operator $\ln\hat\rho$ may be represented by a contour
integral,
\beq
\ln\hat\rho=\frac{1}{2\pi i}\oint_C
\ln z\,\Bigl(z\hat{\openone}-\hat\rho\Bigr)^{-1}dz\;,
\eeq
where the contour $C$ encloses all the eigenvalues of $\hat\rho$ (for
all possible values of $t$).  Then
\beq
\frac{d}{dt}\ln\hat\rho=\frac{1}{2\pi i}\oint_C
\ln z\,\Bigl(z\hat{\openone}-\hat\rho\Bigr)^{-1}\hat\Delta\Bigl(z\hat{\openone}-
\hat\rho\Bigr)^{-1}dz\;,
\eeq
so that
\beqa
\lim_{t\rightarrow1}\,\frac{d}{dt}\,\tr(\hat\rho_1\ln\hat\rho)
&=&
\frac{1}{2\pi i}\oint_C
\ln z\;\tr\!\left(\Bigl(z\hat{\openone}-\hat\rho_1\Bigr)^{-1}\hat\rho_1
\Bigl(z\hat{\openone}-\hat\rho_1\Bigr)^{-1}\hat\Delta\right)dz
\nonumber\\
&=&
\sum_{k=1}^D\lambda_k\Delta_{kk}\,\frac{1}{2\pi i}\oint_C\frac{\ln z}
{(z-\lambda_k)^2}\,dz\;,
\rule{0mm}{10mm}
\eeqa
where $\lambda_k$ are the eigenvalues of $\hat\rho_1$ and $\Delta_{kk}$ are
the matrix elements of $\hat\Delta$ in a basis that diagonalizes $\hat\rho_1$.
By the Cauchy integral theorem, however,
\beq
\frac{1}{2\pi i}\oint_C\frac{\ln z}{(z-\lambda_k)^2}\,dz=\frac{1}{\lambda_k}\;.
\eeq
Therefore
\beqa
\lim_{t\rightarrow1}\,\frac{d}{dt}\,\tr(\hat\rho_1\ln\hat\rho)
&=&
\sum_{k=1}^D\Delta_{kk}
\nonumber\\
&=&
\tr(\hat\Delta)\;\,=\;\,0\;.
\rule{0mm}{6mm}
\eeqa
This proves that the Kullback-Leibler relative information for any measurement
is bounded above by the Umegaki relative information.  In particular, this
places an upper bound on the quantum Kullback information,
\beq
K(\hat\rho_0/\hat\rho_1)\le K_{\rm U}(\hat\rho_0/\hat\rho_1)\;.
\eeq
Moreover, since the Holevo bound on mutual information is achievable if and
only if $\hat\rho_0$ and $\hat\rho_1$ commute, it follows that there is equality
in this bound if and only the density operators commute.

\subsection{One Million One Lower Bounds}
\label{MTS}

It appears to be productive to ask, among other things, if there is 
any systematic procedure for generating successively tighter lower bounds to 
$K(\hat\rho_0/\hat\rho_1)$.  In particular, one would like to know if there is
a procedure for finding lower bounds of the form
\beq
{\rm tr}\!\left(\hat\rho_0\ln\hat\Lambda(\hat\rho_0/\hat\rho_1)\right)\;,
\label{DesiredForm}
\eeq
where $\hat\Lambda(\hat\rho_0/\hat\rho_1)$ is a Hermitian operator that depends
(asymmetrically) on $\hat\rho_0$ and $\hat\rho_1$.

We already know of two such bounds.  The first \cite{Fuchs94,Fuchs95a} is
given by Eq.~(\ref{FisherKullback}),
\beq
K_{\rm F}(\hat\rho_0/\hat\rho_1)\equiv{\rm tr}\biggl(\hat\rho_0\ln
\Bigl({\cal L}_{\hat\rho_1}(\hat\rho_0)\Bigr)\biggr)\;,
\label{FormulaicSpinster}
\eeq
where ${\cal L}_{\hat\rho_1}(\hat\rho_0)$ \cite{Braunstein94a} is an
operator $\hat X$ satisfying the equation
\beq
\hat X\hat\rho_1+\hat\rho_1\hat X=2\hat\rho_0\;.
\eeq
The second \cite{Fuchs95b,Braunstein94c} is given by Eq.~(\ref{KullTrace}),
\beq
K_{\rm B}(\hat\rho_0/\hat\rho_1)\equiv
2\:{\rm tr}\Biggl(\hat\rho_0\,\ln\!\left(
\hat\rho_1^{-1/2}
\sqrt{\hat\rho_1^{1/2}\hat\rho_0\hat\rho_1^{1/2}}
\hat\rho_1^{-1/2}\right)\Biggr)\;.
\label{BhattaKullback}
\eeq
Both these bounds come quite close to the exact answer defined by 
Eq.~(\ref{QKullback}) when $\hat\rho_0$ and $\hat\rho_1$ are $2\times2$ density 
operators \cite{Braunstein95b}.  Nevertheless, these approximations may not
fare so well for density operators of higher dimensionality.

The trick in finding expressions~(\ref{FormulaicSpinster}) and 
(\ref{BhattaKullback}) was in noting that the eigenvalues $\lambda_b$ of
the operators
\beq
\hat\Lambda_1\equiv{\cal L}_{\hat\rho_1}(\hat\rho_0)
\eeq
and
\beq
\hat\Lambda_2\equiv
\hat\rho_1^{-1/2}\sqrt{\hat\rho_1^{1/2}\hat\rho_0\hat\rho_1^{1/2}}
\hat\rho_1^{-1/2}
\eeq
can be written in the form
\beq
\lambda_b^p=\frac{{\rm tr}(\hat\rho_0\hat E_b)}{{\rm tr}(\hat\rho_1\hat E_b)}\;,
\label{EigenvalueForm}
\eeq
where the $\hat E_b=|b\rangle\langle b|$ are projectors onto the one-dimensional
subspaces spanned by eigenvectors $|b\rangle$ of $\hat\Lambda_p$, $p=1$ or 2 respectively.  Using this fact, one simply notes that
\beqa
{\rm tr}\!\left(\hat\rho_0\,\ln\hat\Lambda_p\right)
&=&
{\rm tr}\!\left(\hat\rho_0\sum_b\ln\lambda^p_b\hat E_b\right)
\nonumber\\
&=&
{\rm tr}\!\left(\rule{0mm}{8mm}\hat\rho_0\sum_b\,\ln\!\left(
\frac{{\rm tr}(\hat\rho_0\hat E_b)}{{\rm tr}(\hat\rho_1\hat E_b)}\right)\!
\hat E_b\right)
\rule{0mm}{9mm}
\nonumber\\
&=&
\sum_b\,{\rm tr}(\hat\rho_0\hat E_b)\,\ln\!\left({
{\rm tr}(\hat\rho_0\hat E_b)\over{\rm tr}(\hat\rho_1\hat E_b)}\right)
\rule{0mm}{9mm}
\label{LittleDeriv}
\eeqa

Eq.~(\ref{EigenvalueForm}) is derived easily
from the defining equation for $\hat\Lambda_1$; one just notes that for an
eigenvector $|b\rangle$ of $\hat\Lambda_1$
\beq
\langle b|\hat\Lambda_1\hat\rho_1|b\rangle+
\langle b|\hat\rho_1\hat\Lambda_1|b\rangle=2\langle b|\hat\rho_0|b\rangle
\;\;\;\;\;\Longrightarrow\;\;\;\;\;
2\lambda_b\langle b|\hat\rho_1|b\rangle=2\langle b|\hat\rho_0|b\rangle\;.
\eeq

The corresponding fact for $\hat\Lambda_2$ was derived from more complex considerations in Section~\ref{MachoMan}.  A much simpler way to see it is by
noting that $\hat\Lambda_2$ satisfies the matrix quadratic equation
\cite{Nagaoka94}
\beq
\hat X\hat\rho_1\hat X=\hat\rho_0\;.
\label{NagaokaEq}
\eeq
(This can be seen by inspection.)  Then along similar lines as above, if one
takes $|b\rangle$ to be an eigenvector of $\hat\Lambda_2$, one gets
\beq
\langle b|\hat\Lambda_2\hat\rho_1\hat\Lambda_2|b\rangle=
\langle b|\hat\rho_0|b\rangle
\;\;\;\;\;\Longrightarrow\;\;\;\;\;
\lambda_b^2\langle b|\hat\rho_1|b\rangle=\langle b|\hat\rho_0|b\rangle\;.
\eeq

Now that the common foundation for both lower bounds (\ref{FormulaicSpinster})
and (\ref{BhattaKullback}) is plain, we are led to ask whether there are any
other operators $\hat\Lambda(\hat\rho_0/\hat\rho_1)$
whose eigenvalues have a similar form, i.e.,
\beq
\lambda_b\!\left(\hat\Lambda(\hat\rho_0/\hat\rho_1)\right)\equiv
{{\rm tr}(\hat\rho_0\hat E_b)\over{\rm tr}(\hat\rho_1\hat E_b)}\;,
\eeq
where $\lambda_b(\hat X)$ denotes the $b$'th eigenvalue of the operator
$\hat X$.  If so, then
\beq
\ln\hat\Lambda(\hat\rho_0/\hat\rho_1)=\sum_b\,\ln\!\left(
\frac{{\rm tr}(\hat\rho_0\hat E_b)}{{\rm tr}(\hat\rho_1\hat E_b)}\right)\!
\hat E_b\;,
\eeq
and we get the desired result, Eq.~(\ref{DesiredForm}), via the steps
in Eq.~(\ref{LittleDeriv}).

Posed in this way, the solution to the question becomes quickly apparent.  It
is found by generalizing the defining equations for the
operators $\hat\Lambda_1$ and $\hat\Lambda_2$.  For instance, consider the
Hermitian operator $\hat X$
defined by
\beq
\frac{1}{2}(\hat\rho_1\hat X+\hat X\hat\rho_1)+\hat X\hat\rho_1
\hat X=\hat\rho_0\;.
\eeq
If its eigenvectors and eigenvalues are $|b\rangle$ and $\lambda_b$,
one obtains (by the same method as before)
\beq
{\langle b|\hat\rho_0|b\rangle\over\langle b|\hat\rho_1|b\rangle}=\lambda_b^2
+\lambda_b\;.
\eeq
Therefore, the operator
\beq
\hat\Lambda=\hat X^2+\hat X
\eeq
has eigenvalues
$({\rm tr}\,\hat\rho_0\hat E_b)/({\rm tr}\,\hat\rho_1\hat E_b)$, and we obtain
another lower bound
\beq
{\rm tr}\!\left(\hat\rho_0\ln(\hat X^2+\hat X)\right)
\eeq
to the quantum Kullback information.

More interestingly, however, is that we now have a method for inserting
parameters into the bound which can be varied to obtain an
``optimal'' bound.  For instance, we could instead consider the solutions
$\hat X_\alpha$ to the (parameterized) operator equation
\beq
\frac{1}{2}\alpha\!\left(\hat\rho_1\hat X_\alpha+
\hat X_\alpha\,\hat\rho_1\right)\,+\,(1-\alpha)
\hat X_\alpha\,\hat\rho_1\hat X_\alpha\,=\,\hat\rho_0\;,
\label{ParamEqua}
\eeq
and thus get the best bound of this form by
\beq
\max_\alpha\,{\rm tr}\biggl(\hat\rho_0\ln\Bigl((1-\alpha)\hat X_\alpha^2+
\alpha\hat X_\alpha\Bigr)\biggr)\;.
\eeq
This bound has no choice but to be at least as good or better than the bounds
$K_{\rm F}(\hat\rho_0/\hat\rho_1)$ and $K_{\rm B}(\hat\rho_0/\hat\rho_1)$ simply
because Eq.~(\ref{ParamEqua}) interpolates between the measurements defining 
them in the first place.  Moreover Eq.~(\ref{ParamEqua}) is still within the
realm of equations known to the mathematical community; methods for its solution
exist \cite{Potter66,Coppel74,Faibusovich86,Gohberg86}.

This pretty much builds the picture. More generally, one has measurements 
defined by
\beq
\frac{1}{2}\sum_{ij}\,\alpha_{ij}\!\left(\hat X^i\hat\rho_1\hat X^j+
\hat X^j\hat\rho_1\hat X^i\right)=\hat\rho_0
\label{GenParamEq}
\eeq
giving rise to lower bounds to the Quantum Kullback of the form
\beq
{\rm tr}\!\left(\hat\rho_0\,\ln\biggl(\sum_{ij}\alpha_{ij}\hat X^{i+j}\biggr)
\right).
\eeq
(Here $i$ and $j$ may range anywhere from 0 up to values for which the 
$\alpha_{ij}$ are no longer freely specifiable.)  These may then be varied over 
all the parameters $\alpha_{ij}$ to find the best bound allowable at that order.
To the extent that solutions to Eq.~(\ref{GenParamEq}) can be found, even
numerically, better lower bounds to the Quantum Kullback information can be 
generated.

\subsection{Upper Bound Based on Ando's Inequality and Other Bounds from
the Literature}

Another way to get an upper bound on the quantum Kullback information is by
examining Eq.~(\ref{UncleWalter}), the bound on the quantum R\'enyi overlap
due to Ando's inequality.  With this, one immediately has
\beq
\frac{1}{\,\alpha-1\,}\,\ln\!\left(\,\sum_{b=1}^n\,
p_0(b)^\alpha p_1(b)^{1-\alpha}\right)\,\le\,
\frac{1}{\,\alpha-1\,}\,\ln\tr\!\left(
\hat\rho_1^{1/2}\!\left(\hat\rho_1^{-1/2}\hat\rho_0\hat\rho_1^{-1/2}
\right)^{\!\alpha}\!\hat\rho_1^{1/2}\rule{0mm}{5mm}\right)\;.
\eeq
However, as $\alpha\rightarrow1$, the left hand side of this inequality
converges to the Kullback-Leibler information.  Therefore if we can evaluate
the right hand side of this in the limit, we will have generated a new bound.
Using l'Hospital's rule, we get
\beqa
&\mbox{}&\!\!\!\!\!\!\!\!\!\!
\lim_{\alpha\rightarrow1}\mbox{RHS}\,=
\nonumber\\
&\mbox{}&\!\!\!\!\!\!\!\!=
\lim_{\alpha\rightarrow1}\!
\left[\tr\!\left(\hat\rho_1^{1/2}\!\left(\hat\rho_1^{-1/2}\hat\rho_0
\hat\rho_1^{-1/2}\right)^{\!\alpha}\!\hat\rho_1^{1/2}\rule{0mm}{5mm}\right)
\right]^{\!-1}\!
\tr\!\left(
\hat\rho_1^{1/2}\!\left(\hat\rho_1^{-1/2}\hat\rho_0\hat\rho_1^{-1/2}
\right)^{\!\alpha}\!\ln\!\left(\hat\rho_1^{-1/2}\hat\rho_0\hat\rho_1^{-1/2}
\right)\!\hat\rho_1^{1/2}\rule{0mm}{5mm}\right).
\nonumber\\
\eeqa
Therefore, one arrives at the relatively asymmetric upper bound to the
quantum Kullback information given by
\beq
K(\hat\rho_0/\hat\rho_1)\,\le\,\tr\!\left(
\!\left(\hat\rho_1^{1/2}\hat\rho_0\hat\rho_1^{-1/2}
\right)\!\ln\!\left(\hat\rho_1^{-1/2}\hat\rho_0\hat\rho_1^{-1/2}
\right)\rule{0mm}{5mm}\!\right)\;.
\eeq
Note that when $\rho_0$ and $\rho_1$ commute, this reduces to the Umegaki
relative information.

Finally, we note that the last property of reducing to the Umegaki relative
entropy is not an uncommon property of many known upper bounds to it.  For
instance it is known \cite{Hiai93} that for every $p>0$,
\beq
K_{\rm U}(\hat\rho_0/\hat\rho_1)\,\le\,
\frac{1}{p}\,\tr\!\left(\hat\rho_0\ln\!\left(\hat\rho_0^{p/2}\hat\rho_1^{-p}
\hat\rho_0^{p/2}\right)\rule{0mm}{5mm}\!\right)\;,
\eeq
and all of these have this property.  Alternatively, so do the lower bounds 
Eqs.~(\ref{FisherKullback}) and (\ref{KullTrace}) to the quantum Kullback 
information derived earlier, as well as the lower bound \cite{Hiai93} given by
\beq
K(\hat\rho_0/\hat\rho_1)\,\ge\,
\tr\!\left(\hat\rho_0\ln\!\left(\hat\rho_1^{-1/2}\hat\rho_0
\hat\rho_1^{-1/2}\right)\rule{0mm}{5mm}\!\right)\;.
\eeq
This may or may not lessen the importance of the Umegaki upper bound, depending
upon one's taste.


\chapter{Distinguishability in Action}

\begin{flushright}
\baselineskip=13pt
\parbox{2.7in}{\baselineskip=13pt
``Something only really happens when an observation is being made \ldots\,.  
Between the observations nothing at all happens, only time has, `in the 
interval,' irreversibly progressed on the mathematical papers!''}\medskip\\
---{\it Wolfgang Pauli}\\
Letter to Markus Fierz\\30 March 1947
\end{flushright}

\section{Introduction}

In the preceding chapters, we went to great lengths to define and calculate
various notions of distinguishability for quantum mechanical states.  These
notions are of intrinsic interest for their associated statistical 
problems.  However, we might still wish for a larger payoff on the work
invested.  This is what this Chapter is about.  Here, we present the briefest
of sketches of what can be accomplished by using some of the measures
of distinguishability already encountered.  The applications concern two 
questions that lie much closer to the foundations of quantum theory than
was our previous concern.

``Quantum mechanical measurements disturb the states of quantum systems in
uncontrollable ways.''  Statements like this are uttered in
almost every beginning quantum mechanics course---it is part of the
folklore of the theory.  But what does this really mean?  How is it to be
quantified?  The next two sections outline steps toward answers to these
questions.

\section{Inference vs.~Disturbance of Quantum States: Extended Abstract}

Suppose an observer obtains a quantum system secretly prepared in one of two
standard but nonorthogonal quantum states.  Quantum theory dictates that there
is no measurement he can use to certify which of the two states was actually
prepared.  This is well known and has already been discussed many times in the
preceding chapters.  A simple, but less recognized, corollary to this is that
no interaction used for performing such an information-gathering measurement
can leave both states unchanged in the process. If the observer could
completely regenerate the unknown quantum state after measurement, then---by
making further nondisturbing information-gathering measurements on it---he would
be able eventually to infer the state's identity after all.

This consistency argument is enough to establish a tension between inference
and disturbance in quantum theory.  What it does not capture, however, is the
extent of the tradeoff between these two quantities.
In this Section, we shall lay the groundwork for a quantitative study that goes
beyond the qualitative nature of this tension.\footnote{This Section is based
on a manuscript disseminated during the ``Quantum Computation 1995'' workshop
held at the Institute for Scientific Interchange (Turin, Italy); as such, it 
contains a small redundancy with the previous Chapters.}
Namely, we will show how
to capture in a formal way the idea that, depending upon the particular
measurement interaction, there can be a tradeoff between the disturbance
of the quantum states and the acquired ability to make inferences about their
identity.  The formalism so developed should have applications to quantum
cryptography on noisy channels and to error correction and stabilization in
quantum computing.

\subsection{The Model}
\label{Model}

\begin{figure}
\begin{center}
\leavevmode
\epsfig{figure=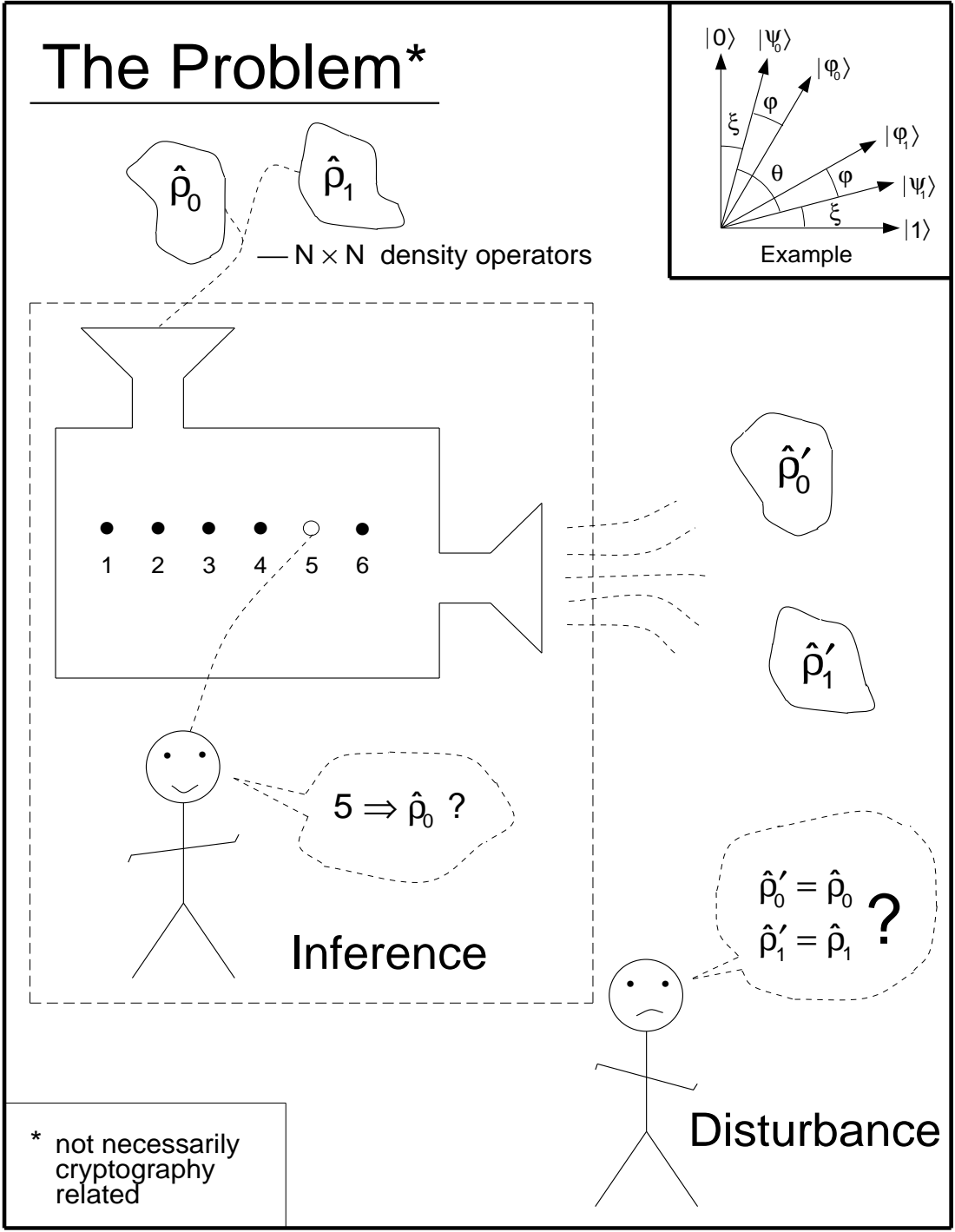,width=5.738in,height=7.425in}
\end{center}
\caption{Set-up for Inference--Disturbance Tradeoff}
\label{FlyBoy}
\end{figure}

The model we shall base our considerations on is most easily described in terms 
borrowed from quantum cryptography, though this problem should not be identified
with the cryptographic one.  Alice randomly prepares a quantum system to be in
either a state $\hat\rho_0$ or a state $\hat\rho_1$.  These states will
be described by $N\!\times\!N$ density operators on an $N$-dimensional Hilbert
space, $N$ arbitrary; 
there is no restriction that they be pure states or orthogonal for that matter.  After the preparation, the 
quantum system is passed into a ``black box'' where it may be probed by an 
eavesdropper Eve in any way allowed by the laws of quantum mechanics.  That is 
to say, Eve may first allow the system to interact with an auxiliary system, or 
ancilla, and then perform quantum mechanical measurements on the ancilla 
itself \cite{Kraus83,Mayers94}. The outcome of such a measurement may provide 
Eve with some information about the quantum state and may even provide her a 
basis on which to make an inference as to the state's identity.  Upon this 
manhandling by Eve, the quantum system is passed out of the ``black box'' and 
into the possession of a third person Bob.  (See related Figure~\ref{FlyBoy}
depicting Eve and Bob only.)

A crucial aspect of this model is that even if Bob knows the state
actually prepared by Alice and, furthermore, the manner
in which Eve operates and the exact measurement she performs,
without knowledge of the answer she actually obtains, he will have to resort to
a new description of the quantum system after it emerges from the ``black 
box''---say some $\hat\rho_0^\prime$ or $\hat\rho_1^\prime$.  This
is where the detail of our work takes its start.  Eve has gathered information
and the state of the quantum system has changed in the process.

The ingredients required formally to pose the question of the Introduction 
follow from the details of the model.  We shall need:
\begin{description}
\item[  A. ]
a convenient description of the most general kind of quantum measurement,
\item[  B. ]
a similarly convenient description of all the possible physical interactions 
that could give rise to that measurement,
\item[  C. ]
a measure of the information or inference power provided by any given 
measurement,
\item[  D. ]
a good notion by which to measure the distinguishability of mixed quantum
states and a measure of disturbance based on it, and finally
\item[  E. ]
a ``figure of merit'' by which to compare the disturbance with the inference.
\end{description}
The way these ingredients are stirred together to make a good soup---very
sche\-matically---is the 
following.  We first imagine some fixed measurement ${\cal M}$ on the
part of Eve and some particular physical interaction ${\cal I}$ used to carry 
out that measurement.  (Note that ${\cal I}$ uniquely determines ${\cal M}$, 
whereas in general ${\cal M}$ says very little about ${\cal I}$.)
Using the measures from ingredients {\bf C} and {\bf D}, we can then
quantify the inference power
\beq
\mbox{Inf}(\hat\rho_0,\hat\rho_1;{\cal M})
\eeq
accorded Eve and the necessary disturbance
\beq
\mbox{Dist}(\hat\rho_0,\hat\rho_1;{\cal I})
\eeq
apparent to Bob.  As
the notation indicates, besides depending on $\hat\rho_0$ and $\hat\rho_1$, the
inference power otherwise depends only on ${\cal M}$ and the disturbance only on
${\cal I}$.  Using the agreed upon {\it figure of merit\/} FOM, we arrive at a
number that describes the tradeoff between inference and disturbance for the 
fixed measurement and interaction:
\begin{equation}
\mbox{FOM}\Bigl[\,\mbox{Inf}(\hat\rho_0,\hat\rho_1;{\cal M}),\,
\mbox{Dist}(\hat\rho_0,\hat\rho_1;{\cal I})\,\Bigr].
\label{FigOMerit}
\end{equation}
Now the idea is to remove the constraint that the measurement ${\cal M}$ and
the interaction ${\cal I}$ be fixed, to re\"express ${\cal M}$ in terms of
${\cal I}$, and to optimize
\begin{equation}
\mbox{FOM}\Bigl[\,\mbox{Inf}(\hat\rho_0,\hat\rho_1;
{\cal M}({\cal I})),\,\mbox{Dist}(\hat\rho_0,\hat\rho_1;{\cal I})\,\Bigr]
\end{equation}
over all measurement interactions ${\cal I}$. This is the whole story.  With
the optimal such expression in hand, one automatically arrives at a tradeoff 
relation between inference and disturbance:  for arbitrary ${\cal M}$ and
${\cal I}$, expression~(\ref{FigOMerit}) will be less or greater than the
optimal such one, depending on its exact definition.

\subsection{The Formalism}
\label{RumpSteak}

The first difficulty in our task is in finding a suitably convenient and useful
formalism for describing Ingredients {\bf A} and {\bf B}.  The most general
measurement procedure allowed by the laws of quantum mechanics is, as already
stated, first to have the system of interest interact with an ancilla and then
to perform a standard (von Neumann) quantum measurement on the ancilla itself.
Taken at face value, this description can be transcribed into mathematical
terms as follows.  The system of interest, starting out in some quantum state
$\hat\rho_{\rm s}$, is placed in conjunction with an ancilla prepared
in a standard state $\hat\rho_{\rm a}$.  The conjunction of these two
systems is described by the initial quantum state
\begin{equation}
\hat\rho_{\rm sa}=\hat\rho_{\rm s}\otimes\hat\rho_{\rm a}\;.
\label{FirstM}
\end{equation}
The interaction of the two systems leads to a unitary time evolution,
\begin{equation}
\hat\rho_{\rm sa}\;\longrightarrow\;\hat U^\dagger\hat\rho_{\rm sa}\hat U\;.
\end{equation}
(Note that the state of the system-plus-ancilla, by way of this interaction,
will generally not remain in its original tensor-product form; rather the states
of the two systems will become inextricably entangled and correlated.)
Finally, a reproducible measurement on the ancilla is described via a set of
orthogonal projection operators $\hat{\openone}\otimes\hat\Pi_b$ acting on the 
ancilla's Hilbert space: any particular outcome $b$ is found with probability
\begin{equation}
p(b)={\rm tr}\!\left((\hat{\openone}\otimes\hat\Pi_b)\hat U^\dagger
(\hat\rho_{\rm s}\otimes\hat\rho_{\rm a})\hat U\right)\;,
\label{ProbForm1}
\end{equation}
and the description of the system-plus-ancilla after the finding of this
outcome must be updated according to the standard rules to
\begin{equation}
\hat\rho_{{\rm sa}|b}^\prime=\frac{1}{p(b)}
(\hat{\openone}\otimes\hat\Pi_b)\hat U^\dagger
(\hat\rho_{\rm s}\otimes\hat\rho_{\rm a})\hat U
(\hat{\openone}\otimes\hat\Pi_b)\;.
\end{equation}
Thus the quantum state describing the system alone after finding outcome $b$ is
\begin{equation}
\hat\rho_{{\rm s}|b}^\prime={\rm tr}_{\rm a}\!\left(\hat
\rho_{{\rm sa}|b}^\prime\right)\;,
\label{CondState}
\end{equation}
where ${\rm tr}_{\rm a}$ denotes a partial trace over the ancilla's Hilbert
space.  If one knows this measurement was performed but does not know
the actual outcome, then the description given the quantum system will be
rather
\begin{equation}
\hat\rho_{\rm s}^\prime=\sum_b p(b)\hat\rho_{{\rm s}|b}^\prime=
{\rm tr}_{\rm a}\!\left(\hat U^\dagger
(\hat\rho_{\rm s}\otimes\hat\rho_{\rm a})\hat U\right)\;.
\label{UnCondState}
\end{equation}

This face-value description of a general measurement
gives---in prin\-ciple---every\-thing required of it.  Namely it gives a formal
description of the probabilities of the measurement outcomes and it gives a 
formal description of the system's state evolution arising from this 
measurement.  The problem with this for the purpose at hand is that it focuses 
attention away from the quantum system itself, placing undue emphasis on the 
fact that there is an ancilla in the background.  Unfortunately, it is this
sort of thing that can make the formulation of optimization problems over
measurements and interactions more difficult than it need be.  On the
brighter side, there is a way of getting around this particular
deficiency of notation.  This is accomplished by introducing the formalism of
``effects and operations''
\cite{Stinespring55,Hellwig69,Hellwig70,Kraus71,Srinivas75,Kraus83}, which we
shall attempt to sketch presently.  (An alternative formalism that may also
be of use in this context is that of Mayers \cite{Mayers94}.)

Recently, more and more attention has been given to the (long known) fact that 
the probability formula, Eq.~(\ref{ProbForm1}), can be written in a way that 
relies on the system Hilbert space alone with no overt reference to the ancilla
\cite{Peres90}.  Namely, one can write
\begin{equation}
p(b)={\rm tr}_{\rm s}(\hat\rho_{\rm s}\hat E_b)\;,
\end{equation}
where ${\rm tr}_{\rm s}$ denotes a partial trace over the system's Hilbert
space, by simply taking the operator $\hat E_b$ to be
\begin{equation}
\hat E_b={\rm tr}_{\rm a}\!\left(
(\hat{\openone}\otimes\hat\rho_{\rm a})
\hat U(\hat{\openone}\otimes\hat\Pi_b)\hat U^\dagger\right)\;.
\label{POMcorr}
\end{equation}
The reason this can be viewed as making no direct reference to the ancilla is
that it has been noted that {\it any\/} set of positive semi-definite
operators $\{\hat E_b\}$ satisfying
\begin{equation}
\sum_b\hat E_b=\hat{\openone}
\label{Completeness}
\end{equation}
can be written in a form specified by Eq.~(\ref{POMcorr}).  Therefore these
sets of operators, known as positive operator-valued measures or POVMs, stand
in one-to-one correspondence with the set of generalized quantum mechanical
measurements.  This correspondence gives us the freedom to exclude or to
include explicitly the ancilla in our description of a measurement's
statistics, whichever is the more convenient.

A quick example illustrating why this particular representation of a 
measurement's outcomes statistics can be useful is the following.  Consider the 
problem of trying to distinguish the quantum states $\hat\rho_0$ and 
$\hat\rho_1$ from
each other by performing some measurement whose outcome probabilities are
$p_0(b)$ if the state is actually $\hat\rho_0$ and $p_1(b)$ if the state is
actually $\hat\rho_1$.  A nice measure of the distinguishability of the 
probability distributions generated by this measurement is their ``statistical
overlap''
\cite{Bhattacharyya43,Wootters81}
\begin{equation}
F(p_0,p_1)=\sum_b\sqrt{p_0(b)}\sqrt{p_1(b)}\;.
\label{BhattaOverlap}
\end{equation}
This quantity is equal to unity whenever the probability distributions are
identical and equal to zero when there is no overlap at all between them.
To get at a notion of how distinct $\hat\rho_0$ and 
$\hat\rho_1$ can be made with respect to this measure, one would like 
to minimize expression~(\ref{BhattaOverlap}) over all possible quantum
measurements---that is to say, over all possible measurement interactions with 
all possible ancilla.  Without the formalism of POVMs this would be quite a
difficult task to pull off.  With the formalism of POVMs, however, we
can pose the problem as that of finding the quantity
\begin{equation}
F(\hat\rho_0,\hat\rho_1)\,\equiv\,
\min_{\{\hat E_b\}}\sum_b
\sqrt{{\rm tr}\hat\rho_0\hat E_b}\sqrt{{\rm tr}\hat\rho_1\hat E_b}\;
\label{QWootters}
\end{equation}
where the minimization is over all sets of operators $\{\hat E_b\}$ satisfying
$\hat E_b\ge0$ and Eq.~(\ref{Completeness}).  This rendition of the problem
makes it tractable.  In fact, it can be shown \cite{Fuchs95b} that
\begin{equation}
F(\hat\rho_0,\hat\rho_1)={\rm 
tr}\,\sqrt{\hat\rho_1^{1/2}\hat\rho_0\hat\rho_1^{1/2}\,}\;,
\label{Fidelity}
\end{equation}
a quantity known in other contexts as (the square root of) Uhlmann's
``transition probability'' or ``fidelity'' for general quantum states 
\cite{Uhlmann76,Jozsa94a}.  The key to the proof of this is in relying as
heavily as one can on the defining characteristic Eq.~(\ref{Completeness})
of all POVMs.  Namely, one uses the standard Schwarz inequality to lower bound
Eq.~(\ref{QWootters}) by an expression linear in the $\hat E_b$, and then uses
the completeness property Eq.~(\ref{Completeness}) to sum those operators out of
this bound. Finally, checking that there is a way of satisfying equality in each step of the process, Eq.~(\ref{Fidelity}) follows.

To restate the point of this example: the tractability of our 
optimization problems may be greatly enhanced by the 
use of mathematical tools that focus on the essential formal characteristics of 
quantum measurements and not on their imagined implementation.  (Other examples 
where great headway has been made by relying on the abstract defining 
properties of POVMs are the maximizing of Fisher information for quantum 
parameter estimation \cite{Braunstein94a} and the bounding of quantum
mutual information \cite{Holevo73d,Fuchs94,Fuchs95c}.)

All that said, the use of POVMs only moves us partially toward our goal.  This
formalism in and of itself has nothing to say about the post-measurement states
given by Eqs.~(\ref{CondState}) and (\ref{UnCondState}); we still do not have a 
particularly convenient (or ancilla-independent) way of representing these
state evolutions.  It takes the formalism of ``effects and operations'' to
complete the story.  This fact is encapsulated by the following
``representation'' theorem of Kraus \cite{Kraus83}.
\begin{theorem}
Let $\{\hat E_b\}$ be the POVM derived from the measurement procedure described
via Eqs.~(\ref{FirstM}) through (\ref{UnCondState}).  Then there exists a set
of (generally non-Her\-mitian) operators $\{\hat A_{bi}\}$ acting on the
system's Hilbert space such that
\begin{equation}
\hat E_b\,=\,\sum_i\hat A^\dagger_{bi}\hat A_{bi}\;,
\label{Stat1}
\end{equation}
and the conditional and unconditional state evolutions under this measurement
can be written as
\begin{equation}
\hat\rho_{{\rm s}|b}^\prime=\frac{1}{{\rm tr}_{\rm s}(\hat\rho_{\rm s}\hat E_b)}
\sum_{i}\hat A_{bi}\hat\rho_{\rm s}\hat A^\dagger_{bi}\;,
\label{Evol1}
\end{equation}
and
\begin{equation}
\hat\rho_{\rm s}^\prime=
\sum_{b,i}\hat A_{bi}\hat\rho_{\rm s}\hat A^\dagger_{bi}\;.
\label{Evol2}
\end{equation}
Moreover, for any set of operators $\{\hat A_{bi}\}$ such that
\begin{equation}
\sum_{b,i}\hat A^\dagger_{bi}\hat A_{bi}=\hat{\openone}\;,
\end{equation}
there exists a measurement interaction whose statistics are described by the
POVM in Eq.~(\ref{Stat1}) and gives rise to the conditional and unconditional
state evolutions of Eqs.~(\ref{Evol1}) and (\ref{Evol2}).
\end{theorem}
(To justify the term ``effects and operations'' we should note that Kraus calls
a POVM $\{\hat E_b\}$ an {\it effect\/} and the state transformations described
by the operators $\{\hat A_{bi}\}$ {\it operations}---the one given by 
Eq.~(\ref{Evol1}) a {\it selective operation\/} and the one given by 
Eq.~(\ref{Evol2}) a {\it nonselective operation}.  In these notes we shall
not make use of the term {\it effect}, but---lacking a more common term---will 
call the operator set $\{\hat A_{bi}\}$ an {\it operation}.)

The way this theorem comes about is seen easily enough from 
Eq.~(\ref{CondState}).  Let the eigenvalues of $\hat\rho_{\rm a}$ be denoted
by $\lambda_\alpha$ and suppose it has an associated eigenbasis
$|a_\alpha\rangle$.  Then $\hat\rho_{\rm s}\otimes\hat\rho_{\rm a}$ can
be written as
\beq
\hat\rho_{\rm s}\otimes\hat\rho_{\rm a}=\sum_\alpha
\sqrt{\lambda_\alpha}\,|a_\alpha\rangle\hat\rho_{\rm s}
\langle a_\alpha|\sqrt{\lambda_\alpha}\;,
\eeq
and, just expanding Eq.~(\ref{CondState}), we have
\beqa
\hat\rho_{{\rm sa}|b}^\prime
&=&
\frac{1}{p(b)}\sum_\beta\langle a_\beta|
(\hat{\openone}\otimes\hat\Pi_b)\hat U^\dagger
(\hat\rho_{\rm s}\otimes\hat\rho_{\rm a})\hat U
(\hat{\openone}\otimes\hat\Pi_b)|a_\beta\rangle
\nonumber\\
&=&
\frac{1}{p(b)}\sum_{\alpha\beta}\left(\sqrt{\lambda_\alpha}\,\langle a_\beta|
(\hat{\openone}\otimes\hat\Pi_b)\hat U^\dagger
|a_\alpha\rangle\right)\hat\rho_{\rm s}\left(\langle a_\alpha|
\hat U(\hat{\openone}\otimes\hat\Pi_b)|a_\beta\rangle\sqrt{\lambda_\alpha}\,
\right)\;.
\rule{0mm}{8mm}
\nonumber\\
\eeqa
Equation~(\ref{Evol1}) comes about by taking
\beq
\hat A_{b\alpha\beta}=\sqrt{\lambda_\alpha}\,\langle a_\alpha|
\hat U(\hat{\openone}\otimes\hat\Pi_b)|a_\beta\rangle
\eeq
and lumping $\alpha$ and $\beta$ into the single index $i$.  Filling in the
remainder of the theorem is relatively easy once this is realized.

Kraus's theorem is the essential new input required to define the 
inference--distur\-bance tradeoff in such a way that it may have a tractable 
solution.  For now, in our recipe Eq.~(\ref{FigOMerit}), we may replace the 
vague symbol ${\cal M}$ (standing for a measurement) by a set of operators
$\{\hat E_b\}$ and we may replace the symbol ${\cal I}$ (standing for a 
measurement interaction) by a set of operators $\{\hat A_{bi}\}$.  Moreover,
we know how to connect these two sets, namely through Eq.~(\ref{Stat1}).  This
reduces our problem to choosing a figure of merit FOM for the tradeoff and
calculating the optimal quantity
\begin{equation}
\mathop{\mbox{optimum}}_{\{\hat A_{bi}\}}\left\{
\mbox{FOM}\!\left[\mbox{Inf}\!\left(\hat\rho_0,\hat\rho_1;
\{\hat A_{bi}\}\right),\mbox{Dist}\!\left(\hat\rho_0,\hat\rho_1;
\{\hat A_{bi}\}\right)\rule{0mm}{5mm}\right]\rule{0mm}{6mm}
\right\}\;,
\end{equation}
where, again, ``optimum'' means either {\it minimum\/} or {\it maximum\/}
depending upon the precise definition of FOM.

\subsection{Tradeoff Relations}
\label{NoonTide}

We finally come to the point where we may attempt to build up various
inference-distur\-bance relations.  To this end we shall satisfy ourselves with
the preliminary work of writing down a fairly arbitrary relation.  We do this
mainly because it is somewhat easier to formulate than other more meaningful 
relations, but also because it appears to be useful for testing out certain 
simple cases.

Before going on to details, however, perhaps we should say a little more about
the significance of these ideas.  It is often said that it is the Heisenberg
uncertainty relations that dictate that quantum mechanical measurements
necessarily disturb the measured system.  That, though, is really not the case.
The Heisenberg relations concern the inability to get hold of two classical 
observables simultaneously, and thus the inability to ascribe classical states 
of motion to quantum mechanical systems.  This is a concern that has very little
to do with the ultimate limits on what can happen to the quantum states
themselves when information is gathered about their identity.
The foundation of this approach differs from that of the standard Heisenberg 
relations in that it makes no reference to conjugate or complementary variables; the only elements entering these considerations are related to the quantum 
states themselves.  In this way one can get at a notion of state disturbance
that is purely quantum mechanical, making no reference to classical
considerations.

What does it really mean to say that the states are disturbed in and of
themselves without reference to variables such as might appear in the
Heisenberg relations?  It means quite literally that Alice faces a loss of
predictability about the outcomes of Bob's measurements whenever an information
gathering eavesdropper intervenes.  Take as an example the case where
$\hat\rho_0$ and $\hat\rho_1$ are nonorthogonal pure states.  Then for each
of these there exists at least one observable for which Alice can predict the
outcome with complete certainty, namely the projectors parallel to $\hat\rho_0$
and $\hat\rho_1$, respectively.  However, after Alice's quantum states pass
into the ``black box'' occupied by Eve, neither Alice nor Bob will any longer
be able to predict with complete certainty the outcomes of both those 
measurements.  This is the real content of these ideas.

\subsubsection{A First ``Trial'' Relation}

The tradeoff relation to be described in this subsection is literally
based on a simple inference problem---that of performing a single quantum 
measurement on one of two unknown quantum states and then using the outcome of 
that measurement to guess the identity of the state.  The criterion of a good
inference is that its expected probability of success be as high as it can
possibly be.  The criterion of a small disturbance is that the expected fidelity
between the initial and final quantum states be as large as it can possibly
be.  There are, of course, many other quantities that we might have used
to gauge inference power, e.g. mutual information, just as there are many other
quantities that we might have used to gauge the disturbance.  We fix our
attention on the ones described here to get the ball rolling.
Let us set up this problem in detail.

Going back to the basic model introduced in Section~\ref{Model}, for this 
scheme, any measurement Eve performs can always be viewed as the measurement
of a two-outcome POVM $\{\hat E_0,\hat E_1\}$.  If the outcome 
corresponds to $\hat E_0$, she guesses the true state to be $\hat\rho_0$; if
it corresponds to $\hat E_1$, she guesses the state to be $\hat\rho_1$.

So, first consider a fixed POVM $\{\hat E_0,\hat E_1\}$ and a fixed operation
$\{\hat A_{bi}\}$ $(b=0,1)$ consistent with it in the sense of 
Eq.~(\ref{Stat1}).  This measurement gives rise to an expected probability 
of success quantified by
\begin{equation}
P_s\,=\,\frac{1}{2}{\rm tr}(\hat\rho_0\hat E_0)\,+\,
\frac{1}{2}{\rm tr}(\hat\rho_1\hat E_1)\;.
\label{SuccessProb}
\end{equation}
That is to say, the expected probability of success for this measurement is the
probability that
$\hat\rho_0$ is the true state times the conditional probability that the 
decision will be right when this is the case {\it plus\/} a similar term for
$\hat\rho_1$.  (Here we have assumed the prior probabilities for the two states
precisely equal.)  Using Eq.~(\ref{Stat1}) and the cyclic property of the
trace, the success probability can be re\"expressed as
\begin{equation}
P_s\,=\,\frac{1}{2}\sum_{b,i}{\rm tr}\!\left(\hat A_{bi}\hat\rho_b
\hat A^\dagger_{bi}\right)\;.
\end{equation}
We shall identify this quantity as $\mbox{Inf}\left(\{\hat A_{bi}\}\right)$,
the measure of the inference power given by this measurement.

Now consider the quantum state Bob gets as the system passes out 
of the black box.  If the original state was $\hat\rho_0$, he obtains
\begin{equation}
\hat\rho_0^\prime=\sum_{b,i}\hat A_{bi}\hat\rho_0\hat A^\dagger_{bi}\;.
\end{equation}
If the original state was $\hat\rho_1$, he obtains
\begin{equation}
\hat\rho_1^\prime=\sum_{b,i}\hat A_{bi}\hat\rho_1\hat A^\dagger_{bi}\;.
\end{equation}
(Eq.~(\ref{Evol2}) is made use of rather than Eq.~(\ref{Evol1}) because it is 
assumed that Bob has no knowledge of Eve's measurement outcome.)
The overall state disturbance by this interaction can be quantified in terms of
any of a number of distinguishability 
measures for quantum states explored in Chapter 3.  Here we choose
to make use of Uhlmann's ``transition probability'' or ``fidelity'',
Eq.~(\ref{Fidelity}), to define a measure of {\it clonability\/},
\begin{equation}
C\,=\,\frac{1}{2}\Big(F(\hat\rho_0,\hat\rho_0^\prime)\Big)^{\!2}\,+\,
\frac{1}{2}\Big(F(\hat\rho_1,\hat\rho_1^\prime)\Big)^{\!2}\;.
\label{Clonability}
\end{equation}
This quantity measures the extent to which the output quantum states ``clone''
the input states.  In particular, the quantity in
Eq.~(\ref{Clonability})---bounded between zero and one---attains a maximum
value only when {\it both\/} states are left completely
undisturbed by the measurement interaction.

Re\"expressing the clonability explicitly in terms of the operation
$\{\hat A_{bi}\}$, we get
\begin{equation}
C\,=\,\frac{1}{2}\sum_{s=0}^1\!\left({\rm tr}
\sqrt{\hat\rho_s^{1/2}\!\left(\sum_{b,i}\hat A_{bi}\hat\rho_s\hat A^\dagger_{bi}
\right)\!\hat\rho_s^{1/2}}\;\rule{0mm}{10mm}\right)^{\!\! 2}
\end{equation}
(Here we have used the fact that $F(\hat\rho_0,\hat\rho_1)$ is
symmetric in its arguments.)  We shall identify this quantity as
$\mbox{Dist}\Bigl(\{\hat A_{bi}\}\Bigr)$, the measure of (non)disturbance given
by this measurement.

Now all that is left is to put the inference and disturbance into a common
figure of merit by which to compare the two.  There are a couple of obvious
ways to do this.  Since the idea is that, as the probability of success in the 
inference increases, the clonability decreases, and vice versa, we know that 
there should be nontrivial upper limits to both the sum and the products of 
$P_s$ and $C$. Indeed the same must be true for an infinite number of monotonic
functions of $P_s$ and $C$.  Here we shall be happy to focus on the sum as an
appropriate figure of merit.  Why?  For no real reason other than that this 
combination looks relatively simple and is enough to demonstrate the
principles involved.  So, following through, we simply write down the tradeoff
relation
\begin{equation}
P_s+C\,\,\le\,\,\max_{\{\hat A_{bi}\}}\,\frac{1}{2}\sum_{s=0}^1\!\left[\,
\sum_{i}{\rm tr}\!\left(\hat A_{si}\hat\rho_s
\hat A^\dagger_{si}\right)+
\left({\rm tr}
\sqrt{\hat\rho_s^{1/2}\left(\sum_{b,i}\hat A_{bi}\hat\rho_s\hat A^\dagger_{bi}
\right)\hat\rho_s^{1/2}}\;\rule{0mm}{10mm}\right)^{\!\! 2}\;\right]\;.
\label{TradeOff1}
\end{equation}

Finally we are in possession of a well-defined mathematical problem awaiting
a solution.  If the right-hand side of Eq.~(\ref{TradeOff1}) can be given an
explicit expression, then the problem will have been solved.  Techniques for
getting at this, however, must be a subject for future research.  Presently we 
have no general solutions.

We should point out that, though Eq.~(\ref{TradeOff1}) is the tightest
bound of the form
\begin{equation}
P_s+C\,\le\,f(\hat\rho_0,\hat\rho_1)\;,
\end{equation}
other, looser, bounds may also be of some utility---for instance, simply because
they may be easier to derive.
This comes about because the right-hand side of Eq.~(\ref{TradeOff1}) is the
actual maximum value of $P_s+C$; often it is easier to bound a quantity than 
explicitly to maximize it.  The only requirement in the game is that the bound
be nontrivial, i.e., smaller than the one that comes about by maximizing both
$P_s$ and $C$ simultaneously,
\beqa
f(\hat\rho_0,\hat\rho_1)
&\le&
\max_{\{\hat A_{bi}\}}C\,+\,
\max_{\{\hat E_b\}}P_s
\nonumber\\
&=&
1\,+\,\max_{\{\hat E_b\}}P_s
\rule{0mm}{6mm}
\nonumber\\
&=&
1\,+\,
\frac{1}{2}\!\left[1+{\sum_j}^\prime\lambda_j(\hat\Gamma)\right]\;,
\rule{0mm}{9mm}
\label{LooseBound}
\eeqa
where $\lambda_j(\hat\Gamma)$ denotes the eigenvalues of the
operator
\beq
\hat\Gamma=\hat\rho_1-\hat\rho_0\;,
\eeq
and the prime on the summation sign signifies that the sum is taken only over 
the positive eigenvalues.  (See Section~\ref{GiveEmHell} and 
Refs.~\cite{Helstrom67,Helstrom76}.)  The right-hand side of 
Eq.~(\ref{LooseBound}) would be the actual solution of
Eq.~(\ref{TradeOff1}) if and only if an inference measurement entailed
{\it no\/} disturbance.  In Eq.~(\ref{LooseBound}),
\beq
\max_{\{\hat A_{bi}\}}C\,=\,1
\eeq
follows from the fact that there is a zero-disturbance measurement, namely the
identity operation.

\subsubsection{An Example}

Let us work out a restricted example of this tradeoff relation, just to 
hint at the interesting insights it can give for concrete problems.\footnote{A 
slight variation of this example is done in much greater detail and generality 
by Fuchs and Peres in Ref.~\protect\cite{Fuchs95d}.}
In this example, the two initial states of interest are pure states
\beq
\hat\rho_0=|\psi_0\rangle\langle\psi_0|
\;\;\;\;\;\;\;\;\mbox{and}\;\;\;\;\;\;\;\;
\hat\rho_1=|\psi_1\rangle\langle\psi_1|
\eeq
separated in Hilbert space by an angle $\theta$.    (See inset to
Figure~\ref{FlyBoy}.)  Eve bases her inference on the outcome of the POVM
\beq
\Big\{\hat\Pi_0\!=\!|0\rangle\langle 0|,\;\hat\Pi_1\!=\!|1\rangle\langle 1|
\Big\}\;,
\eeq
the projection operators onto the basis vectors symmetrically straddling the
states.  This measurement leads to the maximal probability of success for this 
inference problem 
\cite{Helstrom67,Helstrom76}.  The measurement interaction giving rise to this
POVM will be assumed to be of the {\it restricted} class described by the
operation
\beq
\hat A_b=\hat U_b\hat\Pi_b\;,\;\;\;\;\;\;b=0,1\;,
\eeq
where the $\hat U_b$ are arbitrary unitary operators.  Note that
$\hat A_b^\dagger\hat A_b=\hat\Pi_b$ for these operators, as must be the case
by the definition of an operation.

With this operation, the two states evolve
to post-measurement states according to
\begin{eqnarray}
|\psi_0\rangle\;\longrightarrow\;\hat\rho_0^\prime
&=&
\sum_b\hat U_b\hat\Pi_b\hat\rho_0\hat\Pi_b\hat U_b^\dagger
\nonumber\\
&=&
|\langle 0|\psi_0\rangle|^2\left(\hat U_0|0\rangle\langle0|\hat U_0^\dagger
\right)+ 
|\langle 1|\psi_0\rangle|^2\left(\hat U_1|1\rangle\langle1|\hat U_1^\dagger
\right)
\nonumber\\
&=&
\cos^2\xi\left(\hat U_0|0\rangle\langle0|\hat U_0^\dagger\right)+ 
\cos^2(\xi+\theta)\left(\hat U_1|1\rangle\langle1|\hat U_1^\dagger\right)\;,
\rule{0mm}{7mm}
\label{Bob0}
\end{eqnarray}
and
\begin{eqnarray}
|\psi_1\rangle\;\longrightarrow\;\hat\rho_1^\prime
&=&
|\langle 0|\psi_1\rangle|^2\left(\hat U_0|0\rangle\langle0|\hat U_0^\dagger
\right)+ 
|\langle 1|\psi_1\rangle|^2\left(\hat U_1|1\rangle\langle1|\hat U_1^\dagger
\right)
\nonumber\\
&=&
\cos^2(\xi+\theta)\left(\hat U_0|0\rangle\langle0|\hat U_0^\dagger\right)+ 
\cos^2\xi\left(\hat U_1|1\rangle\langle1|\hat U_1^\dagger\right)\;,
\rule{0mm}{7mm}
\label{Bob1}
\end{eqnarray}
where $\xi$ is the angle between $|b\rangle$ and $|\psi_b\rangle$, $b=0,1$.
This evolution has a simple interpretation: if Eve finds outcome $b=0$, she
sends on the quantum system in state
\beq
|\phi_0\rangle\equiv\hat U_0|0\rangle\;;
\eeq
if Eve finds outcome $b=1$, she sends on the state
\beq
|\phi_1\rangle\equiv\hat U_1|1\rangle\;.
\eeq
The (mixed) states---according to Bob's description---appearing in the outside 
world are then those given by Eqs.~(\ref{Bob0}) and (\ref{Bob1}).

To calculate the disturbance given by these operations, we note a
simplification to expressions for $F(\hat\rho_0,\hat\rho_0^\prime)$ 
and $F(\hat\rho_1,\hat\rho_1^\prime)$ due to the fact that the
initial states are pure.  Namely,
\beq
F(\hat\rho_b,\hat\rho_b^\prime)\,=\,{\rm tr}
\sqrt{\hat\Pi_b\hat\rho_b^\prime\hat\Pi_b}\,=\,
\sqrt{\langle b|\hat\rho_b^\prime|b\rangle}\;.
\eeq
If we further restrict $\hat U_0$ and $\hat U_1$ to be
such that $\{|\phi_0\rangle,\,|\phi_1\rangle\}$ lie in the plane spanned by
$\{|0\rangle,\,|1\rangle\}$ and determine equal angles with these basis vectors
(see Figure~\ref{FlyBoy}), then the clonability under this interaction works
out to be
\begin{eqnarray}
C
&=&
\frac{1}{2}\Bigl(\langle 0|\hat\rho_0^\prime|0\rangle\,+\,
\langle 1|\hat\rho_1^\prime|1\rangle\Bigr)
\nonumber\\
&=&
\cos^2\xi\cos^2\phi\,+\,\cos^2(\xi+\theta)\cos^2(\theta-\phi)\;,
\rule{0mm}{6mm}
\end{eqnarray}
where $\phi$ is the angle between $|\psi_b\rangle$ and $|\phi_b\rangle$.

The problem is, of course, to find the optimal tradeoff between inference
and disturbance for this measurement and interaction.  Since the measurement
POVM is fixed, this boils down to determining the angle $\phi$ such that the
clonability $C$ is maximized.  Setting $dC/d\phi$ equal to zero and solving for
$\phi$, we find the least disturbing final states to be specified
by the angle $\phi_{\rm o}$, where
\begin{equation}
\phi_{\rm o}=\frac{1}{2}\arctan\!\left[
\left({1+\sin\theta\over1-\sin\theta}+\cos2\theta\right)^{\!\!-1}\!\sin2\theta
\right]\;.
\end{equation}
This angle ranges from $0^\circ$ at $\theta=0^\circ$ to its maximum value 
$6.99^\circ$ at $\theta=27.73^\circ$, returning to $0^\circ$ at 
$\theta=90^\circ$.  This means that, under the restrictions imposed here, the
best strategy on the part of Eve for minimizing disturbance is {\it not\/} to
send on the quantum states guessed in the inference, but rather a set
of states with slightly higher overlap than the original states.  This result
points out that the (a priori reasonable) strategy of simply sending on the
inferred states will only propagate the error in that inference; it is much 
smarter on Eve's part to attempt to hide that error by decreasing the 
probability that a wrong guess can lead to a detection of itself outside the
boundaries of the black box.

\section{Noncommuting Quantum States Cannot Be Broadcast}

The fledgling field of quantum information theory 
\cite{Bennett95} serves
perhaps its most important role in delimiting wholly new classes of what is
and is not physically possible.  A particularly elegant example of this is the
theorem \cite{Wootters82,Dieks82} that there are no physical means with which
an unknown {\it pure\/} quantum state can be reproduced or copied.  This 
situation is often summarized with the phrase, ``quantum states cannot be 
cloned.''  Here, we demonstrate an impossibility theorem that extends and 
generalizes the pure-state no-cloning theorem to mixed quantum
states.\footnote{This Section represents a collaboration with Howard 
Barnum, Carlton~M. Caves, Richard Jozsa, and Benjamin Schumacher.  The
presentation here is based largely on a manuscript submitted
to Physical Review Letters; as such, it contains a small redundancy with the 
previous Chapters.  Also note that this Section breaks from the notation of the 
rest of the dissertation in that operators are not distinguished by hats; for
instance we now write $\rho$ instead of $\hat\rho$.}
This theorem strikes very close to the heart of the distinction between the
classical and quantum theories, because it provides a nontrivial physical 
classification of {\it commuting\/} versus {\it noncommuting\/} states.

In this Section we ask whether there are any physical means---fixed 
independently of the identity of a quantum state---for {\it broadcasting\/} 
that quantum state onto two separate quantum systems.  By broadcasting
we mean that the marginal density operator of each of the separate 
systems is the same as the state to be broadcast.  

The pure-state ``no-cloning'' theorem \cite{Wootters82,Dieks82} 
prohibits broadcasting pure states.  This is because the only way to broadcast 
a pure state $|\psi\rangle$ is to put the two systems in the product 
state $|\psi\rangle\otimes|\psi\rangle$, i.e., to clone $|\psi\rangle$.
Things are more complicated when the states are mixed.  A mixed-state 
no-cloning theorem is not sufficient to demonstrate no-broadcasting, 
for there are many conceivable ways to broadcast a mixed state $\rho$ 
without the joint state being in the product form $\rho\otimes\rho$, 
the mixed-state analog of cloning; the systems might be correlated 
or entangled in such a way as to give the right marginal density 
operators.  For instance, if the density operator has the spectral 
decomposition
\beq
\rho=\sum_b\lambda_b|b\rangle\langle b|\;,
\eeq
a potential broadcasting state is the highly correlated joint state
\beq
\tilde\rho=\sum_b\lambda_b|b\rangle|b\rangle\langle b|\langle b|\;,
\eeq
which, though not of the product form $\rho\otimes\rho$, reproduces 
the correct marginal density operators.

The general problem, posed formally, is this.  A quantum system AB is 
composed of two parts, A and B, each having an $N$-dimensional Hilbert 
space.  System A is secretly prepared in one state from a set 
${\cal A}\!=\!\{\rho_0,\rho_1\!\}$ of two quantum states.  System B,
slated to receive the unknown state, is in a standard
quantum state $\Sigma$.  The initial state of the composite system AB
is the product state $\rho_s\otimes\Sigma$, where $s=0$ or 1 specifies
which state is to be broadcast.  We ask whether there is any physical 
process $\cal E$, consistent with the laws of quantum theory, that leads 
to an evolution of the form
\beq
\rho_s\otimes\Sigma\rightarrow{\cal E}(\rho_s\otimes\Sigma)=\tilde\rho_s\;,
\eeq 
where $\tilde\rho_s$ is {\it any\/} state on the $N^2$-dimensional 
Hilbert space AB such that
\begin{equation}
{\rm tr}_{\scriptscriptstyle {\rm A}}(\tilde\rho_s)=\rho_s
\;\;\;\;\;\;\;\;\;\;\;\mbox{and}\;\;\;\;\;\;\;\;\;\;\;
{\rm tr}_{\scriptscriptstyle {\rm B}}(\tilde\rho_s)=\rho_s\;.
\label{require}
\end{equation}
Here ${\rm tr}_{\scriptscriptstyle {\rm A}}$ and
${\rm tr}_{\scriptscriptstyle {\rm B}}$ denote partial traces over A and 
B.  If there is an ${\cal E}$ that satisfies Eq.~(\ref{require}) for both 
$\rho_0$ and $\rho_1$, then the set ${\cal A}$ can be {\it broadcast}.  
A special case of broadcasting is the evolution specified by
\beq
{\cal E}(\rho_s\otimes\Sigma)=\rho_s\otimes\rho_s\;.
\eeq
We reserve the word {\it cloning\/} for this strong form of broadcasting.

The most general action $\cal E$ on AB consistent with quantum theory 
is to allow AB to interact unitarily with an auxiliary quantum system 
C in some standard state and thereafter to ignore the auxiliary 
system \cite{Kraus83}; that is,  
\begin{equation}
{\cal E}(\rho_s\otimes\Sigma)=
{\rm tr}_{\scriptscriptstyle {\rm C}}\!\left(U(
\rho_s\otimes\Sigma\otimes\Upsilon) U^\dagger\right),
\label{process}
\end{equation}
for some auxiliary system C, some standard state $\Upsilon$ on C, and 
some unitary operator $U$ on ABC.  We show that such an evolution can 
lead to broadcasting if and only if $\rho_0$ and $\rho_1$ commute.
(In this way the concept of broadcasting makes a communication theoretic cut 
between commuting and noncommuting density operators, and thus between
classical and quantum state descriptions.)
We further show that $\cal A$ is clonable if and only if $\rho_0$ and
$\rho_1$ are identical or orthogonal, i.e.,
\beq
\rho_0\rho_1=0\;.
\eeq

To see that the set $\cal A$ can be broadcast when the states commute, 
we do not have to go to the extra trouble of attaching an auxiliary system.  
Since orthogonal pure states can be cloned, broadcasting can be obtained
by cloning the simultaneous eigenstates of $\rho_0$ and $\rho_1$.
Let $|b\rangle$, $b=1,\ldots ,N$, be an orthonormal basis for A in 
which both $\rho_0$ and $\rho_1$ are diagonal, and let their spectral
decompositions be
\beq
\rho_s=\sum_b\lambda_{sb}|b\rangle\langle b|\;.
\eeq
Consider any unitary operator $U$ on AB consistent with
\beq
U|b\rangle|1\rangle=|b\rangle|b\rangle\;.
\eeq
If we choose $\Sigma=|1\rangle\langle1|$ and let
\begin{equation}
\tilde\rho_s=
U(\rho_s\otimes\Sigma)U^\dagger=
\sum_b\lambda_{sb}|b\rangle|b\rangle\langle b|\langle b|\;,
\label{waldo}
\end{equation}
we immediately have that $\tilde\rho_0$ and $\tilde\rho_1$ satisfy 
Eq.~(\ref{require}).

The converse of this statement---that if $\cal A$ can be broadcast,
$\rho_0$ and $\rho_1$ commute---is more difficult to prove.  Our
proof is couched in  terms of the concept of {\it fidelity\/} between 
two density operators.  The fidelity $F(\rho_0,\rho_1)$ is defined by
\begin{equation}
F(\rho_0,\rho_1)={\rm tr}\sqrt{\rho_0^{1/2}\rho_1\rho_0^{1/2}\,}\;,
\label{fidel}
\end{equation}
where for any positive operator $O$, i.e., any Hermitian operator with {\it nonnegative\/} eigenvalues, $O^{1/2}$ denotes its unique 
positive square root. (Note that Ref.~\cite{Jozsa94a} defines fidelity
to be the square of the present quantity.)  Fidelity is an analogue of the 
modulus of the inner product for pure states \cite{Uhlmann76,Jozsa94a}
and can be interpreted as a measure of distinguishability for quantum 
states: it ranges between 0 and 1, reaching 0 if and only 
if the states are orthogonal and reaching 1 if and only 
if $\rho_0=\rho_1$.  It is  invariant under the interchange 
$0\leftrightarrow1$ and under the transformation
\beq
\rho_0\rightarrow U\rho_0 U^\dagger
\;\;\;\;\;\;\;\;\;\;\;\mbox{\it and}\;\;\;\;\;\;\;\;\;\;\;
\rho_1\rightarrow U\rho_1 U^\dagger\;
\eeq
for any unitary operator $U$ \cite{Jozsa94a,Fuchs95b}.  Also, from the 
properties of the direct product, one has that
\beq
F(\rho_0\otimes\sigma_0,\rho_1\otimes\sigma_1)=
F(\rho_0,\rho_1)F(\sigma_0,\sigma_1)\;.
\eeq

Another reason $F(\rho_0,\rho_1)$ defines a good notion of 
distinguishability \cite{Wootters81} is that it equals the minimal 
overlap between the probability distributions $p_0(b)={\rm tr}(\rho_0 E_b)$ 
and $p_1(b)={\rm tr}(\rho_1 E_b)$ generated by a generalized 
measurement or {\it positive operator-valued measure\/} (POVM) 
$\{E_b\}$ \cite{Kraus83}. That is \cite{Fuchs95b},
\begin{equation}
F(\rho_0,\rho_1)=\min_{\{E_b\}}\sum_b\sqrt{{\rm tr}(\rho_0 E_b)}
\sqrt{{\rm tr}(\rho_1 E_b)}\;,
\label{uncle}
\end{equation}
where the minimum is taken over all sets of positive operators $\{E_b\}$
such that
\beq
\sum_b E_b=\openone\;.
\eeq
This representation of fidelity
has the advantage of being defined operationally in terms of measurements.
We call a POVM that achieves the minimum in Eq.~(\ref{uncle}) an 
{\it optimal\/} POVM.

One way to see the equivalence of Eqs.~(\ref{uncle}) and (\ref{fidel}) 
is through the Schwarz inequality for the operator inner product
${\rm tr}(AB^\dagger)$:
\beq
{\rm tr}(AA^\dagger)\,{\rm tr}(BB^\dagger)\ge|{\rm tr}(AB^\dagger)|^2\;,
\eeq
with equality if and only if
\beq
A=\alpha B
\eeq
for some constant $\alpha$.  
Going through this exercise is useful because it leads directly to the 
proof of the no-broadcasting theorem.  Let $\{E_b\}$ be any POVM and 
let $U$ be any unitary operator.  Using the cyclic property of the trace 
and the Schwarz inequality, we have that
\begin{eqnarray}
\sum_b\sqrt{{\rm tr}(\rho_0 E_b)}\sqrt{{\rm tr}(\rho_1 E_b)}
&=&
\sum_b\sqrt{{\rm tr}\!
\left(U\rho_0^{1/2}E_b\,\rho_0^{1/2}U^\dagger\right)}
\sqrt{{\rm tr}\!\left(\rho_1^{1/2}E_b\,\rho_1^{1/2}\right)}
\nonumber\\
&\ge&
\sum_b\left|{\rm tr}\!
\left(U\rho_0^{1/2}E_b^{1/2}E_b^{1/2}\rho_1^{1/2}\right)\right|
\rule{0mm}{8mm}
\label{nordle}
\\
&\ge&
\left|\sum_b{\rm tr}\!
\left(U\rho_0^{1/2}E_b\rho_1^{1/2}\right)\right|
\rule{0mm}{8mm}
\nonumber\\
&=&
\Bigl|\,{\rm tr}\!\left(U\rho_0^{1/2}\rho_1^{1/2}\right)\Bigr|\;.
\rule{0mm}{8mm}
\label{Quebecois}
\end{eqnarray}
We can use the freedom in $U$ to make the inequality as tight as 
possible.  To do this, we recall \cite{Jozsa94a,Schatten60} that
\beq 
\max_V|{\rm tr}(V\!O)|={\rm tr}\sqrt{O^\dagger O}\;,
\eeq
where $O$ is any operator and the maximum is taken over all unitary operators 
$V$.  The maximum is achieved only by those $V$ such that
\beq
V\!O=\sqrt{O^\dagger O}e^{-i\phi}\;,
\eeq
where $\phi$ is an arbitrary phase.  That there exists at least one such $V$ 
is insured by the operator polar decomposition theorem \cite{Schatten60}.
Therefore, by choosing
\begin{equation}
e^{i\phi}U\rho_0^{1/2}\rho_1^{1/2}=\sqrt{\rho_1^{1/2}\rho_0\rho_1^{1/2}\,}\;,
\label{badland}
\end{equation}
we get that
\beq
\sum_b\!\sqrt{{\rm tr}(\rho_0 E_b)}\sqrt{{\rm tr}(\rho_1 E_b)}\ge 
F(\rho_0,\rho_1)\;.
\eeq

To find optimal POVMs, we consult the conditions for equality in 
Eq.~(\ref{Quebecois}).  These arise from Step~(\ref{nordle}) and the one 
following it: a POVM $\{E_b\}$ is optimal if and only if 
\begin{equation}
U\rho_0^{1/2}E_b^{1/2}=\mu_b\rho_1^{1/2}E_b^{1/2}
\label{rasbo}
\end{equation}
and $U$ is rephased such that
\begin{equation}
{\rm tr}\!\left(U\rho_0^{1/2}E_b\rho_1^{1/2}\right)=\mu_b\,{\rm tr}
(\rho_1 E_b)\ge0\;\;\;\;\;\Longleftrightarrow\;\;\;\;\;\mu_b\ge0\;.
\label{mangoroots}
\end{equation}
When $\rho_1$ is invertible, Eq.~(\ref{rasbo}) becomes
\begin{equation}
M E_b^{1/2}=\mu_b E_b^{1/2}\;,
\label{Fontaine}
\end{equation}
where
\beqa
M
&=&
\rho_1^{-1/2}U\rho_0^{1/2}
\nonumber\\
&=&
\rho_1^{-1/2}\sqrt{\rho_1^{1/2}\rho_0\rho_1^{1/2}\,}\rho_1^{-1/2}
\rule{0mm}{8mm}
\label{rimbo}
\eeqa
is a positive operator. Therefore one way to satisfy Eq.~(\ref{rasbo}) 
with $\mu_b\ge0$ is to take $E_b=|b\rangle\langle b|$,
where the vectors $|b\rangle$ are an orthonormal eigenbasis for $M$,
with $\mu_b$ chosen to be the eigenvalue of $|b\rangle$.  When $\rho_1$
is noninvertible, there are still optimal POVMs.  One can choose the
first $E_b$ to be the projector onto the null subspace of $\rho_1$; in 
the support of $\rho_1$, i.e., the orthocomplement of the null subspace, 
$\rho_1$ is invertible, so one can construct the analogue of $M$ and 
proceed as for an invertible $\rho_1$.  Note that if both $\rho_0$ and
$\rho_1$ are in\-vert\-ible, $M$ is invertible.  

We begin the proof of the no-broadcasting theorem by using Eq.~(\ref{uncle})
to show that fidelity cannot decrease under the operation of partial
trace; this gives rise to an elementary constraint on all potential
broadcasting processes $\cal E$.  Suppose Eq.~(\ref{require}) is 
satisfied for the process $\cal E$ of Eq.~(\ref{process}), and let 
$\{E_b\}$ denote an optimal POVM for distinguishing $\rho_0$
and $\rho_1$.  Then, for each $s$,
\beqa
{\rm tr}\Bigl(\tilde\rho_s(E_b\otimes\openone)\Bigr)
&=&
{\rm tr}_{\scriptscriptstyle{\rm A}}\Bigl(
{\rm tr}_{\scriptscriptstyle{\rm B}}(\tilde\rho_s)E_b\Bigr)
\nonumber\\
&=&
{\rm tr}_{\scriptscriptstyle{\rm A}}(\rho_s E_b)\;;
\eeqa
it follows that 
\begin{eqnarray}
F_{\scriptscriptstyle{\rm A}}(\rho_0,\rho_1)
&\equiv&
\sum_b\sqrt{{\rm tr}\Bigl(\tilde\rho_0(E_b\otimes\openone)\Bigr)}
\sqrt{{\rm tr}\Bigl(\tilde\rho_1(E_b\otimes\openone)\Bigr)}
\nonumber\\
&\ge&
\min_{\{\tilde E_c\}}\,\sum_c\sqrt{{\rm tr}(\tilde\rho_0\tilde E_c)}
\sqrt{{\rm tr}(\tilde\rho_1\tilde E_c)}
\rule{0mm}{6mm}\nonumber\\
&=&
F(\tilde\rho_0,\tilde\rho_1)\rule{0mm}{6mm}\;.
\label{gaggle}
\end{eqnarray}
Here $F_{\scriptscriptstyle{\rm A}}(\rho_0,\rho_1)$ denotes simply
the fidelity $F(\rho_0,\rho_1)$, but the subscript A emphasizes that
$F_{\scriptscriptstyle{\rm A}}(\rho_0,\rho_1)$ stands for the
particular representation on the first line.  The inequality in 
Eq.~(\ref{gaggle}) comes from the fact that $\{E_b\otimes\openone\}$ 
might not be an optimal POVM for distinguishing $\tilde\rho_0$ and 
$\tilde\rho_1$; this demonstrates the said partial trace property.  Similarly
since
\beqa
{\rm tr}\Bigl(\tilde\rho_s(\openone\otimes E_b)\Bigr)
&=&
{\rm tr}_{\scriptscriptstyle{\rm B}}\Bigl(
{\rm tr}_{\scriptscriptstyle{\rm A}}(\tilde\rho_s)E_b\Bigr)
\nonumber\\
&=&
{\rm tr}_{\scriptscriptstyle{\rm B}}(\rho_s E_b)\;,
\eeqa
it follows that
\begin{eqnarray}
F_{\scriptscriptstyle {\rm B}}(\rho_0,\rho_1)
&\equiv&
\sum_b\sqrt{{\rm tr}\Bigl(\tilde\rho_0(\openone\otimes E_b)\Bigr)}
\sqrt{{\rm tr}\Bigl(\tilde\rho_1(\openone\otimes E_b)\Bigr)}
\nonumber\\
&\ge&
F(\tilde\rho_0,\tilde\rho_1)\;,\rule{0mm}{6mm}
\label{jojo}
\end{eqnarray}
where the subscript B emphasizes that 
$F_{\scriptscriptstyle {\rm B}}(\rho_0,\rho_1)$ stands for the 
representation on the first line.  

On the other hand, we can just as easily derive an inequality that is opposite
to Eqs.~(\ref{gaggle}) and (\ref{jojo}).  By the direct product formula and
the invariance of fidelity under unitary transformations,
\begin{eqnarray}
F(\rho_0,\rho_1)
&=&
F(\rho_0\otimes\Sigma\otimes\Upsilon,\rho_1\otimes\Sigma\otimes\Upsilon)
\nonumber\\
&=&
F\Bigl(U(\rho_0\otimes\Sigma\otimes\Upsilon)U^\dagger, 
U(\rho_1\otimes\Sigma\otimes\Upsilon)U^\dagger\Bigr)\;.
\rule{0mm}{7mm}
\end{eqnarray}
Therefore, by the partial-trace property, 
\beqa
F(\rho_0,\rho_1)\,\le\,
F\!\left({\rm tr}_{\scriptscriptstyle {\rm C}}\!
\left(U(\rho_0\otimes\Sigma\otimes\Upsilon)U^\dagger\right), 
{\rm tr}_{\scriptscriptstyle {\rm C}}\!
\left(U(\rho_1\otimes\Sigma\otimes\Upsilon)U^\dagger\right)\rule{0mm}{5mm}\!
\right)\;,
\eeqa
or, more succinctly,
\begin{equation}
F(\rho_0,\rho_1)\,\le\,
F\Bigl({\cal E}(\rho_0\otimes\Sigma),
{\cal E}(\rho_1\otimes\Sigma)\Bigr)\,=\,
F(\tilde\rho_0,\tilde\rho_1)\;.
\label{Wilma}
\end{equation}

The elementary constraint now follows: the only way to 
maintain Eqs.~(\ref{gaggle}), (\ref{jojo}), {\it and\/} 
(\ref{Wilma}) is with strict equality.  In other words, 
we have that if the set $\cal A$ can be broadcast, then there are 
density operators $\tilde\rho_0$ and $\tilde\rho_1$ on AB satisfying
Eq.~(\ref{require}) {\it and}
\begin{equation}
F_{\scriptscriptstyle{\rm A}}(\rho_0,\rho_1)=
F(\tilde\rho_0,\tilde\rho_1)=F_{\scriptscriptstyle{\rm B}}(\rho_0,\rho_1)\;.
\label{SmootGibson}
\end{equation}

Let us pause at this point to consider the restricted question of
cloning. If $\cal A$ is to be clonable, there must exist a process
$\cal E$ such that $\tilde\rho_s=\rho_s\otimes\rho_s$ for $s=0,1$.
But then, by Eq.~(\ref{SmootGibson}), we must have 
\begin{equation}
F(\rho_0,\rho_1)\,=\,F(\rho_0\otimes\rho_0,\rho_1\otimes\rho_1)\,=\,
\Bigl(F(\rho_0,\rho_1)\Bigr)^{\!2},
\label{Spanky}
\end{equation}
which means that $F(\rho_0,\rho_1)=1$ or 0, i.e., $\rho_0$ and $\rho_1$ 
are identical or orthogonal.  There can be no cloning for density operators 
with nontrivial fidelity.  The converse, that orthogonal and identical
density operators can be cloned, follows, in the first case, from the fact
that they can be distinguished by measurement and, in the second case,
because they need not be distinguished at all.

Like the pure-state no-cloning theorem \cite{Wootters82,Dieks82}, this 
no-cloning result for mixed states is a consistency requirement for 
the axiom that quantum measurements cannot distinguish nonorthogonal 
states with perfect reliability.  If nonorthogonal quantum states 
could be cloned, there would exist a measurement procedure for 
distinguishing those states with arbitrarily high reliability: one 
could make measurements on enough copies of the quantum state to
make the probability of a correct inference of its identity 
arbitrarily high.  That this consistency requirement, as expressed 
in Eq.~(\ref{SmootGibson}), should also exclude more general kinds 
of broadcasting problems is not immediately obvious.  Nevertheless, 
this is the content of our claim that Eq.~(\ref{SmootGibson}) generally 
cannot be satisfied; any broadcasting process can be viewed as 
creating distinguishability {\it ex nihilo\/} with respect to 
measurements on the larger Hilbert space AB.  Only for the case 
of commuting density operators does broadcasting not create 
any extra distinguishability.

We now show that Eq.~(\ref{SmootGibson}) implies that $\rho_0$ and 
$\rho_1$ commute.  To simplify the exposition, we assume that 
$\rho_0$ and $\rho_1$ are invertible.  We proceed by studying the 
conditions necessary for the representations  
$F_{\scriptscriptstyle{\rm A}}(\rho_0,\rho_1)$ and
$F_{\scriptscriptstyle{\rm B}}(\rho_0,\rho_1)$ in
Eqs.~(\ref{gaggle}) and (\ref{jojo}) to equal
$F(\tilde\rho_0,\tilde\rho_1)$.  Recall that the optimal POVM
$\{E_b\}$ for distinguishing $\rho_0$ and $\rho_1$ can be chosen
so that the POVM elements $E_b=|b\rangle\langle b|$ are a complete
set of orthogonal one-dimensional projectors onto orthonormal 
eigenstates of $M$.  Then, repeating the steps leading from 
Eqs.~(\ref{Quebecois}) to (\ref{mangoroots}), one finds that 
the necessary conditions for equality in Eq.~(\ref{SmootGibson}) 
are that each
\beq
E_b\otimes\openone\,=\,(E_b\otimes\openone)^{1/2}
\eeq
and each
\beq
\openone\otimes E_b\,=\,(\openone\otimes E_b)^{1/2}
\eeq
satisfy
\beq
\tilde U\tilde\rho_0^{1/2}(\openone\otimes E_b)\,=\,
\alpha_b\,\tilde\rho_1^{1/2}(\openone\otimes E_b)\;,
\label{parceltree}
\eeq
and
\beq
\tilde V\tilde\rho_0^{1/2}(E_b\otimes\openone)\,=\,
\beta_b\,\tilde\rho_1^{1/2}(E_b\otimes\openone)\;,
\label{skypost}
\eeq
where $\alpha_b$ and $\beta_b$ are nonnegative numbers and $\tilde U$ 
and $\tilde V$ are unitary operators satisfying
\begin{equation}
\tilde U\tilde\rho_0^{1/2}\tilde\rho_1^{1/2}\,=\,
\tilde V\tilde\rho_0^{1/2}\tilde\rho_1^{1/2}\,=\,
\sqrt{\tilde\rho_1^{1/2}\tilde\rho_0\tilde\rho_1^{1/2}\,}\;.
\label{temptation}
\end{equation}
Although $\rho_0$ and $\rho_1$ are assumed invertible, one cannot 
demand that $\tilde\rho_0$ and $\tilde\rho_1$ be invertible---a glance 
at Eq.~(\ref{waldo}) shows that to be too restrictive.  This means
that $\tilde U$ and $\tilde V$ need not be the same.  Also we cannot
assume that there is any relation between $\alpha_b$ and $\beta_b$.

The remainder of the proof consists in showing that Eqs.~(\ref{parceltree}) 
through (\ref{temptation}), which are necessary (though perhaps not
sufficient) for broadcasting, are nevertheless restrictive enough to imply
that $\rho_0$ and $\rho_1$ commute.  The first step is to sum over $b$ in 
Eqs.~(\ref{parceltree}) and (\ref{skypost}).  Defining the positive 
operators 
\beq
G\,=\,\sum_b\alpha_b|b\rangle\langle b| 
\eeq
and
\beq
H\,=\,\sum_b\beta_b|b\rangle\langle b|\;,
\end{equation}
we obtain
\begin{equation}
\tilde U\tilde\rho_0^{1/2}\,=\,\tilde\rho_1^{1/2}(\openone\otimes G)
\label{Kunkel}
\eeq
and
\beq
\tilde V\tilde\rho_0^{1/2}\,=\,\tilde\rho_1^{1/2}(H\otimes\openone)\;.
\label{vituperative}
\end{equation}

The next step is to demonstrate that $G$ and $H$ are invertible and, 
in fact, equal to each other.  Multiplying the two equations in 
Eq.~(\ref{vituperative}) from the left by 
$\tilde\rho_0^{1/2}\tilde U^\dagger$ and 
$\tilde\rho_0^{1/2}\tilde V^\dagger$, respectively, and partial tracing
the first over A and the second over B, we get
\begin{equation}
\rho_0\,=\,{\rm tr}_{\scriptscriptstyle{\rm A}}
\Bigl(\tilde\rho_0^{1/2}\tilde U^\dagger\tilde\rho_1^{1/2}\Bigr)G
\eeq
and
\beq
\rho_0\,=\,{\rm tr}_{\scriptscriptstyle{\rm B}}
\Bigl(\tilde\rho_0^{1/2}\tilde V^\dagger\tilde\rho_1^{1/2}\Bigr)H\;.
\label{Lewis}
\end{equation}
Since, by assumption, $\rho_0$ is invertible, it follows that 
$G$ and $H$ are invertible.  Returning to Eq.~(\ref{vituperative}), 
multiplying both parts from the left by $\tilde \rho_1^{1/2}$ and 
tracing over A and B, respectively, we obtain
\begin{equation}
{\rm tr}_{\scriptscriptstyle{\rm A}}
\Bigl(\tilde\rho_1^{1/2}\tilde U\tilde\rho_0^{1/2}\Bigr)\,=\,\rho_1G
\label{stuckle}
\eeq
and
\beq
{\rm tr}_{\scriptscriptstyle{\rm B}}
\Bigl(\tilde\rho_1^{1/2}\tilde V\tilde\rho_0^{1/2}\Bigr)\,=\,\rho_1H\;.
\label{miasma}
\end{equation}
Conjugating Eqs.~(\ref{stuckle}) and (\ref{miasma}) and inserting
the results into the two parts of Eq.~(\ref{Lewis}) yields
\begin{equation}
\rho_0=G\rho_1G
\;\;\;\;\;\;\;\;\mbox{and}\;\;\;\;\;\;\;\;
\rho_0=H\rho_1H\;.
\label{yellowbelly}
\end{equation}
This shows that $\mbox{$G=H$}$, because these equations have a 
unique positive solution, namely the operator $M$ of Eq.~(\ref{rimbo}).  
This can be seen by multiplying Eq.~(\ref{yellowbelly})
from the left and right by $\rho_1^{1/2}$ to get
\beq
\rho_1^{1/2}\rho_0\rho_1^{1/2}\,=\,
\Bigl(\rho_1^{1/2}G\rho_1^{1/2}\Bigr)^{\!2}\;.
\eeq
The positive operator $\rho_1^{1/2}G\rho_1^{1/2}$ is thus
the unique positive square root of $\rho_1^{1/2}\rho_0\rho_1^{1/2}$.

Knowing that
\beq
G=H=M\;,
\eeq
we return to Eqs.~(\ref{Kunkel}) and
(\ref{vituperative}).  The two, taken together, imply that
\begin{equation}
\tilde V^\dagger\tilde U\tilde\rho_0^{1/2}\,=\,\tilde\rho_0^{1/2}
(M^{-1}\!\otimes M)\;.
\label{BiancaJ}
\end{equation}
If $|b\rangle$ and $|c\rangle$ are eigenvectors of $M$, with eigenvalues
$\mu_b$ and $\mu_c$, Eq.~(\ref{BiancaJ}) implies that
\begin{equation}
\tilde V^\dagger\tilde U\Bigl(\tilde\rho_0^{1/2}|b\rangle|c\rangle\Bigr)\,=\,
\frac{\mu_c}{\mu_b}\Bigl(\tilde\rho_0^{1/2}|b\rangle|c\rangle\Bigr)\;.
\label{salsa}
\end{equation}
This means that $\tilde\rho_0^{1/2}|b\rangle|c\rangle$ 
is zero or it is an eigenvector of the unitary operator 
$\tilde V^\dagger\tilde U$.  In the latter case, since the eigenvalues 
of a unitary operator have modulus 1, it must be true that 
$\mu_b=\mu_c$.  Hence we can conclude that 
\begin{equation}
\tilde\rho_0^{1/2}|b\rangle|c\rangle=0
\mbox{\qquad when\qquad}\mu_b\ne\mu_c\;.
\label{Benjamin}
\end{equation}
This is enough to show that $M$ and $\rho_0$ commute and hence
\beq
[\rho_0,\rho_1]=0\;.
\eeq
To see this, consider the matrix element
\begin{eqnarray}
\langle b'|(M\rho_0-\rho_0M)|b\rangle
&=&
(\mu_{b'}-\mu_b)\langle b'|\rho_0|b\rangle\nonumber\\
&=&
(\mu_{b'}-\mu_b)\langle b'|{\rm tr}_{\scriptscriptstyle
{\rm A}}(\tilde\rho_0)|b\rangle\rule{0mm}{7mm}\nonumber\\
&=&
(\mu_{b'}-\mu_b)\sum_c
\langle b'|\langle c|\,\tilde\rho_0|c\rangle|b\rangle\;.\rule{0mm}{7mm}
\end{eqnarray}
If $\mu_b=\mu_{b'}$, this is automatically zero.  If, on the other hand,
$\mu_b\ne\mu_{b'}$, then the sum over $c$ must vanish by Eq.~(\ref{Benjamin}).
It follows that $\rho_0$ and $M$ commute.  Hence, using
Eq.~(\ref{yellowbelly}),
\beqa
\rho_1\rho_0
&=&
M^{-1}\rho_0M^{-1}\rho_0\nonumber\\
&=&
\rho_0M^{-1}\rho_0M^{-1}\nonumber\\
&=&
\rho_0\rho_1\;.
\eeqa
This completes the proof that noncommuting quantum states cannot be broadcast.

Note that, by the same method as above,
\beq
\tilde\rho_1^{1/2}|b\rangle|c\rangle=0
\mbox{\qquad when\qquad}\mu_b\ne\mu_c\;.
\eeq
This condition, along with Eq.~(\ref{Benjamin}), determines the conceivable
broadcasting states, in which the correlations between the systems A
and B range from purely classical to purely quantum.  For example,
since $\rho_0$ and $\rho_1$ commute, the states of Eq.~(\ref{waldo}) satisfy
these conditions, but so do the perfectly entangled pure states
\beq
|\tilde\psi_s\rangle=\sum_b\sqrt{\lambda_{sb}}|b\rangle|b\rangle\;.
\label{HorseMeat}
\eeq
However, not all such potential broadcasting states can be realized by a 
physical process $\cal E$.  The reason for this is quite intuitive: since the
states $\rho_0$ and $\rho_1$ commute, the eigenvalues $\lambda_{0b}$ and 
$\lambda_{1b}$ correspond to two different probability distributions for the
eigenvectors.  Any device that could produce the states in Eq.~(\ref{HorseMeat})
would have to essentially read the mind of the person who set the (subjective)
probability assignments---clearly this cannot be done.

Nevertheless, this can be seen in a more formal way with a simple example.
Suppose $S(\rho_0)\ne S(\rho_1)$, where
\beq
S(\rho)=-\tr(\rho\ln\rho)
\eeq
denotes the von Neumann entropy.  In order for the states in 
Eq.~(\ref{HorseMeat}) to come about, the unitary operator in
Eq.~(\ref{process}) must be such that
\beq
U(\rho_s\otimes\Sigma\otimes\Upsilon)U^\dagger=|\tilde\psi_s\rangle\langle
\tilde\psi_s|\otimes\Upsilon_s\;.
\label{Meal-O-Matic}
\eeq
It then follows, by the unitary invariance of the von Neumann entropy and the
fact that entropies add across independent subsystems, that
\beq
S(\rho_0)-S(\rho_1)=S(\Upsilon_0)-S(\Upsilon_1)\;.
\label{FullBelly}
\eeq
However,
\beq
F(\rho_0,\rho_1)=|\langle\tilde\psi_0|\tilde\psi_1\rangle|
\eeq
by construction.  Therefore, by Eq.~(\ref{Meal-O-Matic}),
$F(\Upsilon_0,\Upsilon_1)=1$. Hence $\Upsilon_0=\Upsilon_1$ and it follows
that Eq.~(\ref{FullBelly}) cannot be satisfied.

In closing, we mention an application of this result.  In some versions of 
quantum cryptography \cite{Bennett92b}, the legitimate users of a communication 
channel encode the bits 0 and 1 into nonorthogonal pure states.  This is done
to ensure that any eavesdropping is detectable, since eavesdropping necessarily 
disturbs the states sent to the legitimate receiver \cite{Bennett92a}.  If 
the channel is noisy, however, causing the bits to evolve to noncommuting mixed
states, the detectability of eavesdropping is no longer a given.  The result 
presented here shows that there are no means available for an eavesdropper to 
obtain the signal, noise and all, intended for the legitimate receiver without 
in some way changing the states sent to the receiver.  Because the
dimensionality of the density operators in the no-broadcasting theorem are
completely arbitrary, this conclusion holds for all possible eavesdropping
attacks. This includes those schemes where measurements are made on whole
strings of quantum systems rather than the individual ones.


\chapter{References for Research in Quantum Distinguishability and State
Disturbance}

\begin{flushright}
\baselineskip=13pt
\parbox{2.8in}{\baselineskip=13pt
``Of course, serendipity played its role---some of the liveliest specimens
\ldots\ were found while looking for something else.''}\medskip\\
---{\it Nicolas Slonimsky}\\
Lexicon of Musical Invective
\end{flushright}
\bigskip

This Chapter contains 528 references that may be useful in answering the
following questions in all their varied contexts: ``How statistically 
distinguishable are quantum states?'' and ``What is the best tradeoff between 
disturbance and inference in quantum measurement?''  References are grouped under three major headings:  Progress Toward the Quantum Problem; Information
Theory and Classical Distinguishability; and Matrix Inequalities, 
Operator Relations, and Mathematical Techniques.  

\sloppy
\baselineskip=14pt

\section{Progress Toward the Quantum Problem}
\bigskip

\begin{enumerate}

\item
P.~M. Alberti, ``A note on the transition probability over ${C}^*$-algebras,''
  {\em Letters in Mathematical Physics}, vol.~7, pp.~25--32, 1983.

\item
P.~M. Alberti and A.~Uhlmann, ``Stochastic linear maps and transition
  probability,'' {\em Letters in Mathematical Physics}, vol.~7, pp.~107--112,
  1983.

\item
A.~Albrecht, ``Locating relative information in quantum systems,'' {\em
  Imperial College preprint}, April 1994.

\item
H.~Araki and G.~Raggio, ``A remark on transition probability,'' {\em Letters in
  Mathematical Physics}, vol.~6, p.~237, 1982.

\item
H.~Araki, ``Recent progress on entropy and relative entropy,'' in {\em VIII'th
  International Congress on Mathematical Physics} (M.~Melokhout and
  R.~S\'{e}n\'{e}or, eds.), (Singapore), pp.~354--365, World Scientific, 1987.

\item
R.~Balian, Y.~Alhassid, and H.~Reinhardt, ``Dissipation in many-body systems:
  {A} geometric approach based on information theory,'' {\em Physics Reports},
  vol.~131(1,2), pp.~1--146, 1986.

\item
L.~E. Ballentine, ``Can the statistical postulate of quantum theory be
  derived?---{A} critique of the many-universes interpretation,'' {\em
  Foundations of Physics}, vol.~3(2), pp.~229--240, 1973.

\item
L.~E. Ballentine, ``Probability theory in quantum mechanics,'' {\em American
  Journal of Physics}, vol.~54(10), pp.~883--889, 1986.

\item
W.~Band and J.~L. Park, ``The empirical determination of quantum states,'' {\em
  Foundations of Physics}, vol.~1(2), pp.~133--144, 1970.

\item
W.~Band and J.~L. Park, ``A general method of empirical state determination in
  quantum physics: {P}art {II},'' {\em Foundations of Physics}, vol.~1(4),
  pp.~339--357, 1971.

\item
W.~Band and J.~L. Park, ``Quantum state determination: {Q}uorum for a particle
  in one dimension,'' {\em American Journal of Physics}, vol.~47(2),
  pp.~188--191, 1979.

\item
S.~M. Barnett and S.~J.~D. Phoenix, ``Information-theoretic limits to quantum
  cryptography,'' {\em Physical Review A}, vol.~48(1), pp.~R5--R8, 1993.

\item
S.~M. Barnett, R.~Loudon, D.~T. Pegg, and S.~J.~D. Phoenix, ``Communication
  using quantum states,'' {\em Journal of Modern Optics}, vol.~41(12),
  pp.~2351--2373, 1994.

\item
P.~A. Benioff, ``Finite and infinite measurement sequences in quantum mechanics
  and randomness: {T}he {E}verett interpretation,'' {\em Journal of
  Mathematical Physics}, vol.~18(12), pp.~2289--2295, 1977.

\item
P.~Benioff, ``A note on the {E}verett interpretation of quantum mechanics,''
  {\em Foundations of Physics}, vol.~8(9/10), pp.~709--720, 1978.

\item
C.~H. Bennett, G.~Brassard, and N.~D. Mermin, ``Quantum cryptography without
  {B}ell's theorem,'' {\em Physical Review Letters}, vol.~68(5), pp.~557--559,
  1992.

\item
C.~H. Bennett, ``Quantum cryptography using any two nonorthogonal states,''
  {\em Physical Review Letters}, vol.~68(21), pp.~3121--3124, 1992.

\item
C.~H. Bennett and S.~J. Wiesner, ``Communication via one- and two-particle
  operators on {E}instein-{P}odolsky-{R}osen states,'' {\em Physical Review
  Letters}, vol.~69(20), pp.~2881--2884, 1992.

\item
C.~H. Bennett, G.~Brassard, C.~Cr\'{e}peau, R.~Jozsa, A.~Peres, and W.~K.
  Wootters, ``Teleporting an unknown quantum state via dual classical and
  {E}instein-{P}odolsky-{R}osen channels,'' {\em Physical Review Letters},
  vol.~70(13), pp.~1895--1899, 1993.

\item
C.~H. Bennett, G.~Brassard, R.~Jozsa, D.~Mayers, A.~Peres, B.~Schumacher, and
  W.~K. Wootters, ``Reduction of quantum entropy by reversible extraction of
  classical information,'' {\em Journal of Modern Optics}, vol.~41(12),
  pp.~2307--2314, 1994.

\item
C.~H. Bennett, ``Quantum information and computation,'' {\em Physics Today},
  vol.~48(10), pp.~24--30, 1995.

\item
G.~Brassard, ``A bibliography of quantum cryptography,'' {\em Universit\'{e} de
  Montr\'{e}al preprint}, pp.~1--10, 3 December 1993.
 A preliminary version of this appeared in {\it Sigact News}, vol.\
  24(3), 1993, pp.\ 16-20.

\item
S.~L. Braunstein and C.~M. Caves, ``Quantum rules: {A}n effect can have more
  than one operation,'' {\em Foundations of Physics Letters}, vol.~1,
  pp.~3--12, 1988.

\item
S.~L. Braunstein and C.~M. Caves, ``Information-theoretic {B}ell
  inequalities,'' {\em Physical Review Letters}, vol.~61(6), pp.~662--665,
  1988.

\item
S.~L. Braunstein, {\em Novel Quantum States and Measurements}.
 PhD thesis, California Institute of Technology, Pasadena, California,
  1988.

\item
S.~L. Braunstein and C.~M. Caves, ``Wringing out better {B}ell inequalities,''
  {\em Annals of Physics}, vol.~202, pp.~22--56, 1990.

\item
S.~L. Braunstein, ``Fundamental limits to precision measurements,'' in {\em
  Symposium on the Foundation of Modern Physics 1993} (P.~Busch, P.~Lahti, and
  P.~Mittelstaedt, eds.), (Singapore), pp.~106--117, World Scientific, 1993.

\item
S.~L. Braunstein and C.~M. Caves, ``Statistical distance and the geometry of
  quantum states,'' {\em Physical Review Letters}, vol.~72, pp.~3439--3443,
  1994.

\item
S.~L. Braunstein, C.~M. Caves, and G.~J. Milburn, ``Generalized uncertainty
  relations: Theory, examples, and {L}orentz invariance,'' {\em University of
  New Mexico preprint}, 1995.

\item
S.~L. Braunstein and C.~M. Caves, ``Geometry of quantum states,'' in {\em
  Quantum Communications and Measurement} (V.~P. Belavkin, O.~Hirota, and R.~L.
  Hudson, eds.), (New York), pp.~21--30, Plenum Press, 1995.

\item
S.~L. Braunstein and C.~M. Caves, ``Geometry of quantum states,'' in {\em
  Fundamental Problems in Quantum Theory: A Conference Held in Honor of John A.
  Wheeler} (D.~M. Greenberger and A.~Zeilinger, eds.), (New York),
  pp.~786--797, New York Academy of Sciences, 1995.
 [{\it Annals of the New York Academy of Sciences\/}, vol.\ 755].

\item
S.~L. Braunstein, ``Geometry of quantum inference,'' in {\em Sixty Years of
  EPR} (A.~Mann and M.~Revzen, eds.), (Israel), Ann. Phys. Soc., 1995.
 To be published.

\item
S.~L. Braunstein, ``Ignorance in the service of uncertainty,'' {\em Weizmann
  Inst.\ preprint}, 1995.

\item
D.~Brody and B.~Meister, ``Minimum decision cost for quantum ensembles,'' {\em
  Imperal College preprint}, pp.~1--11, 1995.

\item
J.~Bub, ``{von} {N}eumann's projection postulate as a probability
  conditionalization rule in quantum mechanics,'' {\em Journal of Philosophical
  Logic}, vol.~6, pp.~381--390, 1977.

\item
D.~Bures, ``An extension of {K}akutani's theorem on infinite product measures
  to the tensor product of semifinite {$w^*$}-algebras,'' {\em Transactions of
  the American Mathematical Society}, vol.~135, pp.~199--212, 1969.

\item
V.~Cantoni, ``Generalized `transition probability','' {\em Communications in
  Mathematical Physics}, vol.~44, pp.~125--128, 1975.

\item
C.~M. Caves, ``Entropy and information: {H}ow much information is needed to
  assign a probability?,'' in {\em Complexity, Entropy and the Physics of
  Information} (W.~H. Zurek, ed.), (Redwood City, CA), pp.~91--115,
  Addison-Wesley, 1990.
 Santa Fe Institute Studies in the Sciences of Complexity, vol. VIII.

\item
C.~M. Caves, ``Information and entropy,'' {\em Physical Review E}, vol.~47(6),
  pp.~4010--4017, 1993.

\item
C.~M. Caves and P.~D. Drummond, ``Quantum limits on bosonic communication
  rates,'' {\em Reviews of Modern Physics}, vol.~66(2), pp.~481--537, 1994.

\item
C.~M. Caves, ``Information, entropy, and chaos,'' in {\em Physical Origins of
  Time Asymmetry} (J.~J. Halliwell, J.~P\'{e}rez-Mercader, and W.~H. Zurek,
  eds.), (Cambridge), pp.~47--89, Cambridge University Press, 1994.

\item
J.~F. Clauser, ``{von} {N}eumann's informal hidden-variable argument,'' {\em
  American Journal of Physics}, vol.~39, pp.~1095--1096, 1971.

\item
E.~B. Davies and J.~T. Lewis, ``An operational approach to quantum
  probability,'' {\em Communications in Mathematical Physics}, vol.~17,
  pp.~239--260, 1970.

\item
E.~B. Davies, {\em Quantum Theory of Open Systems}.
 London: Academic Press, 1976.

\item
E.~B. Davies, ``Information and quantum measurement,'' {\em IEEE Transactions
  on Information Theory}, vol.~IT-24(5), pp.~596--599, 1978.

\item
B.~S. DeWitt and N.~Graham, eds., {\em The Many-Worlds Interpretation of
  Quantum Mechanics}.
 Princeton Series in Physics, Princeton, NJ: Princeton University
  Press, 1973.

\item
D.~Dieks, ``Communication by {EPR} devices,'' {\em Physics Letters A},
  vol.~92(6), pp.~271--272, 1982.

\item
D.~Dieks and P.~Veltkamp, ``Distance between quantum states, statistical
  inference and the projection postulate,'' {\em Physics Letters A},
  vol.~97(1,2), pp.~24--28, 1983.

\item
J.~Dittmann and G.~Rudolph, ``On a connection governing parallel transport
  along $2\times2$ density matrices,'' {\em Journal of Geometry and Physics},
  vol.~10, pp.~93--106, 1992.

\item
J.~Dittmann, ``Some properties of the {R}iemannian {B}ures metric on mixed
  states,'' {\em Journal of Geometry and Physics}, vol.~13, pp.~203--206, 1994.

\item
M.~J. Donald, ``On the relative entropy,'' {\em Communications in Mathematical
  Physics}, vol.~105, pp.~13--34, 1986.

\item
M.~J. Donald, ``Free energy and the relative entropy,'' {\em Journal of
  Statistical Physics}, vol.~49(1/2), pp.~81--87, 1987.

\item
M.~J. Donald, ``Further results on the relative entropy,'' {\em Mathematical
  Proceedings of the Cambridge Philosophical Society}, vol.~101, pp.~363--373,
  1987.

\item
M.~J. Donald, ``{\it A priori} probability and localized observers,'' {\em
  Foundations of Physics}, vol.~22(9), pp.~1111--1172, 1992.

\item
M.~J. Donald, ``Probabilities for observing mixed quantum states given limited
  prior information,'' in {\em Quantum Communications and Measurement} (V.~P.
  Belavkin, O.~Hirota, and R.~L. Hudson, eds.), (New York), pp.~411--418,
  Plenum Press, 1995.

\item
C.~M. Edwards, ``The operational approach to algebraic quantum theory {I},''
  {\em Communications in Mathematical Physics}, vol.~16(3), pp.~207--230, 1970.

\item
C.~M. Edwards, ``Classes of operations in quantum theory,'' {\em Communications
  in Mathematical Physics}, vol.~20(1), pp.~26--56, 1971.

\item
A.~K. Ekert, ``Quantum cryptography based on {B}ell's theorem,'' {\em Physical
  Review Letters}, vol.~67(6), pp.~661--663, 1991.

\item
A.~K. Ekert, B.~Huttner, G.~M. Palma, and A.~Peres, ``Eavesdropping on
  quantum-cryptographical systems,'' {\em Physical Review A}, vol.~50(2),
  pp.~1047--1056, 1994.

\item
E.~Farhi, J.~Goldstone, and S.~Gutman, ``How probability arises in quantum
  mechanics,'' {\em Annals of Physics}, vol.~192, pp.~368--382, 1989.

\item
R.~P. Feynman, ``The concept of probability in quantum mechanics,'' in {\em
  Proceedings of the Second Berkeley Symposium on Mathematical Statistics and
  Probability} (J.~Neyman, ed.), (Berkeley, CA), pp.~533--541, University of
  California Press, 1951.

\item
S.~K. Foong and S.~Kanno, ``Proof of {P}age's conjecture on the average entropy
  of a subsystem,'' {\em Physical Review Letters}, vol.~72(8), pp.~1148--1151,
  1994.

\item
G.~D. {Forney, Jr.}, ``Upper bound to the capacity of a linear time-varying
  channel with additive {G}aussian noise,'' Master's thesis, Massachusetts
  Institute of Technology, Lincoln Laboratory, Cambridge, MA, September 26
  1962.

\item
C.~A. Fuchs and C.~M. Caves, ``Ensemble-dependent bounds for accessible
  information in quantum mechanics,'' {\em Physical Review Letters},
  vol.~73(23), pp.~3047--3050, 1994.

\item
C.~A. Fuchs and C.~M. Caves, ``Bounds for accessible information in quantum
  mechanics,'' in {\em Fundamental Problems in Quantum Theory: A Conference
  Held in Honor of John A. Wheeler} (D.~M. Greenberger and A.~Zeilinger, eds.),
  (New York), pp.~706--715, New York Academy of Sciences, 1995.
 [{\it Annals of the New York Academy of Sciences\/}, vol.\ 755].

\item
C.~A. Fuchs and C.~M. Caves, ``Mathematical techniques for quantum
  communication theory,'' {\em Open Systems and Information Dynamics},
  vol.~3(3), pp.~1--12, to be published.

\item
W.~Gale, E.~Guth, and G.~T. Trammell, ``Determination of the quantum state by
  measurements,'' {\em Physical Review}, vol.~165(6), pp.~1434--1436, 1968.

\item
G.~W. Gibbons, ``Typical states and density matrices,'' {\em Journal of
  Geometry and Physics}, vol.~8, pp.~147--162, 1992.

\item
R.~Giles, ``Foundations for quantum mechanics,'' {\em Journal of Mathematical
  Physics}, vol.~11(7), pp.~2139--2160, 1970.

\item
A.~M. Gleason, ``Measures on the closed subspaces of a {H}ilbert space,'' {\em
  Journal of Mathematics and Mechanics}, vol.~6(6), pp.~885--893, 1957.

\item
J.~P. Gordon, ``Noise at optical frequencies; information theory,'' in {\em
  Quantum Electronics and Coherent Light; Proceedings of the International
  School of Physics ``Enrico Fermi,'' Course XXXI} (P.~A. Miles, ed.), (New
  York), pp.~156--181, Academic Press, 1964.

\item
N.~Graham, {\em The {E}verett Interpretation of Quantum Mechanics}.
 PhD thesis, University of North Carolina at Chapel Hill, Chapel Hill,
  NC, 1970.

\item
S.~Gudder, J.-P. Marchand, and W.~Wyss, ``Bures distance and relative
  entropy,'' {\em Journal of Mathematical Physics}, vol.~20(9), pp.~1963--1966,
  1979.

\item
S.~P. Gudder, ``Expectation and transition probability,'' {\em International
  Journal of Theoretical Physics}, vol.~20(5), pp.~383--395, 1981.

\item
R.~Haag and D.~Kastler, ``An algebraic approach to quantum field theory,'' {\em
  Journal of Mathematical Physics}, vol.~5(7), pp.~848--861, 1964.

\item
N.~Hadjisavvas, ``Distance between states and statistical inference in quantum
  theory,'' {\em Annales de l'Institut Henri Poincar\'{e}--Section A},
  vol.~35(4), pp.~287--309, 1981.

\item
N.~Hadjisavvas, ``On {C}antoni's generalized transition probability,'' {\em
  Communications in Mathematical Physics}, vol.~83, pp.~43--48, 1982.

\item
N.~Hadjisavvas, ``Metrics on the set of states of a {$W^*$}-algebra,'' {\em
  Linear Algebra and Its Applications}, vol.~84, pp.~281--287, 1986.

\item
M.~J.~W. Hall and M.~J. O'Rourke, ``Realistic performance of the maximum
  information channel,'' {\em Quantum Optics}, vol.~5, pp.~161--180, 1993.

\item
M.~J.~W. Hall, ``Information exclusion principle for complementary
  observables,'' {\em Physical Review Letters}, vol.~74(17), pp.~3307--3311,
  1995.

\item
J.~E. Harriman, ``Geometry of density matrices.\ {I}.\ {D}efinitions {$N$} and
  {$1$} matrices,'' {\em Physical Review A}, vol.~17(4), pp.~1249--1256, 1978.

\item
J.~E. Harriman, ``Geometry of density matrices.\ {II}.\ {R}educed density
  matrices and {$N$} representability,'' {\em Physical Review A}, vol.~17(4),
  pp.~1257--1268, 1978.

\item
J.~B. Hartle, ``Quantum mechanics of individual systems,'' {\em American
  Journal of Physics}, vol.~36(8), pp.~704--712, 1968.

\item
H.~Hasegawa, ``$\alpha$-divergence of the non-commutative information
  geometry,'' {\em Reports on Mathematical Physics}, vol.~33(1/2), pp.~87--93,
  1993.

\item
H.~Hasegawa, ``Non-commutative extension of the information geometry,'' in {\em
  Quantum Communications and Measurement} (V.~P. Belavkin, O.~Hirota, and R.~L.
  Hudson, eds.), (New York), pp.~327--337, Plenum Press, 1995.

\item
P.~Hausladen and W.~K. Wootters, ``A `pretty good' measurement for
  distinguishing quantum states,'' {\em Journal of Modern Optics}, vol.~41(12),
  pp.~2385--2390, 1994.

\item
P.~Hausladen, B.~Schumacher, M.~Westmoreland, and W.~K. Wootters, ``Sending
  classical bits via quantum its,'' in {\em Fundamental Problems in Quantum
  Theory: A Conference Held in Honor of John A. Wheeler} (D.~M. Greenberger and
  A.~Zeilinger, eds.), (New York), pp.~698--705, New York Academy of Sciences,
  1995.
 [{\it Annals of the New York Academy of Sciences\/}, vol.\ 755].

\item
K.~Hellwig and K.~Kraus, ``Pure operations and measurements,'' {\em
  Communications in Mathematical Physics}, vol.~11(3), pp.~214--220, 1969.

\item
K.~Hellwig and K.~Kraus, ``Operations and measurements.\ {II},'' {\em
  Communications in Mathematical Physics}, vol.~16(2), pp.~142--147, 1970.

\item
K.~Hellwig and W.~Stulpe, ``A classical reformulation of finite-dimensional
  quantum mechanics,'' in {\em Symposium on the Foundation of Modern Physics
  1993} (P.~Busch, P.~Lahti, and P.~Mittelstaedt, eds.), (Singapore),
  pp.~209--214, World Scientific, 1993.

\item
C.~W. Helstrom, ``Detection theory and quantum mechanics,'' {\em Information
  and Control}, vol.~10, pp.~254--291, 1967.

\item
C.~W. Helstrom, {\em Quantum Detection and Estimation Theory}.
 Mathematics in Science and Engineering, vol.\ 123, New York: Academic
  Press, 1976.

\item
F.~Hiai and D.~Petz, ``The proper formula for relative entropy and its
  asymptotics in quantum probability,'' {\em Communications in Mathematical
  Physics}, vol.~143, pp.~99--114, 1991.

\item
A.~S. Holevo, ``Statistical problems in quantum physics,'' in {\em Proceedings
  of the Second Japan--USSR Symposium on Probability Theory} (G.~Maruyama and
  J.~V. Prokhorov, eds.), (Berlin), pp.~104--119, Springer-Verlag, 1973.
 Lecture Notes in Mathematics, vol.\ 330.

\item
A.~S. Holevo, ``Statistical decisions in quantum theory,'' {\em Theory of
  Probability and Its Applications}, vol.~18, pp.~418--420, 1973.

\item
A.~S. Holevo, ``Information-theoretical aspects of quantum measurement,'' {\em
  Problemy Pere\-dachi Informatsii}, vol.~9(2), pp.~31--42, 1973.
 [A. S. Kholevo, {\it Problems of Information Transmission\/}, vol.\
  9, pp.\ 110-118 (1973)].

\item
A.~S. Holevo, ``Bounds for the quantity of information transmitted by a quantum
  communication channel,'' {\em Problemy Pere\-dachi Informatsii}, vol.~9(3),
  pp.~3--11, 1973.
 [A. S. Kholevo, {\it Problems of Information Transmission\/}, vol.\
  9, pp.\ 177-183 (1973)].

\item
A.~S. Holevo, ``Statistical decision theory for quantum systems,'' {\em Journal
  of Multivariate Analysis}, vol.~3, pp.~337--394, 1973.

\item
A.~S. Holevo, ``Some statistical problems for quantum {G}aussian states,'' {\em
  IEEE Transactions on Information Theory}, vol.~IT-21(5), pp.~533--543, 1975.

\item
A.~S. Holevo, ``Noncommutative analogues of the {C}ram\'{e}r-{R}ao inequality
  in the quantum theory of measurement,'' in {\em Proceedings of the Third
  Japan--USSR Symposium on Probability Theory} (G.~Maruyama and J.~V.
  Prokhorov, eds.), (Berlin), pp.~194--222, Springer-Verlag, 1976.
 Lecture Notes in Mathematics, vol.\ 550.

\item
A.~S. Holevo, ``Commutation superoperator of a state and its applications to
  the noncommutative statistics,'' {\em Reports on Mathematical Physics},
  vol.~12(2), pp.~251--271, 1977.

\item
A.~S. Holevo, ``Problems in the mathematical theory of quantum communication
  channels,'' {\em Reports on Mathematical Physics}, vol.~12(2), pp.~273--278,
  1977.

\item
A.~S. Holevo, {\em Probabilistic and Statistical Aspects of Quantum Theory}.
 North-Holland Series in Statistics and Probability, vol.\ 1,
  Amsterdam: North-Holland, 1982.

\item
M.~H\"{u}bner, ``Explicit computation of the {B}ures distance for density
  matrices,'' {\em Physics Letters A}, vol.~163, pp.~239--242, 1992.

\item
M.~H\"{u}bner, ``Computation of {U}hlmann's parallel transport for density
  matrices and the {B}ures metric on three-dimensional {H}ilbert space,'' {\em
  Physics Letters A}, vol.~179, pp.~226--230, 1993.

\item
L.~P. Hughston, R.~Jozsa, and W.~K. Wootters, ``A complete classification of
  quantum ensembles having a given density matrix,'' {\em Physics Letters A},
  vol.~183, pp.~14--18, 1993.

\item
L.~P. Hughston, ``Geometric aspects of quantum mechanics,'' {\em Merrill Lynch
  International Limited preprint}, pp.~1--21, 1994.

\item
B.~Huttner and A.~Peres, ``Quantum cryptography with photon pairs,'' {\em
  Journal of Modern Optics}, vol.~41(12), pp.~2397--2403, 1994.

\item
B.~Huttner and A.~K. Ekert, ``Information gain in quantum eavesdropping,'' {\em
  Journal of Modern Optics}, vol.~41(12), pp.~2455--2466, 1994.

\item
R.~S. Ingarden, ``Information geometry in functional spaces of classical and
  quantum finite statistical systems,'' {\em International Journal of
  Engineering Science}, vol.~19(12), pp.~1609--1633, 1981.

\item
I.~D. Ivanovi\'{c}, ``Geometrical description of quantal state determination,''
  {\em Journal of Physics A}, vol.~14(1), pp.~3241--3245, 1981.

\item
I.~D. Ivanovi\'{c}, ``Formal state determination,'' {\em Journal of
  Mathematical Physics}, vol.~24(5), pp.~1199--1205, 1983.

\item
G.~Jaeger and A.~Shimony, ``Optimal distinction between two non-orthogonal
  quantum states,'' {\em Physics Letters A}, vol.~197, pp.~83--87, 1995.

\item
J.~M. Jauch and C.~Piron, ``Generalized localizability,'' {\em Helvetica
  Physica Acta}, vol.~40, pp.~559--570, 1967.

\item
K.~R.~W. Jones, {\em Quantum Inference and the Optimal Determination of Quantum
  States}.
 PhD thesis, University of Bristol, Bristol, England, 1989.

\item
K.~R.~W. Jones, ``Entropy of random quantum states,'' {\em Journal of Physics
  A}, vol.~23(23), pp.~L1247--L1250, 1990.

\item
K.~R.~W. Jones, ``Principles of quantum inference,'' {\em Annals of Physics},
  vol.~207(1), pp.~140--170, 1991.

\item
K.~R.~W. Jones, ``Quantum limits to information about states for finite
  dimensional {H}ilbert space,'' {\em Journal of Physics A}, vol.~24,
  pp.~121--130, 1991.

\item
K.~R.~W. Jones, ``Towards a proof of two conjectures from quantum inference
  concerning quantum limits to knowledge of states,'' {\em Journal of Physics
  A}, vol.~24(8), pp.~L415--L419, 1991.

\item
K.~R.~W. Jones, ``Riemann-{L}iouville fractional integration and reduced
  distributions on hyperspheres,'' {\em Journal of Physics A}, vol.~24,
  pp.~1237--1244, 1991.

\item
K.~R.~W. Jones, ``Fractional integration and uniform densities in quantum
  mechanics,'' in {\em Recent Advances in Fractional Calculus} (R.~N. Kalia,
  ed.), (Sauk Rapids, MN), pp.~203--218, Global Publishing, 1993.

\item
K.~R.~W. Jones, ``Fundamental limits upon the measurement of state vectors,''
  {\em Physical Review A}, vol.~50(5), pp.~3682--3699, 1994.

\item
R.~Jozsa, ``Fidelity for mixed quantum states,'' {\em Journal of Modern
  Optics}, vol.~41(12), pp.~2315--2323, 1994.

\item
R.~Jozsa, D.~Robb, and W.~K. Wootters, ``Lower bound for accessible information
  in quantum mechanics,'' {\em Physical Review A}, vol.~49, pp.~668--677, 1994.

\item
R.~Jozsa and B.~Schumacher, ``A new proof of the quantum noiseless coding
  theorem,'' {\em Journal of Modern Optics}, vol.~41(12), pp.~2343--2349, 1994.

\item
R.~V. Kadison, ``Isometries of operator algebras,'' {\em Annals of
  Mathematics}, vol.~54(2), pp.~325--338, 1951.

\item
R.~V. Kadison, ``Transformations of states in operator theory and dynamics,''
  {\em Topology}, vol.~3, suppl.\ 2, pp.~177--198, 1965.

\item
H.~Kosaki, ``On the {B}ures distance and the {U}hlmann transition probability
  of states on a von {N}eumann algebra,'' {\em Proceedings of the American
  Mathematical Society}, vol.~89(2), pp.~285--288, 1983.

\item
K.~Kraus, ``General state changes in quantum theory,'' {\em Annals of Physics},
  vol.~64, pp.~311--335, 1971.

\item
K.~Kraus, {\em States, Effects, and Operations:\ Fundamental Notions of Quantum
  Theory}.
 Lecture Notes in Physics, vol.\ 190, Berlin: Springer-Verlag, 1983.

\item
L.~B. Levitin, ``On the quantum measure of information,'' in {\em Proceedings
  of the Fourth All-Union Conference on Information and Coding Theory}, Sec.\
  II, Tashkent, 1969.
 Translation by A. Bezinger and S. L. Braunstein available.

\item
L.~B. Levitin, ``Physical information theory part {II}: Quantum systems,'' in
  {\em Workshop on Physics and Computation: PhysComp '92} (D.~Matzke, ed.),
  (Los Alamitos, CA), pp.~215--219, IEEE Computer Society, 1993.

\item
L.~B. Levitin, ``Optimal quantum measurements for two pure and mixed states,''
  in {\em Quantum Communications and Measurement} (V.~P. Belavkin, O.~Hirota,
  and R.~L. Hudson, eds.), (New York), pp.~439--448, Plenum Press, 1995.

\item
G.~Lindblad, ``An entropy inequality for quantum measurements,'' {\em
  Communications in Mathematical Physics}, vol.~28, pp.~245--249, 1972.

\item
G.~Lindblad, ``Entropy, information and quantum measurements,'' {\em
  Communications in Mathematical Physics}, vol.~33, pp.~305--322, 1973.

\item
G.~Lindblad, ``Expectations and entropy inequalities for finite quantum
  systems,'' {\em Communications in Mathematical Physics}, vol.~39,
  pp.~111--119, 1974.

\item
G.~Lindblad, ``On the generators of quantum dynamical semigroups,'' {\em
  Communications in Mathematical Physics}, vol.~48, pp.~119--130, 1976.

\item
G.~Lindblad, {\em Non-Equilibrium Entropy and Irreversibility}.
 Mathematical Physics Studies, vol.\ 5, Dordrecht: D. Reidel, 1983.

\item
G.~Lindblad, ``Quantum entropy and quantum measurements,'' in {\em Quantum
  Aspects of Optical Communications} (C.~Bendjaballah, O.~Hirota, and
  S.~Reynaud, eds.), Lecture Notes in Physics, vol.\ 378, (Berlin), pp.~71--80,
  Springer-Verlag, 1991.

\item
H.-K. Lo, ``Quantum coding theorem for mixed states,'' {\em Optics
  Communications}, vol.~119, pp.~552--556, 1995.

\item
J.~D. Malley and J.~Hornstein, ``Quantum statistical inference,'' {\em
  Statistical Sciences}, vol.~8(4), pp.~433--457, 1993.

\item
S.~Massar and S.~Popescu, ``Optimal extraction of information from finite
  quantum ensembles,'' {\em Physical Review Letters}, vol.~74(8),
  pp.~1259--1263, 1995.

\item
D.~Mayers, ``Generalized measurements and quantum transformations,'' {\em
  Universit\'{e} de Mon\-tr\'{e}al preprint}, pp.~1--13, 1994.

\item
H.~Nagaoka, ``On {F}isher information of quantum statistical models,'' in {\em
  The 10th Symposium on Information Theory and Its Applications} (??, ed.),
  (??), pp.~241--246, ??, 1987.
 Enoshima Island, Japan, Nov.\ 19-21, 1987. In Japanese.

\item
H.~Nagaoka, ``On the parameter estimation problem for quantum statistical
  models,'' in {\em The 12th Symposium on Information Theory and Its
  Applications (SITA '89)} (??, ed.), (??), pp.~577--582, ??, 1989.
 Inuyama, Japan, Dec.\ 6-9, 1989.

\item
H.~Nagaoka, ``Differential geometrical aspects of quantum state estimation and
  relative entropy,'' {\em University of Tokyo preprint}, September 1994.

\item
H.~Nagaoka, ``Differential geometrical aspects of quantum state estimation and
  relative entropy,'' in {\em Quantum Communications and Measurement} (V.~P.
  Belavkin, O.~Hirota, and R.~L. Hudson, eds.), (New York), pp.~449--452,
  Plenum Press, 1995.

\item
M.~Nakamura and T.~Turumaru, ``Expectations in an operator algebra,'' {\em
  Tohoku Mathematics Journal}, vol.~6, pp.~182--188, 1954.

\item
M.~Ohya, ``On compound state and mutual information in quantum information
  theory,'' {\em IEEE Transactions on Information Theory}, vol.~IT-29(5),
  pp.~770--774, 1983.

\item
M.~Ohya and D.~Petz, {\em Quantum Entropy and Its Use}.
 Texts and Monographs in Physics, Berlin: Springer-Verlag, 1993.

\item
M.~Ohya, ``State change, complexity and fractal in quantum systems,'' in {\em
  Quantum Communications and Measurement} (V.~P. Belavkin, O.~Hirota, and R.~L.
  Hudson, eds.), (New York), pp.~309--320, Plenum Press, 1995.

\item
M.~Ohya and N.~Watanabe, ``A mathematical study of information transmission in
  quantum communication processes,'' in {\em Quantum Communications and
  Measurement} (V.~P. Belavkin, O.~Hirota, and R.~L. Hudson, eds.), (New York),
  pp.~371--378, Plenum Press, 1995.

\item
M.~Osaki and O.~Hirota, ``On an effectiveness of quantum mini-max formula in
  quantum communication,'' in {\em Quantum Communications and Measurement}
  (V.~P. Belavkin, O.~Hirota, and R.~L. Hudson, eds.), (New York),
  pp.~401--409, Plenum Press, 1995.

\item
M.~Ozawa, ``Mathematical characterization of measurement statistics,'' in {\em
  Quantum Communications and Measurement} (V.~P. Belavkin, O.~Hirota, and R.~L.
  Hudson, eds.), (New York), pp.~109--117, Plenum Press, 1995.

\item
D.~N. Page, ``Average entropy of a subsystem,'' {\em Physical Review Letters},
  vol.~71(9), pp.~1291--1294, 1993.

\item
J.~L. Park and W.~Band, ``A general method of empirical state determination in
  quantum physics: {P}art {I},'' {\em Foundations of Physics}, vol.~1(3),
  pp.~211--226, 1971.

\item
M.~Pavi\u{c}i\'{c}, ``Bibliography on quantum logics and related structures,''
  {\em International Journal of Theoretical Physics}, vol.~31(3), pp.~373--460,
  1992.
 This bibliography contains more than 1600 items.

\item
A.~Peres, ``What is a state vector?,'' {\em American Journal of Physics},
  vol.~52(7), pp.~644--650, 1984.

\item
A.~Peres, ``Neumark's theorem and quantum inseparability,'' {\em Foundations of
  Physics}, vol.~20(12), pp.~1441--1453, 1990.

\item
A.~Peres and W.~K. Wootters, ``Optimal detection of quantum information,'' {\em
  Physical Review Letters}, vol.~66, pp.~1119--1122, 1991.

\item
A.~Peres, ``Storage and retrieval of quantum information,'' in {\em Workshop on
  Physics and Computation: PhysComp '92} (D.~Matzke, ed.), (Los Alamitos, CA),
  pp.~155--158, IEEE Computer Society, 1993.

\item
A.~Peres, {\em Quantum Theory: Concepts and Methods}.
 Fundamental Theories of Physics, Dordrecht: Kluwer Academic
  Publishers, 1993.

\item
D.~Petz and G.~Toth, ``The {B}ogoliubov inner product in quantum statistics,''
  {\em Letters in Mathematical Physics}, vol.~27, pp.~205--216, 1993.

\item
D.~Petz, ``Geometry of canonical correlation on the state space of a quantum
  system,'' {\em Journal of Mathematical Physics}, vol.~35(2), pp.~780--795,
  1994.

\item
S.~Popescu, ``Bell's inequalities versus teleportation: What is nonlocality?,''
  {\em Physical Review Letters}, vol.~72(6), pp.~797--799, 1994.

\item
S.~Popescu, ``Bell's inequalities and density matrices: Revealing `hidden'
  nonlocality,'' {\em Physical Review Letters}, vol.~74(14), pp.~2619--2622,
  1995.

\item
J.~P. Provost and G.~Vallee, ``Riemannian structure on manifolds of quantum
  states,'' {\em Communications in Mathematical Physics}, vol.~76,
  pp.~289--301, 1980.

\item
S.~Pulmannov\'{a} and B.~Stehl\'{\i}kov\'{a}, ``Strong law of large numbers and
  central limit theorem on a {H}ilbert space logic,'' {\em Reports on
  Mathematical Physics}, vol.~23(1), pp.~99--107, 1986.

\item
H.~P. Robertson, ``The uncertainty principle,'' {\em Physical Review}, vol.~34,
  pp.~163--164, 1929.

\item
R.~Schack, G.~M. {D'A}riano, and C.~M. Caves, ``Hypersensitivity to
  perturbation in the quantum kicked top,'' {\em Physical Review E},
  vol.~50(2), pp.~972--987, 1994.

\item
B.~Schumacher, ``Information from quantum measurements,'' in {\em Complexity,
  Entropy and the Physics of Information} (W.~H. Zurek, ed.), (Redwood City,
  CA), pp.~29--37, Addison-Wesley, 1990.
 Santa Fe Institute Studies in the Sciences of Complexity, vol. VIII.

\item
B.~W. Schumacher, {\em Communication, Correlation and Complementarity}.
 PhD thesis, The University of Texas at Austin, Austin, TX, 1990.

\item
B.~W. Schumacher, ``Information and quantum nonseparability,'' {\em Physical
  Review A}, vol.~44, pp.~7074--7079, 1991.

\item
B.~Schumacher, ``Quantum coding,'' {\em Physical Review A}, vol.~51(4),
  pp.~2738--2747, 1995.

\item
H.~Scutaru, ``Lower bound for mutual information of a quantum channel,'' {\em
  Physical Review Letters}, vol.~75(5), pp.~773--776, 1995.

\item
M.~D. Srinivas, ``Foundations of a quantum probability theory,'' {\em Journal
  of Mathematical Physics}, vol.~16(8), pp.~1672--1685, 1975.

\item
H.~P. Stapp, ``The {C}openhagen interpretation,'' {\em American Journal of
  Physics}, vol.~60, pp.~1098--1116, 1972.

\item
O.~Steuernagel and J.~A. Vaccaro, ``Reconstructing the density operator via
  simple projectors,'' {\em Physical Review Letters}, vol.~75(18),
  pp.~3201--3205, 1995.

\item
W.~F. Stinespring, ``Positive functions on {$C^*$}-algebras,'' {\em Proceedings
  of the American Mathematical Society}, vol.~6, pp.~211--216, 1955.

\item
E.~St{\o}rmer, ``Positive linear maps of operator algebras,'' {\em Acta
  Mathematica}, vol.~110, pp.~233--278, 1963.

\item
E.~St{\o}rmer, ``On projection maps on von {N}eumann algebras,'' {\em
  Mathematica Scandinavica}, vol.~30(1), pp.~46--50, 1972.

\item
M.~Strauss, {\em Modern Physics and Its Philosophy: Selected Papers in the
  Logic, History and Philosophy of Science}.
 Synthese Library, Dordrecht: D. Reidel, 1972.

\item
S.~S\'{y}kora, ``Quantum theory and the {B}ayesian inference problems,'' {\em
  Journal of Statistical Physics}, vol.~11(1), pp.~17--27, 1974.

\item
Y.~Tikochinsky, ``On the generalized multiplication and addition of complex
  numbers,'' {\em Journal of Mathematical Physics}, vol.~29(2), pp.~398--399,
  1988.

\item
Y.~Tikochinsky, ``Feynman rules for probability amplitudes,'' {\em
  International Journal of Theoretical Physics}, vol.~27(5), pp.~543--549,
  1988.

\item
A.~Uhlmann, ``The `transition probability' in the state space of a
  $\,^*$-algebra,'' {\em Reports on Mathematical Physics}, vol.~9,
  pp.~273--279, 1976.

\item
A.~Uhlmann, ``Parallel transport and `quantum holonomy' along density
  operators,'' {\em Reports on Mathematical Physics}, vol.~24(2), pp.~229--240,
  1986.

\item
A.~Uhlmann, ``Density operators as an arena for differential geometry,'' {\em
  Reports on Mathematical Physics}, vol.~33, pp.~253--263, 1993.

\item
A.~Uhlmann, ``The n-sphere as a quantal states space,'' in {\em Symposium on
  the Foundation of Modern Physics 1993} (P.~Busch, P.~Lahti, and
  P.~Mittelstaedt, eds.), (Singapore), pp.~390--397, World Scientific, 1993.

\item
H.~Umegaki, ``Conditional expectations in an operator algebra,'' {\em Tohoku
  Mathematics Journal}, vol.~6, pp.~177--181, 1954.

\item
H.~Umegaki, ``Conditional expectations in an operator algebra {IV} (entropy and
  information),'' {\em K\={o}dai Mathematical Seminar Reports}, vol.~14,
  pp.~59--85, 1962.

\item
R.~Urigu, ``On the physical meaning of the {S}hannon information entropy,'' in
  {\em Symposium on the Foundation of Modern Physics 1993} (P.~Busch, P.~Lahti,
  and P.~Mittelstaedt, eds.), (Singapore), pp.~398--405, World Scientific,
  1993.

\item
V.~S. Varadarajan, {\em Geometry of Quantum Theory}, vol.~1.
 Princeton, NJ: van Nostrand, 1968.

\item
V.~S. Varadarajan, {\em Geometry of Quantum Theory}, vol.~2.
 Princeton, NJ: van Nostrand, 1970.

\item
J.~{von Neumann}, {\em Mathematische Grundlagen der Quantenmechanik}.
 Berlin: Springer, 1932.
 Translated by E.\ T.\ Beyer, {\it Mathematical Foundations of Quantum
  Mechanics\/} (Princeton University Press, Princeton, NJ, 1955).

\item
A.~Wehrl, ``General properties of entropy,'' {\em Reviews of Modern Physics},
  vol.~50(2), pp.~221--260, 1978.

\item
E.~P. Wigner, ``On hidden variables and quantum mechanical probabilities,''
  {\em American Journal of Physics}, vol.~38(8), pp.~1005--1009, 1970.

\item
W.~K. Wootters and W.~H. Zurek, ``Complementarity in the double-slit
  experiment: {Q}uantum nonseparability and a quantitative statement of
  {B}ohr's principle,'' {\em Physical Review D}, vol.~19, pp.~473--484, 1979.

\item
W.~K. Wootters, ``Information is maximized in photon polarization
  measurements,'' in {\em Quantum Theory and Gravitation} (A.~R. Marlow, ed.),
  (Boston), pp.~13--26, Academic Press, 1980.

\item
W.~K. Wootters, {\em The Acquisition of Information from Quantum Measurements}.
 PhD thesis, The University of Texas at Austin,, Austin, TX, 1980.

\item
W.~K. Wootters, ``Statistical distance and {H}ilbert space,'' {\em Physical
  Review D}, vol.~23, pp.~357--362, 1981.

\item
W.~K. Wootters and W.~H. Zurek, ``A single quantum cannot be cloned,'' {\em
  Nature}, vol.~299, pp.~802--803, 1982.

\item
W.~Wootters, ``A measure of the distinguishability of quantum states,'' in {\em
  Quantum Optics, General Relativity, and Measurement Theory} (M.~O. Scully and
  P.~Meystre, eds.), (New York), pp.~145--154, Plenum Press, 1983.
 NATO Advanced Science Institute Series.

\item
W.~K. Wootters, ``Quantum mechanics without probability amplitudes,'' {\em
  Foundations of Physics}, vol.~16(4), pp.~391--405, 1986.

\item
W.~K. Wootters, ``A {W}igner-function formulation of finite-state quantum
  mechanics,'' {\em Annals of Physics}, vol.~176(1), pp.~1--21, 1987.

\item
W.~K. Wootters and B.~D. Fields, ``Optimal state-determination by mutually
  unbiased measurements,'' {\em Annals of Physics}, vol.~191(2), pp.~363--381,
  1989.

\item
W.~K. Wootters and B.~D. Fields, ``Searching for mutually unbiased
  observables,'' in {\em Bell's Theorem, Quantum Theory and Conceptions of the
  Universe} (M.~C. Kafatos, ed.), (Dordrecht), pp.~65--67, Kluwer Academic
  Publishers, 1989.

\item
W.~K. Wootters, ``Random quantum states,'' {\em Foundations of Physics},
  vol.~20(11), pp.~1365--1378, 1990.

\item
W.~K. Wootters, ``Two extremes of information in quantum mechanics,'' in {\em
  Workshop on Physics and Computation: PhysComp '92} (D.~Matzke, ed.), (Los
  Alamitos, CA), pp.~181--183, IEEE Computer Society, 1993.

\item
T.~Y. Young, ``Asymptotically efficient approaches to quantum-mechanical
  parameter estimation,'' {\em Information Sciences}, vol.~9, pp.~25--42, 1975.

\item
H.~P. Yuen and M.~Ozawa, ``Ultimate information carrying limit of quantum
  systems,'' {\em Physical Review Letters}, vol.~70(4), pp.~363--366, 1993.

\item
S.~Zanzinger, ``Coherent superposability of states,'' in {\em Symposium on the
  Foundation of Modern Physics 1993} (P.~Busch, P.~Lahti, and P.~Mittelstaedt,
  eds.), (Singapore), pp.~450--457, World Scientific, 1993.

\end{enumerate}


\section{Classical Distinguishability Measures and Information Theory}
\bigskip

\begin{enumerate}
\setcounter{enumi}{212}

\item
J.~Acz\'{e}l, ``A solution of some problems of {K. Borsuk} and {L.
  J\'{a}nossy},'' {\em Acta Physica Academiae Scientiarum Hungaricae}, vol.~4,
  pp.~351--362, 1955.

\item
J.~Acz\'{e}l and J.~Pfanzagl, ``Remarks on the measurement of subjective
  probability and information,'' {\em Metrika}, vol.~11, pp.~91--105, 1966.

\item
J.~Acz\'{e}l and M.~Ostrowski, ``On the characterization of {S}hannon's entropy
  by {S}hannon's inequality,'' {\em The Journal of the Australian Mathematical
  Society}, vol.~16, pp.~368--374, 1973.

\item
J.~Acz\'{e}l, B.~Forte, and C.~T. Ng, ``Why the {S}hannon and {H}artley
  entropies are `natural','' {\em Advances in Applied Probability}, vol.~6,
  pp.~131--146, 1974.

\item
J.~Acz\'{e}l and Z.~Dar\'{o}czy, {\em On Measures of Information and Their
  Characterizations}.
 Mathematics in Science and Engineering, vol.\ 115, New York: Academic
  Press, 1975.

\item
J.~Acz\'{e}l, ``Some recent results on characterizations of measures of
  information related to coding,'' {\em IEEE Transactions on Information
  Theory}, vol.~IT-24(5), pp.~592--595, 1978.

\item
J.~Acz\'{e}l, ``A mixed theory of information.\ {V}.\ {H}ow to keep the (inset)
  expert honest,'' {\em Journal of Mathematical Analysis and Applications},
  vol.~75, pp.~447--453, 1980.

\item
J.~Acz\'{e}l, ``Derivations and information functions ({A} tale of two
  surprises and a half),'' in {\em Contributions to Probability: A Collection
  of Papers Dedicated to Eugene Lukacs} (J.~M. Gani and V.~K. Rohatgi, eds.),
  (New York), pp.~191--200, Academic Press, 1981.

\item
J.~Acz\'{e}l, ``Some recent results on information measures, a new
  generalization and some `real life' interpretations of the `old' and new
  measures,'' in {\em Functional Equations: History, Applications and Theory}
  (J.~Acz\'{e}l, ed.), (Dordrecht), pp.~175--189, D. Reidel, 1984.

\item
S.-I. Amari, ``Differential geometry of curved exponential
  families---curvatures and information loss,'' {\em The Annals of Statistics},
  vol.~10(2), pp.~357--385, 1982.

\item
S.-I. Amari and T.~S. Han, ``Statistical inference under multiterminal rate
  restrictions: {A} differential geometric approach,'' {\em IEEE Transactions
  on Information Theory}, vol.~35(2), pp.~217--227, 1989.

\item
R.~B. Ash, {\em Information Theory}.
 New York: Dover Publications, 1965.

\item
C.~Atkinson and A.~F.~S. Mitchell, ``Rao's distance measure,'' {\em
  Sankhy\={a}, The Indian Journal of Statistics}, vol.~43, pp.~345--365, 1981.
 Series A, Pt.\ 3.

\item
M.~Ben-Bassat and J.~Raviv, ``R\'{e}nyi's entropy and the probability of
  error,'' {\em IEEE Transactions on Information Theory}, vol.~IT-24(3),
  pp.~324--331, 1978.

\item
M.~Ben-Bassat, ``Use of distance measures, information measures and error
  bounds in feature evaluation,'' in {\em Classification, Pattern Recognition,
  and Reduction of Dimensionality} (P.~R. Krishnaiah and L.~N. Kanal, eds.),
  Handbook of Statistics, vol.\ 2, (Amsterdam), pp.~773--791, North-Holland,
  1982.

\item
J.~O. Berger, {\em Statistical Decision Theory and Bayesian Analysis}.
 Springer Series in Statistics, New York: Springer-Verlag,
  {S}econd~ed., 1985.

\item
J.~M. Bernardo and A.~F. Smith, {\em Bayesian Theory}.
 Wiley Series in Probability and Mathematical Statistics, Chichester:
  Wiley, 1994.

\item
A.~Bhattacharyya, ``On a measure of divergence between two statistical
  populations defined by their probability distributions,'' {\em Bulletin of
  the Calcutta Mathematical Society}, vol.~35, pp.~99--109, 1943.

\item
A.~Bhattacharyya, ``On a measure of divergence between two multinomial
  populations,'' {\em Sankhy\={a}, The Indian Journal of Statistics}, vol.~7,
  pp.~401--406, 1946.

\item
S.~L. Braunstein, ``How large a sample is needed for the maximum likelihood
  estimator to be approximately {G}aussian,'' {\em Journal of Physics A},
  vol.~25, pp.~3813--3826, 1992.

\item
R.~J. Buehler, ``Measuring information and uncertainty,'' in {\em Foundations
  of Statistical Inference} (V.~P. Godambe and D.~A. Sprott, eds.), (Toronto),
  pp.~330--341, Holt, Rinehart and Winston of Canada, 1971.

\item
J.~Burbea and C.~R. Rao, ``On the convexity of some divergence measures based
  on entropy functions,'' {\em IEEE Transactions on Information Theory},
  vol.~IT-28(3), pp.~489--495, 1982.

\item
J.~Burbea and C.~R. Rao, ``Entropy differential metric, distance and divergence
  measures in probability spaces: {A} unified approach,'' {\em Journal of
  Multivariate Analysis}, vol.~12, pp.~575--596, 1982.

\item
L.~L. Campbell, ``A coding theorem and {R}\'{e}nyi's entropy,'' {\em
  Information and Control}, vol.~8(4), pp.~423--429, 1965.

\item
L.~L. Campbell, ``Definition of entropy by means of a coding problem,'' {\em
  Zeitschrift {f\"{u}r} Wahrscheinlichkeitstheorie und verwandte Gebiete},
  vol.~6, pp.~113--118, 1966.

\item
L.~L. Campbell, ``The relation between information theory and the differential
  geometry approach to statistics,'' {\em Information Sciences}, vol.~35,
  pp.~199--210, 1985.

\item
L.~L. Campbell, ``An extended \v{C}encov characterization of the information
  metric,'' {\em Proceedings of the American Mathematical Society}, vol.~98(1),
  pp.~135--141, 1986.

\item
R.~M. Capocelli and A.~{De Santis}, ``New bounds on the redundancy of {H}uffman
  codes,'' {\em IEEE Transactions on Information Theory}, vol.~37(4),
  pp.~1095--1104, 1991.

\item
N.~N. \v{C}encov, {\em Statistical Decision Rules and Optimal Inference}.
 Translations of Mathematical Monographs, vol.\ 53, Providence, RI:
  American Mathematical Society, 1982.
 Translation edited by Lev J. Leifman.

\item
C.~H. Chen, ``Theoretical comparison of a class of feature selection criteria
  in pattern recognition,'' {\em IEEE Transactions on Computers}, vol.~20,
  pp.~1054--1056, 1971.

\item
C.~H. Chen, ``On information and distance measures, error bounds, and feature
  selection,'' {\em Information Sciences}, vol.~10, pp.~159--173, 1976.

\item
H.~Chernoff, ``A measure of asymptotic efficiency for tests of a hypothesis
  based on the sum of observations,'' {\em The Annals of Mathematical
  Statistics}, vol.~23, pp.~493--507, 1952.

\item
H.~Chernoff and L.~E. Moses, {\em Elementary Decision Theory}.
 New York: Dover, 1959.

\item
K.~L. Chung, ``On the probability of the occurrence of at least $m$ events
  among $n$ arbitrary events,'' {\em The Annals of Mathematical Statistics},
  vol.~12, pp.~328--338, 1941.

\item
W.~S. Cleveland and P.~A. Lachenbruch, ``A measure of divergence among several
  populations,'' {\em Communications in Statistics}, vol.~3(3), pp.~201--211,
  1974.

\item
T.~M. Cover and P.~E. Hart, ``Nearest neighbor pattern classification,'' {\em
  IEEE Transactions on Information Theory}, vol.~IT-13(1), pp.~278--284, 1967.

\item
T.~M. Cover, ``The best two independent measurements are not the two best,''
  {\em IEEE Transactions on Systems, Man, and Cybernetics}, vol.~SMC-4(1),
  pp.~116--117, 1974.

\item
T.~M. Cover and J.~A. Thomas, {\em Elements of Information Theory}.
 Wiley Series in Telecommunications, New York: John Wiley \& Sons,
  1991.

\item
R.~T. Cox, ``Probability, frequency, and reasonable expectation,'' {\em
  American Journal of Physics}, vol.~14, pp.~1--13, 1946.

\item
R.~T. Cox, {\em The Algebra of Probable Inference}.
 Baltimore, MD: Johns Hopkins Press, 1961.

\item
H.~Cram\'{e}r, {\em Mathematical Methods of Statistics}.
 Princeton Mathematical Series, vol.\ 9, Princeton, NJ: Princeton
  University Press, 1946.

\item
I.~Csisz\'{a}r and J.~Fischer, ``Informationsentfernungen im {R}aum der
  {W}ahr\-schein\-lich\-keits\-ver\-teil\-ung\-en,'' {\em A Magyar
  Tudom\'{a}nyos Akad\'{e}mia Matematikai Kutat\'{o} Int\'{e}zet\'{e}nek
  K\"{o}zlem\'{e}nyei}, vol.~7, pp.~159--179, 1962.
 Pub. of the Mathematical Inst. of the Hungarian Acad. of Sciences.

\item
I.~Csisz\'{a}r, ``\"{U}ber topologische und metrische {E}igenschaften der
  relativen {I}nformation der {O}rdnung $\alpha$,'' in {\em Transactions of the
  Third Prague Conference on Information Theory, Statistical Decision Functions
  and Random Processes}, (Prague), pp.~63--73, Academia, 1964.
 (Translation by R. Schack and C. A. Fuchs in preparation).

\item
I.~Csisz\'{a}r, ``Information-type measures of difference of probability
  distributions and indirect observations,'' {\em Studia Scientiarum
  Mathematicarum Hungarica}, vol.~2, pp.~299--318, 1967.

\item
I.~Csisz\'{a}r, ``On topological properties of $f$-divergences,'' {\em Studia
  Scientiarum Mathematicarum Hungarica}, vol.~2, pp.~329--339, 1967.

\item
I.~Csisz\'{a}r, ``{$I$}-divergence geometry of probability distributions and
  minimization problems,'' {\em The Annals of Probability}, vol.~3(1),
  pp.~146--158, 1975.

\item
I.~Csisz\'{a}r, ``Information measures: {A} critical survey,'' in {\em
  Transactions of the Seventh Prague Conference on Information Theory,
  Statistical Decision Functions, and Random Processes and the 1974 European
  Meeting of Statisticians, vol. B}, (Boston), pp.~73--86, D. Reidel, 1978.

\item
I.~Csisz\'{a}r and J.~K\"orner, {\em Information Theory: Coding Theorems for
  Discrete Memoryless Systems}.
 Probability and Mathematical Statistics, New York: Academic Press,
  1981.

\item
Z.~Dar\'{o}czy, ``Generalized information functions,'' {\em Information and
  Control}, vol.~16, pp.~36--51, 1970.

\item
Z.~Dar\'{o}czy and G.~Maksa, ``Nonnegative information functions,'' in {\em
  Analytic Function Methods in Probability Theory} (B.~Gyires, ed.), Colloquia
  Mathematica Societatis J\'{a}nos Bolyai, vol. 21, (Amsterdam), pp.~67--78,
  North Holland, 1980.

\item
A.~P. Dawid, ``Further comments on some comments on a paper by {B}radley
  {E}fron,'' {\em The Annals of Statistics}, vol.~5(6), p.~1249, 1977.

\item
A.~Dembo, T.~M. Cover, and J.~A. Thomas, ``Information theoretic
  inequalities,'' {\em IEEE Transactions on Information Theory}, vol.~37(6),
  pp.~1501--1518, 1991.

\item
P.~A. Devijver, ``On a new class of bounds on {B}ayes risk in multihypothesis
  pattern recognition,'' {\em IEEE Transactions on Computers}, vol.~C-23(1),
  pp.~70--80, 1974.

\item
G.~T. Diderrich, ``The role of boundedness in characterizing {S}hannon
  entropy,'' {\em Information and Control}, vol.~29, pp.~149--161, 1975.

\item
G.~T. Diderrich, ``Local boundedness and the {S}hannon entropy,'' {\em
  Information and Control}, vol.~36, pp.~292--308, 1978.

\item
J.~Earman, {\em Bayes or Bust? A Critical Examination of Bayesian Confirmation
  Theory}.
 Cambridge, MA: MIT Pess, 1992.

\item
B.~Efron, ``Defining the curvature of a statistical problem (with applications
  to second order efficiency),'' {\em The Annals of Statistics}, vol.~3(6),
  pp.~1189--1242, 1975.

\item
W.~Feller, {\em An Introduction to Probability Theory and Its Applications,
  vol. 1}.
 Wiley Series in Probability and Mathematical Statistics, New York:
  John Wiley \& Sons, {T}hird~ed., 1968.

\item
P.~Fischer, ``On the inequality $\sum p_if(p_i)\ge\sum p_if(q_i)$,'' {\em
  Metrika}, vol.~18, pp.~199--208, 1972.

\item
R.~A. Fisher, ``On the dominance ratio,'' {\em Proceedings of the Royal Society
  of Edinburgh}, vol.~42, pp.~321--341, 1922.

\item
B.~Forte and C.~T. Ng, ``On a characterization of the entropies of degree
  $\beta$,'' {\em Utilitas Mathematica}, vol.~4, pp.~193--205, 1973.

\item
B.~Forte and C.~T. Ng, ``Derivation of a class of entropies including those of
  degree $\beta$,'' {\em Information and Control}, vol.~28, pp.~335--351, 1975.

\item
R.~G. Gallager, {\em Information Theory and Reliable Communication}.
 New York: John Wiley and Sons, 1968.

\item
R.~G. Gallager, ``Variations on a theme by {H}uffman,'' {\em IEEE Transactions
  on Information Theory}, vol.~IT-24(6), pp.~668--674, 1978.

\item
F.~E. Glave, ``A new look at the {B}arankin lower bound,'' {\em IEEE
  Transactions on Information Theory}, vol.~IT-18(3), pp.~349--356, 1972.

\item
N.~Glick, ``Separation and probability of correct classification among two or
  more distributions,'' {\em Annals of the Institute of Statistical
  Mathematics}, vol.~25, pp.~373--382, 1973.

\item
I.~J. Good, ``Rational decisions,'' {\em Journal of the Royal Statistical
  Society, Series B}, vol.~14, pp.~107--114, 1952.

\item
T.~L. Grettenberg, ``Signal selection in communication and radar systems,''
  {\em IEEE Transactions on Information Theory}, vol.~9, pp.~265--275, 1963.

\item
T.~S. Han and S.~Verd\'{u}, ``Generalizing the {F}ano inequality,'' {\em IEEE
  Transactions on Information Theory}, vol.~40(4), pp.~1247--1248, 1994.

\item
M.~E. Hellman and J.~Raviv, ``Probability of error, equivocation, and the
  {C}hernoff bound,'' {\em IEEE Transactions on Information Theory},
  vol.~IT-16(4), pp.~368--372, 1970.

\item
H.~Heyer, {\em Theory of Statistical Experiments}.
 Springer Series in Statistics, New York: Springer-Verlag, 1982.

\item
A.~Hobson, ``A new theorem of information theory,'' {\em Journal of Statistical
  Physics}, vol.~1(3), pp.~383--391, 1969.

\item
A.~Hobson, {\em Concepts in Statistical Physics}.
 New York: Gordon and Breach, 1971.

\item
A.~Hobson and B.-K. Cheng, ``A comparison of the {S}hannon and {K}ullback
  information measures,'' {\em Journal of Statistical Physics}, vol.~7(4),
  pp.~301--310, 1973.

\item
Y.~Horibe, ``A note on entropy metrics,'' {\em Information and Control},
  vol.~22, pp.~403--404, 1973.

\item
Y.~Horibe, ``Entropy and correlation,'' {\em IEEE Transactions on Systems, Man,
  and Cybernetics}, vol.~SMC-15(5), pp.~641--642, 1985.

\item
H.~Hudimoto, ``A note on the probability of correct classification when the
  distributions are not specified,'' {\em Annals of the Institute of
  Statistical Mathematics}, vol.~9(1), pp.~31--36, 1957.

\item
D.~A. Huffman, ``A method for the construction of minimum redundancy codes,''
  {\em Proceedings of the Institute of Radio Engineers}, vol.~40(2),
  pp.~1098--1101, 1952.

\item
L.~J\'{a}nossy, ``Remarks on the foundation of probability calculus,'' {\em
  Acta Physica Academiae Scientiarum Hungaricae}, vol.~4, pp.~333--349, 1955.

\item
E.~T. Jaynes, {\em Probability Theory: The Logic of Science}.
 Electronic-mail preprint, 1993.
 Book in preparation---fragmentary edition of September 1993. 30
  Chapters, 9 Appendices.

\item
H.~Jeffreys, ``An invariant form for the prior probability in estimation
  problems,'' {\em Proceedings of the Royal Society of London, Series A},
  vol.~186, pp.~453--461, 1946.

\item
T.~Kailath, ``The divergence and {B}hattacharyya distance measures in signal
  selection,'' {\em IEEE Transactions on Communication Technology},
  vol.~COM-15(1), pp.~52--60, 1967.

\item
P.~Kannappan and P.~N. Rathie, ``On various characterizations of
  directed-divergence,'' in {\em Transactions of the Sixth Prague Conference on
  Information Theory, Statistical Decision Functions, Random Processes},
  (Prague), pp.~331--339, Academia, 1973.

\item
H.~Kaufman and A.~M. Mathai, ``An axiomatic foundation for a multivariate
  measure of affinity among a number of distributions,'' {\em Journal of
  Multivariate Analysis}, vol.~3(2), pp.~236--242, 1973.

\item
S.~N. U.~A. Kirmani, ``Some limiting properties of {M}atusita's measure of
  distance,'' {\em Annals of the Institute of Statistical Mathematics},
  vol.~23, pp.~157--162, 1971.

\item
L.~G. Kraft, ``A device for quantizing, grouping and coding amplitude modulated
  pulses,'' Master's thesis, Massachusetts Institute of Technology, Cambridge,
  MA, 1949.

\item
S.~Kullback and R.~A. Leibler, ``On information and sufficiency,'' {\em The
  Annals of Mathematical Statistics}, vol.~22, pp.~79--86, 1951.

\item
S.~Kullback, {\em Information Theory and Statistics}.
 Wiley Publication in Mathematical Statistics, New York: John Wiley \&
  Sons, 1959.

\item
S.~Kullback, ``A lower bound for discrimination information in terms of
  variation,'' {\em IEEE Transactions on Information Theory}, vol.~IT-13,
  pp.~126--127, 1967.

\item
S.~Kullback, ``Correction to a lower bound for discrimination information in
  terms of variation,'' {\em IEEE Transactions on Information Theory},
  vol.~IT-16, p.~652, 1970.

\item
D.~G. Lainiotis, ``A class of upper bounds on probability of error for
  multihypothesis pattern recognition,'' {\em IEEE Transactions on Information
  Theory}, vol.~15, pp.~730--731, 1969.

\item
P.~M. Lee, ``On the axioms of information theory,'' {\em The Annals of
  Mathematical Statistics}, vol.~35, pp.~415--418, 1964.

\item
D.~V. Lindley, ``On a measure of the information provided by an experiment,''
  {\em The Annals of Mathematical Statistics}, vol.~27, pp.~986--1005, 1956.

\item
A.~M. Mathai, ``Characterization of some statistical concepts,'' in {\em
  Conference on Measures of Information and Their Applications} (P.~N. Rathie,
  ed.), (Bombay), pp.~1--10, Indian Institute of Technology, 1974.

\item
K.~Matusita, ``On the estimation by the minimum distance method,'' {\em Annals
  of the Institute of Statistical Mathematics}, vol.~5, pp.~59--65, 1954.

\item
K.~Matusita, ``Decision rules, based on the distance, for problems of fit, two
  samples, and estimation,'' {\em The Annals of Mathematical Statistics},
  vol.~26, pp.~631--640, 1955.

\item
K.~Matusita and H.~Akaike, ``Decision rules, based on the distance, for the
  problems of independence, invariance and two samples,'' {\em Annals of the
  Institute of Statistical Mathematics}, vol.~7(2), pp.~67--80, 1956.

\item
K.~Matusita, ``Decision rule, based on the distance, for the classification
  problem,'' {\em Annals of the Institute of Statistical Mathematics},
  vol.~7(3), pp.~67--77, 1956.

\item
K.~Matusita and M.~Motoo, ``On the fundamental theorem for the decision rule
  based on distance $\left\|\;\right\|$,'' {\em Annals of the Institute of
  Statistical Mathematics}, vol.~8(2), pp.~137--142, 1956.

\item
K.~Matusita, ``Some properties of affinity and applications,'' {\em Annals of
  the Institute of Statistical Mathematics}, vol.~23, pp.~137--155, 1971.

\item
K.~Matusita, ``Discrimination and the affinity of distributions,'' in {\em
  Discriminant Analysis and Applications} (T.~Cacoullos, ed.), Academic Press
  Rapid Manuscript Reproduction, (New York), pp.~213--223, Academic Press,
  1973.

\item
G.~Musz\'{e}ly, ``On continuous solutions of a functional inequality,'' {\em
  Metrika}, vol.~20, pp.~65--69, 1973.

\item
P.~Nath, ``On a coding theorem connected with {R}\'{e}nyi's entropy,'' {\em
  Information and Control}, vol.~29, pp.~234--242, 1975.

\item
T.~Nemetz, ``Information theory and the testing of a hypothesis,'' in {\em
  Proceedings of the Colloquium on Information Theory, vol.\ II} (A.~R\'{e}nyi,
  ed.), (Budapest), pp.~283--294, J\'{a}nos Bolyai Mathematical Society, 1968.

\item
E.~J.~G. Pitman, {\em Some Basic Theory for Statistical Inference}.
 Monographs on Applied Probability and Statistics, London: Chapman and
  Hall, 1979.

\item
C.~Rajski, ``A metric space of discrete probability distributions,'' {\em
  Information and Control}, vol.~4, pp.~371--377, 1961.

\item
C.~Rajski, ``Entropy and metric spaces,'' in {\em Information Theory: Fourth
  London Symposium} (C.~Cherry, ed.), (Washington), pp.~41--45, Butterworths,
  1961.

\item
C.~Rajski, ``On the normed information rate of discrete random variables,''
  {\em Zastosowania Matematyki}, vol.~6, pp.~459--461, 1963.

\item
C.~R. Rao, ``The problem of classification and distance between two
  populations,'' {\em Nature}, vol.~159, pp.~30--31, 1947.

\item
C.~R. Rao, ``On the distance between two populations,'' {\em Sankhy\={a}, The
  Indian Journal of Statistics}, vol.~9, pp.~246--248, 1949.

\item
C.~R. Rao, ``Efficient estimates and optimum inference procedures in large
  samples,'' {\em Journal of the Royal Statistical Society, Series B},
  vol.~24(1), pp.~46--72, 1962.

\item
P.~N. Rathie and P.~Kannappan, ``A directed-divergence function of type
  $\beta$,'' {\em Information and Control}, vol.~20, pp.~28--45, 1972.

\item
P.~N. Rathie and P.~Kannappan, ``On a new characterization of
  directed-divergence in information theory,'' in {\em Transactions of the
  Sixth Prague Conference on Information Theory, Statistical Decision
  Functions, Random Processes}, (Prague), pp.~733--745, Academia, 1973.

\item
P.~N. Rathie, ``Characterizations of the harmonic mean and the associated
  distance measure useful in pattern recognition,'' in {\em Conference on
  Measures of Information and Their Applications} (P.~N. Rathie, ed.),
  (Bombay), pp.~81--88, Indian Institute of Technology, 1974.

\item
A.~R\'{e}nyi, ``On measures of entropy and information,'' in {\em Proceedings
  of the Fourth Berkeley Symposium on Mathematical Statistics and Probability,
  vol.\ 1} (J.~Neyman, ed.), (Berkeley, CA), pp.~547--561, University of
  California Press, 1961.

\item
A.~R\'{e}nyi, ``On the amount of information concerning an unknown parameter in
  a sequence of observations,'' {\em A Magyar Tudom\'{a}nyos Akad\'{e}mia
  Matematikai Kutat\'{o} Int\'{e}zet\'{e}nek K\"{o}zlem\'{e}nyei}, vol.~9,
  pp.~617--624, 1965.
 Publications of the Mathematical Institute of the Hungarian Academy
  of Sciences.

\item
A.~R\'{e}nyi, ``On the amount of missing information and the {N}eyman-{P}earson
  lemma,'' in {\em Research Papers in Statistics: Festschrift for J. Neyman}
  (F.~N. David, ed.), (New York), pp.~281--288, Wiley, 1966.

\item
A.~R\'{e}nyi, ``On some basic problems of statistics from the point of view of
  information theory,'' in {\em Proceedings of the Fifth Berkeley Symposium on
  Mathematical Statistics and Probability, vol.\ 1} (L.~M. {LeCam} and
  J.~Neyman, eds.), (Berkeley, CA), pp.~531--543, University of California
  Press, 1967.

\item
A.~R\'{e}nyi, ``Statistics and information theory,'' {\em Studia Scientiarum
  Mathematicarum Hungarica}, vol.~2, pp.~249--256, 1967.

\item
A.~R\'{e}nyi, ``On some problems of statistics from the point of view of
  information theory,'' in {\em Proceedings of the Colloquium on Information
  Theory, vol.\ II} (A.~R\'{e}nyi, ed.), (Budapest), pp.~343--357, J\'{a}nos
  Bolyai Mathematical Society, 1968.

\item
A.~R\'{e}nyi, {\em Probability Theory}.
 North-Holland Series in Applied Mathematics and Mechanics, vol.\ 10,
  Amsterdam: North-Holland and Akad\'{e}miai Kiad\'{o}, 1970.
 translated by L. Vekerdi from {\it Wahrscheinlichkeitsrechnung. Mit
  einem Anhang \"uber Informationstheorie\/} (VEB Deutscher Verlag der
  Wissenschaften, Berlin, 1962).

\item
E.~Schr\"{o}dinger, ``The foundation of the theory of probability---{I},'' {\em
  Proceedings of the Royal Irish Academy, Section A}, vol.~51, pp.~51--66,
  1947.

\item
E.~Schr\"{o}dinger, ``The foundation of the theory of probability---{II},''
  {\em Proceedings of the Royal Irish Academy, Section A}, vol.~51,
  pp.~141--146, 1947.

\item
C.~E. Shannon, ``A mathematical theory of communication,'' {\em The Bell System
  Technical Journal}, vol.~27, pp.~379--423, 623--656, 1948.
 Reprinted in book form, with postscript by Warren Weaver: C. E.
  Shannon and W. Weaver, {\it The Mathematical Theory of Communication},
  (University of Illinois Press, Urbana, IL, 1949).

\item
C.~E. Shannon, ``The lattice theory of information,'' in {\em Symposium on
  Information Theory (Report of Proceedings)} (W.~Jackson, ed.), (London),
  pp.~105--107, The Ministry of Supply, 1950.
 Held in the Lecture Theatre of the Royal Society, Burlington House,
  September 1950.

\item
B.~D. Sharma, ``On amount of information of type-$\beta$ and other measures,''
  {\em Metrika}, vol.~19, pp.~1--10, 1972.

\item
B.~D. Sharma and R.~Autar, ``On characterization of a generalized inaccuracy
  measure in information theory,'' {\em Journal of Applied Probability},
  vol.~10, pp.~464--468, 1973.

\item
J.~E. Shore and R.~W. Johnson, ``Axiomatic derivation of the principle of
  maximum entropy and the principle of minimum cross-entropy,'' {\em IEEE
  Transactions on Information Theory}, vol.~IT-26(1), pp.~26--37, 1980.

\item
J.~E. Shore and R.~W. Johnson, ``Properties of cross-entropy minimization,''
  {\em IEEE Transactions on Information Theory}, vol.~IT-27(4), pp.~472--482,
  1981.

\item
D.~Slepian, ed., {\em Key Papers in the Development of Information Theory}.
 IEEE Press Selected Reprint Series, New York: IEEE Press, 1974.

\item
C.~R. Smith and G.~J. Erickson, ``From rationality and consistency to
  {B}ayesian probability,'' in {\em Maximum Entropy and Bayesian Methods,
  Cambridge, England, 1988} (J.~Skilling, ed.), Fundamental Theories of
  Physics, (Dordrecht), pp.~29--44, Kluwer Academic Publishers, 1989.

\item
C.~R. Smith and G.~J. Erickson, ``Probability theory and the associativity
  equation,'' in {\em Maximum Entropy and Bayesian Methods, Dartmouth, U.S.A.,
  1989} (P.~F. Foug\`{e}re, ed.), Fundamental Theories of Physics, (Dordrecht),
  pp.~17--29, Kluwer Academic Publishers, 1990.

\item
G.~T. Toussaint, ``Note on optimal selection of independent binary-valued
  features for pattern recognition,'' {\em IEEE Transactions on Information
  Theory}, vol.~IT-17(5), p.~618, 1971.

\item
G.~T. Toussaint, ``Some upper bounds on error probability for multiclass
  pattern recognition,'' {\em IEEE Transactions on Computers}, vol.~C-20,
  pp.~943--944, 1971.

\item
G.~T. Toussaint, ``Some functional lower bounds on the expected divergence for
  multihypothesis pattern recognition, communication, and radar systems,'' {\em
  IEEE Transactions on Systems, Man, and Cybernetics}, vol.~SMC-1,
  pp.~384--385, 1971.

\item
G.~T. Toussaint, ``Feature evaluation with quadratic mutual information,'' {\em
  Information Processing Letters}, vol.~1, pp.~153--156, 1972.

\item
G.~T. Toussaint, ``Some inequalities between distance measures for feature
  evaluation,'' {\em IEEE Transactions on Computers}, vol.~C-21(4),
  pp.~409--410, 1972.

\item
G.~T. Toussaint, ``Comments on `{T}he divergence and {B}hattacharyya distance
  measures in signal selection','' {\em IEEE Transactions on Communication
  Technology}, vol.~COM-20, p.~485, 1972.

\item
G.~T. Toussaint, ``Some properties of {M}atusita's measure of affinity of
  several distributions,'' {\em Annals of the Institute of Statistical
  Mathematics}, vol.~26, pp.~389--394, 1974.

\item
G.~T. Toussaint, ``On some measures of information and their application to
  pattern recognition,'' in {\em Conference on Measures of Information and
  Their Applications} (P.~N. Rathie, ed.), (Bombay), pp.~21--28, Indian
  Institute of Technology, 1974.

\item
G.~T. Toussaint, ``Sharper lower bounds for discrimination information in terms
  of variation,'' {\em IEEE Transactions on Information Theory}, vol.~IT-21(1),
  pp.~99--100, 1975.

\item
G.~T. Toussaint, ``Comments on `{O}n a new class of bounds on {B}ayes' risk in
  multihypothesis pattern recognition','' {\em IEEE Transactions on Computers},
  vol.~C-24(8), pp.~855--856, 1975.

\item
G.~T. Toussaint, ``An upper bound on the probability of misclassification in
  terms of affinity,'' {\em Proceedings of the IEEE}, vol.~65, pp.~275--276,
  1977.

\item
G.~T. Toussaint, ``Probability of error, expected divergence, and the affinity
  of several distributions,'' {\em IEEE Transactions on Systems, Man, and
  Cybernetics}, vol.~SMC-8(6), pp.~482--485, 1978.

\item
I.~Vajda, ``On the convergence of information contained in a sequence of
  observations,'' in {\em Proceedings of the Colloquium on Information Theory,
  vol.\ II} (A.~R\'{e}nyi, ed.), (Budapest), pp.~489--501, J\'{a}nos Bolyai
  Mathematical Society, 1968.

\item
I.~Vajda, ``$\chi^\alpha$-divergence and generalized {F}isher's information,''
  in {\em Transactions of the Sixth Prague Conference on Information Theory,
  Statistical Decision Functions, Random Processes}, (Prague), pp.~873--886,
  Academia, 1973.

\item
I.~Vajda, {\em Theory of Statistical Inference and Information}.
 Theory and Decision Library, Series B: Mathematical and Statistical
  Methods, Dordrecht: Kluwer Academic Publishers, 1989.

\item
T.~R. Vilmansen, ``On dependence and discrimination in pattern recognition,''
  {\em IEEE Transactions on Computers}, vol.~C-21(9), pp.~1029--1031, 1972.

\item
T.~R. Vilmansen, ``Feature evaluation with measures of probabilistic
  dependence,'' {\em IEEE Transactions on Computers}, vol.~C-22(4),
  pp.~381--388, 1973.

\item
P.~M. Woodward, {\em Probability and Information Theory, with Applications to
  Radar}.
 International Series of Monographs on Electronics and
  Instrumentation, vol. 3, Oxford: Pergamon Press, {S}econd~ed., 1953.

\item
J.~Zv\'{a}rov\'{a}, ``On asymptotic behavior of a sample estimator of
  {R}\'{e}nyi's information of order $\alpha$,'' in {\em Transactions of the
  Sixth Prague Conference on Information Theory, Statistical Decision
  Functions, Random Processes}, (Prague), pp.~919--924, Academia, 1973.

\end{enumerate}


\section{Matrix Inequalities, Operator Relations, and Mathematical Techniques}
\bigskip

\begin{enumerate}
\setcounter{enumi}{363}

\item
J.~Acz\'{e}l, {\em Lectures on Functional Equations and Their Applications}.
 Mathematics in Science and Engineering, vol.\ 19, New York: Academic
  Press, 1966.

\item
K.~Aizu, ``Parameter differentiation of quantum-mechanical linear operators,''
  {\em Journal of Mathematical Physics}, vol.~4(6), pp.~762--775, 1963.

\item
A.~R. Amir-Moez, {\em Singular Values of Linear Transformations}.
 Mathematics Series, vol.\ 11, Lubbock, TX: Texas Technological
  College, 1968.

\item
T.~Ando, ``Concavity of certain maps on positive definite matrices and
  applications to {H}adamard products,'' {\em Linear Algebra and Its
  Applications}, vol.~26, pp.~203--241, 1979.

\item
T.~Ando, ``Hua-{M}arcus inequalities,'' {\em Linear and Multilinear Algebra},
  vol.~8, pp.~347--352, 1979.

\item
T.~Ando, ``On some operator inequalities,'' {\em Mathematische Annalen},
  vol.~279, pp.~157--159, 1987.

\item
T.~Ando, ``Comparison of norms $|\!|\!|f({{\rm A}})-f({{\rm B}})|\!|\!|$ and
  $|\!|\!|\,f(|{{\rm A}}-{{\rm B}}|)\,|\!|\!|$,'' {\em Mathematische
  Zeitschrift}, vol.~197, pp.~403--409, 1988.

\item
T.~Ando, ``Majorization, doubly stochastic matrices, and comparison of
  eigenvalues,'' {\em Linear Algebra and Its Applications}, vol.~118,
  pp.~163--248, 1989.

\item
T.~Ando and F.~Hiai, ``Log majorization and complementary {G}olden-{T}hompson
  type inequalities,'' {\em Linear Algebra and Its Applications}, vol.~197,198,
  pp.~113--131, 1994.

\item
T.~Ando, ``Majorizations and inequalities in matrix theory,'' {\em Linear
  Algebra and Its Applications}, vol.~199, pp.~17--67, 1994.

\item
T.~Ando and F.~Hiai, ``Inequality between powers of positive semidefinite
  matrices,'' {\em Linear Algebra and Its Applications}, vol.~208/209,
  pp.~65--71, 1994.

\item
H.~Araki and S.~Yamagami, ``An inequality for the {H}ilbert-{S}chmidt norm,''
  {\em Communications in Mathematical Physics}, vol.~81, pp.~89--91, 1981.

\item
H.~Araki, ``On an inequality of {L}ieb and {T}hirring,'' {\em Letters in
  Mathematical Physics}, vol.~19, pp.~167--170, 1990.

\item
R.~B. Bapat, ``Majorization and singular values {II},'' {\em SIAM Journal on
  Matrix Analysis and Applications}, vol.~10(4), pp.~429--434, 1989.

\item
E.~R. Barnes and A.~J. Hoffman, ``Bounds for the spectrum of normal matrices,''
  {\em Linear Algebra and Its Applications}, vol.~201, pp.~79--90, 1994.

\item
S.~Barnett and C.~Storey, ``Analysis and synthesis of stability matrices,''
  {\em Journal of Differential Equations}, vol.~3, pp.~414--422, 1967.

\item
S.~Barnett and C.~Storey, {\em Matrix Methods in Stability Theory}.
 Applications of Mathematics Series, New York: Barnes \& Noble, Inc.,
  1970.

\item
S.~Barnett, ``Matrix differential equations and {K}ronecker products,'' {\em
  SIAM Journal on Applied Mathematics}, vol.~24(1), pp.~1--5, 1973.

\item
N.~Bebiano and M.~E. Miranda, ``On a recent determinantal inequality,'' {\em
  Linear Algebra and Its Applications}, vol.~201, pp.~99--102, 1994.

\item
E.~F. Beckenbach and R.~Bellman, {\em Inequalities}.
 Ergebnisse der Mathematik und ihrer Grenzgebiete, Heft 30, Berlin:
  Springer-Verlag, 1961.

\item
R.~Bellman, {\em Introduction to Matrix Analysis}.
 New York: McGraw-Hill, {S}econd~ed., 1970.

\item
R.~Bellman, ``Some inequalities for positive definite matrices,'' in {\em
  General Inequalities 2: Proceedings of the Second International Conference on
  General Inequalities} (E.~F. Beckenbach, ed.), International Series on
  Numerical Mathematics, vol.\ 47, (Basel, Switzerland), pp.~89--90,
  Birkh\"{a}user Verlag, 1980.
 Held in the Mathematical Institute at Oberwolfach, Black Forest, July
  30--August 5, 1978.

\item
D.~S. Bernstein, ``Inequalities for the trace of matrix exponentials,'' {\em
  SIAM Journal on Matrix Analysis and Applications}, vol.~9(2), pp.~156--158,
  1988.

\item
J.~H. Bevis, F.~J. Hall, and R.~E. Hartwig, ``The matrix equation
  {$A\overline{X}-XB=C$} and its special cases,'' {\em SIAM Journal on Matrix
  Analysis and Applications}, vol.~9(3), pp.~348--359, 1988.

\item
K.~V. Bhagwat and R.~Subramanian, ``Inequalities between means of positive
  operators,'' {\em Mathematical Proceedings of the Cambridge Philosophical
  Society}, vol.~83, pp.~393--401, 1978.

\item
R.~Bhatia, ``Some inequalities for norm ideals,'' {\em Communications in
  Mathematical Physics}, vol.~111, pp.~33--39, 1987.

\item
R.~Bhatia, ``Perturbation inequalities for the absolute value map in norm
  ideals of operator,'' {\em Journal of Operator Theory}, vol.~19,
  pp.~129--136, 1988.

\item
R.~Bhatia and J.~A. Holbrook, ``A softer, stronger {L}idskii theorem,'' {\em
  Proceedings of the Indian Academy of Sciences (Mathematical Sciences)},
  vol.~99, pp.~75--83, 1989.

\item
R.~Bhatia and F.~Kittaneh, ``On the singular values of a product of
  operators,'' {\em SIAM Journal on Matrix Analysis and Applications},
  vol.~11(2), pp.~272--277, 1990.

\item
R.~Bhatia and C.~Davis, ``More matrix forms of the arithmetic-geometric mean
  inequality,'' {\em SIAM Journal on Matrix Analysis and Applications},
  vol.~14(1), pp.~132--136, 1993.

\item
R.~Bhatia and C.~Davis, ``A {C}auchy-{S}chwarz inequality for operators with
  applications,'' {\em Linear Algebra and Its Applications}, vol.~223/224,
  pp.~119--129, 1995.

\item
P.~S. Bullen, D.~S. Mitrinovi\'{c}, and P.~M. Vasi\'{c}, {\em Means and Their
  Inequalities}.
 Mathematics and Its Applications (East European Series), Dordrecht:
  D. Reidel, 1988.

\item
E.~A. Carlen and E.~H. Lieb, ``Optimal hypercontractivity for {F}ermi fields
  and related non-commutative integration inequalities,'' {\em Communications
  in Mathematical Physics}, vol.~155, pp.~27--46, 1993.

\item
D.~H. Carlson and B.~N. Datta, ``The {L}yapunov matrix equation
  {$SA+A^*S=S^*B^*BS$},'' {\em Linear Algebra and Its Applications}, vol.~28,
  pp.~43--52, 1979.

\item
J.~E. Cohen, ``Spectral inequalities for matrix exponentials,'' {\em Linear
  Algebra and Its Applications}, vol.~111, pp.~25--28, 1988.

\item
W.~A. Coppel, ``Matrix quadratic equations,'' {\em Bulletin of the Australian
  Mathematical Society}, vol.~10, pp.~377--401, 1974.

\item
G.~Corach, H.~Porta, and L.~Recht, ``An operator inequality,'' {\em Linear
  Algebra and Its Applications}, vol.~142, pp.~153--158, 1990.

\item
J.~L. Daleck\u{}\hglue-5.2pt\i\i\ and S.~G. Kre\u{\i}n, ``Integration and
  differentiation of functions of {H}ermitian operators and applications to the
  theory of perturbations,'' {\em American Mathematical Society Translations,
  Series 2}, vol.~47, pp.~1--30, 1965.

\item
J.~L. Daleck\u{}\hglue-5.2pt\i\i\, ``Differentiation of non-{H}ermitian matrix
  functions depending on a parameter,'' {\em American Mathematical Society
  Translations, Series 2}, vol.~47, pp.~73--87, 1965.

\item
E.~B. Davies, ``Lipschitz continuity of functions of operators in the
  {S}chatten classes,'' {\em Journal of the London Mathematical Society},
  vol.~37, pp.~148--157, 1988.

\item
C.~Davis, ``All convex invariant functions of {H}ermitian matrices,'' {\em
  Archiv der Mathematik}, vol.~8, pp.~276--278, 1957.

\item
C.~Davis, ``Operator-valued entropy of a quantum mechanical measurement,'' {\em
  Proceedings of the Japan Academy}, vol.~37(9), pp.~533--538, 1961.

\item
C.~Davis, ``Notions generalizing convexity for functions defined on spaces of
  matrices,'' in {\em Convexity, Proceedings of Symposia in Pure Mathematics,
  vol.\ VII} (V.~L. Klee, ed.), (Providence, RI), pp.~187--201, American
  Mathematical Society, 1963.
  Held at University of Washington, Seattle, June 13-15, 1961.

\item
A.~S. Deif, {\em Advanced Matrix Theory for Scientists and Engineers}.
 Kent, England: Abacus Press, 1982.

\item
W.~F. {Donoghue, Jr.}, {\em Monotone Matrix Functions and Analytic
  Continuation}.
 Die Grundlehren der mathematischen Wissenschaften in
  Einzeldarstellungen, Band 207, New York: Springer-Verlag, 1974.

\item
L.~E. Faibusovich, ``Algebraic {R}iccati equation and symplectic algebra,''
  {\em International Journal of Control}, vol.~43(3), pp.~781--792, 1986.

\item
H.~Falk, ``Inequalities of {J. W. Gibbs},'' {\em American Journal of Physics},
  vol.~38(7), pp.~858--869, 1970.

\item
K.~Fan, ``On a theorem of {W}eyl concerning eigenvalues of linear
  transformations. {I},'' {\em Proceedings of the National Academy of Sciences
  (U.S.A.)}, vol.~35, pp.~652--655, 1949.

\item
K.~Fan, ``Maximum properties and inequalities for the eigenvalues of completely
  continuous operators,'' {\em Proceedings of the National Academy of Sciences
  (U.S.A.)}, vol.~37, pp.~760--766, 1951.

\item
R.~P. Feynman, ``An operator calculus having applications in quantum
  electrodynamics,'' {\em Physical Review}, vol.~84(1), pp.~108--128, 1951.

\item
J.~I. Fujii and M.~Fujii, ``A norm inequality for operator monotone
  functions,'' {\em Mathematica Japonica}, vol.~35(2), pp.~249--252, 1990.

\item
M.~Fujii, T.~Furuta, and E.~Kamei, ``Operator functions associated with
  {F}uruta's inequality,'' {\em Linear Algebra and Its Applications}, vol.~149,
  pp.~91--96, 1991.

\item
M.~Fujii, T.~Furuta, and E.~Kamei, ``Furuta's inequality and its applications
  to {A}ndo's theorem,'' {\em Linear Algebra and Its Applications}, vol.~179,
  pp.~161--169, 1993.

\item
M.~Fujii and R.~Nakamoto, ``Rota's theorem and {H}einz inequalities,'' {\em
  Linear Algebra and Its Applications}, vol.~214, pp.~271--275, 1995.

\item
T.~Furuta, ``A proof via operator means of an order preserving inequality,''
  {\em Linear Algebra and Its Applications}, vol.~113, pp.~129--130, 1989.

\item
T.~Furuta, ``Two operator functions with monotone property,'' {\em Proceedings
  of the American Mathematical Society}, vol.~111(2), pp.~511--516, 1991.

\item
T.~Furuta, ``A note on the arithmetic--geometric mean inequality for every
  unitarily invariant matrix norm,'' {\em Linear Algebra and Its Applications},
  vol.~208/209, pp.~223--228, 1994.

\item
T.~Furuta, ``Extension of the {F}uruta inequality and {A}ndo-{H}iai
  log-majorization,'' {\em Linear Algebra and Its Applications}, vol.~219,
  pp.~139--155, 1995.

\item
M.~I. Gil, ``On inequalities for eigenvalues of matrices,'' {\em Linear Algebra
  and Its Applications}, vol.~184, pp.~201--206, 1993.

\item
I.~C. Gohberg and M.~G. Kre\u{\i}n, {\em Introduction to the Theory of Linear
  Nonselfadjoint Operators}.
 Translations of Mathematical Monographs, vol.\ 18, Providence, RI:
  American Mathematical Society, 1969.
 Translated by A. Feinstein.

\item
I.~Gohberg, P.~Lancaster, and L.~Rodman, ``On {H}ermitian solutions of the
  symmetric algebraic {R}iccati equation,'' {\em SIAM Journal on Control and
  Optimization}, vol.~24(6), pp.~1323--1334, 1986.

\item
A.~Graham, {\em Kronecker Products and Matrix Calculus: with Applications}.
 Ellis Horwood Series in Mathematics and Its Applications, Chichester,
  England: Ellis Horwood Limited, 1981.

\item
P.~R. Halmos, ``Finite dimensional vector spaces,'' {\em Annals of Mathematics
  Studies}, vol.~7, pp.~1--196, 1942.

\item
F.~Hansen, ``An operator inequality,'' {\em Mathematische Annalen}, vol.~246,
  pp.~249--250, 1980.

\item
F.~Hansen and G.~K. Pedersen, ``Jensen's inequality for operators and
  {L}oewner's theorem,'' {\em Mathematische Annalen}, vol.~258, pp.~229--241,
  1981/1982.

\item
G.~H. Hardy, J.~E. Littlewood, and G.~P\'{o}lya, {\em Inequalities}.
 Cambridge: Cambridge University Press, {S}econd~ed., 1952.

\item
E.~Heinz, ``Beitr\"{a}ge zur {S}t\"{o}rungstheorie der {S}pektralzerlegung,''
  {\em Mathematische Annalen}, vol.~123, pp.~415--438, 1951.

\item
F.~Hiai and D.~Petz, ``The {G}olden-{T}hompson trace inequality is
  complemented,'' {\em Linear Algebra and Its Applications}, vol.~181,
  pp.~153--185, 1993.

\item
F.~Hiai, ``Trace norm convergence of exponential product formula,'' {\em
  Letters in Mathematical Physics}, vol.~33, pp.~147--158, 1995.

\item
A.~S. Hodel and K.~Poolla, ``Parallel solution of large {L}yapunov equations,''
  {\em SIAM Journal on Matrix Analysis and Applications}, vol.~13(4),
  pp.~1189--1203, 1992.

\item
R.~A. Horn and C.~R. Johnson, {\em Matrix Analysis}.
 Cambridge: Cambridge University Press, 1985.

\item
R.~A. Horn and R.~Mathias, ``An analog of the {C}auchy-{S}chwarz inequality for
  {H}adamard products and unitarily invariant norms,'' {\em SIAM Journal on
  Matrix Analysis and Applications}, vol.~11(4), pp.~481--498, 1990.

\item
R.~A. Horn and C.~R. Johnson, {\em Topics in Matrix Analysis}.
 Cambridge: Cambridge University Press, 1994.

\item
K.~D. Ikramov, ``A simple proof of the generalized {S}chur inequality,'' {\em
  Linear Algebra and Its Applications}, vol.~199, pp.~143--149, 1994.

\item
R.~Jackiw, ``Minimum uncertainty product, number-phase uncertainty product, and
  coherent states,'' {\em Journal of Mathematical Physics}, vol.~9(3),
  pp.~339--346, 1968.

\item
G.~W. Johnson and M.~L. Lapidus, ``Generalized {D}yson series, generalized
  {F}eynman diagrams, the {F}eynman integral and {F}eynman's operational
  calculus,'' {\em Memoirs of the American Mathematical Society}, vol.~62(351),
  pp.~1--78, 1986.

\item
G.~W. Johnson and M.~L. Lapidus, ``Noncommutative operations on {W}iener
  functionals and {F}eynman's operational calculus,'' {\em Journal of
  Functional Analysis}, vol.~81, pp.~74--99, 1988.

\item
R.~V. Kadison, ``A generalized {S}chwarz inequality and algebraic invariants
  for operator algebras,'' {\em Annals of Mathematics}, vol.~56(3),
  pp.~494--503, 1952.

\item
T.~Kato, ``Continuity of the map {$S\rightarrow|S|$} for linear operators,''
  {\em Proceedings of the Japan Academy}, vol.~49, pp.~157--160, 1973.

\item
F.~Kittaneh, ``Inequalities for the {S}chatten $p$-norm,'' {\em Glasgow
  Mathematical Journal}, vol.~26, pp.~141--143, 1985.

\item
F.~Kittaneh, ``Inequalities for the {S}chatten $p$-norm. {III},'' {\em
  Communications in Mathematical Physics}, vol.~104, pp.~307--310, 1986.

\item
F.~Kittaneh, ``Inequalities for the {S}chatten $p$-norm. {IV},'' {\em
  Communications in Mathematical Physics}, vol.~106, pp.~581--585, 1986.

\item
F.~Kittaneh, ``Inequalities for the {S}chatten $p$-norm. {II},'' {\em Glasgow
  Mathematical Journal}, vol.~29, pp.~99--104, 1987.

\item
F.~Kittaneh, ``On the continuity of the absolute value map in the {S}chatten
  classes,'' {\em Linear Algebra and Its Applications}, vol.~118, pp.~61--68,
  1989.

\item
F.~Kittaneh, ``Norm inequalities for fractional powers of positive operators,''
  {\em Letters in Mathematical Physics}, vol.~27, pp.~279--285, 1993.

\item
F.~Kittaneh, ``On some operator inequalities,'' {\em Linear Algebra and Its
  Applications}, vol.~208/ 209, pp.~19--28, 1994.

\item
H.~Kosaki, ``An inequality of {A}raki-{L}ieb-{T}hirring (von {N}eumnann algebra
  case),'' {\em Proceedings of the American Mathematical Society}, vol.~114(2),
  pp.~477--481, 1992.

\item
C.~Kourkoumelis and S.~Nettel, ``Operator functionals and path integrals,''
  {\em American Journal of Physics}, vol.~45(1), pp.~26--30, 1977.

\item
F.~Kubo and T.~Ando, ``Means of positive linear operators,'' {\em Mathematische
  Annalen}, vol.~246, pp.~205--224, 1980.

\item
M.~K. Kwong, ``Some results on matrix monotone functions,'' {\em Linear Algebra
  and Its Applications}, vol.~118, pp.~129--153, 1989.

\item
P.~Lancaster, ``Explicit solutions of linear matrix equations,'' {\em SIAM
  Review}, vol.~12(4), pp.~544--566, 1970.

\item
L.~J. Landau and R.~F. Streater, ``On {B}irkhoff's theorem for doubly
  stochastic completely positive maps of matrix algebras,'' {\em Linear Algebra
  and Its Applications}, vol.~193, pp.~107--127, 1993.

\item
O.~E. {Lanford III} and D.~W. Robinson, ``Mean entropy of states in
  quantum-statistical mechanics,'' {\em Journal of Mathematical Physics},
  vol.~9(7), pp.~1120--1125, 1968.

\item
M.~L. Lapidus, ``Quantification, calcul op\'{e}rationnel de {F}eynman
  axiomatique et int\'{e}grale fonctionnelle g\'{e}n\'{e}ralis\'{e}e,'' {\em
  Comptes Rendus de l'Acad\'{e}mie des Sciences, S\'{e}rie I}, vol.~308,
  pp.~133--138, 1989.

\item
L.~L. Lee, ``Continuum calculus and {F}eynman's path integrals,'' {\em Journal
  of Mathematical Physics}, vol.~17(11), pp.~1988--1997, 1976.

\item
A.~Lenard, ``Generalization of the {G}olden-{T}hompson inequality {${\rm
  Tr}(e^Ae^B)\ge{\rm Tr}\,e^{A+B}$},'' {\em Indiana University Mathematics
  Journal}, vol.~21(5), pp.~457--467, 1971.

\item
C.~K. Li, ``A generalization of spectral radius, numerical radius, and spectral
  norm,'' {\em Linear Algebra and Its Applications}, vol.~90, pp.~105--118,
  1987.

\item
C.~K. Li and N.~K. Tsing, ``Distance to the convex hull of unitary orbits,''
  {\em Linear and Multilinear Algebra}, vol.~25, pp.~93--103, 1989.

\item
E.~H. Lieb, ``Convex trace functions and the {W}igner-{Y}anase-{D}yson
  conjecture,'' {\em Advances in Mathematics}, vol.~11, pp.~267--288, 1973.

\item
E.~H. Lieb and M.~B. Ruskai, ``Some operator inequalities of the {S}chwarz
  type,'' {\em Advances in Mathematics}, vol.~12, pp.~269--273, 1974.

\item
E.~H. Lieb, ``Inequalities for some operator and matrix functions,'' {\em
  Advances in Mathematics}, vol.~20, pp.~174--178, 1976.

\item
E.~R. Ma, ``A finite series solution of the matrix equation {$AX-XB=C$},'' {\em
  SIAM Journal on Applied Mathematics}, vol.~14(3), pp.~490--495, 1966.

\item
J.~H. Maclagan-Wedderburn, ``Note on the linear matrix equation,'' {\em
  Proceedings of the Edinburgh Mathematical Society}, vol.~22, pp.~49--53,
  1904.

\item
M.~Marcus and B.~N. Moyls, ``On the maximum principle of {K}y {F}an,'' {\em
  Canadian Journal of Mathematics}, vol.~9, pp.~313--320, 1957.

\item
M.~Marcus, ``Convex functions of quadratic forms,'' {\em Canadian Journal of
  Mathematics}, vol.~9, pp.~321--326, 1957.

\item
M.~Marcus and H.~Minc, {\em A Survey of Matrix Theory and Matrix Inequalities}.
 New York: Dover Publications, 1992.

\item
A.~S. Markus, ``The eigen- and singular values of the sum and product of linear
  operators,'' {\em Russian Mathematical Surveys}, vol.~19(4), pp.~91--120,
  1964.

\item
A.~W. Marshall and I.~Olkin, {\em Inequalities: Theory of Majorization and Its
  Applications}.
 Mathematics in Science and Engineering, vol.\ 143, New York: Academic
  Press, 1979.

\item
V.~P. Maslov, {\em Operational Methods}.
 Moscow: Mir Publishers, 1976.
 Translated by V. Golo, N. Kulman, and G. Voropaeva.

\item
R.~Mathias, ``Concavity of monotone matrix function of finite order,'' {\em
  Linear and Multilinear Algebra}, vol.~27, pp.~129--138, 1990.

\item
R.~Mathias, ``Matrices with positive definite {H}ermitian part: {I}nequalities
  and linear systems,'' {\em SIAM Journal on Matrix Analysis and Applications},
  vol.~13(2), pp.~640--654, 1992.

\item
R.~Mathias, ``Perturbation bounds for the polar decomposition,'' {\em SIAM
  Journal on Matrix Analysis and Applications}, vol.~14(2), pp.~588--597, 1993.

\item
R.~Mathias, ``Approximation of matrix-valued functions,'' {\em SIAM Journal on
  Matrix Analysis and Applications}, vol.~14(4), pp.~1061--1063, 1993.

\item
R.~Mathias, ``The {H}adamard operator norm of a circulant and applications,''
  {\em SIAM Journal on Matrix Analysis and Applications}, vol.~14(4),
  pp.~1152--1167, 1993.

\item
R.~McEachin, ``Analyzing specific cases of an operator inequality,'' {\em
  Linear Algebra and Its Applications}, vol.~208/209, pp.~343--365, 1994.

\item
C.~L. Mehta, ``Some inequalities involving traces of operators,'' {\em Journal
  of Mathematical Physics}, vol.~9(5), pp.~693--697, 1968.

\item
A.~D. Michal, {\em Matrix and Tensor Calculus, with Applications to Mechanics,
  Elasticity, and Aeronautics}.
 Galcit Aeronautical Series, New York: John Wiley \& Sons, 1947.

\item
L.~Mirsky, ``Maximum principles in matrix theory,'' {\em Proceedings of the
  Glasgow Mathematical Association}, vol.~4, pp.~34--37, 1958.

\item
L.~Mirsky, ``On the trace of matrix products,'' {\em Mathematische
  Nachrichten}, vol.~20, pp.~171--174, 1959.

\item
L.~Mirsky, ``A trace inequality of {J}ohn von {N}eumann,'' {\em Monatshefte
  {f\"{u}r} Mathematik}, vol.~79, pp.~303--306, 1975.

\item
D.~S. Mitrinovi\'{c}, {\em Elementary Inequalities}.
 Groningen: ??, 1964.

\item
D.~S. Mitrinovi\'{c}, {\em Analytic Inequalities}.
 Die Grundlehren der mathematischen Wissenschaften in
  Einzeldarstellungen, Band 165, Berlin: Springer-Verlag, 1970.
 In cooperation with Petar M. Vasi\'{c}.

\item
B.~Mond and J.~E. Pe\v{c}ari\'{c}, ``Inequalities involving powers of
  generalized inverses,'' {\em Linear Algebra and Its Applications}, vol.~199,
  pp.~293--303, 1994.

\item
P.~C. M\"{u}ller, ``Solution of the matrix equations {$AX+XB=-Q$} and
  {$S^{T}X+XS=-Q$},'' {\em SIAM Journal on Applied Mathematics}, vol.~18(3),
  pp.~682--687, 1970.

\item
M.~Nakamura and H.~Umegaki, ``A note on the entropy for operator algebras,''
  {\em Proceedings of the Japan Academy}, vol.~37(3), pp.~149--154, 1961.

\item
E.~Nelson, ``Operants: {A} functional calculus for non-commuting operators,''
  in {\em Functional Analysis and Related Fields: Proceedings of a Conference
  in Honor of Professor Marshall Stone} (F.~E. Browder, ed.), (Berlin),
  pp.~172--187, Springer-Verlag, 1970.

\item
H.~Neudecker, ``Some theorems on matrix differentiation with special reference
  to {K}ronecker matrix products,'' {\em Journal of the American Statistical
  Association}, vol.~64, pp.~953--963, 1969.

\item
M.~L. Overton and R.~S. Womersley, ``On the sum of the largest eigenvalues of a
  symmetric matrix,'' {\em SIAM Journal on Matrix Analysis and Applications},
  vol.~13(1), pp.~41--45, 1992.

\item
P.~A. Pearce and C.~J. Thompson, ``The anisotropic {H}eisenberg model in the
  long-range interaction limit,'' {\em Communications in Mathematical Physics},
  vol.~41, pp.~191--201, 1975.

\item
D.~Petz, ``Monotone metrics on matrix spaces,'' in {\em Quantum Communications
  and Measurement} (R.~Hudson, V.~P. Belavkin, and O.~Hirota, eds.), (New
  York), Plenum Press, 1995.
 to appear.

\item
H.~Poincar\'{e}, ``Sur les groupes continus,'' {\em Transactions. Cambridge
  Philosophical Society}, vol.~18, p.~220, 1899.

\item
J.~E. Potter, ``Matrix quadratic solutions,'' {\em SIAM Journal on Applied
  Mathematics}, vol.~14(3), pp.~496--501, 1966.

\item
W.~Pusz and S.~L. Woronowicz, ``Functional calculus for sesquilinear forms and
  the purification map,'' {\em Reports on Mathematical Physics}, vol.~8(2),
  pp.~159--170, 1975.

\item
G.~S. Rogers, {\em Matrix Derivatives}.
 Lecture Notes in Statistics, vol.\ 2, New York: Marcel Dekker, Inc.,
  1980.

\item
M.~Rosenblum, ``On the operator equation {$BX-XA=Q$},'' {\em Duke Mathematical
  Journal}, vol.~23, pp.~263--269, 1956.

\item
H.~L. Royden, {\em Real Analysis}.
 New York: Macmillan, {S}econd~ed., 1963.

\item
D.~Ruelle, {\em Statistical Mechanics: Rigorous Results}.
 The Mathematical Physics Monograph Series, Reading, MA: W. A.
  Benjamin, Inc., 1969.

\item
M.~B. Ruskai, ``Inequalities for traces on von {N}eumann algebras,'' {\em
  Communications in Mathematical Physics}, vol.~26, pp.~280--289, 1972.

\item
D.~E. Rutherford, ``On the solution of the matrix equation {$AX+XB=C$},'' {\em
  Koninklijke Nederlandse Akademie van Wetenschappen, Proceedings Series A},
  vol.~35, pp.~54--59, 1932.

\item
R.~Schatten, ``A theory of cross-spaces,'' {\em Annals of Mathematics Studies},
  vol.~26, pp.~1--153, 1950.

\item
R.~Schatten, {\em Norm Ideals of Completely Continuous Operators}.
 Ergebnisse der Mathematik und ihrer Grenzgebiete, Heft 27, Berlin:
  Springer-Verlag, 1960.

\item
B.~Simon, {\em Trace Ideals and Their Applications}.
 London Mathematical Society Lecture Note Series, vol.\ 35, Cambridge:
  Cambridge University Press, 11979.

\item
M.~F. Smiley, ``Inequalities related to {L}idskii's,'' {\em Proceedings of the
  American Mathematical Society}, vol.~19, pp.~1029--1034, 1968.

\item
R.~A. Smith, ``Bounds for quadratic {L}yapunov functions,'' {\em Journal of
  Mathematical Analysis and Applications}, vol.~12, pp.~425--435, 1965.

\item
R.~A. Smith, ``Matrix calculations for {L}iapunov quadratic forms,'' {\em
  Journal of Differential Equations}, vol.~2, pp.~208--217, 1966.

\item
R.~A. Smith, ``Matrix equation {$XA+BX=C$},'' {\em SIAM Journal on Applied
  Mathematics}, vol.~16(1), pp.~198--201, 1968.

\item
W.~So, ``Equality cases in matrix exponential inequalities,'' {\em SIAM Journal
  on Matrix Analysis and Applications}, vol.~13(4), pp.~1154--1158, 1992.

\item
W.~H. Steeb, ``A lower bound for the free energy of the {H}ubbard model,'' {\em
  Journal of Physics C}, vol.~8, pp.~L103--L106, 1975.

\item
C.~M. Theobald, ``An inequality for the trace of the product of two symmetric
  matrices,'' {\em Mathematical Proceedings of the Cambridge Philosophical
  Society}, vol.~77, pp.~265--267, 1975.

\item
C.~J. Thompson, ``Inequality with applications in statistical mechanics,'' {\em
  Journal of Mathematical Physics}, vol.~6(11), pp.~1812--1813, 1965.

\item
C.~J. Thompson, ``Inequalities and partial orders on matrix spaces,'' {\em
  Indiana University Mathematics Journal}, vol.~21(5), pp.~469--479, 1971.

\item
R.~C. Thompson, ``Matrix type metric inequalities,'' {\em Linear and
  Multilinear Algebra}, vol.~5, pp.~303--319, 1978.

\item
R.~C. Thompson, ``The case of equality in the matrix valued triangle
  inequality,'' {\em Pacific Journal of Mathematics}, vol.~82, pp.~279--280,
  1979.

\item
P.~A.~J. Tindemans and H.~W. Capel, ``On the free energy in systems with
  separable interactions. {III},'' {\em Physica}, vol.~79A, pp.~478--502, 1975.

\item
R.~Vaidyanathaswamy, {\em Set Topology}.
 New York: Chelsea Publishing Company, {S}econd~ed., 1960.

\item
J.~L. {van Hemmen} and T.~Ando, ``An inequality for trace ideals,'' {\em
  Communications in Mathematical Physics}, vol.~76, pp.~143--148, 1980.

\item
W.~J. Vetter, ``Vector structures and solutions of linear matrix equations,''
  {\em Linear Algebra and Its Applications}, vol.~10, pp.~181--188, 1975.

\item
J.~{von Neumann}, ``Some matrix-inequalities and metrization of matrix-space,''
  {\em Tomsk Univ.\ Rev.}, vol.~1, pp.~286--300, 1937.
 Reprinted in {\it John von Neumann: Collected Works\/}, vol. IV,
  (Macmillan, New York, 1962), edited by A. H. Taub.

\item
B.-Y. Wang and M.-P. Gong, ``Some eigenvalue inequalities for positive
  semidefinite matrix power products,'' {\em Linear Algebra and Its
  Applications}, vol.~184, pp.~249--260, 1993.

\item
R.~M. Wilcox, ``Exponential operators and parameter differentiation in quantum
  physics,'' {\em Journal of Mathematical Physics}, vol.~8(4), pp.~962--982,
  1967.

\item
H.~K. Wimmer, ``Extremal problems for {H}\"older norms of matrices and
  realizations of linear systems,'' {\em SIAM Journal on Matrix Analysis and
  Applications}, vol.~9(3), pp.~314--322, 1988.

\item
H.~K. Wimmer, ``Linear matrix equations, controllability and observability, and
  the rank of solutions,'' {\em SIAM Journal on Matrix Analysis and
  Applications}, vol.~9(4), pp.~570--578, 1988.

\item
H.~K. Wimmer, ``Explicit solutions of the matrix equation {$\sum A^iXD_i=C$},''
  {\em SIAM Journal on Matrix Analysis and Applications}, vol.~13(4),
  pp.~1123--1130, 1992.

\item
F.~J. Yeadon and P.~E. Kopp, ``Inequalities for non-commutative {$L^p$}-spaces
  and an application,'' {\em Journal of the London Mathematical Society},
  vol.~19, pp.~123--128, 1979.

\item
N.~J. Young, ``Formulae for the solution of {L}yapunov matrix equations,'' {\em
  International Journal of Control}, vol.~31(1), pp.~159--179, 1980.

\end{enumerate}


\end{document}